\documentclass[12pt,english,a4paper]{book}

\pdfoutput=1 

\usepackage[T1]{fontenc}
\usepackage[latin9]{inputenc}
\usepackage{amsmath}
\usepackage{amssymb}
\usepackage[only,llbracket,rrbracket]{stmaryrd}
\usepackage{amsfonts}
\usepackage{pifont}
\usepackage[usenames,dvipsnames]{color}

\usepackage{graphicx}
\usepackage[margin = 1cm]{caption}
\usepackage{subcaption}

\newlength{\figwidth}
\newlength{\figheight}
\usepackage{layout}

\usepackage{fancyhdr}
\usepackage{calc}
\fancyheadoffset[LE,RO]{0cm}

\fancypagestyle{plain}{%
\fancyhead{} 
}

\numberwithin{equation}{chapter}
\numberwithin{figure}{chapter}

\usepackage{babel}
\usepackage{hyperref}
\hypersetup{
    unicode=false,          
    pdftoolbar=true,        
    pdfmenubar=true,        
    pdffitwindow=false,     
    pdfstartview={FitH},    
    pdftitle={Stability and properties of electron-driven fishbones in tokamaks},    
    pdfauthor={A. Merle},     
    pdfsubject={PhD},   
    pdfkeywords={keywords}, 
    pdfnewwindow=true,      
    colorlinks=true,       	
    linkcolor=magenta,       	
    citecolor=blue,        	
    filecolor=green,      
    urlcolor=cyan           
}

\usepackage[toc,nonumberlist]{glossaries}
\newglossary[nlg]{notation}{not}{ntn}{List of Notations}
\makeglossaries

\newcommand{\mm}{\mathrm{m}}
\newcommand{\nn}{\mathrm{n}}

\renewcommand{\Re}{\mathrm{Re}\:}
\renewcommand{\Im}{\mathrm{Im}\:}

\newcommand{\dW}{\delta\hat{W}}
\newcommand{\dWk}{\delta\hat{W}_k}
\newcommand{\dWh}{\delta\hat{W}_h}
\newcommand{\dWf}{\delta\hat{W}_f}
\newcommand{\wi}{\omega_{*i}}

\newcommand{\dd}[1]{\mathrm{d} #1}
\newcommand{\ddr}[2]{\frac{\mathrm{d} #1}{\mathrm{d} #2}}
\newcommand{\mdd}[3]{\frac{\mathrm{d}^{#3} #1}{\mathrm{d} #2^{#3}}}
\newcommand{\pdd}[2]{\frac{\partial #1}{\partial #2}}
\newcommand{\pmdd}[3]{\frac{\partial^{#3} #1}{\partial #2^{#3}}}

\newcommand{\vc}[1]{\mathbf{#1}}
\newcommand{\vb}[1]{\boldsymbol{#1}}

\renewcommand{\O}[2]{O(#1^{#2})}
\newcommand{\Oe}[1]{\O{\varepsilon}{#1}}

\newcommand{\comment}[1]{}

\newenvironment{prof}[0]{\vspace{2mm}\noindent\hspace{0.05\textwidth}\begin{minipage}{0.9\textwidth} \small $\square$}{\end{minipage}\normalsize\vspace{2mm}}
\newenvironment{proof}[0]{\begin{prof}}{\end{prof}}

\newglossaryentry{not:magn}{type=notation,name=$B$,description={magnetic field},sort=AB}
\newglossaryentry{not:elec}{type=notation,name=$E$,description={electric field},sort=AE}
\newglossaryentry{not:epot}{type=notation,name=$\Phi$,description={electrostatic potential},sort=APhi}
\newglossaryentry{not:vpot}{type=notation,name=$A$,description={vector potential},sort=AA}
\newglossaryentry{not:Rmaj}{type=notation,name=$R$,description={distance to the vertical axis},sort=AR}
\newglossaryentry{not:Zele}{type=notation,name=$Z$,description={vertical coordinate},sort=AZ}
\newglossaryentry{not:phig}{type=notation,name=$\varphi$,description={toroidal angle such that $\tan \varphi = Y/X$},sort=Aphi}
\newglossaryentry{not:psil}{type=notation,name=$\psi$,description={flux-label},sort=Apsi}
\newglossaryentry{not:thta}{type=notation,name=$\theta$,description={poloidal angle},sort=Atheta}
\newglossaryentry{not:zeta}{type=notation,name=$\zeta$,description={toroidal angle},sort=Azeta}
\newglossaryentry{not:psip}{type=notation,name=$\psi_p$,description={poloidal magnetic flux (normalized by $2\pi$)},sort=Apsip}
\newglossaryentry{not:psia}{type=notation,name=$\psi_a$,description={value of $\psi_p$ on the last closed flux-surface},sort=Apsia,parent=not:psip}
\newglossaryentry{not:psit}{type=notation,name=$\psi_t$,description={toroidal magnetic flux (normalized by $2\pi$)},sort=Apsit}
\newglossaryentry{not:rmag}{type=notation,name=$r$,description={radial coordinate},sort=Armin}
\newglossaryentry{not:qsaf}{type=notation,name=$q$,description={safety factor},sort=Aq}
\newglossaryentry{not:smag}{type=notation,name=$s$,description={magnetic shear},sort=As}
\newglossaryentry{not:Jcur}{type=notation,name=$J$,description={plasma current},sort=AJcur}
\newglossaryentry{not:Bax0}{type=notation,name=$B_0$,description={magnetic field amplitude on axis},sort=ABax}
\newglossaryentry{not:Rax0}{type=notation,name=$R_0$,description={major radius on axis},sort=ARax}
\newglossaryentry{not:amin}{type=notation,name=$a$,description={plasma minor radius},sort=Aamin}
\newglossaryentry{not:epsa}{type=notation,name=$\varepsilon$,description={inverse aspect ratio $\varepsilon = a/R_0$, sometimes used as a local variable $\varepsilon = r/R_0$},sort=Aeps}
\newglossaryentry{not:Bmin}{type=notation,name=$B_m$,description={minimum amplitude of $B$ on a given flux-surface},sort=ABmin}
\newglossaryentry{not:kapc}{type=notation,name=$\kappa$,description={curvature of magnetic field lines},sort=Akapc}

\newglossaryentry{not:lagr}{type=notation,name=$\mathcal{L}$,description={particle Lagrangian},sort=Blag}
\newglossaryentry{not:hami}{type=notation,name=$\mathcal{H}$,description={particle Hamiltonian},sort=Bham}
\newglossaryentry{not:mass}{type=notation,name=$m$,description={particle mass},sort=Bm}
\newglossaryentry{not:chrg}{type=notation,name=$e$,description={particle charge},sort=Be}
\newglossaryentry{not:posi}{type=notation,name=$x$,description={particle position},sort=Bx}
\newglossaryentry{not:velo}{type=notation,name=$v$,description={particle velocity},sort=Bv}
\newglossaryentry{not:momt}{type=notation,name=$p$,description={particle momentum},sort=Bp}
\newglossaryentry{not:rhoL}{type=notation,name=$\rho$,description={particle Larmor radius or gyroradius},sort=Brho}
\newglossaryentry{not:mumg}{type=notation,name=$\mu$,description={particle magnetic momentum},sort=Bmu}
\newglossaryentry{not:Pphi}{type=notation,name=$P_\zeta$,description={particle toroidal angular momentum},sort=BPphi}
\newglossaryentry{not:engy}{type=notation,name=$E$,description={particle energy},sort=BEn}
\newglossaryentry{not:xipa}{type=notation,name=$\xi$,description={cosine of the particle's pitch-angle such that $v_\Vert = v \xi$},sort=Bxipa}
\newglossaryentry{not:xi0p}{type=notation,name=$\xi_0$,description={value of $\xi$ at the position of minimum magnetic field amplitude},sort=Bxi0p,parent=not:xipa}
\newglossaryentry{not:xi0T}{type=notation,name=$\xi_{0T}$,description={value of $\xi_0$ at the trapped/passing boundary},sort=Bxi0T,parent=not:xipa}
\newglossaryentry{not:dist}{type=notation,name=$F$,description={particle distribution function},sort=BFdist}
\newglossaryentry{not:wcyc}{type=notation,name=$\omega_c$,description={Larmor frequency or cyclotron frequency or gyrofrequency},sort=Bwcyc}
\newglossaryentry{not:wbnc}{type=notation,name=$\omega_b$,description={bounce/transit frequency of trapped/passing particles},sort=Bwbnc}
\newglossaryentry{not:wdrf}{type=notation,name=$\omega_d$,description={toroidal drift precession frequency},sort=Bwdrf}
\newglossaryentry{not:tbnc}{type=notation,name=$\tau_b$,description={bounce/transit time of trapped/passing particles},sort=Btbnc}

\newglossaryentry{not:dens}{type=notation,name=$n$,description={plasma density},sort=Cn}
\newglossaryentry{not:masd}{type=notation,name=$\rho$,description={plasma mass density},sort=Crho}
\newglossaryentry{not:temp}{type=notation,name=$T$,description={plasma temperature},sort=CT}
\newglossaryentry{not:pres}{type=notation,name=$p$,description={plasma pressure},sort=Cp}
\newglossaryentry{not:velp}{type=notation,name=$v$,description={plasma velocity},sort=Cv}
\newglossaryentry{not:visc}{type=notation,name=$\Pi$,description={plasma viscous tensor},sort=CPvisc}
\newglossaryentry{not:resi}{type=notation,name=$\eta$,description={plasma electrical resistivity},sort=Ceresi}
\newglossaryentry{not:gamf}{type=notation,name=$\gamma$,description={ratio of specific heats},sort=Cgamf}
\newglossaryentry{not:ximd}{type=notation,name=$\xi$,description={MHD displacement},sort=Cximd}
\newglossaryentry{not:xick}{type=notation,name=$\xi_c$,description={radial MHD displacement amplitude for the internal kink mode},sort=Cxick,parent=not:ximd}
\newglossaryentry{not:rsin}{type=notation,name=$r_s$,description={radial location of the $q=1$ surface},sort=Crsin}
\newglossaryentry{not:beta}{type=notation,name=$\beta$,description={ratio of plasma kinetic energy and magnetic energy},sort=Cbeta}
\newglossaryentry{not:mpol}{type=notation,name=$\mm$,description={poloidal mode number},sort=Cmpol}
\newglossaryentry{not:ntor}{type=notation,name=$\nn$,description={toroidal mode number},sort=Cntor}
\newglossaryentry{not:kpar}{type=notation,name=$k_\Vert$,description={parallel wave number},sort=Ckpar}
\newglossaryentry{not:Fmag}{type=notation,name=$F$,description={toroidal covariant component of $B$},sort=CFbtor}
\newglossaryentry{not:shft}{type=notation,name=$\Delta$,description={Shafranov shift},sort=CDshft}
\newglossaryentry{not:Etot}{type=notation,name=$E$,description={plasma perturbed energy},sort=CEMHD}
\newglossaryentry{not:dWma}{type=notation,name=$\delta W$,description={plasma perturbed potential energy},sort=CdW,parent=not:Etot}
\newglossaryentry{not:Ekin}{type=notation,name=$K$,description={plasma perturbed kinetic energy},sort=CK,parent=not:Etot}
\newglossaryentry{not:dWha}{type=notation,name=$\dW$,description={normalized value of $\delta W$},sort=CdW,parent=not:dWma}
\newglossaryentry{not:Minr}{type=notation,name=$M$,description={inertial enhancement factor},sort=CMinr}
\newglossaryentry{not:omst}{type=notation,name=$\omega_*$,description={diamagnetic frequency},sort=Comst}

\newglossaryentry{not:dIne}{type=notation,name=$\delta I$,description={Inertial term of the fishbone dispersion relation (FDR)},sort=DdIne}
\newglossaryentry{not:dWff}{type=notation,name=$\dWf$,description={Fluid contribution to the FDR},sort=DdWff}
\newglossaryentry{not:dWhh}{type=notation,name=$\dWh$,description={Total contribution of energetic particles to the FDR},sort=DdWhh}
\newglossaryentry{not:dWkh}{type=notation,name=$\dWk$,description={Kinetic contribution of energetic particles to the FDR},sort=DdWkh,parent=not:dWhh}
\newglossaryentry{not:dWfh}{type=notation,name=$\dW_{f,h}$,description={Fluid contribution of energetic particles to the FDR},sort=DdWfh,parent=not:dWhh}

\newglossaryentry{not:echa}{type=notation,name=$e$,description={elementary charge ($1.60217646 \, 10^{-19}\ \mathrm{C}$)},sort=echa}
\newglossaryentry{not:mu_0}{type=notation,name=$\mu_0$,description={vacuum permeability ($1.25663706 \, 10^{-6}\ \mathrm{m\, kg\, s^{-2}\, A^{-2}}$)},sort=Emu0}
\newglossaryentry{not:ep_0}{type=notation,name=$\varepsilon_0$,description={vacuum permittivity ($8.85418782 \, 10^{-12}\ \mathrm{m^{-3}\, kg^{-1}\, s^4\, A^2}$)},sort=Eep0}
\newglossaryentry{not:clum}{type=notation,name=$c$,description={speed of light in vaccum ($2.99792458 \, 10^8\ \mathrm{m\, s^{-1}}$)},sort=Eclum}
\newglossaryentry{not:kBol}{type=notation,name=$k_B$,description={Boltzmann's constant ($1.3806503 \, 10^{-23}\ \mathrm{m^2\, kg\, s^{-2}\, K^{-1}}$)},sort=EkBol}
\newglossaryentry{not:Zdis}{type=notation,name=$Z$,description={plasma dispersion function},sort=EZdisp}

\begin{document}

\glsaddall



\title{Stability and properties of \\electron-driven fishbones in tokamaks}

\author{Antoine Merle \\ \\ CEA Supervisor: Joan Decker \\ PhD Advisor: Xavier Garbet}

\date{\today}

\pagenumbering{roman}

\begin{titlepage}

\includegraphics*[width=3cm]{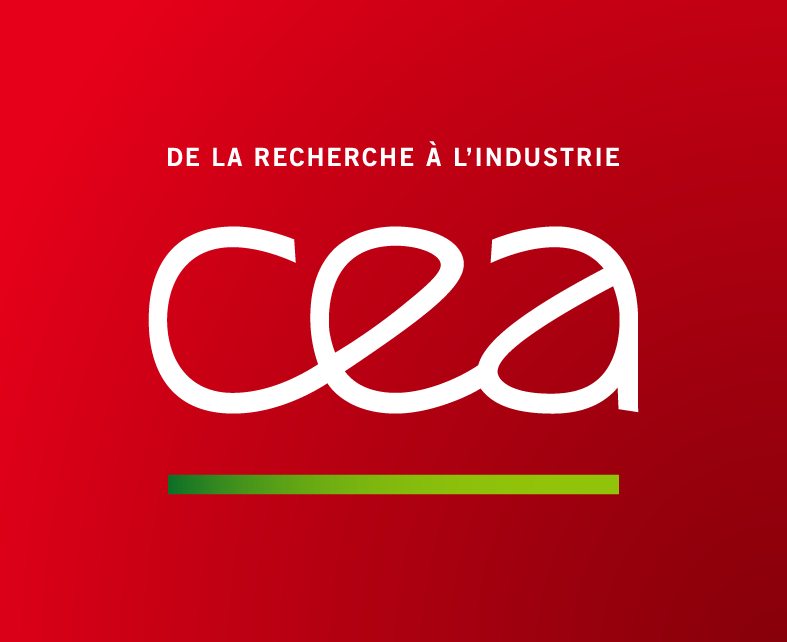}
\hfill
\includegraphics*[width=3cm]{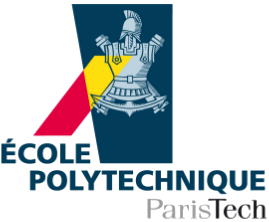}

\vspace{2.5cm}

\rule{\linewidth}{0.2mm}

\vspace{-5mm}

\begin{center}
\huge
\bf
Stability and properties of electron-driven fishbones in tokamaks
\end{center}

\vspace{-5mm}

\rule{\linewidth}{0.2mm}

\begin{center}
\Large
Th\`{e}se de doctorat
\end{center}

\vspace{0mm}

\begin{center}
soutenue le 29/11/2012 par 
\Large

Antoine Merle
\end{center}

\vspace{0mm}

\begin{center}

en vue d'obtenir le grade de 

\Large
Docteur de l'\'{E}cole Polytechnique

Sp\'{e}cialit\'{e} : Physique
\end{center}

\vspace{0mm}

\begin{center}
\begin{tabular}{lll}
\hline
 & \\
Responsable CEA & Joan Decker & Ing\'{e}nieur de recherches CEA  \\
Directeur de th\`{e}se & Xavier Garbet & Directeur de recherches CEA \\
Rapporteur & Jonathan Graves & Professeur \`{a} l'EPFL \\
Rapporteur & Fulvio Zonca & Professeur \`{a} l'ENEA \\
Examinateur & Sadruddin Benkadda & Directeur de recherches CNRS\\
Examinateur & Jean-Marcel Rax & Professeur \`{a} l'\'{E}cole Polytechnique\\
 & \\
\hline
\end{tabular}
\end{center}

\end{titlepage}

\cleardoublepage
\section*{Abstract}

In tokamaks, the stability of magneto-hydrodynamic modes can be modified by populations of energetic particles. In ITER-type fusion reactors, such populations can be generated by fusion reactions or auxiliary heating. The electron-driven fishbone mode belongs to this category of instabilities. It results from the resonant interaction of the internal kink mode with the slow toroidal precessional motion of energetic electrons and is frequently observed in present-day tokamaks with Electron Cyclotron Resonance Heating or Lower Hybrid Current Drive. These modes provide a good test bed for the linear theory of fast-particle driven instabilities as they exhibit a very high sensitivity to the details of both the equilibrium and the electronic distribution function.

In Tore Supra, electron-driven fishbones are observed during LHCD-powered discharges in which a high-energy tail of the electronic distribution function is created. Although the destabilization of those modes is related to the existence of a fast particle population, the modes are observed at a frequency that is lower than expected. Indeed, the corresponding energy assuming resonance with the toroidal precession frequency of barely trapped electrons falls in the thermal range.

The linear stability analysis of electron-driven fishbone modes is the main focus of this thesis. The fishbone dispersion relation is derived in a form that accounts for the contribution of the parallel motion of passing particles to the resonance condition. The MIKE code is developed to compute and solve the dispersion relation of electron-driven fishbones. The code is successfully benchmarked against theory using simple analytical distributions. When coupled to the relativistic Fokker-Planck code LUKE and to the integrated modeling platform CRONOS, it is used to compute the stability of electron-driven fishbones using reconstructed data from tokamak experiments. Using the code MIKE with parametric distributions and equilibria, we show that both barely trapped and barely passing electrons resonate with the mode and can drive it unstable. More deeply trapped and passing electrons have a non-resonant effect on the mode that is, respectively, stabilizing and destabilizing. MIKE simulations using complete ECRH-like distribution functions show that energetic barely passing electrons can contribute to drive a mode unstable at a relatively low frequency. This observation could provide some insight to the understanding of Tore Supra experiments.

\cleardoublepage
\section*{R\'{e}sum\'{e}}

La stabilit\'{e} des modes magn\'{e}to-hydrodynamiques dans les plasmas de tokamaks est modifi\'{e}e par la pr\'{e}sence de particules rapides. Dans un tokamak tel qu'ITER ces particules rapides peuvent \^{e}tre soit les particules alpha cr\'{e}\'{e}es par les r\'{e}actions de fusion, soit les ions et \'{e}lectrons acc\'{e}l\'{e}r\'{e}s par les dispositifs de chauffage additionnel et de g\'{e}n\'{e}ration de courant. Les modes appel\'{e}s \emph{fishbones \'{e}lectroniques} correspondent \`{a} la d\'{e}stabilisation du mode de kink interne due \`{a} la r\'{e}sonance avec le lent mouvement de pr\'{e}cession toroidale des \'{e}lectrons rapides. Ces modes sont fr\'{e}quemment observ\'{e}s dans les plasmas des tokamaks actuels en pr\'{e}sence de chauffage par onde cyclotronique \'{e}lectronique (ECRH) ou de g\'{e}n\'{e}ration de courant par onde hybride basse (LHCD). La stabilit\'{e} de ces modes est particuli\`{e}rement sensible aux d\'{e}tails de la fonction de distribution \'{e}lectronique et du facteur de s\'{e}curit\'{e}, ce qui fait des fishbones \'{e}lectroniques un excellent candidat pour tester la th\'{e}orie lin\'{e}aire des instabilit\'{e}s li\'{e}es aux particules rapides.

Dans le tokamak Tore Supra, des fishbones \'{e}lectroniques sont couramment observ\'{e}s lors de d\'{e}charges o\`{u} l'utilisation de l'onde hybride basse cr\'{e}e une importante queue de particules rapides dans la fonction de distribution \'{e}lectronique. Bien que ces modes soit clairement li\'{e}s \`{a} la pr\'{e}sence de particules rapides, la fr\'{e}quence observ\'{e}e de ces modes est plus basse que celle pr\'{e}vue par la th\'{e}orie. En effet, si on estime l'\'{e}nergie des \'{e}lectrons r\'{e}sonants en faisant correspondre la fr\'{e}quence du mode avec la fr\'{e}quence de pr\'{e}cession toroidale des \'{e}lectrons faiblement pi\'{e}g\'{e}s, on obtient une valeur comparable \`{a} celle des \'{e}lectrons thermiques.

L'objet principal de cette th\`{e}se est l'analyse lin\'{e}aire de la stabilit\'{e} des fishbones \'{e}lectroniques. La relation de dispersion de ces modes est d\'{e}riv\'{e}e et la forme obtenue prend en compte, dans la condition de r\'{e}sonance, la contribution du mouvement parall\`{e}le des particules passantes. Cette relation de dispersion est impl\'{e}ment\'{e}e dans le code MIKE qui est ensuite test\'{e} avec succ\'{e}s en utilisant des fonctions de distributions analytiques. En le couplant au code Fokker-Planck relativiste LUKE et \`{a} la plate-forme de simulation int\'{e}gr\'{e}e CRONOS, MIKE peut estimer la stabilit\'{e} des fishbones \'{e}lectroniques en utilisant les donn\'{e}es reconstruites de l'exp\'{e}rience. En utilisant des fonctions de distributions et des \'{e}quilibres analytiques dans le code MIKE nous montrons que les \'{e}lectrons faiblement pi\'{e}g\'{e}s ou faiblement passants peuvent d\'{e}stabiliser le mode de kink interne en r\'{e}sonant avec lui. Si l'on s'\'{e}loigne de la fronti\`{e}re entre \'{e}lectrons passants et pi\'{e}g\'{e}s, les effets r\'{e}sonants s'affaiblissent. Cependant les \'{e}lectrons passants conservent une influence d\'{e}stabilisante alors que les \'{e}lectrons pi\'{e}g\'{e}s tendent \`{a} stabiliser le mode. D'autres simulations avec MIKE, utilisant cette fois des distributions compl\`{e}tes similaires \`{a} celles obtenues en pr\'{e}sence de chauffage de type ECRH, montrent que l'interaction avec les \'{e}lectrons faiblement passants peut entra\^{i}ner une d\'{e}stabilisation du mode \`{a} une fr\'{e}quence relativement basse ce qui pourrait permettre d'expliquer les observations sur le tokamak Tore Supra.

\hypersetup{linkcolor=black}

\cleardoublepage
\phantomsection
\addcontentsline{toc}{chapter}{Table of Contents}
\tableofcontents

\listoffigures
\addcontentsline{toc}{chapter}{List of Figures}

\listoftables
\addcontentsline{toc}{chapter}{List of Tables}

\hypersetup{linkcolor=magenta}


\cleardoublepage


\pagenumbering{arabic}

\chapter{Introduction}
\label{cha:introduction}

\section{Nuclear fusion}

In a \emph{nuclear fusion} reaction, two light nuclei are brought together to form one heavier element. If the mass of the products of the reaction is smaller than the total mass of the initial elements, the reaction releases energy. The source of this energy is the strong nuclear interaction which binds the protons and neutrons inside the nucleus. This process of nuclear fusion is very efficient in terms of energy production per mass of the reactants, far above processes involving chemical reaction like the oil combustion. But this tremendous energy comes at a price, indeed in order to fuse the reactants must overcome their mutual repulsion due to the Coulomb interaction between the two positively charged nuclei.

Today nuclear fusion is studied as a potential energy source. The most accessible reaction is the one involving \emph{deuterium} ${}^2_1\mathrm{D}$ and \emph{tritium} ${}^3_1\mathrm{T}$, two heavy isotopes of hydrogen, and producing one Helium nucleus ${}^4_2\mathrm{He}$ (also named $\alpha$-\emph{particle}) and a \emph{neutron} $n$,
\begin{equation}
  \label{eq:DT_reaction}
  {}^2_1\mathrm{D} + {}^3_1\mathrm{T}\  \longrightarrow \ {}^4_2\mathrm{He} \  (3.56 \ \mathrm{MeV}) + \mathrm{n} \  (14.03 \ \mathrm{MeV}).
\end{equation}
The numbers between parenthesis are the amount of kinetic energy carried by the fusion products, such that the total energy released per reaction is $17.6 \ \mathrm{MeV}$. The \emph{temperature} of the reactants plays an important role in reaching an efficient energy production, since they must carry enough kinetic energy to overcome the Coulomb barrier. The reaction rate reaches a maximum when the thermal energy is about $k_B T \sim 25 \ \mathrm{keV}$ ($T \sim 2. \, 10^8 \  \mathrm{K}$). At this level the deuterium and the tritium form a fully ionized gas or \emph{plasma}.

Until the plasma can be self-heated by fusion reactions, one has to inject energy into the plasma to bring and maintain the fuel at the required temperature due to energy losses. The Lawson criterion \cite{law57} states that the fusion power overcomes the power losses when the product $n\,k_B T\, \tau_E$ reaches a certain value, where $n$ is the fuel density and $\tau_E = W/P_{loss}$ is named the energy confinement time and is defined in a steady-state regime as the ratio of the energy content of the plasma $W$ and the level of power losses $P_{loss}$. At $k_B T = 25 \ \mathrm{keV}$ the product $n \tau_E$ must reach the value of $1.5 \, 10^{20} \ \mathrm{m}^{-3}\,\mathrm{s}$. Two different approaches can be considered to satisfy this criterion.
\begin{itemize}
\item Achieve a very high density plasma ($n \sim 10^{31} \ \mathrm{m}^{-3}$) for a short time ($\tau_E \sim 10^{-11} \ \mathrm{s}$). In inertial confinement devices, these conditions are achieved by compressing D-T targets with powerful lasers.
\item Maintain a low density plasma ($n \sim 10^{20} \ \mathrm{m}^{-3}$) for a longer time ($\tau_E \sim 1 \ \mathrm{s}$). In magnetic confinement devices, the plasma is confined by a strong magnetic field which keeps the plasma from cooling down on the wall of the reactor. 
\end{itemize}

\section{Magnetic confinement fusion}

Charged particles in a magnetic field follow trajectories which are helically wound around magnetic field lines. The extent $\rho_s$  of the helix perpendicular to the magnetic field line is called the Larmor radius or gyration radius and is inversely proportional to the amplitude of the magnetic field. For a particle of mass $m_s$ and charge $e_s$ and with a velocity $v_\perp$ in the direction perpendicular to the magnetic fiel of amplitude $B$, the Larmor radius is
\begin{equation}
  \label{eq:Larmor_radius}
  \rho_s = \frac{m_s v_\perp}{e_s B}.
\end{equation}
Thus a stronger magnetic field will provide better confinement properties. In present day magnetic confinement machines, the magnetic field amplitude is typically of several teslas ($\mathrm{T}$), while the earth magnetic field has an amplitude of a few $10^{-4} \ \mathrm{T}$. The $\beta$ parameter measures the ratio of the plasma kinetic energy and the magnetic energy
\begin{equation}
  \label{eq:plasma_beta}
  \beta = 2\mu_0 p/B_T^2;
\end{equation}
in magnetic confinement devices $\beta$ is generally of the order of $1 \%$.

The confinement properties depend also on the geometry of the magnetic field. Initially linear devices with open field lines were tested but the energy confinement times measured were not compatible with a sustainable production of energy due to the important particle and energy losses at both ends of the machines. The simplest configuration with closed magnetic field lines (or at least closed magnetic surfaces) is when the magnetic field lines form a torus. Unfortunately if the magnetic field is purely toroidal, the charged particles suffer a vertical drift due to the curvature of the magnetic field lines and are not confined. But if one adds a poloidal component to the magnetic field so as to make the field lines wind helically around the torus, then the particles orbits are periodic and are confined to the reactor chamber.

\section{The tokamak configuration}

\begin{figure}[!ht]
  \centering
  \includegraphics*[width = 1.3\figwidth]{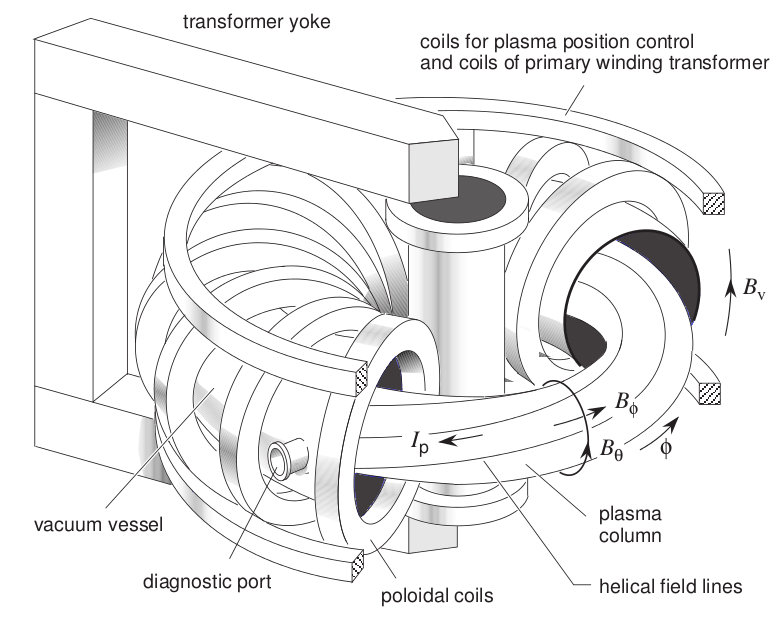}
  \caption{The tokamak configuration coil system.}
  \label{fig:Tokamak_conf}
\end{figure}

The \emph{tokamak} is currently the most successful configuration based on this idea. In this configuration, the toroidal field is produced by vertical coils surrounding the torus while the poloidal magnetic field is produced by an intense \emph{toroidal electric current} which flows inside the torus. The magnetic system of a tokamak is presented in figure \ref{fig:Tokamak_conf}, where additional coils needed for the plasma shape and stability control have been added. 

The level of performance of the plasma can be measured by introducing the enhancement factor $Q$ which corresponds to the ratio of the power released into the plasma by fusion reactions $P_{fus}$ and the level of power injected into the plasma $P_{inj}$. The $Q=1$ limit is called the \emph{break-even}, it corresponds to the state where the plasma is sustained to equal parts by the fusion energy power and by the external power input. This has been achieved in the JET (for Joint European Torus) tokamak \cite{jac99}. The ITER tokamak actually in construction is designed to operate routinely at $Q=10$ when operating with D-T fuel. The predicted fusion power output is of the order of $500 \ \mathrm{MW}$ well above the present record of $16 \ \mathrm{MW}$ with the JET tokamak, this level should be maintained for as long as $400 \ \mathrm{s}$. The characteristics of the ITER machine can be found in table \ref{tab:Tok_Parameters} along with those of the JET and of the Tore Supra tokamak.
\begin{table}[!ht]
  \centering
  \begin{tabular}{l*{3}c}
    & Tore Supra & JET & ITER \\ \hline\hline
    Major radius ($\mathrm{m}$) & $2.5$ & $3$ & $6.2$ \\
    Minor radius ($\mathrm{m}$) & $0.7$ & $1$ & $2.0$ \\
    Plasma volume ($\mathrm{m^3}$) & $25$ & $125$ & $830$ \\
    Plasma current ($\mathrm{MA}$) & $1.5$ & $6$ & $15$ \\
    Magnetic field amplitude ($\mathrm{T}$) & $3.8$ & $3.4$ & $5.3$ \\
    Pulse Duration ($\mathrm{s}$) & $~100$ & $~10$ & $400$ \\
    Fuel mix & D-D & D-D / D-T & D-T\\
    Fusion Power ($\mathrm{MW}$) & $~10^{-3}$ & $5.0\, 10^{-2}\ /\ 10$ & $500$ \\
    Amplification factor $Q$ & $\ll 1$ & $> 1$ & $>10$ \\ \hline\hline
  \end{tabular}
  \caption{Principal parameters for the Tore Supra, JET and ITER tokamaks.}
  \label{tab:Tok_Parameters}
\end{table}
A commercial fusion reactor should operate around $Q= 40$ while the $Q=+\infty$ limit is called \emph{ignition}. 

The large plasma current necessary for the plasma stability is usually induced by a secondary set of electromagnets which create an inductive toroidal electric field inside the plasma which in turn creates an electric current due to the finite resistivity of the plasma. Simultaneously the plasma is heated by \emph{Joule effect}, this process is the principal source of plasma heating and current drive in most present day machines. But at high temperatures the resistivity and the efficiency of the Joule heating drop and additional heating techniques have been developed.
\begin{itemize}
\item The Neutral Beam Injection (NBI) system: since charged particles cannot enter the plasma due to the magnetic field, deuterium ions are accelerated to an energy of about $1 \ \mathrm{MeV}$ before being neutralized. Once inside the plasma the atoms are stripped from their electrons. The energy of the energetic ions is then transferred to the background plasma by successive collisions.
\item Ion  Cyclotron Resonance Heating (ICRH): electromagnetic waves are sent into the plasma at the ion cyclotron frequency. The resonant interaction between the particles and the waves results in a net transfer of energy from the waves to the particles and therefore heats the plasma. The ion cyclotron frequency is $\omega_{ci} = e_i B / m_i$ is of the order of $50 \ \mathrm{MHz}$ and lies in the radio-frequency part of the electromagnetic spectrum. Note that in a tokamak, the magnetic field amplitude is typically inversely proportional to the major radius $R$ such that the region where the particles can resonate with the wave is limited to a layer near a given value of the major radius.
\item Electron Cyclotron Resonance Heating (ECRH) follows the same principle as ICRH. The frequency of the waves matches the electron cyclotron frequency $\omega_{ce} = e_e B / m_e$ which is of the order of $150 \ \mathrm{GHz}$ (microwaves).
\end{itemize}
For the ITER tokamak the amount of power available is of about $73 \ \mathrm{MW}$, $33\ \mathrm{MW}$ of deuterium neutral beams and $40\ \mathrm{MW}$ of radio-frequency heating \cite{iterweb}.

When aiming at long discharges the problem of maintaining the plasma current for a long period of time arises. Since the amount of flux which can be varied through the secondary circuit formed by the plasma is finite, the plasma current cannot be sustained solely by the transformer for an infinite time. In ITER an important part of the total plasma current will be driven non-inductively (not relying on the transformer). Non-inductive current-drive can be achieved by
\begin{itemize}
\item Radiofrequency waves. This technique relies once again on the resonant absorption of electron-magnetic waves. The spatial structure of the waves generates an electric field which accelerates the electrons primarily in the parallel direction. Currently two different techniques have been used, the first one uses the electron cyclotron resonance and is called Electron Cyclotron Current-Drive or ECCD. The second one uses the lower-hybrid resonance, one then speaks of Lower-Hybrid Current-Drive or LHCD. The efficiency of the current-drive is measured by the ratio of the driven current and the amount of power injected by the waves. Typical values for the efficiency are around $0.1 \ \mathrm{A}.\mathrm{W}^{-1}$.
\item Bootstrap current. Due to the inhomogeneity of the magnetic field, some particles are trapped in the region of low magnetic field (or low field side noted LFS, the region of high magnetic field on the inboard side is called the high field side and is noted HFS). When the collision frequency is lower than the bounce frequency (the orbit frequency of trapped particles) ,the existence of these trapped particles and of a radial pressure gradient produces a parallel current called the \emph{bootstrap current}. This current is important in the regions of strong pressure gradients such as transport barriers. During the ITER tokamak operation, the level of bootstrap current is expected to be around $30 \ \%$.

\end{itemize}

The good confinement properties of the plasma are guarantied by the existence of nested flux-surfaces which means that surfaces exist such that the magnetic field is everywhere tangent to those surfaces. Let $\psi$ be a flux-label, we define the safety factor $q$ by
\begin{equation}
  \label{eq:safety_factor}
  q(\psi) = \frac{1}{2\pi} \int_{0}^{2\pi} \frac{B^\varphi}{B^\theta} \dd{\theta},
\end{equation}
where $B^\varphi$ and $B^\theta$ are the contravariant components of the magnetic field in the toroidal and poloidal directions. The value of $q$ corresponds to the average field-line pitch and can be interpreted as the number of toroidal turns done by a field line for every poloidal turn. Depending on the value of $q$ two type of surfaces exist.
\begin{itemize}
\item If $q(\psi)$ is an irrational number then each field line on this surface fills the surface ergodically. Due to the large parallel heat conductivity of electrons, the pressure is homogenized over the whole flux-surface and is a quantity which depends only on the flux label $\psi$.
\item If $q(\psi)$ is a rational number then the field lines are closed. These surfaces are prone to instabilities since perturbations with structures aligned with the magnetic field and minimize the bending of the magnetic field lines will grow more easily. If $q=\mm/\nn$ these structures have the following spatial dependence $(\mm \theta - \nn \varphi)$.
\end{itemize}

In this thesis, one of these instabilities is studied, the electron-driven fishbone mode.

\section{Introduction to the electron-driven fishbone mode}

The electron-driven fishbone mode belongs to the so-called category of energetic particle driven instabilities. Energetic particles correspond to particles with a velocity higher than the thermal velocity $v_{Ts} = \sqrt{k_B T/m_s}$. In tokamaks the additional heating and current-drive systems such as NBI, ICRH, ECRH/ECCD or LHCD provide large sources of energetic ions and electrons such that the population of those particles is generally higher than the level in a purely maxwellian distribution. The fusion reactions produce $\alpha$-particles at an energy of $3.5 \ \mathrm{MeV}$ and are therefore another source of energetic particles. The characteristic frequencies of the motion of those particles, in particular the slow toroidal precession motion due to the magnetic drifts, are in the same range as the frequencies of the Magneto-HydroDynamic (MHD) instabilities allowing a resonant interaction between the particles and the waves. These instabilities have a radial extent corresponding to a large fraction of the minor radius and are well described by the MHD model which considers the plasma as a magnetized fluid.

\subsection{The fishbone instability}

The fishbone instability was first observed in the PDX tokamak during high-$\beta$ experiments using NBI in near-perpendicular injection, meaning that the initial velocity of the injected ions is nearly perpendicular to the magnetic field lines \cite{mcg83}.

\begin{figure}[!ht]
  \centering
  \includegraphics*[width = 1.5\figwidth]{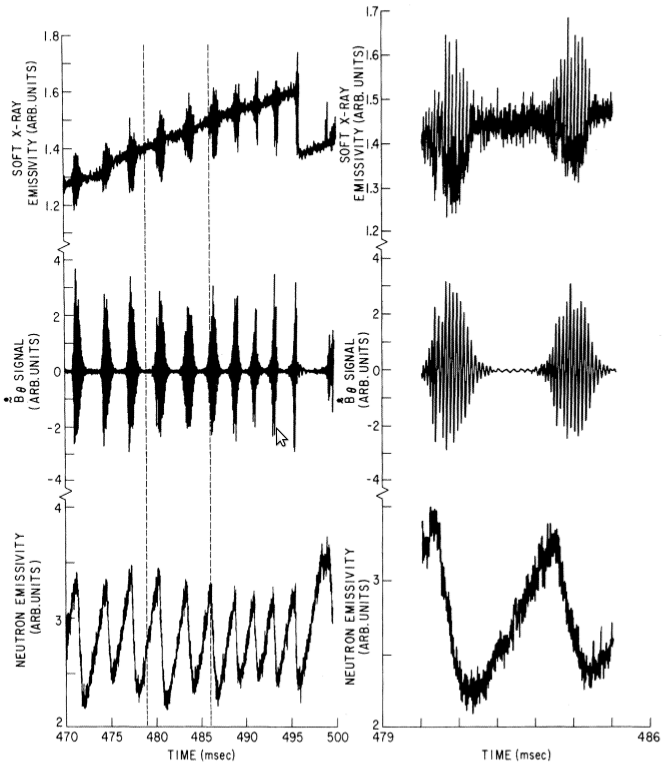}
  \caption[Original report of the fishbone instabilityin ]{Original report of the fishbone instability with traces of the soft X-ray emissivity (top), of the poloidal magnetic field variations (middle) and neutron emissivity (bottom). Taken from \cite{mcg83}.}
  \label{fig:PDX_FB}
\end{figure}

Figure \ref{fig:PDX_FB} reproduces the original figure describing the fishbone instability from reference \cite{mcg83}.
The instability appears at high levels of neutral beam power in the form of successive bursts of $\mm = 1$ activity on the central soft X-ray signals and Mirnov coils measurements (see the top and middle panels of figure \ref{fig:PDX_FB}). The instability is located in the plasma core $r< 20 \ \mathrm{cm}$. The name \emph{fishbone} was given because of the shape of the signal on the Mirnov coils. The $\mm=1$ activity was correlated with a drop in the measured neutron emissivity (see the bottom panel of figure \ref{fig:PDX_FB}) indicating a loss of the energetic ion content of the plasma. This was correlated with measures of the energetic ion distribution with the charge-exchange diagnostic indicating that the population of ions with energies between $E_{inj}$ and $E_{inj}/2$ (where $E_{inj}$ is the injection energy of the beam ions) drops immediately after the $\mm=1$ bursts. The measured frequency of the instability $f \sim 20 \ \mathrm{kHz}$ was in the ion diamagnetic direction and was compatible with the precession drift frequency of deeply trapped ions with energies close to $E_{inj}$ which are abundantly produced by near-perpendicular beam injection. It was then suggested that both the energetic ion losses and the mechanism of the instability growth were  linked to the resonant interaction of the particles with the $\mm=1$ mode. 

Later, ion-driven fishbone instabilities were reported on other machines such as TFTR \cite{kai90}, JET \cite{nav91}, JT-60 \cite{nin88} or DIII-D \cite{hei90}. The instabilities were observed during NBI heating with or without additional ICRF heating, the measured frequencies were situated close to the precession frequency of energetic ions or to the ion diamagnetic frequency or in-between those two frequencies.

\subsection{Theoretical interpretation}

Two theoretical models were proposed to interpret these instabilities \cite{che84,cop86}. Both rely on the modification of the ideal stability of the $\mm=1,\nn=1$ internal kink by resonance with a population of energetic ions. The resonance occurred at the precession drift frequency of trapped ions. The source of the instability is the radial gradient of the distribution function of ions. A negative radial gradient, which corresponds to a central deposition of beam ions is necessary for the growth of the instability. The difference between the two models being that; in the model proposed by Chen et al. \cite{che84} the frequency is fixed by the precession frequency of deeply trapped ions such that the mode is a continuum resonant mode, while in the Model of Coppi et al. \cite{cop86} the frequency is close to the ion diamagnetic frequency such that the mode is described as a discrete gap mode. In fact these 2 models can be described using a single formalism \cite{big87}.

\subsection{Sawtooth stabilization}

\begin{figure}[!ht]
  \centering
  \includegraphics*[width = 0.9\figwidth]{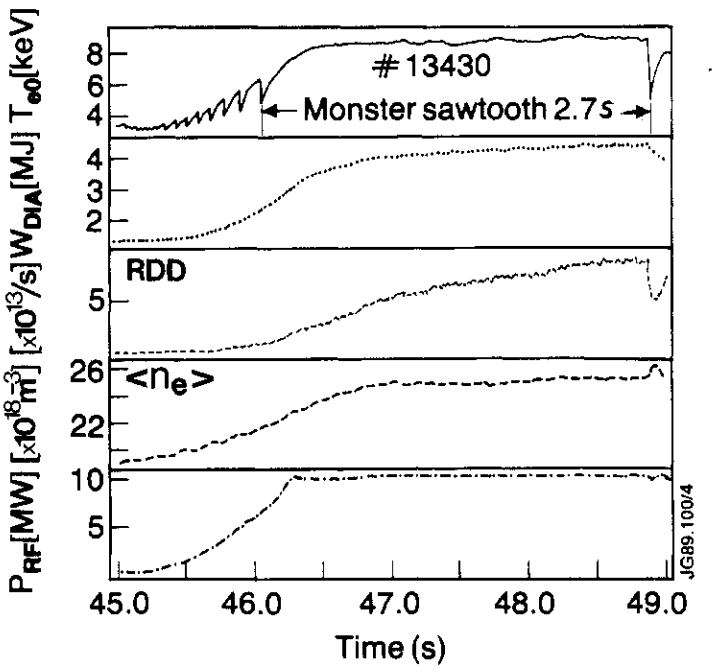}
  \caption[A characteristic monster sawtooth discharge in JET]{A characteristic monster sawtooth discharge in JET during ICRF minority heating. $T_{e0}$ is the central electron temperature, $W_{DIA}$ is the plasma stored energy, $R_{DD}$ is the D-D fusion reaction rate, $\langle n_e \rangle $ is the line-averaged electronic density and $P_{RF}$ is the level of ICRH power \cite{bha89}.}
  \label{fig:JETMonsterST}
\end{figure}

This formalism was also used to explain the stabilization of the sawtooth instabilities by energetic ions and the apparition of \emph{monster sawteeth} such as the ones observed on the JET tokamak \cite{cam88}. An example of a monster sawtooth is shown on figure \ref{fig:JETMonsterST}.

The sawtooth instability is a periodic relaxation of the plasma core which is constituted of a ramp-up phase where the core plasma temperature rises followed by the apparition of an $\mm = 1$ precursor and by a sudden crash of the temperature profile over the whole central region where $q \leq 1$. The $\mm=1$ precursor has the same structure as the one of the fishbone mode, which is the one of the internal kink mode.

In the JET experiments, the sawtooth instability is stabilized by an input of ICRH power and the temperature crash can be triggered by cutting the ICRH power, in this case the crash occurs $30$ to $40 \ \mathrm{ms}$ after the end of the RF pulse which is consistent with the slowing down time of the energetic ions. This confirms the assumption of a stabilization by energetic particles.

White et al. showed, using analytical trapped fast-ion distributions, that a window of values for the fast-ion beta parameter existed in which both the sawtooth and the fishbone instability were stable \cite{whi89}. 

\subsection{Electron-driven fishbones}

The initial theory of Chen et al. only considered the modification of the internal kink stability by energetic trapped ions. But it was later showed \cite{sun05,wan06a,zon07,wan07} that it could be applied to the influence of energetic electrons since the precession frequency has the same absolute value for ions and electrons at the same energy. Yet the drift motion of electrons has to be reversed and the distribution function has to have an inverted radial gradient for a transfer of energy from the electrons to the mode, due to the fact that the internal kink mode rotates preferably in the ion diamagnetic direction. In this case the resonant drive is mostly provided by barely trapped electrons.

\subsubsection*{Electron-driven fishbones in DIII-D}

\begin{figure}[!ht]
  \centering
  \includegraphics*[width = 0.8\figwidth]{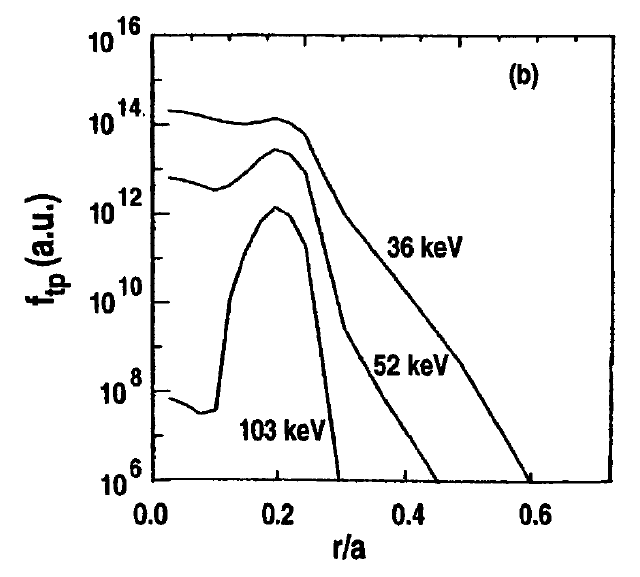}
  \caption[Radial gradient inversion of the energetic electron population in DIII-D]{Spatial profile of $f_{tp}$ the population of electrons at the trapped-passing boundary for specified energies. The data comes from the reconstructed suprathermal electron distribution in the DIII-D shot 96163 \cite{won00}.}
  \label{fig:Wong_DIIID_invgrad}
\end{figure}

The first report of electron-driven fishbones was published by Wong et al. \cite{won00} for the DIII-D tokamak. The fishbones were observed in discharges where off-axis ECCD was used to obtain negative magnetic shear (a region where $q$ decreases with radius) in the central region. Bursts of fishbone activity appeared when the ECCD power was deposited just outside the $q=1$ surface. The inversion of the radial gradient of energetic barely trapped electrons was confirmed by numerical reconstruction of the electronic distribution function as can be seen on figure \ref{fig:Wong_DIIID_invgrad}.
The influence of barely trapped electrons was also confirmed by varying the poloidal angle of the position of the peak in power deposition (but keeping its radial position outside of the $q=1$ surface), the fishbone activity was maximum when the power deposition peaked on the inboard midplane corresponding to optimal conditions for the production of energetic barely trapped electrons. It should be noted that energetic ions were present in the plasma due to NBI heating but their influence was ruled out by the authors.

\subsubsection*{Electron-driven fishbones in FTU}

\begin{figure}[!ht]
  \centering
  \includegraphics*[width = 1.3\figwidth]{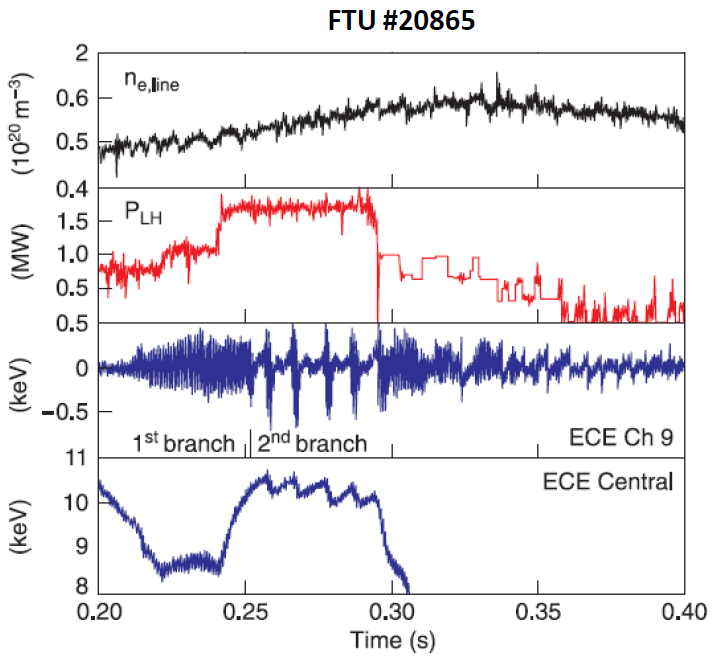}
  \caption[Electron fishbone observation in FTU]{Electron fishbone observation in FTU, $n_{e,line}$ is the line-averaged density, $P_{LH}$ is the level of LHCD power. The two bottom panels use the ECE radiometer data and show the fast electron temperature fluctuations and the central radiation temperature respectively \cite{zon07}.}
  \label{fig:EFB_FTU_Zonca}
\end{figure}

Fishbone instabilities driven by suprathermal electrons have also been observed in FTU using LHCD only \cite{rom03,zon07,ces09}. In FTU two different regimes were obtained as can be seen on figure \ref{fig:EFB_FTU_Zonca}.
At a moderate level of LHCD power the growth of an instability is observed on ECE radiation fluctuation measurements until this instability reaches a saturated level. A simultaneous diminution of the central radiation temperature is observed indicating the loss of energetic electrons. At higher levels of LHCD power the typical bursts of $\mm=1$ activity appear in conjunction with drops of the central radiation temperature similar to the drops in neutron rate measurements in the case of ion-driven fishbones. Using a linear stability analysis \cite{zon07}, it has been established that in the case of the saturated mode the fast particle beta is just above marginal stability whereas it is well above marginal stability in the bursting regime.

\subsubsection*{Electron-driven fishbones in Tore Supra}

Electron-driven fishbones are also observed on the Tore Supra tokamak during LHCD discharges \cite{gon08,mac09}. The modes are observed during the so-called oscillating regime or O-regime where the equilibrium profiles such as $T_e$ and $q$ experience periodic oscillations \cite{gir03,imb06}, slow frequency chirping but also frequency jumps are observed corresponding to a modification of the structure of the mode (modification of $\mm$ and $\nn$). The modes are also seen during steady-state discharges with fixed equilibrium profiles, the modes frequency and structure is similar to those occurring in oscillating discharges. More recently similar modes were observed in-between sawteeth.

\begin{figure}[!ht]
  \centering
  \begin{subfigure}[b]{0.495\textwidth}
    \includegraphics[width = 0.95\figwidth]{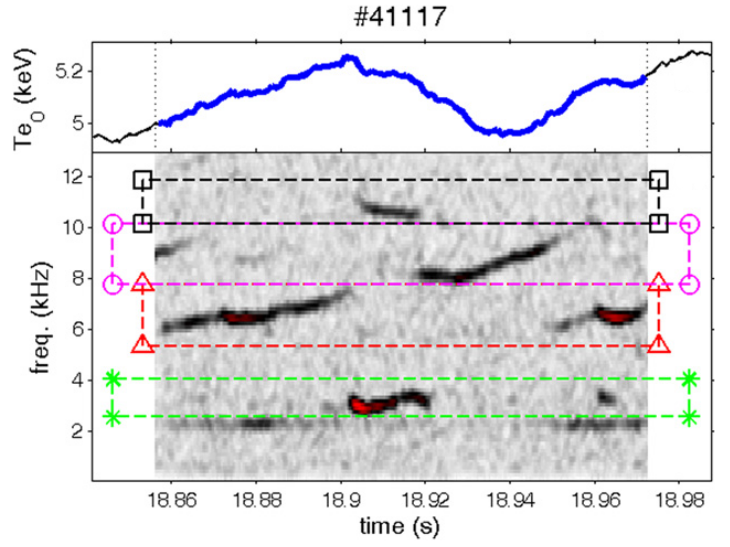}
    \caption{}
    \label{fig:Gui12_41117_Spectro}
  \end{subfigure}
  \begin{subfigure}[b]{0.495\textwidth}
    \includegraphics[width = 0.866\figwidth]{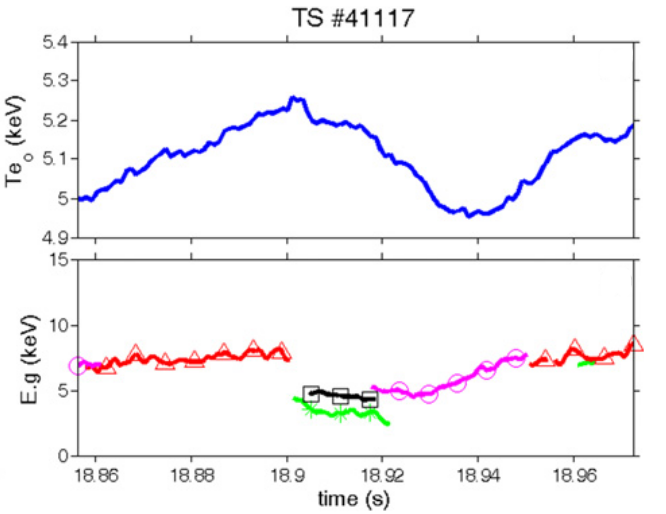}
    \caption{}
    \label{fig:Gui12_41117_Energy}
  \end{subfigure}
  \caption[Electron fishbone observation in Tore Supra]{Central temperature evolution (top a and b), spectrogram of the ECE radiation fluctuations (bottom a) and energy of resonant barely trapped electrons (bottom b) for the Tore Supra discharge \#41117. The black line with squares corresponds to a $\mm/\nn = 4/4$ mode, the magenta line with circles to a $3/3$ mode, the red line with triangles corresponds to a $2/2$ mode and the green line with stars to a $1/1$ mode \cite{gui12}.}
  \label{fig:Gui12_41117}
\end{figure}

Figure \ref{fig:Gui12_41117_Spectro} reproduces the central electron temperature evolution of the Tore Supra discharge number 41117 together with the spectrogram of one of the central ECE channels. If one considers the time where the central electron temperature is maximum as a reference, the frequency decreases at each frequency jump, beginning around $11 \ \mathrm{kHz}$ down to $9 \ \mathrm{kHz}$ then $6 \ \mathrm{kHz}$ and finally $3 \ \mathrm{kHz}$. Each frequency jumps is correlated with a modification of the structure of the mode, the analysis of the radial structures taking into account the vertical extent of the ECE antenna and the alignment of the antenna line of sight with the plasma midplane showed that the poloidal mode numbers are successively $\mm=4$, $\mm=3$, $\mm=2$ and  $\mm=1$ \cite{gui12}. Since the radial position of the modes is consistent with the position of the $q=1$ surface of the reconstructed equilibrium, the toroidal mode numbers are assumed to match the poloidal mode numbers. While the $\mm=2,3,4$ modes have relatively constant radial positions during the cycle, the one of the $\mm=1$ mode drifts slowly inward \cite{gui11}. This observation, together with the fact that both $\mm=1$ and $\mm=4$ modes are present at the same time with the $\mm=4$ mode being located further from the plasma center, indicates that the magnetic shear is low inside the $q=1$ surface. Moreover low-shear profiles are known to be more unstable to modes with high mode numbers \cite{wes86,has88}.

In order to estimate the energy of the resonant electrons, the Doppler shift due to the plasma toroidal rotation $\omega_\varphi$ has to be estimated. This evaluation was made using an $\mm/\nn = 1/1$ diamagnetic mode which does not appear on the spectrogram of figure \ref{fig:Gui12_41117_Spectro}, the measured frequency of this mode is around $2 \ \mathrm{kHz}$ while the ion-diamagnetic frequency in TS is typically of a few hundred kHz. The mode measured frequency was therefore assumed to be dominated by plasma rotation and the value $\omega_{\varphi} = 2 \ \mathrm{kHz}$ was retained \cite{gui12}. The energy of resonant barely trapped electrons was then estimated by matching the frequency of the modes in the plasma rest-frame with the precession frequency of barely trapped electrons $\omega = \nn(\omega_\varphi + \omega_{d,BT})$, the results are shown in figure \ref{fig:Gui12_41117_Energy}. It appears that this energy is comparable to the energy of thermal particles around $5\ \mathrm{keV}$. \comment{This relatively low energy can be explained by the inclusion of finite $k_\Vert$ effects in the resonance condition of passing electrons \cite{mer12} \footnote{This analysis is reproduced in chapter \ref{sec:EFB_stability_circ} of this thesis.}.}

\subsubsection*{Electron-driven fishbones in other tokamaks}

Other machines, such as COMPASS-D \cite{val00}, HL-1M \cite{din02} or HL-2A \cite{che09} also reported observations of electron-driven fishbone modes. In all cases the frequency of the mode is observed to be in or near the ion diamagnetic gap of the  Alfv\'{e}n spectrum except in the case of COMPASS-D where the frequency is higher ($f \sim 400 \ \mathrm{kHz}$) and close to the TAE frequency.

 It should be noted that the theory also allows the existence of fishbones rotating in the electron direction and driven by deeply trapped electrons but these would be more heavily damped by coupling to the MHD continuum and would therefore require a stronger drive than in the case of electron-driven fishbones rotating in the ion direction \cite{zon07}. Some numerical simulations were able to produce such instabilities \cite{vla11,vla12}.

\section{Thesis motivation and outline}

The example of the fishbone instability described in the previous section shows that populations of energetic particles can give rise to macro-scale instabilities through resonant interaction. Simultaneously this interaction affects the confinement of the particles. On the one hand the development of such instabilities could prevent the fusion-born alpha particles from transferring their energy to the plasma bulk \cite{hei94,ITE99,fas07}. On the other hand this phenomenon is considered as a way to control the accumulation of the \emph{helium ash} made up of the slowed-down alpha particles which can affect the fusion reaction rate by diluting the fuel \cite{ITE99}. Energetic-particle driven instabilities can also affect the power deposition profiles of auxiliary heating systems by modifying the spatial distribution of the populations of energetic particles, one corollary being that we can use this phenomenon to control these profiles. Whether to prevent anomalous energetic particle transport or to provide new control mechanisms for the plasma, it is important to understand the mechanisms of the onset of such instabilities.

The study of electron-driven fishbones is directly relevant to the study of the interaction of alpha particles with low frequency MHD instabilities since in this case the resonance would happen at the toroidal precession frequency of energetic particles which depends on the energy of the particles and not their mass. Also energetic electrons have very thin orbits much like fusion-born alphas in ITER \cite{zon07}. Moreover the stability of electron-driven fishbones is very sensitive to the details of both the electronic distribution function and the safety factor profile \cite{dec09a,mer10}. Thus they provide a sensitive test for the linear stability model.

In the Tore Supra tokamak electron-driven fishbones have been observed at a frequency well below the precession frequency of barely trapped energetic electrons which was the one predicted by the theory. The aim of this thesis is to study the stability of electron-driven fishbones to provide a possible explanation for this phenomenon.

\vspace{2mm}

In the first three chapters we introduce some of the tools necessary to our analysis. Chapter \ref{cha:Magn_config} is dedicated to the description of the equilibrium magnetic field configuration in a tokamak, the formalism developed is then used in chapter \ref{sec:GC_motion} where a hamiltonian formalism is used to study the motion of the particles in a tokamak. In chapter \ref{cha:MHD-Energy-Principle} we introduce the framework of the ideal MHD energy principle which is used to study MHD instabilities in magnetized plasmas. The stability of the internal kink mode is investigated in chapter \ref{cha:internal-kink-mode} using this formalism. The final part of this thesis is dedicated to electron-driven fishbones. The modification of the internal kink dispersion relation by resonance with energetic particles is then derived in \ref{cha:FDR_derivation} using a kinetic description of energetic electrons. Special care is given to the resonance with passing particles which are of importance in the case of energetic electrons. The MIKE code which implements this model is introduced in chapter \ref{cha:MIKE_solver} and is used in chapter \ref{sec:EFB_stability_circ} where we show that the resonance with passing electrons lowers not only the density of energetic electrons at the instability threshold but also the frequency of the mode.


\chapter{Magnetic configuration}
\label{cha:Magn_config}

The configuration of the magnetic field in a tokamak is investigated. In the first section different coordinate systems used to describe the magnetic field are introduced. In the second section the Grad-Shafranov equation \cite{gra58,sha58} which describes the equilibrium configurations for a toroidal magnetic field is derived. Finally we introduce some notations which will be used throughout this thesis.

\section{Coordinate system}

The case of an axisymmetric magnetic field with nested flux-surfaces is considered. The innermost flux-surface is degenerate and is called the \emph{magnetic axis}. The axis of symmetry is supposed to be in the vertical direction $Z$. Three different coordinate systems are defined:
\begin{itemize}
\item a right-handed orthonormal Cartesian coordinate system $(X,Y,Z)$, 
\item a right-handed polar coordinate system $(R,Z,\varphi)$ such that $\tan \varphi = X/Y$ ($\varphi$ will be further referred to as the geometrical toroidal angle) and $R$ is the distance to the vertical axis,
\item a general coordinate system $(\psi,\theta,\zeta)$ where $\psi$ is a flux-label ($\psi$ is constant on a given flux-surface and its value on the magnetic axis is chosen to be $0$) $\theta$ is a poloidal angle such that $\theta = 0$ corresponds to the outboard midplane and $\zeta$ a toroidal angle.
\end{itemize}
The standard definitions for the covariant basis are used
\begin{equation}
  \label{eq:cov_bas}
  \vc{e}_\psi = \left.\pdd{\vc{X}}{\psi}\right|_{\theta,\zeta},\, \vc{e}_\theta = \left.\pdd{\vc{X}}{\theta}\right|_{\psi,\zeta},\, \vc{e}_\zeta = \left.\pdd{\vc{X}}{\zeta}\right|_{\psi,\theta},\, 
\end{equation}
the contravariant basis
\begin{equation}
  \label{eq:cont_bas}
  \vc{e}^\psi = \nabla{\psi},\, \vc{e}^\theta = \nabla{\theta},\, \vc{e}^\zeta = \nabla{\zeta},\, 
\end{equation}
the metric tensor elements
\begin{equation}
  \label{eq:metric_tensor}
  g_{ij} = \vc{e}_i \cdot \vc{e}_j,\, g^{ij} = \vc{e}^i \cdot \vc{e}^j,
\end{equation}
and the jacobian $\mathcal{J}$ such that
\begin{equation}
  \label{eq:jacobian_MC}
  \mathcal{J}^2 = \det \left(g_{ij}\right) = \left(\det \left(g^{ij}\right)\right)^{-1}.
\end{equation}
The following identities hold
\begin{align}
  \vc{e}_i \cdot \vc{e}^j &= \delta_i^j, \\
  \left(\vc{e}_i \times \vc{e}_j\right)\cdot \vc{e}_k &= \epsilon_{ijk} \, \mathcal{J} \\
  \left(\vc{e}^i \times \vc{e}^j\right)\cdot \vc{e}^k &= \epsilon^{ijk} \, \mathcal{J}^{-1}
\end{align}
where $\delta$ is the Kronecker symbol and $\epsilon$ is the antisymmetric tensor.

Because an axisymmetric field is considered, one can choose $\zeta$ to be the geometric toroidal angle $\varphi$ and $\theta$ such that $\vc{e}_\zeta$ is orthogonal to both $\vc{e}_\psi$ and $\vc{e}_\theta$. In this way, one has
\begin{equation*}
  \nabla \zeta \cdot \nabla \psi = 0, 
  \nabla \zeta \cdot \nabla \theta = 0, 
  \nabla \zeta \cdot \nabla \zeta = \frac{1}{R^2}.
\end{equation*}

\section{Vector potential and magnetic field}

The vector potential of an axisymmetric magnetic field can be put in the form (see \cite{whi03})
\begin{equation}
  \vc{A} = - \eta(\psi,\theta) \nabla \psi + \psi_t(\psi) \nabla \theta - \psi_p(\psi) \nabla \zeta.
  \label{eq:A_vect_pot_White}
\end{equation}
$\psi_t(\psi)$ is, up to a constant, the flux of the magnetic field through a poloidal surface $S_p(\psi)$ (which is defined by $\psi' \in [0,\psi]$, $\theta \in [0,2\pi]$ and $\zeta$ an arbitrary constant).
\begin{equation*}
\iint_{S_p(\psi)} \vc{B} \cdot \vc{\dd{S}} = \oint \vc{A} \cdot \vc{e}_\theta \dd{\theta} = 2 \pi \psi_t(\psi)
\end{equation*}
In the same way $\psi_p(\psi)$ is $2 \pi$ times the flux of the magnetic field though a toroidal surface $S_t(\psi)$ (which is defined by $\psi' \in [0,\psi]$, $\zeta \in [0,2\pi]$ and $\theta$ an arbitrary constant).
\begin{equation*}
\iint_{S_t(\psi)} \vc{B} \cdot \vc{\dd{S}} = - \oint \vc{A} \cdot \vc{e}_\zeta \dd{\zeta} = 2 \pi \psi_p(\psi)
\end{equation*}

Without loss of generality, one can choose $\psi = \psi_p$ which is a flux-label as the radial variable. The derivative of $\psi_t$ against $\psi_p$ is known as the safety factor and is denoted $q$. It is a function of $\psi_p$ only. Its inverse $\iota$ is known as the rotational transform.
The contravariant representation of the magnetic field is then easily obtained,
\begin{align}
  \vc{B} &= \nabla \times \vc{A} = \left(q + \pdd{\eta}{\theta}\right) \nabla \psi_p \times \nabla \theta - \nabla \psi_p \times \nabla \zeta, \\
  B^\theta &= \vc{B} \cdot \nabla \theta = \nabla \psi_p \times \nabla\theta \cdot \nabla \zeta = \mathcal{J}^{-1}, \\
  B^\zeta &= \vc{B} \cdot \nabla \zeta = \left(q + \pdd{\eta}{\theta}\right)\nabla \psi_p \times \nabla \theta \cdot \nabla \zeta = \left(q + \pdd{\eta}{\theta}\right) \mathcal{J}^{-1}.
\end{align}
where $\mathcal{J}$ is the jacobian of the coordinate system, such that $\mathcal{J}^{-1} = \nabla \psi_p \times \nabla \theta \cdot \nabla \zeta$.

\section{Magnetic field lines}

Consider a magnetic field line $t \rightarrow \vc{X}(t)$, the magnetic field is aligned with the tangent of the field line for all $t$:
\begin{equation}
  \vc{B} \times \ddr{\vc{X}}{t} = 0.
\end{equation}
Solving this ordinary differential equation, one obtains first that $\psi_p$ is constant along the field line, which is not surprising since magnetic field lines are embedded in magnetic flux surfaces, but one obtains also the relationship between $\theta$ and $\zeta$ along the field line, 
\begin{equation}
  \ddr{\zeta}{t} B^{\theta} - \ddr{\theta}{t} B^{\zeta} = 0
\end{equation}
such that
\begin{equation}
  \ddr{\zeta}{\theta}  = \frac{B^{\zeta}}{B^{\theta}} = \left(q + \pdd{\eta}{\theta}\right)
\end{equation}

The physical meaning of $\eta$ is now apparent. If $\eta = 0$, then the pitch of the field lines in the $(\theta,\zeta)$ plane, $\dd{\zeta}/\dd{\theta} = q$, is a function of $\psi_p$ alone and the field lines are straight. Now, if $\eta \neq 0$, the pitch of the field-lines is not constant anymore, but its average on one poloidal period is still $q$.

A set of coordinates $(\psi_p, \theta, \zeta)$ with $\eta = 0$ is called \emph{flux-coordinates} or \emph{straight field line coordinates}. It can be shown that for well-behaved fields, such coordinates always exist, and that they can be found starting from any set of coordinates $(\psi_p,\theta,\zeta)$ where $a$ is a flux label, $\theta$ is a poloidal angle and $\zeta$ a toroidal angle, by only modifying either the poloidal angle or the toroidal angle.

A method to obtain flux coordinates by modifying only the poloidal angle is presented. Let $\theta_F(\psi_p,\theta,\zeta)$ be the new poloidal angle. Without loss of generality $\theta_F$ can be written $\theta_F = \theta + \hat{\theta}_F(\psi_p,\theta,\zeta)$ where $\hat{\theta}_F$ is a periodic function of $\theta$. Then from the contravariant form of $\vc{B}$, one has
\begin{align}
  \pdd{\eta}{\theta} &= q\pdd{\hat{\theta}_F}{\theta}, \\
  0 &= q \pdd{\hat{\theta}_F}{\zeta},
\end{align}
from which a solution is $\hat{\theta}_F(\psi_p,\theta,\zeta) = \eta(\psi_p,\theta)/q(\psi_p)$. \footnote{Similarly, $\zeta_F = \zeta + \eta(\psi_p,\theta)$ is a new toroidal angle such that $(\psi_p,\theta,\zeta_F)$ are flux-coordinates.}

\begin{figure}[!ht]
  \centering
  \includegraphics*[width=\figwidth]{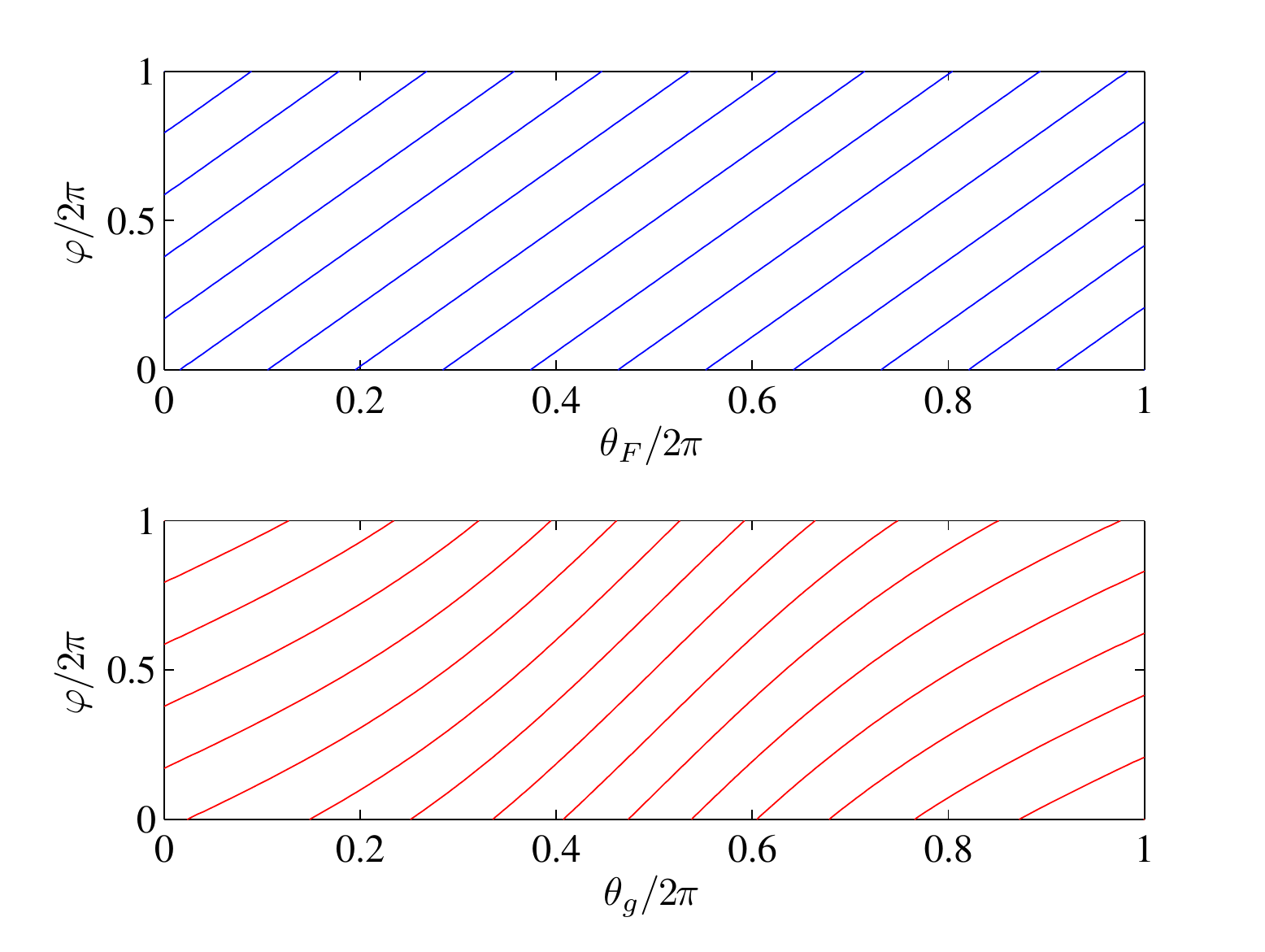}
  \caption[Comparison of geometrical coordinates and flux coordinates]{Comparison of the shape of magnetic field lines for a single flux-surface in the case of geometrical coordinates (bottom) and flux-coordinates (top). Computation based on the reconstructed equilibrium of the Tore Supra discharge \#40816, the average field-line pitch for this flux-surface is $q \simeq 2.3$.}
  \label{fig:straight_vs_geo_fl}
\end{figure}

Then $\eta$ is a measure of the distance between the actual coordinate system and a set of field-aligned coordinates. Figure \ref{fig:straight_vs_geo_fl} presents the shape of the magnetic field lines in both geometrical coordinates $(\theta_g,\varphi)$ (the geometrical poloidal angle $\theta_g$ is defined such that $\tan \theta_g = (Z - Z_0)/(R - R_0)$) and flux coordinates $(\theta_F,\varphi)$, based on a reconstructed equilibrium of a discharge in the Tore Supra tokamak.

\begin{figure}[!ht]
  \centering
  \includegraphics*[width=\figwidth]{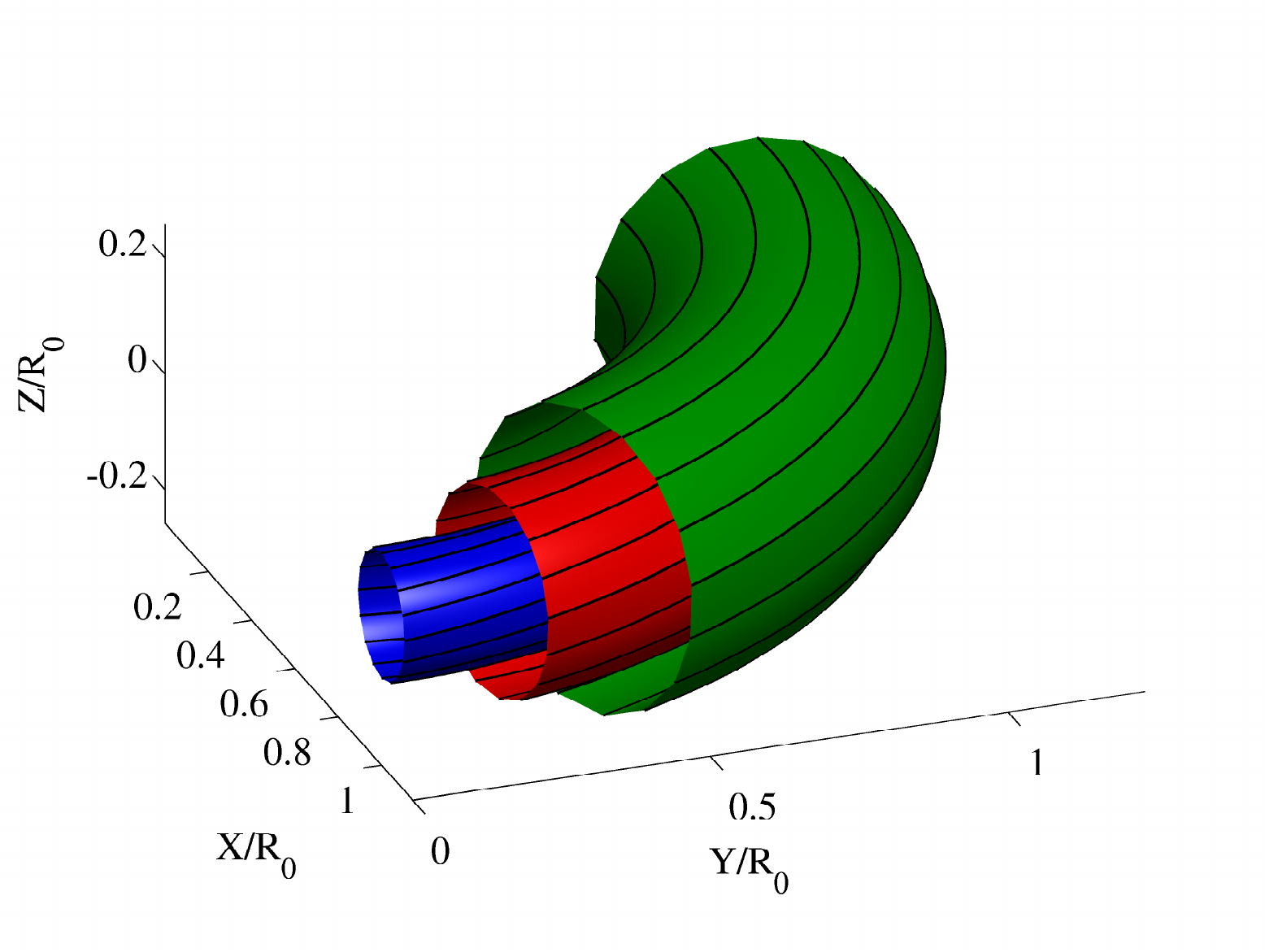}
  \caption[Flux surfaces with different values of $q$]{Different flux surfaces can have different averaged field line pitch.}
  \label{fig:plt_3_surf_shear}
\end{figure}

The variation of $q$ with the flux-surface label $\psi_p$ is called the \emph{magnetic shear}. It is quantified by the quantity $s = r/q\dd{q}/\dd{r}$ which plays a major role in tokamak physics, for example magnetic configurations with a region of \emph{reversed shear} (negative $s$) have been observed to have enhanced confinement properties. Figure \ref{fig:plt_3_surf_shear} illustrates this where 3 different surfaces with different values of $q$ have been drawn.

\section{The Grad-Shafranov equation}

In the absence of toroidal rotation of the plasma or pressure anisotropy, the force balance equation for the equilibrium plasma ($\partial/\partial t \equiv 0$) is written
\begin{equation}
  \vc{J} \times \vc{B} = \nabla p
  \label{eq:Force_balance}
\end{equation}
where $\vc{J} = \mu_0^{-1} \nabla \times \vc{B}$ is the current density. Since $p$ is constant on flux-surfaces due to the large parallel heat conductivity of electrons,and is therefore only a function of $\psi_p$, the $\theta$ and $\zeta$ components of this equation imply that the $\psi_p$ contravariant component of $\vc{J}$ vanishes. For the structure of the magnetic field, this means
\begin{equation}
  \pdd{B_\zeta}{\theta} - \pdd{B_\theta}{\zeta} = 0
\end{equation}
which in the axisymmetric case implies that $B_\zeta$ is a function of $\psi_p$ alone.

The $\psi_p$ component of equation (\ref{eq:Force_balance}) writes
\begin{equation}
  \mathcal{J} \left(J^\theta B^\zeta - J^\zeta B^\theta\right) = p'(\psi_p)
  \label{eq:Force_balance_psi}
\end{equation}
The components of $\vc{J}$ and $\vc{B}$ are expressed as
\begin{align}
  B^\theta &= \mathcal{J}^{-1}, \\
  J^\theta &= \mu_0^{-1} \mathcal{J}^{-1} \left(\pdd{B_{\psi_p}}{\zeta} - \pdd{B_\zeta}{\psi_p}\right) = - \mu_0^{-1} \mathcal{J}^{-1} B_\zeta'(\psi_p), \\
  B^\zeta &= \frac{B_\zeta}{R^2}, \\
  J^\zeta &= \mu_0^{-1} \nabla \times \vc{b} \cdot \nabla \zeta = \mu_0^{-1} \nabla \cdot \left(\vc{B} \times \nabla \zeta\right) = \mu_0^{-1} \nabla \cdot \left(\frac{\nabla \psi_p}{R^2}\right),
\end{align}
the last two identities have been obtained using the fact that the toroidal direction is orthogonal to the other two. These are then injected into equation (\ref{eq:Force_balance_psi}) to obtain the Grad-Shafranov equation,
\begin{equation}
  \nabla \cdot \left(\frac{\nabla \psi_p}{R^2}\right) + \frac{B_\zeta B_\zeta'(\psi_p)}{R^2} + \mu_0\, p'(\psi_p) = 0.
\end{equation}

\vspace{2mm}

Approximate solutions of the Grad-Shafranov equation can be obtained in the case of circular flux-surfaces using an expansion in $\epsilon$ the ratio of the plasma minor radius and major radius \cite{war66,gre71}. See appendix \ref{cha:app_circular} for the case of concentric flux-surfaces (first order in $\epsilon$) and section \ref{sec:high-aspect-ratio} for the case of shifted flux-surfaces (second order in $\epsilon$). In an appendix of reference \cite{whi84}, the Grad-Shafranov equation is extended to geometries with a helicoidal symmetry.

\section{Additional definitions}
\label{sec:eq_add_def}

The value of the poloidal flux on the last closed magnetic surface (LCFS) is noted $\psi_a$.

The coordinates of the magnetic axis in the $(R,Z)$ system are noted $(R_0,Z_0)$ and the amplitude of the magnetic field on the axis is noted $B_0$. The assumption $Z_0 = 0$ will be made unless mentioned otherwise. $R_0$ will be referred to as the plasma \emph{major radius}, and the plasma minor radius $a$ is defined as the distance of the magnetic axis to the LCFS on the outboard midplane $a = R(\psi_a,0) - R_0$.

The poloidal angle corresponding to the minimum of the magnetic field amplitude for one given flux-surface is noted $\theta_m$ and one then defines $B_m(\psi_p) = B(\psi_p,\theta_m(\psi_p))$. Similarly the value of $\theta$ corresponding to the maximum of $B$ is noted $\theta_M$ and one defines $B_M(\psi_p) = B(\psi_p,\theta_M(\psi_p))$.

Two quantities with the same structure as the safety factor $q$ are defined:
\begin{align}
  \label{eq:qtilde_def}
  \tilde{q}(\psi_p) &= \int_0^{2\pi} \frac{\dd{\theta}}{2\pi} \frac{B}{R_0 B^\theta}, \\
  \label{eq:qhat_def}
  \hat{q}(\psi_p) &= \int_0^{2\pi} \frac{\dd{\theta}}{2\pi} \frac{B_m}{R_0 B^\theta}. 
\end{align}


\chapter{Guiding-center motion}
\label{sec:GC_motion}

Guiding-center theory, introduced by Littlejohn \cite{lit79,lit81} deals with the motion of charged particles in magnetic fields. The description of this motion is simplified by averaging over the fast cyclotronic motion of the particles around the field lines. In the first section of this paragraph the basic assumptions and features of guiding-center motion are recalled. Then the equations of motion for particles evolving in static fields are derived following White et al. \cite{whi03}. Using the results from chapter \ref{cha:Magn_config} the case of the tokamak is studied and some basic features of the guiding-center motion applicable to more general magnetic configurations such as the particle drifts are shown. In section \ref{sec:orbits-tokamak}, the motion of charged particles in a tokamak is decomposed in three different motions with well-separated timescales, the expressions of the corresponding frequencies in general flux-surface geometry are recalled as well as their expansion in the case where the Larmor radius of the particles is much smaller than the plasma minor radius (this is called the \emph{zero-orbit width limit}).

\section{Charged particle motion lagrangian}

The standard form of the lagrangian for the motion of a non-relativistic charged particle of mass $m_s$ and charge $e_s$, in an electromagnetic field characterized by the potentials $\vc{A}$ and $\Phi$ is
\begin{align}
  \mathcal{L} &= (e_s\vc{A} + m_s\vc{v}) \cdot \vc{\dot{x}} - \mathcal{H}(\vc{v},\vc{x}) \\
  \mathcal{H} &= \frac{m_s v^2}{2} + e_s\Phi
  \label{eq:GenHamiltonian}
\end{align}
from which we can recover the standard result for the canonical momentum $\vc{p} = m_s \vc{v} + e_s \vc{A}$.

Littlejohn derived an expression for the guiding-center lagrangian correct to first order in the gyro-radius \cite{lit79,lit81}. This expansion is valid under the assumption that the fields evolve slowly compared to the Larmor frequency $\omega_c$ and that the characteristic gradient lengths of the fields are greater than the Larmor radius $\rho_{s} = m_s v_\perp/e_s B$ of the particles (where $v_\perp$ is the magnitude of the velocity component perpendicular to the magnetic field),
\begin{equation}
  \left|\frac{\nabla F}{F}\right| \gg \rho_{s},\quad \left|\frac{1}{F}\pdd{F}{t}\right| \ll \omega_{c,s},
  \label{eq:GCexpansion}
\end{equation}
where $F$ is either one of the electromagnetic field components or of the potentials. We define the parameter $\rho_{*s}$ which corresponds to the ratio of the Larmor radius and the gradient length of the magnetic field amplitude
\begin{equation}
  \label{eq:rho_star}
  \rho_{*s} = \rho_s \frac{\nabla B}{B},
\end{equation}
then the condition for the validity of the guiding-center theory for equilibrium fields is simply $\rho_{*s} \ll 1$.

The velocity is separated into a parallel and a perpendicular velocity $\vc{v} = v_\Vert \vc{b} + v_\perp \vc{c}$ where $\vc{b}$ is the unit vector in the direction of the magnetic field and $\vc{c} = -\sin \xi\, \vc{e}_1 - \cos \xi\, \vc{e}_2$ where $\vc{e}_1,\, \vc{e}_2$ are two unit vectors such that $\vc{e}_1 \times \vc{e}_2 = \vc{b}$ and $\xi$ is the \emph{gyrophase}. The position  is written $\vc{x} = \vc{X} + m_s v_\perp/ e_s B(\vc{X},t)\, \vc{a}(\vc{X},t)$ with $\vc{a}$ defined by $\vc{a} = \cos \xi\, \vc{e}_1 - \sin \xi\, \vc{e}_2$ giving $\vc{c} \times \vc{a} = \vc{b}$. This expression for $\vc{x}$ uniquely defines the quantity $\vc{X}$ which is called the \emph{guiding-center} position, the quantity $\vb{\rho} = \vc{a} m_s v_\perp / e_s B$ is the \emph{gyroradius}.

Under the assumptions (\ref{eq:GCexpansion}) we make the so-called gyrocenter expansion by writing 
\[F(\vc{x},t) = F(\vc{X},t) + \frac{m_s v_\perp}{e_s B}\vc{a}\cdot \nabla F(\vc{X},t),\]
the expression obtained by Littlejohn for the Lagrangian is
\begin{align}
  \mathcal{L} &= e_s \left[ \vc{A} + \rho_\Vert \vc{B}\right] \cdot \dot{\vc{X}} + \frac{m_s\mu}{e_s} \dot{\xi} - \mathcal{H}(\rho_\Vert,\mu,\vc{X},t) 
  \label{eq:GC_lagrangian} \\
  \mathcal{H} &= \frac{e_s^2}{m_s}\frac{\rho_\Vert^2 B^2}{2} + \mu B + e_s \Phi
  \label{eq:GC_hamiltonian} 
\end{align}
The quantity $\mu = m_s v_\perp^2 /2 B$ is the first-order expression of the magnetic momentum which is an adiabatic invariant. The quantity $\rho_\Vert = m_s v_\Vert/e_s B$ is called the parallel gyroradius. The following expressions of the lagrangian and hamiltonian are correct to first-order in the gyroradius, the associated Euler-Lagrange equations describe the motion of the gyrocenter position.

\section{Equilibrium motion}

\subsection{Phase-space variables}

For the gyrocenter's position, the same variables are used as for the description of the magnetic field. The phase-space lagrangian $\mathcal{L}$, depending on the phase-space variables $\psi_p,\theta,\zeta,\rho_\Vert,\mu,\xi$ and their time-derivatives, is expressed using the covariant representations of $\vc{A}$ and $\vc{B}$, giving
\begin{equation}
  \mathcal{L} = e_s \left( A_{\psi_p} + \rho_\Vert B_{\psi_p}\right) \dot{\psi_p} +  e_s \left( A_\theta + \rho_\Vert B_{\theta}\right) \dot{\theta} +  e_s \left( A_\zeta + \rho_\Vert B_{\zeta}\right) \dot{\zeta} + \frac{m_s\mu}{e_s}\dot{\xi} - \mathcal{H},
  \label{eq:GC_Lag_magn_coord_muxi}
\end{equation}

\subsection{Fast gyromotion}

In expression (\ref{eq:GC_Lag_magn_coord_muxi}), all the fields and potentials are not evaluated at the particle's position but at the gyrocenter's position. The consequence for the Euler-Lagrange equation is 
\begin{equation}
  \ddr{}{t} \left(\frac{m_s}{e_s} \mu\right) = 0,
\end{equation}
which is not a surprise since $\mu$ is an adiabatic invariant. If we now consider the dependence over $\mu$, then one obtains the following equation
\begin{equation}
  \frac{m_s}{e_s}\dot{\xi} = B
\end{equation}
which simply tells us that the gyrophase $\xi$ oscillates at the gyrofrequency $\omega_{c,s} = e_sB/m_s$. In other words, $\xi$ is the angle associated to the gyromotion of the particle and $m_s\mu/e_s$ the action which is canonically conjugate to $\xi$. The $(m_s\mu/e_s)\dot{\xi}$ term in the lagrangian is further dropped for convenience since it will not affect the equations for the $4$ remaining variables.

\subsection{Equations of motion}

The gyrocenter lagrangian can be written
\begin{equation}
  \mathcal{L} = P_{\psi_p} \dot{\psi_p} +  P_\theta \dot{\theta} +  P_\zeta \dot{\zeta} - \mathcal{H}.
  \label{eq:GC_Lag_magn_coord}
\end{equation}
with $P_i = e_s(A_i + \rho_\Vert B_i)$. The extremalization over the remaining variables $(\psi_p,\theta,\zeta,\rho_\Vert)$ yields the following equations
\begin{equation*}
  \begin{pmatrix}
    0 & a_{12} & a_{13} & a_{14} \\
    -a_{12} & 0 & a_{23} & a_{24} \\
    -a_{13} & -a_{23} & 0 & a_{34} \\
    -a_{14} & -a_{24} & -a_{34} & 0 \\
  \end{pmatrix}
  \begin{pmatrix}
    \dot{\psi_p} \\ \dot{\theta} \\ \dot{\zeta} \\ \dot{\rho_\Vert}
  \end{pmatrix}
  = 
  \begin{pmatrix}
    \partial_{\psi_p} \mathcal{H} \\ \partial_{\theta} \mathcal{H} \\ \partial_{\zeta} \mathcal{H} \\ \partial_{\rho_\Vert} \mathcal{H} 
  \end{pmatrix}
\end{equation*}
with $a_{12} = \partial_{\psi_p}P_{\theta} - \partial_{\theta}P_{\psi_p}$, $a_{13} = \partial_{\psi_p}P_{\zeta} - \partial_{\zeta}P_{\psi_p}$, $a_{23} = \partial_{\theta}P_{\zeta} - \partial_{\zeta}P_{\theta}$, $a_{14} = -\partial_{\rho_\Vert}P_{\psi_p}$, $a_{24} = -\partial_{\rho_\Vert}P_{\theta}$ and $a_{34} = - \partial_{\rho_\Vert}P_{\zeta}$. The equations of motion are then obtained by inverting the matrix. 
\begin{equation*}
  \begin{pmatrix}
    \dot{\psi_p} \\ \dot{\theta} \\ \dot{\zeta} \\ \dot{\rho_\Vert}
  \end{pmatrix}
  = \frac{1}{D}
  \begin{pmatrix}
    0 & a_{34} & -a_{24} & a_{23} \\
    -a_{34} & 0 & a_{14} & -a_{13} \\
    a_{24} & -a_{14} & 0 & a_{12} \\
    -a_{23} & a_{13} & -a_{12} & 0 
  \end{pmatrix}
  \begin{pmatrix}
    \partial_{\psi_p} \mathcal{H} \\ \partial_{\theta} \mathcal{H} \\ \partial_{\zeta} \mathcal{H} \\ \partial_{\rho_\Vert} \mathcal{H} 
  \end{pmatrix}
\end{equation*}
with $D = a_{13}a_{24} - a_{12}a_{34} - a_{23}a_{14}$. This result is valid for any equilibrium field with nested flux-surfaces such that conditions (\ref{eq:GCexpansion}) are met.

\section{The tokamak case}
\label{sec:tokamak-case}

\subsection{Equations of motion}

In the case of an axisymmetric tokamak, $\zeta$ is an ignorable coordinate such that all $\zeta$-derivatives vanish. In particular $\partial \mathcal{H}/\partial \zeta = 0$ such that Noether's theorem tells us that $P_\zeta = e_s(- \psi_p + \rho_\Vert B_\zeta)$ is an invariant of motion. Additionally we showed in chapter \ref{cha:Magn_config}, that $\partial B_\varphi/\partial \theta = 0$, such that $a_{23} = \partial_{\theta}P_{\zeta} - \partial_{\zeta}P_{\theta} = e \rho_\Vert \partial_\theta B_\zeta = 0$.

The equations of motion are
\begin{align}
  \dot{\psi_p} &= - \frac{e_s}{D} B_\zeta \left[\left(\frac{e_s^2}{m_s}\rho_\Vert^2 B + \mu\right)\pdd{B}{\theta} + e_s\pdd{\Phi}{\theta}\right], \\
  \dot{\theta} &= \frac{e_s}{D} B_\zeta\left[\left(\frac{e_s^2}{m_s}\rho_\Vert^2 B + \mu\right)\pdd{B}{\psi_p} + e_s\pdd{\Phi}{\psi_p}\right] - \frac{e_s}{D} \left(-1 + \rho_\Vert \pdd{B_\zeta}{\psi_p}\right) \frac{e_s^2}{m_s}\rho_\Vert B^2, \label{eq:GC_dot_theta}\\
  \dot{\zeta} &= -\frac{e_s}{D} B_\theta \left[\left(\frac{e_s^2}{m_s}\rho_\Vert^2 B + \mu\right)\pdd{B}{\psi_p} + e_s\pdd{\Phi}{\psi_p}\right] + \frac{e_s}{D} B_{\psi_p} \left[\left(\frac{e_s^2}{m_s}\rho_\Vert^2 B + \mu\right)\pdd{B}{\theta} + e_s\pdd{\Phi}{\theta}\right] + \ldots \nonumber \\
  &\qquad \qquad \frac{e_s}{D} \left(q+\pdd{\eta}{\theta} + \rho_\Vert \left(\pdd{B_\theta}{\psi_p} - \pdd{B_{\psi_p}}{\theta}\right)\right) \frac{e_s^2}{m_s}\rho_\Vert B^2, \label{eq:GC_dot_zeta}\\
  \dot{\rho_\Vert} &= \frac{e_s}{D} \left(-1 + \rho_\Vert \pdd{B_\zeta}{\psi_p}\right) \left[\left(\frac{e_s^2}{m_s}\rho_\Vert^2 B + \mu\right)\pdd{B}{\theta} + e_s\pdd{\Phi}{\theta}\right],
\end{align}
with $\displaystyle D = e_s^2 \left\{\left(q + \pdd{\eta}{\theta}\right)B_\zeta + B_\theta + \rho_\Vert\left[B_\zeta \left(\pdd{B_\theta}{\psi_p} - \pdd{B_{\psi_p}}{\theta}\right) - B_\theta \pdd{B_\zeta}{\psi_p}\right]\right\}$.

In the following paragraphs, we suppose that the ordering of the static electric field is such that its effect on the motion of the guiding center is only of second order in $\rho_{*s}$.

\subsection{Motion along field lines}

Since $q + \partial_\theta \eta = B^{\zeta}/B_\theta$ and $(q + \partial_\theta \eta)B_\zeta + B_\theta = (B_\zeta B^\zeta + B_\theta B^\theta)/B^\theta = B^2/B^\theta$, the movement of the guiding center is, to first order in $\rho_{*s}$, along the field line,
\begin{align}
  \dot{\psi_p} &= O(\rho_{*s}^2), \\
  \dot{\theta} &= \frac{e_s}{m_s} \rho_\Vert B^\theta + O(\rho_{*s}^2), \\
  \dot{\zeta} &= \frac{e_s}{m_s} \rho_\Vert B^\zeta + O(\rho_{*s}^2),
\end{align}
which is simply $\vc{v} = v_\Vert \vc{b} + O(\rho_{*s}^2)$. 

\subsection{Magnetic and electric drifts}
\label{sec:Mag_elec_drift}

It is also interesting to look at the second order terms in these equations. One obtains
\begin{align}
  \dot{\psi_p} &= - \frac{B^\theta B_\zeta}{e_s B^2} \left[\left(\frac{e_s^2}{m_s}\rho_\Vert^2 B + \mu\right)\pdd{B}{\theta} + e_s\pdd{\Phi}{\theta}\right] + O(\rho_{*s}^3), \\
  \dot{\theta} &= \frac{e_s}{m_s} \rho_\Vert B^\theta + \frac{B^\theta B_\zeta}{e_s B^2} \left[\left(\frac{e_s^2}{m_s}\rho_\Vert^2 B + \mu\right)\pdd{B}{\psi_p} + e_s\pdd{\Phi}{\psi_p}\right] + \ldots \\
  &\qquad\qquad- \frac{e_s}{m_s} \rho_\Vert^2 \frac{B^\theta B_\zeta}{B^2} \left(B^\theta \left(\pdd{B_\theta}{\psi_p} - \pdd{B_{\psi_p}}{\theta}\right) + B^\zeta \pdd{B_\zeta}{\psi_p}\right) + O(\rho_{*s}^3), \\
  \dot{\zeta} &= \frac{e_s}{m_s} \rho_\Vert B^\zeta + \frac{B^\theta}{e_s B^2} \left(\frac{e_s^2}{m_s}\rho_\Vert^2 B + \mu\right)\left(B_{\psi_p}\pdd{B}{\theta} - B_\theta \pdd{B}{\psi_p}\right) + \frac{B^\theta}{B^2}\left(B_{\psi_p}\pdd{\Phi}{\theta} - B_\theta \pdd{\Phi}{\psi_p}\right) + \ldots \nonumber \\
  &\qquad \qquad  \frac{e_s}{m_s} \rho_\Vert^2 \frac{B^\theta B_\theta}{B^2} \left(B^\theta\left(\pdd{B_\theta}{\psi_p} - \pdd{B_{\psi_p}}{\theta}\right) + B^\zeta \pdd{B_\zeta}{\psi_p}\right) + O(\rho_{*s}^3).
\end{align}
If one then separates the terms in $\rho_\Vert^2$ from the terms in $\mu$, then one obtains the decomposition $\vc{v} = v_\Vert \vc{b} + \vc{v}_{E \times B} + \vc{v}_\nabla + \vc{v}_\kappa + O(\rho^3)$, where $\vc{v}_{E \times B}$ is the \emph{electric drift} or \emph{$E \times B$ drift}, $\vc{v}_\nabla$ is the \emph{gradient-B} drift and $\vc{v}_\kappa$ the \emph{curvature} drift defined by 
\begin{align}
  \vc{v}_{E \times B} &= \frac{\vc{E} \times \vc{B}}{B^2} = \frac{\vc{B} \times \nabla \Phi}{B^2} , \\
  \vc{v}_\nabla &= \frac{\mu}{e_s B^2} \left(\vc{B} \times \nabla B\right), \\
  \vc{v}_\kappa &= \frac{e_s}{m_s}\rho_\Vert^2 \left(\vc{B} \times \vb{\kappa}\right)
\end{align}
where $\vb{\kappa} = \vc{b} \cdot \nabla \vc{b} = - \vc{b} \times (\nabla \times \vc{b})$ is the magnetic field curvature, which expressed in terms of the components of the magnetic field is
\begin{multline}
  \vb{\kappa} = \left[\frac{B^\theta}{B}\left(\pdd{}{\theta}\left(\frac{B_{\psi_p}}{B}\right) - \pdd{}{\psi_p}\left(\frac{B_{\theta}}{B}\right)\right) - \frac{B^\zeta}{B}\pdd{}{\psi_p}\left(\frac{B_{\zeta}}{B}\right)\right] \nabla \psi_p + \ldots \\
  - \frac{B^\zeta}{B} \pdd{}{\theta}\left(\frac{B_{\zeta}}{B}\right) \nabla \theta + \frac{B^\theta}{B} \pdd{}{\theta}\left(\frac{B_{\zeta}}{B}\right) \nabla \zeta.
\end{multline}

\subsection{Mirror force}

We wish to obtain the equation ruling the evolution of $v_\Vert$. We first recall the expression of $\dot{\rho_\Vert}$, to order $O(\rho^3)$
\begin{equation}
  \dot{\rho_\Vert} = -\frac{B^\theta}{e_s B^2} \left[\left(\frac{e_s^2}{m_s}\rho_\Vert^2 B + \mu\right)\pdd{B}{\theta} + e_s\pdd{\Phi}{\theta}\right] + O(\rho_{*s}^3),
\end{equation}
then $\dot{v_\Vert}$ is obtained by writing
\begin{equation*}
  m\dot{v_\Vert} = e_s B \dot{\rho_\Vert} + e_s \rho_\Vert \pdd{B}{\psi_p} \dot{\psi_p} + e_s \rho_\Vert \pdd{B}{\theta} \dot{\theta}
\end{equation*}
such that
\begin{equation}
  m\dot{v_\Vert} = e_s \frac{B^\theta}{B} \pdd{\Phi}{\theta} - \mu \frac{B^\theta}{B} \pdd{B}{\theta} + O(\rho_{*s}^3)
\end{equation}
which can be rewritten as $m\dot{v_\Vert} = e_s (\vc{b} \cdot \vc{E}) - \mu \vc{b}\cdot \nabla B + O(\rho_{*s}^3)$.
The first term corresponds to the acceleration by the static parallel electric field. The second term has the dimension of a force and is called the \emph{mirror force} or \emph{$\mu$-gradB force}. This force slows the motion in the parallel direction when the particle goes through regions of increasing magnetic field amplitude.

\section{Orbits in a tokamak}
\label{sec:orbits-tokamak}

\subsection{Particle trapping}

In a tokamak and for a pure MHD equilibrium, the parallel electric field is equal to $0$ and the energy of the guiding-center is conserved along the trajectory. The guiding-center then evolves in a four dimensional space $(\psi_p,\theta,\zeta,\rho_\Vert)$ with two invariants 
\begin{equation}
  P_\zeta = e_s(- \psi_p + \rho_\Vert B_\zeta(\psi_p))
  \label{eq:Pzeta_GC}
\end{equation}
and 
\begin{equation}
E = \frac{e_s^2}{2m_s}\left(\rho_\Vert B(\psi_p,\theta)\right)^2 + \mu B(\psi_p,\theta).
  \label{eq:Energy_GC}
\end{equation}

\subsubsection*{Low energy limit}

At low energies (meaning small Larmor radius), the conservation of $P_\zeta$ indicates that the guiding-center's radial position is almost constant. As a first approximation, the flux-surfaces are circular and the magnetic field strength is inversely proportional to the major radius such that, on a given flux surface it is maximum at $\theta = \pi$ and minimum at $\theta = 0$. Starting from the position $\theta = 0$, with a given $\rho_\Vert$, the guiding-center motion along the field-line will slow down due to the mirror force, two categories of orbits can then be distinguished:
\begin{itemize}
\item If $E - \mu B(\psi_p,\pi) > 0$ then $\rho_\Vert$ never vanishes and since $\dot{\theta} \sim (e_s^2/m_s) \rho_\Vert B^\theta$, $\theta$ increases or decreases monotonically. These orbits are called \emph{passing orbits}.
\item If  $E - \mu B(\psi_p,\pi) < 0$, then $\rho_\Vert$ vanishes and changes its sign. So does $\dot{\theta}$, so that the trajectory is limited to the portion of the torus where $E - \mu B(\psi_p,\theta) > 0$. These orbits are called \emph{trapped orbits}.
\end{itemize}

\subsubsection*{General case}

\begin{figure}[!ht]
  \centering
  \includegraphics*[width=\figwidth]{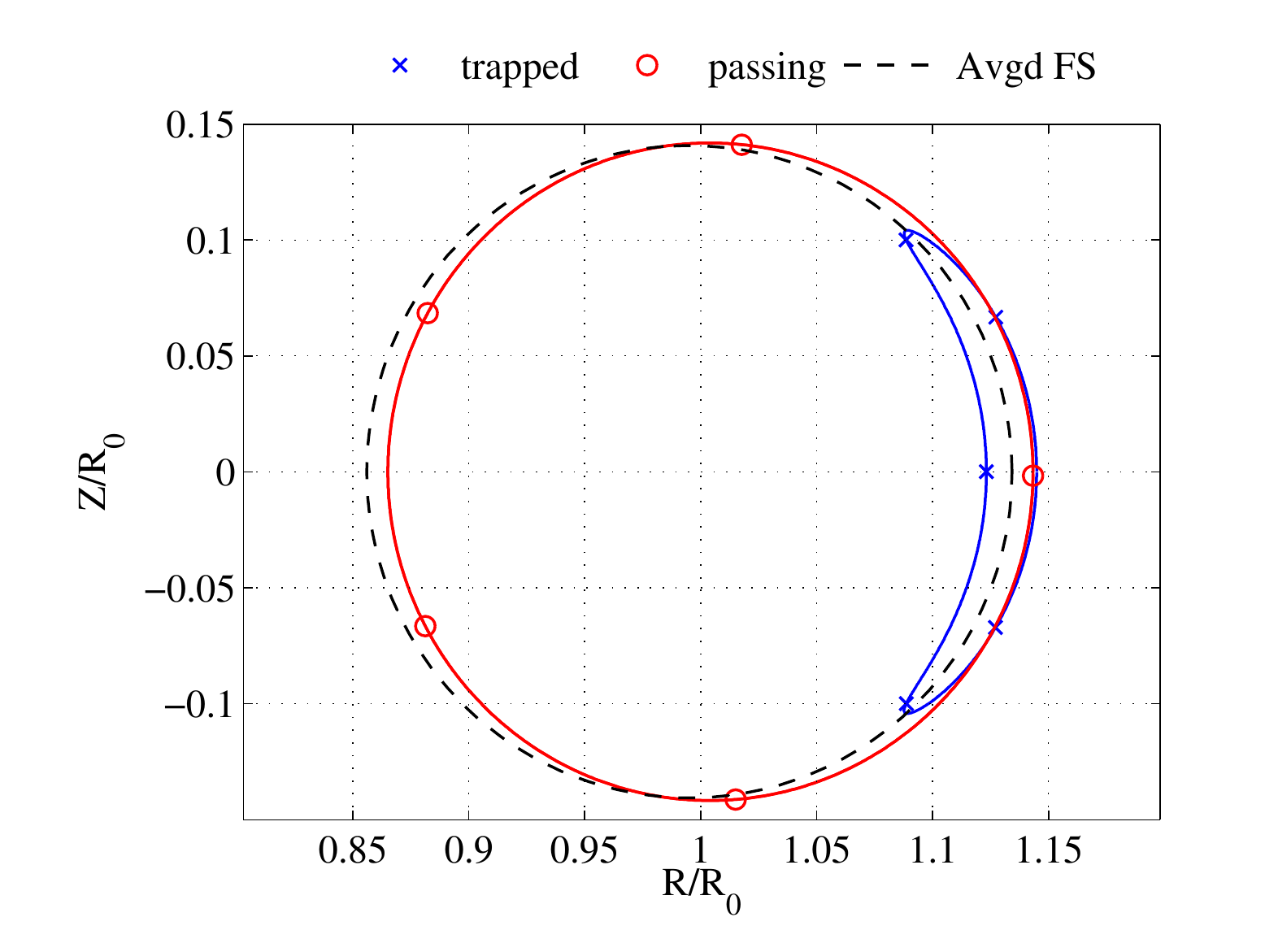}
  \caption[Comparison of trapped and passing orbits]{Projection of the guiding-center trajectory in the poloidal plane of a trapped (blue crosses) and passing (red circles) deuterium ion. The black dashed line indicates the position of the flux-surface corresponding the orbit-averaged poloidal flux.}
  \label{fig:Comp_pass_trap_traj}
\end{figure}

At higher energies, the $\rho_\Vert B_\zeta(\psi_p)$ term in $P_\zeta$ will become non-negligible and this will result in a widening of the orbit in the radial direction. One can still separate \emph{trapped orbits}, for which $\rho_\Vert$ vanishes at some point along the trajectory, and \emph{passing orbits}, for which $\rho_\Vert$ remains of constant sign along the trajectory. The point of zero parallel velocity can be used to compute the equation of the trapped-passing boundary in the invariant space. If $\rho_\Vert = 0$, then $P_\zeta = -e_s \psi_p$ and $E = \mu B(\psi_p,\theta)$, such that the trapped-passing boundary is given by
\begin{equation}
  E - \mu B(-P_\zeta/e_s,\pi) = 0,
  \label{eq:TP_boundary_GC}
\end{equation}
with the condition $E - \mu B(-P_\zeta/e_s,\pi) < 0$ corresponding to trapped orbits and $E - \mu B(-P_\zeta/e_s,\pi) > 0$ to passing orbits. Figure \ref{fig:Comp_pass_trap_traj} features a comparison of the trajectories of two deuterium ions with the same energy and orbit-averaged poloidal flux, one on a trapped orbit and one on a passing orbit. 
Due to their characteristic shape, the trapped orbits are sometimes called \emph{banana orbits}.

\begin{figure}[!ht]
  \centering
  \includegraphics*[width=\figwidth]{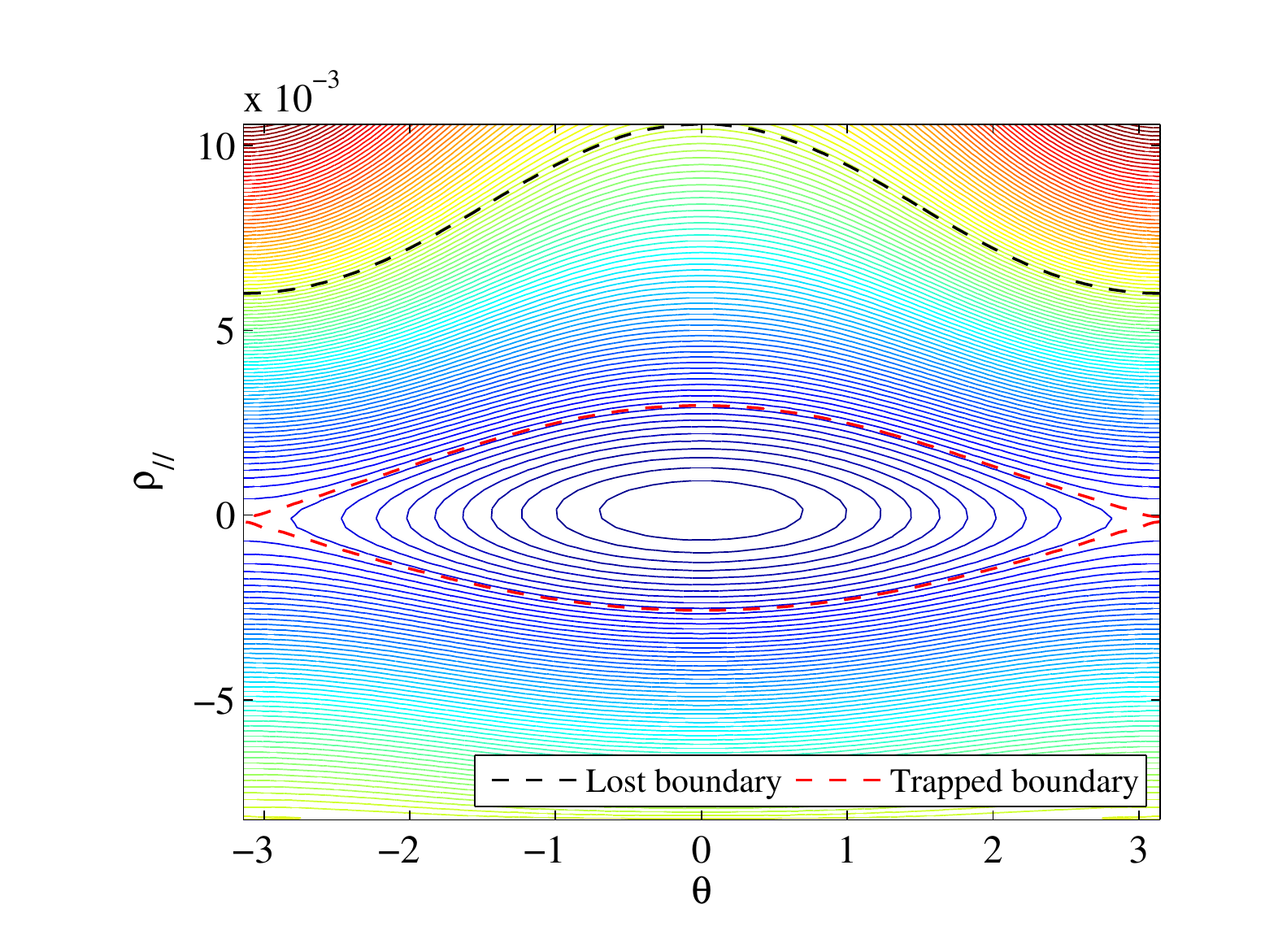}
  \caption[Phase-space island due to the magnetic field inhomogeneity]{Guiding-center trajectories in the $(\rho_\Vert,\theta)$ plane, corresponding to iso-$E$ contours, for $P_\zeta/e_s \simeq - 0. 44 \,\psi_a$ and $\mu B_0 \simeq 27.5 \, \mathrm{keV}$, $m_s = 2 m_p$, $e_s = q_e$. Closed contours correspond to trapped orbits, while open contours correspond to passing orbits. The red dashed-line is the trapped-passing boundary according to equation (\ref{eq:TP_boundary_GC}). The black dashed line is the lost boundary corresponding to $\psi_p = \psi_a$ at some point of the trajectory.}
  \label{fig:rho_par_theta_island_TB}
\end{figure}

For general equilibria, the different orbits and regions can be visualized by using equation (\ref{eq:Pzeta_GC}) to write $\psi_p$ as a function of $P_\zeta$ and $\rho_\Vert$ and injecting this expression into equation (\ref{eq:Energy_GC}) to get $E$ as a function of $(P_\zeta,\mu,\rho_\Vert,\theta)$. If the values of $P_\zeta$ and $\mu$ are kept fixed then the iso-contours of $E$ in the $(\rho_\Vert,\theta)$ plane correspond to the trajectories. This was done for an equilibrium corresponding to the Tore Supra geometry, the result is presented in figure \ref{fig:rho_par_theta_island_TB} where the characteristic phase-space island is recovered.

\vspace{2mm}

\begin{figure}[!ht]
  \centering
  \includegraphics*[height=\figheight]{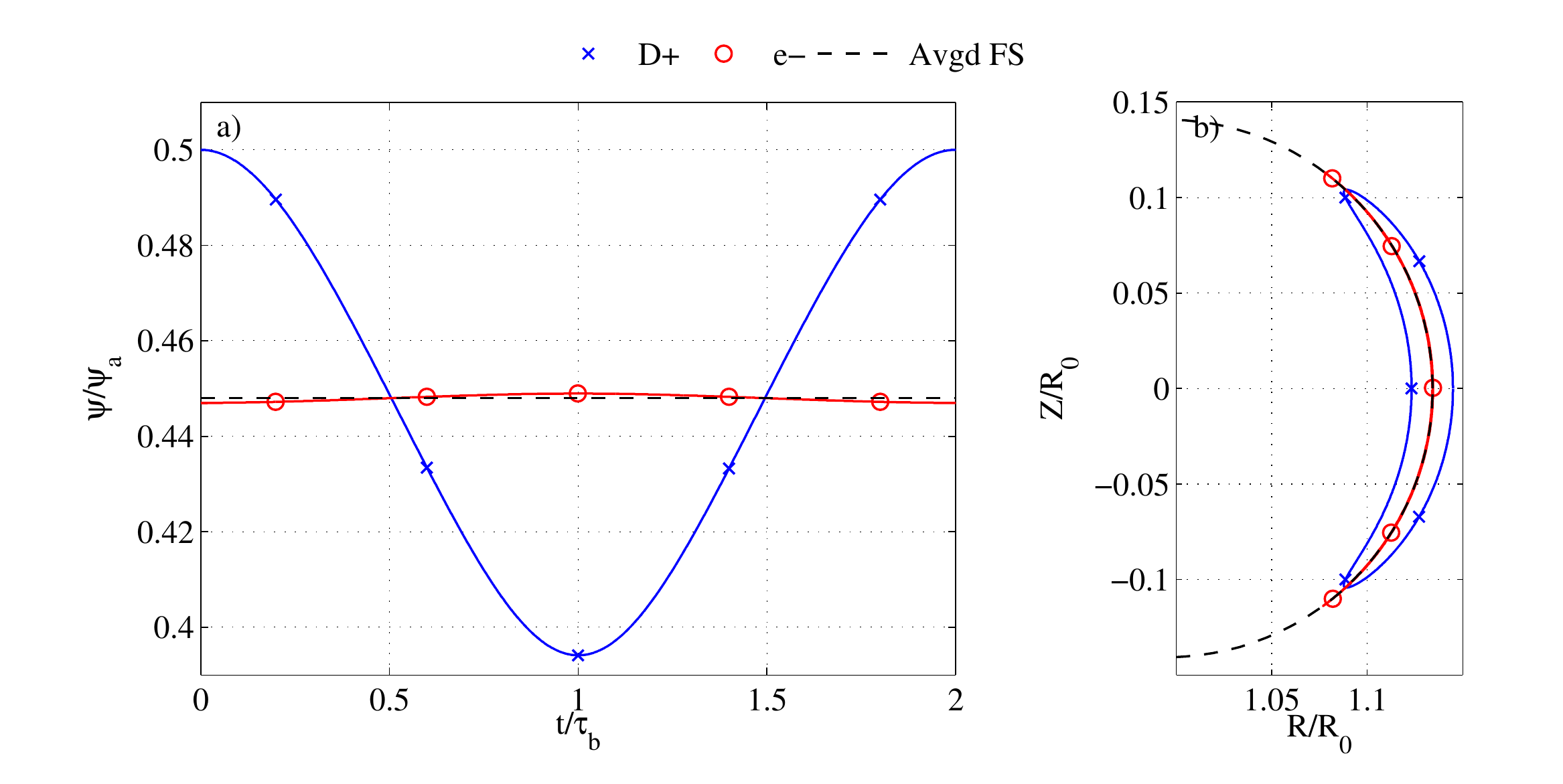}
  \caption[Comparison of the trapped orbits of an electron and a deuterium ion]{Guiding-center radial position (a) and projection of the trajectory in the poloidal plane for a deuterium ion (blue crosses) and an electron (red circles). The black dashed lines indicate the orbit-averaged poloidal flux and corresponding flux-surface for both particles.}
  \label{fig:Comp_d_e_traj}
\end{figure}

The variation of $\psi_p$ along the trajectory are of the order of $\rho_{s} R_0 B_0$ where $R_0$ and $B_0$ are the major radius and magnetic field strength on the magnetic axis. At fixed energy, the Larmor radius is proportional to the square root of the particle's mass, such that electrons have much thinner orbits than ions of the same energy. This can be seen on figure \ref{fig:Comp_d_e_traj}, where we compare the trajectories of a deuterium ion and an electron, with similar parameters (same energy, orbit-averaged poloidal flux and same $\xi_0 = v_\Vert(\theta = 0)/v$).

Other types of orbits can exist, such as \emph{potato orbits} which are passing orbits which do not encircle the magnetic axis, but they are beyond the scope of this thesis. A more complete description of the different types of particle orbits can be found in \cite{whi06}.

\subsection{Toroidal drift}

\begin{figure}[!ht]
  \centering
  \includegraphics*[height=\figheight]{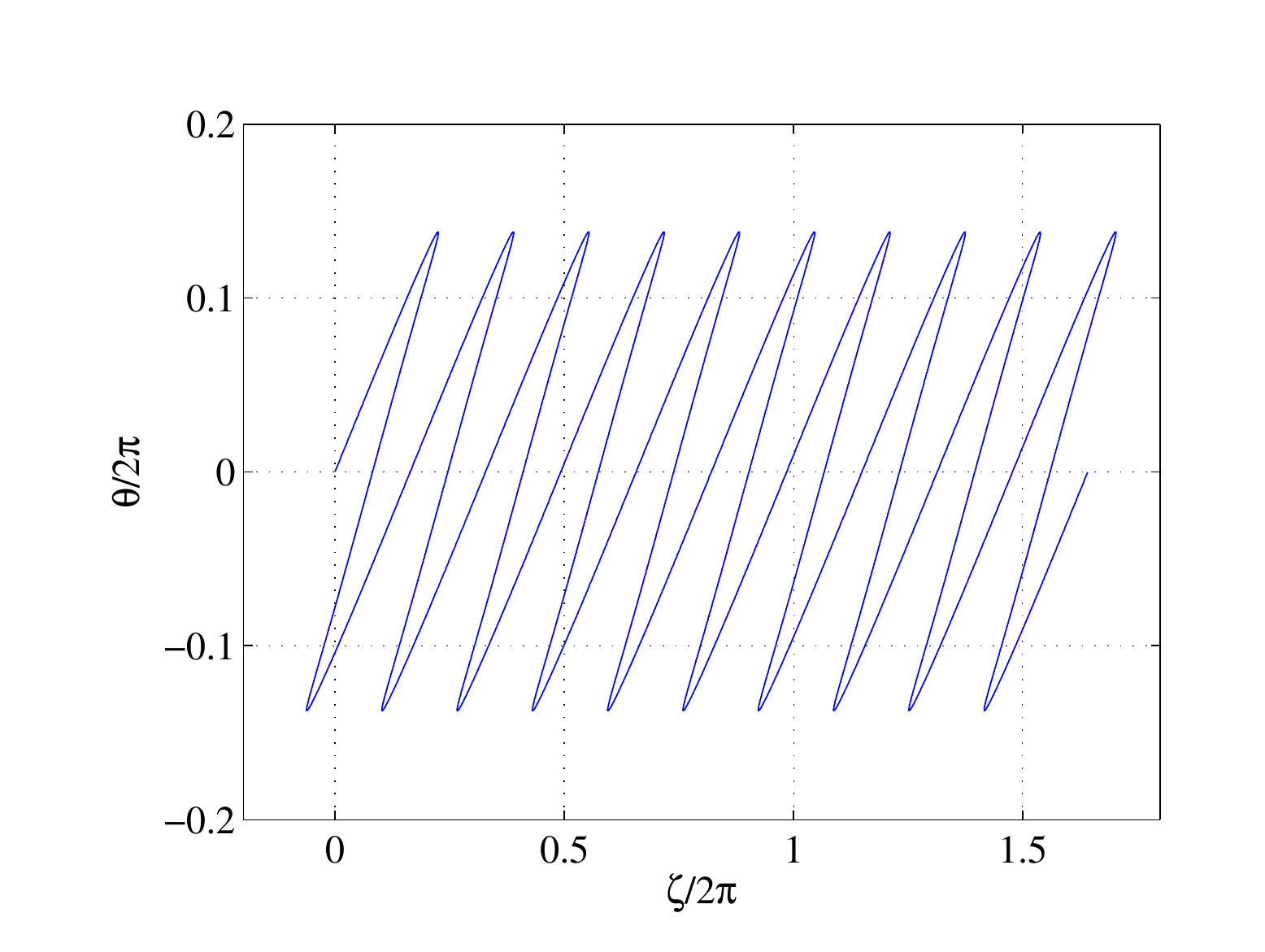}
  \caption{Toroidal drift motion of a trapped deuterium ion.}
  \label{fig:Dion_tor_drift_10}
\end{figure}

Let us consider a trapped particle, after one poloidal orbit the guiding-center's position in the poloidal plane is unchanged. But the orbit is not exactly closed in space because the toroidal angle has changed. This is due to the magnetic drifts introduced in section \ref{sec:Mag_elec_drift}, which projection in the toroidal direction do not average to zero over one complete poloidal orbit. This effect is illustrated in figure \ref{fig:Dion_tor_drift_10}, where we have plotted the guiding-center trajectory of a trapped deuterium ion in the $(\zeta,\theta)$ plane for several poloidal orbits.

For passing particles, the situation is slightly different. If the orbit width is negligible, then the variation of the toroidal angle due to the motion of the particle streaming along the field-line is, after a complete poloidal orbit, $\left(\Delta \zeta\right)_{stream} = q(\psi_p) \Delta \theta$. The total variation of $\zeta$ is then the sum of this $\Delta \zeta$ due to the streaming and of the one due to the magnetic drifts. For arbitrary orbit width, we choose for the streaming part the following definition
\begin{equation}
  \Delta \zeta = q(\bar{\psi}_p)\Delta \theta + \left(\Delta \zeta\right)_{drift},
  \label{eq:dzeta_drift_p}
\end{equation}
where $\bar{\psi}_p$ is the orbit-averaged poloidal flux.

Since the curvature and grad-B drifts are only second order in $\rho_{*s}$, the toroidal drift motion of particles is slower than the bounce motion by a $\rho*$ factor.

\subsection{Orbit characteristic frequencies}
\label{sec:orbit-char-freq}

The motion of a charged particle in a tokamak can then be decomposed into three separate motions. The cyclotronic motion around the field-lines, the bounce or transit motion along the field-lines and the toroidal drift motion across the torus. In tokamaks these motions have well-separated time-scales since the ordering between the characteristic frequencies is $\omega_{c} / \omega_{b} \simeq \omega_{b} / \omega_{d} \simeq \rho_{*} \ll 1$. Note that in a tokamak, $\rho_{*s}$ is the ratio of the Larmor radius and the plasma minor radius.

\subsubsection*{Cyclotron frequency}

Its expression is simply $\omega_{c,s} = e_s B/m_s$. For ions its value is in the range of a few hundreds of megahertz ($\mathrm{MHz}$), while for electrons it is in the range of a few hundreds of gigahertz ($\mathrm{GHz}$).

\subsubsection*{Bounce frequency}

The bounce time corresponds to the time taken by a trapped particle to complete a full poloidal orbit. For continuity reasons, it is extended to passing particles as the time taken for the particle to go twice around the magnetic axis. It can be computed as
\begin{equation}
  \tau_b = \oint \frac{\dd{\theta}}{\dot{\theta}}.
\end{equation}
where $\oint$ stands for the appropriate variation of $\theta$. The bounce frequency is then simply defined as
\begin{equation}
  \omega_b = \frac{2\pi}{\tau_b} = \left(\oint \frac{1}{\dot{\theta}}\frac{\dd{\theta}}{2\pi}\right)^{-1}.
  \label{eq:def_omegab}
\end{equation}

For passing particles with a small orbit width, the bounce time can be approximated by considering that the guiding-center will stream at a velocity $v_\Vert$ along the field line which, for a variation of $\theta$ equal to $4\pi$ has an approximate length of $4\pi q R_0$. For well-passing particle $\mu \sim 0$ and the parallel velocity is almost constant, such that $\tau_b$ is, up to a factor of the order of unity,
\begin{equation}
  \tau_b \sim \frac{4 \pi q R_0}{v_\Vert}
\end{equation}
For thermal ions ($E \sim 5\,\mathrm{keV}$), this approximation gives a bounce frequency of about $10\,\mathrm{kHz}$ while for thermal electrons, it is of the order of $1 \,\mathrm{MHz}$.
\begin{figure}[!ht]
  \centering
  \includegraphics*[height=\figheight]{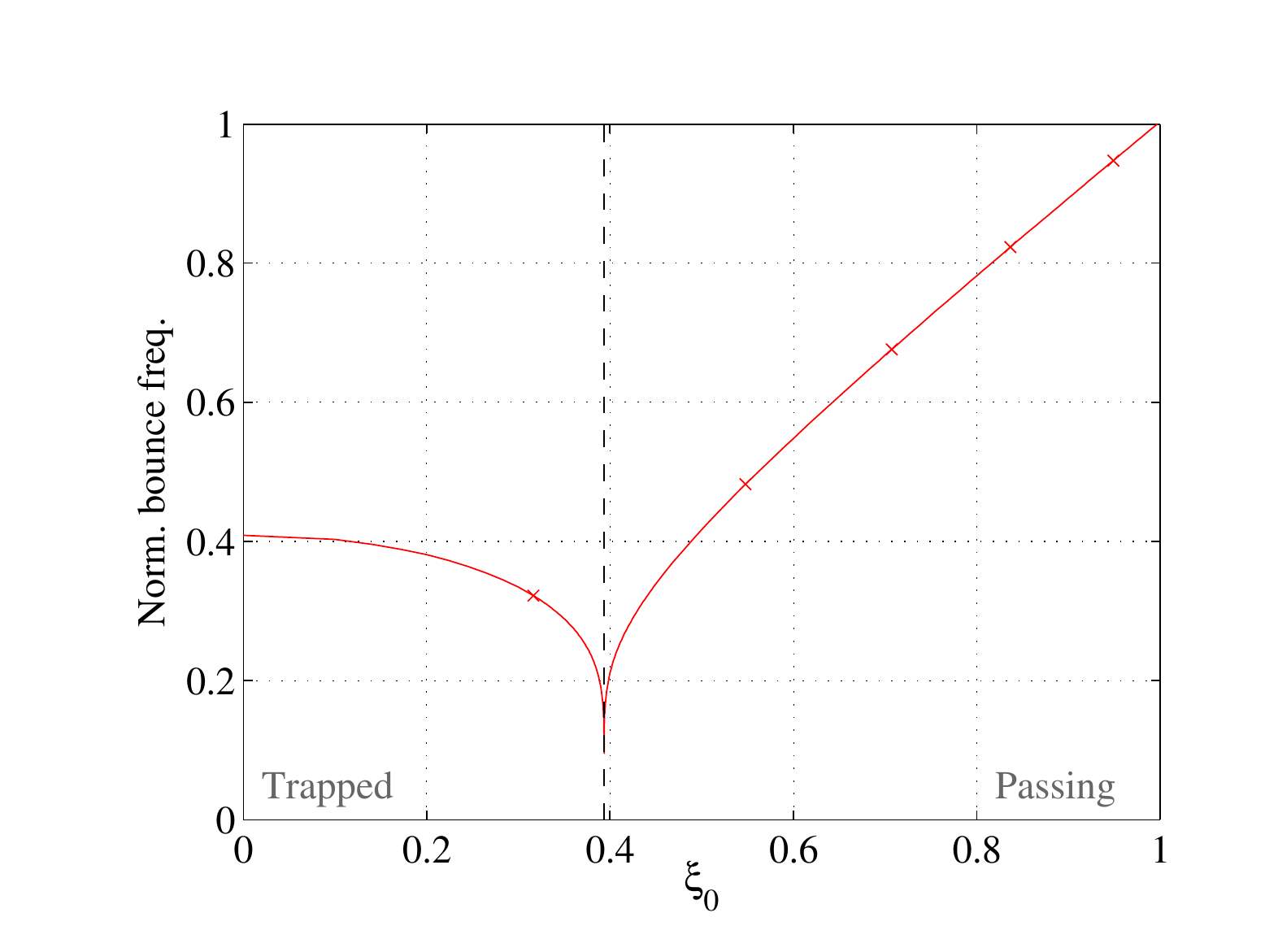}
  \caption[Normalized bounce frequency versus pitch-angle]{Normalized bounce frequency against $\xi_0 = (v_\Vert/v)_{\theta = 0}$ for low energy particles.}
  \label{fig:GCT_omegab_0p2_xi0}
\end{figure}

\subsubsection*{Toroidal drift frequency}

The toroidal drift frequency is the frequency associated to the slow drift motion across the torus of the particles. It can be computed as
\begin{equation}
  \omega_d = \frac{1}{\tau_b} \oint \left(\dot{\zeta} - q(\bar{\psi}_p) \dot{\theta}\right) \dd{t}  = \frac{\omega_b}{2\pi} \oint \left( \frac{\dot{\zeta}}{\dot{\theta}} - q(\bar{\psi}_p)\right)\dd{\theta},
  \label{eq:def_omegad}
\end{equation}
where the term in $q(\bar{\psi}_p)$ ensures that $\omega_d$ is only second order in Larmor radius. This expression then reduces to 
\begin{equation}
  \omega_d = \frac{\omega_b}{2\pi} \oint \frac{\dot{\zeta}}{\dot{\theta}}\dd{\theta},
\end{equation}
for trapped particles and
\begin{equation}
  \omega_d = \frac{\omega_b}{2\pi} \oint \frac{\dot{\zeta}}{\dot{\theta}} \dd{\theta} - 2 q(\bar{\psi}_p) \omega_b
\end{equation}
for passing particles (the factor $2$ comes from the definition of $\tau_b$ as the time to make two complete orbits for passing particles).

As we will see later on, the toroidal drift frequency can be put in the form
\begin{equation}
  \omega_d = \frac{q E}{e_s B_0 R_0 r} \bar{\Omega}_d
\end{equation}
where $\bar{\Omega}_d$ is of the order of unity. Since this expression does not involve the particle's mass, electrons and singly-charged ions at the same energy will have the same absolute value for $\omega_d$. For thermal particles, the typical value is a few $\mathrm{kHz}$.
\begin{figure}[!ht]
  \centering
  \includegraphics*[height=\figheight]{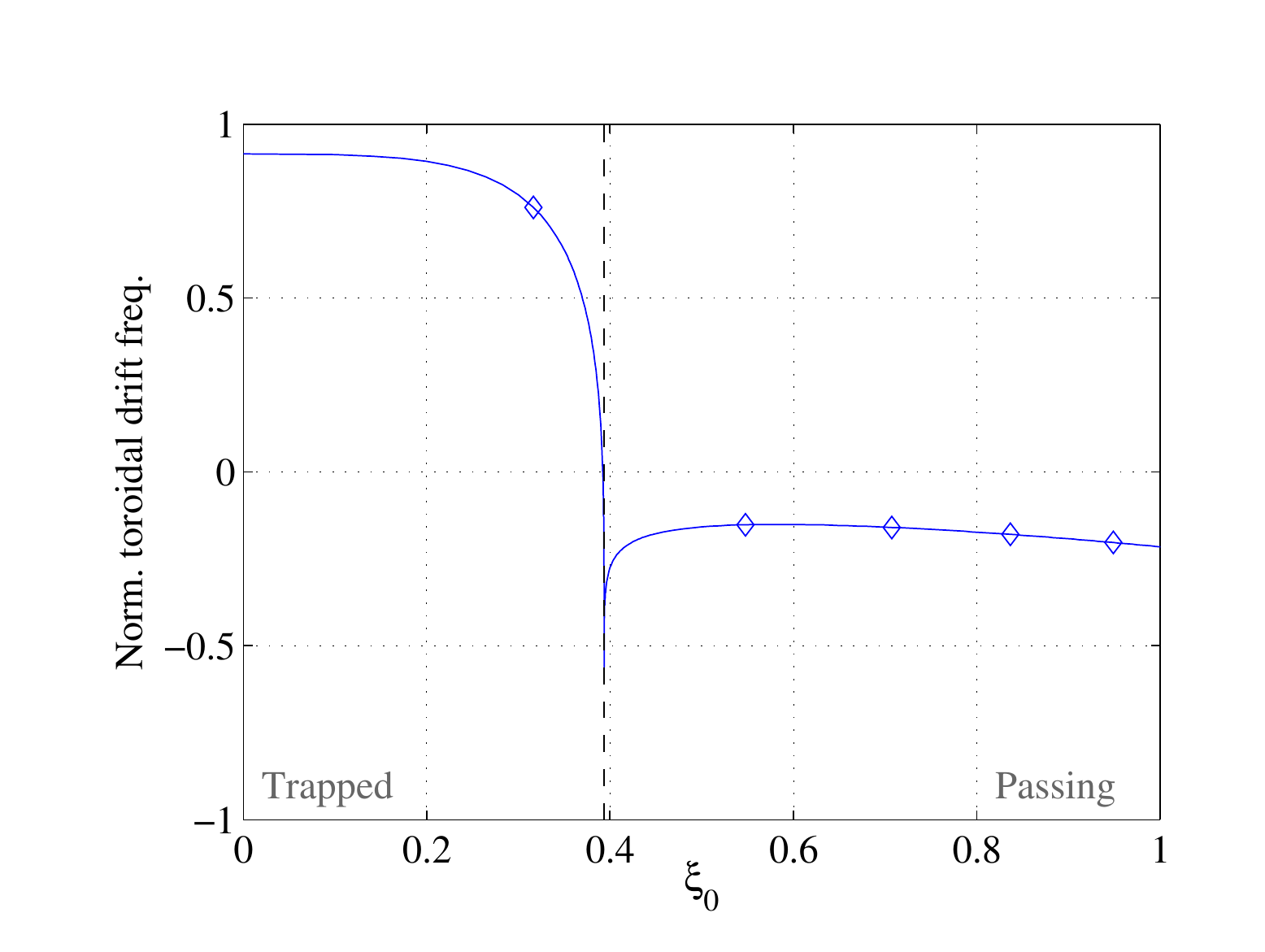}
  \caption[Normalized drift frequency versus pitch-angle]{Normalized drift frequency against $\xi_0 = (v_\Vert/v)_{\theta = 0}$ for low energy particles.}
  \label{fig:GCT_omegad_0p2_xi0}
\end{figure}

\subsection{The thin orbit width limit}
\label{sec:thin-orbit-width}

If the orbit width (or Larmor radius) is much smaller than the plasma minor radius, approximate expressions for $\tau_b$ (or equivalently $\omega_b$) and $\omega_d$ can be obtained. We proceed by injecting equations (\ref{eq:GC_dot_theta}) and (\ref{eq:GC_dot_zeta}) in the integrals (\ref{eq:def_omegab}) and (\ref{eq:def_omegad}) and the radial position $\psi_p$ is replaced by its orbit-averaged value $\bar{\psi}_p$.

\subsubsection*{Bounce frequency}

For the first order approximation of $\omega_b$, one only needs the first order approximation of $\dot{\theta}$. Then $\tau_b$ can be approximated by 
\begin{equation*}
  \tau_b = \oint \frac{\dd{\theta}}{\frac{e_s}{m_s} \rho_\Vert B^\theta}\left(1 + O(\rho_{*s})\right)
\end{equation*}
such that the first-order approximation of $\tau_b$ is
\begin{equation*}
  \tau_b = \oint \frac{B}{B^\theta}\frac{\dd{\theta}}{v_\Vert}.
\end{equation*}

\subsubsection*{Toroidal drift frequency}

For the toroidal drift frequency, one needs the second order approximation of both $\dot{\theta}$ and $\dot{\zeta}$. Writing $\dot{\theta} = \dot{\theta}_1 + \dot{\theta}_2 + \ldots$ and $\dot{\zeta} = \dot{\zeta}_1 + \dot{\zeta}_2 + \ldots$, one has
\begin{equation*}
  \omega_d = \frac{\omega_b}{2\pi} \oint \left( \frac{\dot{\zeta}_1}{\dot{\theta}_1} + \frac{\dot{\zeta}_2}{\dot{\theta}_1} - \frac{\dot{\zeta}_1}{\dot{\theta}_1} \frac{\dot{\theta}_2}{\dot{\theta}_1}- q(\bar{\psi}_p)\right)\dd{\theta} + \ldots
\end{equation*}
which gives
\begin{multline*}
  \omega_d = \frac{\omega_b}{2\pi} \oint \left( \frac{B^\zeta}{B^\theta} + \frac{m_s}{e_s \rho_\Vert B^\theta}\left[\frac{B^\theta}{e_s B^2} \left(\frac{e_s^2}{m_s}\rho_\Vert^2 B + \mu\right)\left(B_{\psi_p}\pdd{B}{\theta} - B_\theta \pdd{B}{\psi_p}\right) + \ldots \right.\right. \\
  \left.\left.\frac{e_s}{m_s} \rho_\Vert^2 \frac{B^\theta B_\theta}{B^2} \left(B^\theta\left(\pdd{B_\theta}{\psi_p} - \pdd{B_{\psi_p}}{\theta}\right) + B^\zeta \pdd{B_\zeta}{\psi_p}\right)\right] + \ldots \right. \\
  \left.- \frac{m_s}{e_s \rho_\Vert B^\theta}\left[ \frac{B^\zeta B_\zeta}{e B^2} \left(\frac{e_s^2}{m_s}\rho_\Vert^2 B + \mu\right)\pdd{B}{\psi_p} + \ldots\right.\right. \\
  \left.\left. - \frac{e_s}{m_s} \rho_\Vert^2 \frac{B^\zeta B_\zeta}{B^2} \left(B^\theta\left(\pdd{B_\theta}{\psi_p} - \pdd{B_{\psi_p}}{\theta}\right) + B^\zeta \pdd{B_\zeta}{\psi_p}\right)\right]- q(\bar{\psi}_p)\right)\dd{\theta} + O(\rho_{*s}^3)
\end{multline*}
and can be simplified to 
\begin{multline*}
  \omega_d = \frac{\omega_b}{2\pi} \oint \left( \frac{B^\zeta}{B^\theta} - q(\bar{\psi}_p)\right)\dd{\theta} + \\ \frac{\omega_b}{2\pi} \oint \left( \frac{m_s}{e_s^2 \rho_\Vert B^\theta}\left[\left(\frac{e_s^2}{m_s}\rho_\Vert^2 B + \mu\right)\left(\frac{B^\theta B_{\psi_p}}{B^2}\pdd{B}{\theta} - \pdd{B}{\psi_p}\right) + \ldots \right.\right. \\
  \left.\left. \frac{e_s^2}{m_s} \rho_\Vert^2 \left(B^\theta\left(\pdd{B_\theta}{\psi_p} - \pdd{B_{\psi_p}}{\theta}\right) + B^\zeta \pdd{B_\zeta}{\psi_p}\right)\right]\right)\dd{\theta} + O(\rho_{*s}^3)
\end{multline*}

Since $q = B^\zeta/B^\theta - \partial \eta/\partial \theta$, the integral of $q(\bar{\psi}_p)$ can be replaced by the integral of $B^\zeta(\bar{\psi}_p,\theta)/B^\theta(\bar{\psi}_p,\theta)$. Such that the first term in $\omega_d$ is of the same order as the other terms, since $\psi_p - \bar{\psi}_p$ is first order in $\rho_{*s}$. One then has
\begin{align*}
  \psi_p - \bar{\psi}_p &= \left(\psi_p + \frac{P_\zeta}{e_s}\right) - \left(\bar{\psi}_p + \frac{P_\zeta}{e_s}\right) \\
  &= \rho_\Vert B_\zeta - \frac{\omega_b}{2\pi} \oint \rho_\Vert B_\zeta \frac{\dd{\theta}}{\dot{\theta}}, \\
  &= \rho_\Vert B_\zeta - \frac{\omega_b}{2\pi} \frac{m_s}{e_s} \oint \frac{B_\zeta}{B^\theta} \dd{\theta} + O(\rho_{*s}^2),
\end{align*}
where the integral $\oint B_\zeta/B^\theta \dd{\theta}$ is equal to zero for trapped particles. Its expression is similar to the one of the safety factor such that we write
\begin{equation}
\psi_p - \bar{\psi}_p = \rho_\Vert B_\zeta - 2 \delta_P \frac{m_s}{e_s} R_0^2 \,\omega_b q_{\zeta}(\bar{\psi}_p) + O(\rho_{*s}^2)
\end{equation}
with $\delta_P = 1$ for passing particles only and 
\begin{equation}
  q_{\zeta}(\psi_p) = \int_0^{2\pi} \frac{1}{R_0^2} \frac{B_\zeta(\psi_p)}{B^\theta(\psi_p,\theta)} \frac{\dd{\theta}}{2 \pi}
\end{equation}
The first term in $\omega_d$ is
\begin{align*}
  &\frac{\omega_b}{2\pi} \oint \left( \frac{B^\zeta}{B^\theta} - q(\bar{\psi}_p)\right)\dd{\theta} = \frac{\omega_b}{2\pi} \oint \left( \rho_\Vert B_\zeta - 2 \delta_P \frac{m_s}{e_s} R_0^2 \,\omega_b q_{\zeta}(\bar{\psi}_p)\right) \pdd{}{\psi_p} \left(\frac{B^\zeta}{B^\theta}\right) \dd{\theta} + O(\rho_{*s}^3) \\
  &= \frac{\omega_b}{2\pi} \oint \rho_\Vert B_\zeta \pdd{}{\psi_p} \left(\frac{B^\zeta}{B^\theta}\right) \dd{\theta} - 2 \delta_P \frac{m_s}{e_s} R_0^2 \,\omega_b^2 q_{\zeta}(\bar{\psi}_p) \left(\oint \pdd{}{\psi_p} \left(\frac{B^\zeta}{B^\theta}\right)\frac{\dd{\theta}}{2\pi}\right) + O(\rho_{*s}^3) \\
  &= \frac{\omega_b}{2\pi} \oint \rho_\Vert B_\zeta \pdd{}{\psi_p} \left(\frac{B^\zeta}{B^\theta}\right) \dd{\theta} - 2 \delta_P \frac{m_s}{e_s} R_0^2 \,\omega_b^2 q_{\zeta}(\bar{\psi}_p) \pdd{q}{\psi_p}(\bar{\psi}_p) + O(\rho_{*s}^3).
\end{align*}
The expression for $\omega_d$ can then be worked out to the following expression, where we have separated the terms in $\rho_\Vert$ from the ones in $\mu/\rho_\Vert$:
\comment{
\begin{multline*}
  \omega_d =  - 2 \delta_P \frac{m_s}{e_s} R_0^2 \,\omega_b^2 q_{\zeta}(\bar{\psi}_p) \pdd{q}{\psi_p}(\bar{\psi}_p) + \ldots \\
\frac{\omega_b}{2\pi} \oint \left[\left(\rho_\Vert + \frac{m_s\mu}{e_s^2\rho_\Vert B}\right)\left(\frac{B_{\psi_p}}{B}\pdd{B}{\theta} - \frac{B}{B^\theta}\pdd{B}{\psi_p}\right) + \ldots \right. \\
  \left. \rho_\Vert \left(B_\zeta\pdd{}{\psi_p} \left(\frac{B^\zeta}{B^\theta}\right) + \pdd{B_\theta}{\psi_p} - \pdd{B_{\psi_p}}{\theta} + \frac{B^\zeta}{B^\theta}\pdd{B_\zeta}{\psi_p}\right)\right]\dd{\theta} + O(\rho_{*s}^3).
\end{multline*}
\begin{multline*}
  \omega_d =  - 2 \delta_P \frac{m_s}{e_s} R_0^2 \,\omega_b^2 q_{\zeta}(\bar{\psi}_p) \pdd{q}{\psi_p}(\bar{\psi}_p) + \ldots \\
\frac{\omega_b}{2\pi} \oint \left[\left(\rho_\Vert + \frac{m_s\mu}{e_s^2\rho_\Vert B}\right)\left(\frac{B_{\psi_p}}{B}\pdd{B}{\theta} - \frac{B}{B^\theta}\pdd{B}{\psi_p}\right) + \ldots \right. \\
  \left. \rho_\Vert \left(\pdd{}{\psi_p}\left(\frac{B^2}{B^\theta}\right) - \pdd{B_{\psi_p}}{\theta}\right)\right]\dd{\theta} + O(\rho_{*s}^3).
\end{multline*}
}
\begin{multline}
  \omega_d =  - 2 \delta_P \frac{m_s}{e_s} R_0^2 \,\omega_b^2 q_{\zeta}(\bar{\psi}_p) \pdd{q}{\psi_p}(\bar{\psi}_p) + \ldots \\
\frac{\omega_b}{2\pi} \oint \left[\frac{m_s\mu}{e_s^2\rho_\Vert B}\left(\frac{B_{\psi_p}}{B}\pdd{B}{\theta} - \frac{B}{B^\theta}\pdd{B}{\psi_p}\right) + \ldots \right. \\
  \left. \rho_\Vert \left(B \pdd{}{\psi_p}\left(\frac{B}{B^\theta}\right) - B\pdd{}{\theta}\left(\frac{B_{\psi_p}}{B}\right)\right)\right]\dd{\theta} + O(\rho_{*s}^3).
\label{eq:omegad_gen}
\end{multline}

Expression \eqref{eq:omegad_gen} can be used to compute the toroidal drift precession frequency in the limit of zero-orbit width and for general flux-surface geometry. The corresponding expression for a low-beta equilibrium with circular concentric flux-surfaces can be found in appendix \ref{cha:app_circular}.

\section{Summary}

The motion of charged particles in a tokamak has been studied using the guiding-center theory. The motion can be decomposed into three motions with well-separated timescales. The fastest of these motions is the cyclotronic motion around the field-lines, then comes the bounce/transit motion of the particles along the field-lines and finally the particles slowly drift toroidally around the torus. Due to the particle drifts described in section \ref{sec:Mag_elec_drift}, the guiding-center orbits have a finite radial width which is of the order of the Larmor radius. The expressions of the characteristic frequencies of motion have been derived in section \ref{sec:orbit-char-freq} for particles with arbitrary energy and in section \ref{sec:thin-orbit-width} in the limit of zero-orbit width. The expressions for circular concentric flux-surfaces are shown in appendix \ref{cha:app_circular}.

The correction of the toroidal precession frequency for circular surfaces with a Shafranov shift can be found in reference \cite{con83} for trapped particles and in reference \cite{zon07} for both passing and trapped particles. A more detailed study of the toroidal drift precession of passing particles can be found in \cite{kol03}. The influence of an anisotropic pressure equilibrium on the guiding center motion and the toroidal drift precession can be found in references \cite{coo06,juc08}.


\chapter{The ideal MHD Energy Principle}
\label{cha:MHD-Energy-Principle}

The aim of this chapter is the derivation of the ideal MHD energy principle \cite{ber58,gre68}. The derivation presented here follows the one from J. P. Freidberg \cite{fre82,fre87}. In the first section the ideal MHD model which considers  the plasma as an ideally conducting magnetized fluid is introduced and its conditions of validity are discussed. The ideal MHD energy principle which deals with the perturbations of the ideal MHD equilibrium is then derived in section \ref{sec:ideal-mhd-pert}.

\section{MHD theory}
\label{sec:mhd-theory}

MHD theory considers the plasma as a magnetized fluid. It is based on the equations ruling quantities relied to the different moments of the distribution function of particles in velocity space. The equations themselves are then obtained by taking moments of the Fokker-Planck equation. The equation ruling a given moment of $F_s$ will involve the moment of the next higher order such that the set of equations is infinite unless a closure relation is chosen.

\subsection{2-fluid MHD}

The case of a plasma with electrons and a single kind of ions is assumed. The ions are singly-charged such that $e_i = - e_e = e$. For each species, the density $n_s$, the mean velocity $\vc{v}_s$ and the kinetic pressure tensor $\vc{P}_s$ are defined by
\begin{equation}
  \left\{n_s, n_s \vc{v}_s,\vc{P}_s\right\}(\vc{x},t) = \iiint \left\{1,\vc{u},m_s (\vc{u} - \vc{v}_s)\otimes(\vc{u} - \vc{v}_s)\right\} F_s(\vc{x},\vc{u},t)\dd^3 \vc{u}
\end{equation}
The scalar pressure $p_s$ is the isotropic part of the pressure tensor, $p_s = (\vb{I} : \vb{P}_s)/3$ and the temperature for each kind of particles is then defined by $p_s = n_s k_B T_s$.

The Vlasov equation for each kind of particles expresses that in the absence of collisions the number of particles along the phase-space flow is conserved. When collisions are included, it becomes the Fokker-Planck equation, namely
\begin{equation}
  \pdd{F_s}{t} + \vc{u} \cdot \pdd{F_s}{\vc{x}} + \frac{e_s}{m_s}\left(\vc{E} + \vc{u}\times\vc{B}\right) \cdot \pdd{F_s}{\vc{u}} = \sum_{s'} C(F_s,F_s'),
\end{equation}
where the right-hand side term is the collision term which contains collisions with particles of the same kind as well as with particles of other kind.

The zero-th moment of the Fokker-Planck equation is the continuity equation and it can be expressed in a conservative form
\begin{equation}
  \label{eq:continuity_s}
  \pdd{n_s}{t} + \nabla \cdot (n_s \vc{v}_s) = 0
\end{equation}
since it is assumed that the different collision types conserve the particles in number and kind. The first moment is the momentum conservation equation
\begin{equation}
  \label{eq:momentum_s}
  m_s n_s \left(\ddr{\vc{v}_s}{t}\right)_s - e_s n_s \left(\vc{E} + \vc{v}_s \times \vc{B}\right) + \nabla p_s + \nabla \cdot \vb{\Pi}_s = \vc{R}_s
\end{equation}
where $\vb{\Pi}_s = \vb{P}_s - p_s \vb{I}$ is the anisotropic part of the pressure tensor and $\vc{R}_s$ is the mean momentum transfer due to collisions (since there can be no net transfer of momentum between particles of the same kind, this term comes only from electron-ion collisions). Finally the second moment equation is the expression of the conservation of energy, and can be written as \cite{fre87}
\begin{equation}
  \label{eq:energy_s}
  \frac{3}{2}n_s \left(\ddr{T_s}{t}\right)_s + \vb{P}_s : \nabla \vc{v}_s + \nabla \cdot \vc{h}_s = Q_s,
\end{equation}
where $h_s$ is a third-order moment of the distribution function and describes the heat flux due to random motion ($\vc{u} - \vc{v}_s$), and $Q_s$ is the heat generated by collisions between particles of different kinds. Note that in these equations the following notation has been used
\begin{equation}
  \left(\ddr{X}{t}\right)_s = \pdd{X}{t} + \vc{v}_s \cdot \nabla X
\end{equation}
which is a derivative along the mean flow of particles of type $s$.

Equations (\ref{eq:continuity_s},\ref{eq:momentum_s},\ref{eq:energy_s}) are then coupled via Maxwell's equations
\begin{align}
  \nabla \times \vc{E} &= - \pdd{\vc{B}}{t}, \\
  \nabla \times \vc{B} &= \mu_0 e (n_i \vc{v}_i - n_e \vc{v}_e) + \frac{1}{c^2} \pdd{\vc{E}}{t}, \\
  \nabla \cdot \vc{B} &= 0, \\
  \nabla \cdot \vc{E} &=  \frac{e}{\varepsilon_0} (n_i - n_e).
\end{align}

At this point, there is still more unknowns than equations since a closure relation has not been chosen, this will be done later. The next section introduces assumptions which leads to the derivation of a new set of equations which describe the evolution of the plasma as a single fluid.

\subsection{Single-fluid MHD}

The basic assumptions leading to single-fluid MHD equations are that the processes of interests are all low-frequency and long-wavelengths processes. The low-frequency, long-wavelength approximation allows to neglect in Maxwell's equation the displacement current $1/\varepsilon_0 \, \partial\vc{E}/\partial t$ as well as the charge separation $n_i - n_e$. Neglecting the displacement current is valid if the phase velocities of interest are small compared to the speed of light $\omega/k \ll c$ and if the same is true for thermal velocities $\sqrt{2 k_B T_{e,i}/m_{e,i}} \ll c$. Concerning the charge separation, the approximation is valid if are considered frequencies much smaller than the electron plasma frequency $\omega \ll \omega_{p,e} = \sqrt{n e^2/m_e \varepsilon_0}$ and length-scales much longer than the Debye length $1/k \gg \lambda_{D} = V_{T,e}/\omega_{p,e} = \sqrt{2 \varepsilon_0 k_B T_e/ n e^2}$. Note that the quasi-neutrality condition does not rule out the presence of an electric field but it implies that the electrostatic potential verifies $\Delta \Phi = 0$.

Additionally, it is assumed that due to the very low mass ratio $m_e/m_i$ the electron inertia can be neglected and that all terms proportional to $m_e$ can be set to $0$. This means that, on the time-scales of interest, the electrons are able to respond instantaneously. For this to be justified, the frequencies considered must be smaller than the electron plasma frequency $\omega_{p,e}$ and cyclotron frequency $\omega_{c,e} = e B /m_e$, and the length-scales must be much larger than the debye length $\lambda_D$ and the electron Larmor radius $\rho_e = m_e v_{T,e} / e B$. 

These two assumptions are usually met in typical fusion plasmas with one noticeable exception, the physics of the drift-waves are not reproduced when the electron inertia is neglected \cite{fre82}.

\subsubsection*{Single-fluid variables}

The particle density $n$ and the mass density $\rho$ are then defined by
\begin{equation}
  n = n_e = n_i, \quad
  \rho = m_e n_e + m_i n_i \simeq m_i n
\end{equation}
where the last equality is obtained by neglecting the electron mass. Another consequence of the very low mass ratio $m_e/m_i$ is that the momentum is mostly carried by ions such that the fluid mean velocity $\vc{v}$ is defined as
\begin{equation}
  \vc{v} = \vc{v}_i,
\end{equation}
The electron mean velocity is accounted for via the plasma current $\vc{J}$ defined by
\begin{equation}
  \label{eq:current}
  \vc{J} = e n \left(\vc{v}_i - \vc{v}_e\right).
\end{equation}
The plasma pressure $p$ is defined as the sum of the kinetic pressure of ions and of electrons. In fusion plasmas, both the temperature of ions and electrons are of the same order such that the two terms in the definition of $p$ are of equal importance,
\begin{equation}
  p = p_e + p_i.
\end{equation}

\subsubsection*{Single-fluid equations}

The single-fluid continuity equation is obtained from the ion continuity equation
\begin{equation}
  \label{eq:MHD_cont}
  \pdd{\rho}{t} + \nabla \cdot (\rho\vc{v}) = 0.
\end{equation}
Combining the electron and ion continuity equation, one then obtains
\begin{equation}
  \label{eq:MHD_charge}
  \nabla \cdot \vc{J} = 0,
\end{equation}
which is consistent with the conservation of charge coming from Maxwell's equation.

The momentum conservation equation is obtained by combining its ion and electron counterparts. This yields, noticing that $\vc{R}_e = -\vc{R}_i$ and neglecting electron inertia,
\begin{equation}
  \label{eq:MHD_momentum}
  \rho \ddr{\vc{v}}{t} - \vc{J} \times \vc{B} + \nabla p = - \nabla \cdot \left(\vc{\Pi}_i + \vc{\Pi}_e\right)
\end{equation}
where 
\begin{equation}
  \ddr{X}{t} = \pdd{X}{t} + \vc{v} \cdot \nabla X.
\end{equation}
The electron momentum conservation equation can be rewritten in the form 
\begin{equation}
  \label{eq:MHD_OhmsLaw}
  \vc{E} + \vc{v} \times \vc{B} = \frac{1}{e n} \left(\vc{J} \times \vc{B} - \nabla p_e - \nabla \cdot \vc{\Pi}_e + \vc{R}_e - \frac{m_e}{e}\left(\ddr{\vc{v_e}}{t}\right)_e\right)
\end{equation}
and is usually referred to as \emph{Ohm's law}. The first term on the right-hand side is Hall's term, next comes the effects of electronic pressure and viscous tensor, $\vc{R}_e$ is dominated by the electrical resistivity effects such that $ \vc{R}_e / e n \simeq \eta \vc{J}$, finally the last term in equation \eqref{eq:MHD_OhmsLaw} corresponds to electron inertia.

The energy conservation equations can be rewritten
\begin{align}
  \ddr{}{t} \left( \frac{p_i}{\rho^\gamma}\right) &= \frac{2}{3\rho^\gamma}\left(Q_i - \nabla \cdot \vc{h}_i - \vc{\Pi}_i : \nabla \vc{v} \right), \label{eq:MHD_energy_i}\\
  \ddr{}{t} \left( \frac{p_e}{\rho^\gamma}\right) &= \frac{2}{3\rho^\gamma}\left[ q_e - \nabla \cdot \vc{h}_e - \vc{\Pi}_e : \nabla \left(\vc{v} - \frac{1}{e n}\vc{J}\right) \right] + \frac{1}{e n}\vc{J} \cdot \nabla \left(\frac{p_e}{\rho^\gamma}\right), \label{eq:MHD_energy_e}
\end{align}
with $\gamma = 5/3$.

Finally, the form of Maxwell's equation in the low-frequency approximation are recalled,
\begin{equation*}
  \nabla \times \vc{E} = - \pdd{\vc{B}}{t}, \quad
  \nabla \times \vc{B} = \mu_0 \vc{J}, \quad
  \nabla \cdot \vc{B} = 0, \quad
  \nabla \cdot \vc{E} = 0.
\end{equation*}

\subsubsection*{Fluid closure}

The closure chosen here is a very common one when considering fluid theories. The main assumption being that the evolution of the distribution functions is dominated by collisions. In this manner, one then expects the collisions to enforce a Maxwellian form on the distribution functions, such that both distribution functions are well-described by their first 3 moments, or equivalently the 3 quantities $n_s,\vc{v}_s,T_s$. With this assumption one is then able to evaluate higher order moments (like $\vc{h}_s$) as a function of those three quantities and in terms of different transport coefficients, see for example Braginskii \cite{bra58}.

If one then evaluates the right hand sides of equations (\ref{eq:MHD_momentum},\ref{eq:MHD_OhmsLaw},\ref{eq:MHD_energy_i},\ref{eq:MHD_energy_e}), it appears that they can all be neglected under certain assumptions. 

\subsection{Ideal MHD}

The model described by the equations obtained is called the \emph{Ideal MHD} model.
\begin{align}
  \pdd{\rho}{t} + \nabla \cdot (\rho \vc{v}) = 0, \label{eq:mass}\\
  \rho \ddr{\vc{v}}{t} - \vc{J}\times \vc{B} + \nabla p= 0,  \label{eq:momt}\\
  \ddr{p}{t} = - \gamma p \nabla \cdot \vc{v},  \label{eq:engy}\\
  \vc{E} + \vc{v} \times \vc{B} = 0,  \label{eq:OhmL}\\
  \nabla \times \vc{B} = \mu_0 \vc{J},  \label{eq:MAmp}\\
  \nabla \cdot \vc{B} = 0, \label{eq:MFlx}\\
  \nabla \times \vc{E} = - \pdd{\vc{B}}{t}. \label{eq:MFar}
\end{align}
Note that equation \eqref{eq:engy} is equivalent to $ \dd/\dd{t}(p/\rho^{\gamma}) = 0$, which is the equation of state for perfect gases, with $\gamma = 5/3$ the ratio of the specific heats.

The details for going from equations (\ref{eq:MHD_momentum},\ref{eq:MHD_OhmsLaw},\ref{eq:MHD_energy_i},\ref{eq:MHD_energy_e}) to equations (\ref{eq:momt},\ref{eq:engy},\ref{eq:OhmL}) are out of the scope of this thesis and can be found in \cite{fre82}. It is however important to mention the conditions for the validity of this derivation, they are listed below \cite{fre82}
\begin{itemize}
\item High collisionality $\omega \tau_{ii} \ll 1$. This is crucial to obtain that the evolution of the distribution functions of ions and electrons are dominated by collisions. This condition is very restrictive.
\item Low resistivity $\omega \mu_0/a^2\eta \gg 1$. This allows us to neglect the effects of resistivity in Ohm's law. In the approximation $\eta = 0$ one consequence is the \emph{frozen-in law}, i.e.\ the magnetic flux is convected by the plasma flow, thus preventing any reconnection of magnetic field-lines.
\item Small gyroradius $\rho_* \ll 1$ and $k_\perp \rho_i \ll 1$.   
\end{itemize}

\begin{figure}[!ht]
  \centering
  \includegraphics*[height=\figheight]{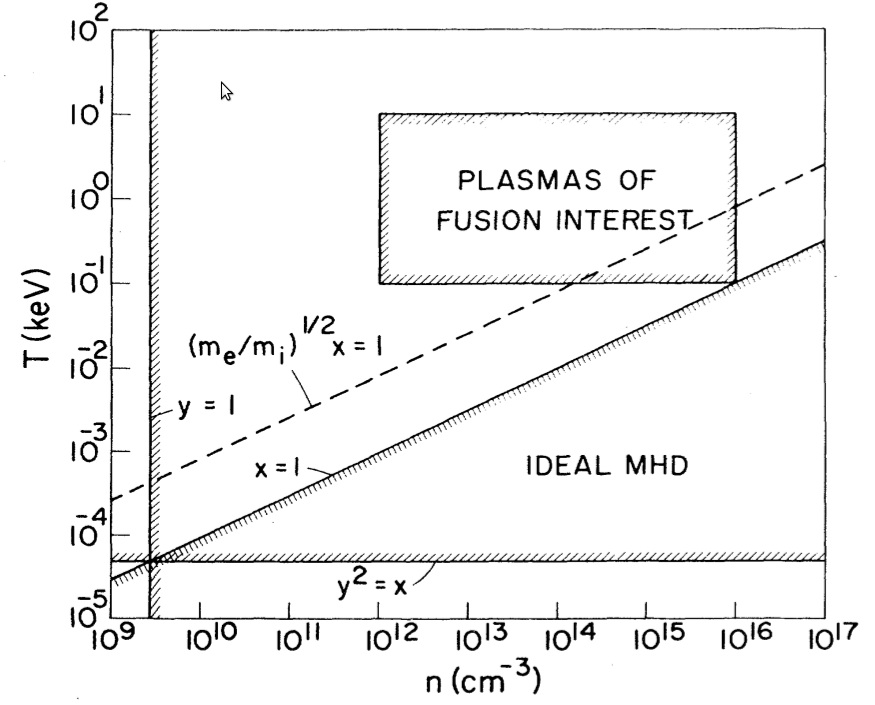}
  \caption[Region of validity of the ideal MHD model]{Region of validity of the ideal MHD model in $(n,T)$ parameter space for fixed $\beta = 0.05$ and $a = 1 \mathrm{m}$ (figure taken from \cite{fre82}).}
  \label{fig:Freidberg_IdealMHD_val}
\end{figure}

These conditions are usually not met in the range of parameters found in plasmas of a typical tokamak (see figure \ref{fig:Freidberg_IdealMHD_val}). 
However the predictions obtained using the ideal MHD model are very often in good agreement with what is observed in the experiments.

\subsection{Collisionless MHD}
\label{sec:collisionless-mhd}

The condition of a collision dominated plasma is released in the collisionless MHD model which can be derived from the drift-kinetic theory \cite{fre82,fre87}. In this model equations (\ref{eq:mass},\ref{eq:momt},\ref{eq:engy}) are replaced by the following equations.
\begin{align}
  &\ddr{\rho}{t} = 0, \label{eq:mass_perp}\\
  \rho &\ddr{\vc{v}_\perp}{t} - \vc{J} \times \vc{B} + \nabla p = 0,  \label{eq:momt_perp}\\
  &\ddr{p}{t} = 0,  \label{eq:engy_perp}
\end{align}
and the parallel momentum equation is replaced by the condition $\nabla \cdot \vc{v} = 0$ so that the plasma flow is incompressible. The conditions of validity of this model are usually met in tokamak plasmas. It is important to mention that the derivation of this model uses the fact that the $\vc{B}\cdot \nabla$ operator is invertible which is generally the case but has some notable exceptions. This point will be discussed later.

\vspace{2mm}

The reader should note that in reference \cite{ede92} Edery et al. show that the potential energy of the ideal MHD energy principle which will be derived in the next section can be recovered from collisionless kinetic theory. The starting point of their analysis was Vlasov's equation without collisions and the conditions for the ideal MHD limit were a low frequency such that no resonance between particles and MHD waves is possible and vanishing parallel components of the perturbed electric field and magnetic field. This analysis was later confirmed in \cite{ngu08,nguPhD} and is reproduced in chapter \ref{cha:FDR_derivation}. This points out that the region of validity of the ideal MHD model, at least in the linear regime, can be extended to low-collisionality plasmas such as fusion plasmas.

\section{Ideal MHD perturbation theory}
\label{sec:ideal-mhd-pert}

\subsection{Linearization near equilibrium}

Considering a static equilibrium, characterized by a current distribution $\vc{J}_0$, a magnetic field $\vc{B}_0$, a pressure $p_0$ (and a density $\rho_0$). Because the equilibrium is static, $\vc{v}_0 = 0$ is set. The relations between $\vc{J}_0$, $\vc{B}_0$ and $p_0$ are 
\begin{align}
  \vc{J}_0 \times \vc{B}_0 = \nabla p_0, \\
  \nabla \times \vc{B}_0 = \vc{J}_0, \\
  \nabla \cdot \vc{B}_0 = 0.
\end{align}
Note that $\vc{E}_0 = 0$ and that $\rho_0$ is not constrained.

Next, let us consider a perturbation of this equilibrium. For each variable $Z$ let $Z = Z_0 + Z_1$ where $Z_0$ is the equilibrium value of $Z$ and $Z_1$ is the perturbed part of $Z$. If the perturbation is small enough, they can be described by linearizing the ideal MHD set of equations.

Keeping only first order terms, the following equations stand.
\begin{align}
  \pdd{\rho_1}{t} + \nabla \cdot (\rho_0 \vc{v}_1) = 0, \label{eq:Lmass}\\
  \rho_0 \pdd{\vc{v}_1}{t} - \vc{J}_1\times \vc{B}_0 - \vc{J}_0\times \vc{B}_1 + \nabla p_1 = 0,  \label{eq:Lmomt}\\
  \pdd{p_1}{t} + \vc{v}_1 \cdot \nabla p_0 = - \gamma p_0 \nabla \cdot \vc{v}_1,  \label{eq:Lengy}\\
  \vc{E}_1 + \vc{v}_1 \times \vc{B}_0 = 0,  \label{eq:LOhmL}\\
  \nabla \times \vc{B}_1 = \mu_0 \vc{J}_1,  \label{eq:LMAmp}\\
  \nabla \cdot \vc{B}_1 = 0,  \label{eq:LMFlx}\\
  \nabla \times \vc{E}_1 = - \pdd{\vc{B}_1}{t}, \label{eq:LMFar}
\end{align}
It is then possible to express all perturbed quantities in terms of the perturbed velocity $\vc{v}_1$ and more precisely to an integral of $\vc{v}_1$, $\vb{\xi}$ defined by $\vc{v}_1 = \partial \vb{\xi}/\partial t$ and called the plasma displacement. At $t=0$, all perturbed quantities are set to $0$ (and $\vb{\xi}$ as well).

Ohm's Law \eqref{eq:LOhmL}, Ampere's Law \eqref{eq:LMAmp} and Faraday's Law \eqref{eq:LMFar} can be combined to express $\vc{E}_1$, $\vc{J}_1$ and $\vc{B}_1$ in terms of $\vb{\xi}$. Conservation of mass \eqref{eq:Lmass} will give the expression of $\rho_1$ and conservation of energy \eqref{eq:Lengy} the expression of $p_1$. 
\begin{multline*}
  \rho_1 = - \nabla \cdot (\rho_0 \vb{\xi}), \quad
  p_1 = - \gamma p_0 (\nabla \cdot \xi) - \xi \cdot \nabla p_0, \\
  \vc{E}_1 = - \pdd{\vb{\xi}}{t} \times \vc{B}_0, \quad
  \vc{B}_1 = \nabla \times \left(\vb{\xi} \times \vc{B}_0\right), \quad
  \vc{J}_1 = \mu_0^{-1} \nabla \times \left(\nabla \times \left(\vb{\xi} \times \vc{B}_0\right)\right).
\end{multline*}

The conservation of momentum \eqref{eq:Lmomt}, then gives the equation for the evolution of $\xi$. If one adopts the following notation $\vc{Q} = \vc{B}_1$ (which will be used throughout the entire chapter) and if the subscript $0$ for equilibrium quantities is dropped, one then has:
\begin{equation}
  \rho \pmdd{\vb{\xi}}{t}{2} = \mu_0^{-1} \left(\nabla \times \vc{Q}\right) \times \vc{B} + \mu_0^{-1} \vc{Q} \times \left(\nabla \times \vc{B}\right) + \nabla \left(\gamma p \nabla\cdot\vb{\xi} - \xi \cdot \nabla p\right) = \vc{F}(\vb{\xi})
\end{equation}
$\vc{F}(\vb{\xi})$ is known as the force operator.

\subsection{The Energy principle}

\subsubsection*{An MHD variational formalism}

The force operator $\vc{F}$ has the interesting property of being self-adjoint \cite{ber58,gre68}. It means that for all fields $\vb{\xi}$ and $\vb{\eta}$, the following holds,
\begin{equation*}
\iiint \vb{\eta}^* \cdot \vc{F}(\vb{\xi}) \mathrm{d}\vc{x} = \iiint \vb{\xi}^* \cdot \vc{F}(\vb{\eta}) \mathrm{d}\vc{x}.
\end{equation*}
One important consequence of this is that the exponential stability of the equilibrium can be expressed in a variational form (see \cite{fre87} for a proof). If one looks at perturbations with a time-dependence corresponding to a single Fourier mode of frequency $\omega$: $\vb{\xi}(\vc{x},t) = \vb{\xi}(\vc{x})\exp(-i \omega t)$, and if one defines
\begin{equation}
  \delta W(\vb{\xi}^*,\vb{\eta}) = - \frac{1}{2} \iiint \vb{\xi}^* \cdot \vc{F}(\vb{\eta})\dd\vc{x}
\end{equation}
and 
\begin{equation}
  K(\vb{\xi}^*,\vb{\eta}) = - \frac{\omega^2}{2} \iiint \rho\, (\vb{\xi}^* \cdot \vb{\eta})\dd\vc{x},
\end{equation}
then the eigen-modes of the force operator are the displacements $\vb{\xi}$ which extremize the quantity $E(\vb{\xi}^*,\vb{\eta}) = \delta W(\vb{\xi}^*,\vb{\eta}) + K(\vb{\xi}^*,\vb{\eta})$ at fixed $\omega^2$. The eigenvalue $\omega^2$ is then obtained by setting $E(\vb{\xi}^*,\vb{\xi}) = 0$. Since $\vc{F}$ is self-adjoint, all its eigenvalues are real. This means that an eigenvalue can either correspond to a solution oscillating at the frequency $\omega = \sqrt{\omega^2}$ (if $\omega^2 > 0$) or an exponentially growing solution with a growth rate $\gamma = i \sqrt{-\omega^2}$ (if $\omega^2 < 0$).

A weaker form of this property is called the Energy Principle and states that an equilibrium is stable if and only if $\delta W(\vb{\xi}^*,\vb{\xi}) \geq 0$ for all allowable displacements ($\vb{\xi}$ bounded in energy and satisfying appropriate boundary conditions).

\subsubsection*{Plasma boundary and vacuum contributions}

The plasma volume $P$ is defined as the region where $p$ is non-vanishing. The plasma is supposed to be surrounded by some vacuum region $V$ which itself is bounded by the wall of the machine. The plasma-vacuum boundary is noted $S$ and the vector $\vc{n}$ is the normal vector to the surface $S$.

Since later only internal modes will be considered, that is modes that have a vanishing displacement outside of the plasma, it can be interesting to distinguish in $\delta W$ the contributions coming from integrals over the plasma from those over the vacuum or from surface terms.

Appropriate boundary conditions can be derived from equations (\ref{eq:mass}-~\ref{eq:MFar}), after some algebra it is found that $\delta W$ can be decomposed into \cite{fre82}
\begin{equation}
  \delta W = \delta W_P + \delta W_S + \delta W_V
  \label{eq:EnPrinc}
\end{equation}
with 
\begin{align}
  \delta W_P &= \frac{1}{2} \int_P \left(\frac{\left|\vc{Q}\right|^2}{\mu_0} + \gamma p \left|\nabla \cdot \vb{\xi}\right|^2- \vb{\xi}_\perp^* \cdot \left(\vc{J} \times \vc{Q}\right) + \left(\vb{\xi}_\perp \cdot \nabla p\right) \nabla \cdot \vb{\xi}_\perp^*\right) \mathrm{d}\tau, \label{eq:dWp} \\
  \delta W_S &= \frac{1}{2} \int_S | \vc{n} \cdot \vb{\xi}_\perp |^2 \left(\vc{n} \cdot \left\llbracket \nabla \left(p+\frac{B^2}{2 \mu_0}\right)\right\rrbracket \right) \mathrm{d}S, \label{eq:dWs}\\
  \delta W_V &= \frac{1}{2} \int_V \frac{|\vc{Q}|^2}{\mu_0} \mathrm{d}\tau. \label{eq:dWv}
\end{align}
Note that for any vector quantity $\vc{X}$ the parallel and perpendicular components have been defined as $\vc{X} = X_\Vert \vc{b} + \vc{X}_\perp$ with $\vc{b}$ the unit vector in the direction of the equilibrium magnetic field and $\vc{b}\cdot \vc{X}_\perp = 0$. The double brackets $\left\llbracket X \right\rrbracket$ indicate the jump of the quantity $X$ at the plasma boundary.

\subsection{Underneath the formula}

To obtain an expression of the Energy Principle where each term will have a simple physical meaning, the expression of $\delta W_P$ can be further modified, by separating the parallel and perpendicular components of $\vc{Q}$ and $\vc{J}$.

The parallel component of the perturbed magnetic field $Q_\Vert$ can be put in the form 
\begin{equation}
  Q_\Vert = -B \left(\nabla \cdot \vb{\xi}_\perp + 2 \vb{\xi}_\perp \cdot \kappa \right) + \frac{\mu_0}{B} \vb{\xi}_\perp \cdot \nabla p,
  \label{eq:Qpar}
\end{equation}
while the perpendicular current $\vc{J}_\perp$ is 
\begin{equation}
  \vc{J}_\perp = \frac{\vc{b} \times \nabla p}{B}.
  \label{eq:Jperp}
\end{equation}

This leads to the following expression for $\delta W_P$:
\begin{multline}
  \delta W_P = \frac{1}{2} \int_P \left( \frac{|\vc{Q}_\perp|^2}{\mu_0} + \frac{B^2}{\mu_0}\left|\nabla \cdot \vb{\xi}_\perp + 2 \vb{\xi}_\perp \cdot \vb{\kappa}\right|^2 - \left(\vb{\xi}_\perp\cdot \nabla p\right) \left(2 \vb{\xi}_\perp^* \cdot \vb{\kappa}\right) \right. \\
  \left. - J_\Vert \left(\vb{\xi}_\perp \times \vc{b} \right) \cdot \vc{Q}_\perp + \gamma p \left| \nabla \cdot \xi \right|^2 \vphantom{\frac{|\vc{Q}_\perp|^2}{\mu_0}} \right) \mathrm{d}\tau
  \label{eq:dWp-int}
\end{multline}

The first term, proportional to $\left|\vc{Q}_\perp \right|^2$ represents the energy associated with the bending of magnetic field lines and it is the dominant term for the shear Alfv\'en wave. The second term represents the energy associated with the compression of the magnetic field and is dominant for the compressional Alfv\'en wave. The last term $\gamma p \left| \nabla \cdot \xi \right|^2$ represents the energy associated with the compression of the plasma, it is dominant for the sound wave. Those three terms are always positive and therefore stabilizing. The remaining two terms have indefinite sign and are the ones that drive the instabilities, one is proportional to the pressure gradient and will be associated with \emph{pressure-driven} modes, the other is proportional to $J_\Vert$ and is associted with \emph{current-driven} modes.

\subsection{The Collisionless MHD energy principle}
\label{sec:coll-mhd-en-princ}

The ideal MHD energy principle can be adapted to the collisionless MHD set of equations. Since $\nabla \cdot \vc{\xi} = 0$, $\delta W(\vb{\xi},\vb{\xi}^*)$ is independent of the parallel component of the MHD displacement. Also, since the momentum equation \eqref{eq:momt_perp} does not include parallel inertia, the kinetic energy $K(\vb{\xi},\vb{\xi}^*)$ is replaced by $K(\vb{\xi}_\perp,\vb{\xi}_\perp^*)$. Consequently for incompressible modes, the stability boundaries are the same for the two models but the growth rates predicted by the collisionless MHD model are bigger than those predict ed by the ideal MHD model.

\section{Summary}

After introducing the ideal MHD model in section \ref{sec:mhd-theory}, the ideal MHD energy principle initially introduced by Bernstein et al. \cite{ber58} and which expresses the exponential stability of the ideal MHD equilibrium has been derived in section \ref{sec:ideal-mhd-pert} following Freidberg \cite{fre82}. It will  be used in the next chapter to study one particular instability called the internal kink mode.

\chapter{The Internal kink mode}
\label{cha:internal-kink-mode}

The MHD energy principle presented in the previous chapter is used to study a particular instability located in the plasma core called the internal kink mode. The method presented here to obtain the dispersion relation of the internal kink mode in sections \ref{sec:high-aspect-ratio} to \ref{sec:dispersion-relation} reproduces the one found in De Blank et al. \cite{deb91} for high aspect ratio equilibria with circular flux-surfaces. The results are comparable to those of Bussac et al. \cite{bus75} or Hastie et al. \cite{has87}. The resistive modification of the internal kink stability has been calculated by Coppie et al. \cite{cop66} and is treated in section \ref{sec:res_disp_rel} following Ara et al. \cite{ara78}. Finally section \ref{sec:bifluid_disp_rel} deals with the modification of the dispersion relation by diamagnetic effects and follows the analysis of \cite{ara78}.

\section{High aspect ratio equilibria with shifted surfaces}
\label{sec:high-aspect-ratio}

In the case of a low-beta equilibrium with a high aspect ratio, the shape of the flux surfaces is circular and the toroidal magnetic field amplitude is almost inversely proportional to the distance to the vertical axis such that one has $\psi_t \simeq B_0 \tilde{r}^2/2$ where $\tilde{r}$ is the geometrical radius of the flux-surfaces and $B_0$ is the magnitude of the magnetic field on the magnetic axis.

One can then define a radial coordinate $r$ by the following equation
\begin{equation}
  \label{eq:def_r_DeB}
  r^2(\psi_p) = 2 R_0 \int_0^{\psi_p} d\psi \frac{q(\psi)}{B_\varphi(\psi)},
\end{equation}
or the opposite if this quantity is negative. The system of coordinates $(r,\theta,\varphi)$ is a set of flux coordinates (see chapter \ref{cha:Magn_config}) with $\varphi$ orthogonal to the other two coordinates. From the definition of $q$, one has
\begin{equation*}
  q = \frac{B^\varphi}{B^\theta} = \frac{B_\varphi}{R^2} \mathcal{J}_{\psi_p,\theta,\varphi}
\end{equation*}
such that the jacobian of the $(r,\theta,\varphi)$ set of coordinates is
\begin{equation}
  \label{eq:Jac_DeB}
  \mathcal{J}_{r,\theta,\varphi} = \frac{r R^2}{R_0}.
\end{equation}

In the next sections one will make use of the notation $F = B_\varphi$ which is a function of $r$ only.

\subsection{Metric tensor}

As pointed out before, for a low inverse aspect ratio equilibrium the flux-surfaces are circular but the centers of the flux-surfaces are shifted and the value of the shift depends on the flux-label. This shift is called the Shafranov shift.

One defines $\varepsilon$ as the ratio of the plasma minor radius $a$ to the plasma major radius $R_0$. The following ordering stands, the minor radii $\tilde{r}$ of the flux-surfaces is of order $\Oe{}$ compared to $R_0$, the shift of the flux surfaces $\Delta$ is of order $\Oe{2}$ compared to $R_0$, the deviation from circular flux-surfaces (like ellipticity or triangularity) are of higher order and are therefore neglected here. Finally the parameter $\beta$ is of order $\Oe{2}$ while the ratio of the poloidal field to the toroidal field is of order $\Oe{}$.

The approximate expressions for the components of the metric tensor as functions of $(r,\theta)$ can then be obtained 
\begin{align}
  R^2 = g_{\varphi\varphi} &= R_0^2\left(1-2\frac{r}{R_0}\cos \theta -2 \frac{\Delta}{R_0} - \frac{r \Delta'}{R_0} - \frac{1}{2}\frac{r^2}{R_0^2} + \Oe{3}\right), \\
  \nabla r \cdot \nabla r = g^{rr} &= 1 - 2 \Delta' \cos \theta + \frac{\Delta}{R_0} + \frac{\Delta'^2}{2} + \frac{3}{4}\frac{r^2}{R_0^2}+ \Oe{3}, \\
  \nabla \theta \cdot \nabla \theta = g^{\theta\theta} &= \frac{1}{r^2}\left(1+2\left(\Delta'+\frac{r}{R_0}\right)\cos\theta + \frac{1}{2} \left(r \Delta'' + \Delta' + \frac{r}{R_0}\right)^2 + \ldots \right. \\
& \hspace{4cm} \left. \frac{3}{2} \left(\Delta' + \frac{r}{R_0}\right)^2 + \frac{\Delta}{R_0} + \frac{1}{4}\frac{r^2}{R_0^2}\Oe{3}\right), \\
  \nabla r \cdot \nabla \theta = g^{r\theta} &= \frac{1}{r}\left(\left(r\Delta'' + \Delta' + \frac{r}{R_0}\right)\sin \theta + \Oe{3}\right),
\end{align}
where prime indicates a derivative against $r$. It is important to note that the radial coordinate $r$ is different from the minor radii of the flux surfaces $\tilde{r}$ (the difference is of order $\Oe{2}$) and $\theta$ is not the geometrical poloidal angle.

\subsection{Solution to the Grad-Shafranov equation}

The $\mm = 0$ component of the Grad-Shafranov equation is
\begin{equation}
  \label{eq:GS_m0_solution}
  F F' + \langle R^2 \rangle \,\mu_0 p' + \frac{F}{q R_0^2}\bigg\langle \frac{r^2 F}{q} g^{rr}\bigg\rangle' = 0
\end{equation}
If one then injects the expressions for the metric tensor elements obtained in the previous section to obtain an approximate solution, one obtains the following conclusions. The lowest order term is the $FF'$ term and all other terms are smaller by a factor of order $\Oe{2}$. Then $F$ can be written $F = F_0 + \Oe{2}$ where $F_0 = B_0 R_0$. Equation \eqref{eq:GS_m0_solution} then yields an approximate expression for $F'$ correct to order $\Oe{4}$,
\begin{equation}
  \label{eq:GS_m0_solution_eps}
  F'  = - \frac{R_0^2}{F_0} \mu_0 p'- \frac{F_0}{q R_0^2} \left(\frac{r^2}{q}\right)'.
\end{equation}

If one then integrates the $\mm=1$ component of the Grad-Shafranov equation, one obtains an expression for the approximate dependence of $\Delta$ over $r$,
\begin{equation}
  \label{eq:Delta_eps}
  \Delta'(r) = \frac{q^2 r }{R_0} \left(\beta_p(r) + \hat{s}(r) + \frac{1}{4}\right),
\end{equation}
with the following definitions of the quantities $\beta_p$ and $\hat{s}$,
\begin{align}
  \beta_p(r) &= -\frac{2 \mu_0 R_0^4}{F_0^2 r^4} \int_0^r \dd{r}\, r^2 p', \\
  \hat{s}(r) &= \frac{1}{r^4} \int_0^r \dd{r}\, r^3\left(\frac{1}{q^2} - 1\right).
\end{align}

\section{Preliminary steps}
\label{sec:preliminary-steps}

\subsection{The perturbed parallel flow}
\label{sec:pert-parall-flow}

The MHD displacement vector $\vb{\xi}$ is decomposed in the following way $\vb{\xi} = \vb{\xi}_p + \alpha \vc{B}$ with $\vb{\xi}_p \cdot \nabla \varphi = 0$ such that $\alpha = (\vb{\xi} \cdot \nabla \varphi)/B^\varphi$ and $\vb{\xi}_p$ has no contravariant component in $\varphi$. The other two contravariant components of $\vb{\xi}_p$ are defined as $\vb{\xi}_p = F_0/F (\xi\, \vc{e}_r + (\chi/r)\, \vc{e}_\theta)$, such that one has
\begin{equation}
  \label{eq:displ_DeB}
  \vb{\xi} = \frac{F_0}{F} \left(\xi\, \vc{e}_r + \frac{\chi}{r}\, \vc{e}_\theta\right) + \alpha \vc{B}.
\end{equation}

Starting from equation \ref{eq:dWp},
\begin{equation}
  \delta W_P = \frac{1}{2} \int_P \left(\frac{\left|Q\right|^2}{\mu_0} + \gamma p \left|\nabla \cdot \vb{\xi}\right|^2 - \vb{\xi}_\perp^* \cdot \left(\vc{J} \times \vc{Q}\right) + \left(\vb{\xi}_\perp \cdot \nabla p\right) \nabla \cdot \vb{\xi}_\perp^*\right) \mathrm{d}\tau.
\end{equation}
One then writes
\begin{equation}
  \delta W_p = \overline{\delta W} + \overline{\overline{\delta W}},
\end{equation}
with
\begin{align}
  \overline{\overline{\delta W}} &= \frac{1}{2} \int_P \gamma p \left|\nabla \cdot \vb{\xi}\right|^2 \mathrm{d}\tau, \\
  \overline{\delta W} &= \frac{1}{2} \int_P \left(\frac{\left|\vc{Q}\right|^2}{\mu_0} - \vb{\xi}_p^* \cdot \left(\vc{J} \times \vc{Q}\right) + \left(\vb{\xi}_p \cdot \nabla p\right) \nabla \cdot \vb{\xi}_p^*\right) \mathrm{d}\tau.
\end{align}
Since $\vc{Q} = \nabla \times (\vb{\xi} \times \vc{B}) = \nabla \times (\vb{\xi}_p \times \vc{B})$, all the dependencies over $\alpha$ are contained in $\overline{\overline{\delta W}}$. Furthermore if both the frequency and the growth rate of the mode are small compared to the Alfvén frequency (and the ratio is of the order of the inverse aspect ratio $\varepsilon$), then the kinetic energy and $\overline{\overline{\delta W}}$ contain terms that are only of order $\Oe{4}$ compared to the leading order terms of $\overline{\delta W}$. This means that the minimization of $E(\vb{\xi},\vb{\xi}^*)$ is carried out by first minimizing $\overline{\delta W}$ to order $\Oe{4}$. The minimization will be carried out by neglecting all terms of order $\Oe{6}$ and higher.

\subsection{A new expression for \texorpdfstring{$\overline{\delta W}$}{dW1}}
\label{sec:new-expr-texorpdfstr}

Here the mode is supposed to be composed of a single toroidal wave number such that the components of $\vb{\xi}_p$ can be put in the form$\{\xi,\chi\}(r,\theta,\varphi) = \{\xi,\chi\}(r,\theta) e^{-i \nn \varphi}$. $F' + \mu_0 (R^2/F) p'$ is expressed in terms of the metric tensor elements through the Grad-Shafranov equation,
\begin{align*}
  F' + \mu_0 (R^2/F) p' &= -\frac{R^2}{F} \ddr{\psi}{r} \left( \nabla \cdot \frac{\nabla \psi}{R^2}\right) \\
  &= -\frac{1}{q R_0^2} \left[\pdd{}{r}\left(\frac{r^2 F}{q} g^{rr}\right) + \frac{r^2 F}{q} \pdd{g^{r\theta}}{\theta}\right]
\end{align*}
with $g^{xy} = \nabla x \cdot \nabla y$. It can then be shown that $\overline{\delta W}$ can be written \cite{deb91}
\begin{multline}
\overline{\delta W} = \pi \frac{F_0^2}{\mu_0 R_0} \int_P r \dd{r}\dd{\theta} \left\{  \frac{1}{R_0^2}\left|r\left(\frac{1}{q}\pdd{\xi}{\theta} - i\nn\xi\right)\nabla \theta + \left(\frac{1}{q} \pdd{(r\xi)}{r} + i\nn \chi \right) \nabla r\right|^2 + \ldots \right. \\ \left. \frac{1}{r^2}\left|\pdd{(r\xi)}{r} + \pdd{\chi}{\theta}\right|^2  - \frac{F'}{rF}\left(\xi^*\left(\pdd{(r\xi)}{r} + \pdd{\chi}{\theta}\right) + c.c. \right) + \mu_0 p' \frac{\xi \xi^*}{F} \pdd{}{r}\left(\frac{R^2}{F}\right) + \ldots \right. \\ \left. -\frac{1}{q R_0^2}\left(\frac{1}{r^2 F} \pdd{(r^2 F g^{rr})}{r} + \pdd{g^{r\theta}}{\theta}\right)\left[ \frac{1}{q}\pdd{}{r}\left(r^2 \xi \xi^*\right) + r \left(i \nn \chi\xi^* +c.c.\right)\right] + \ldots \right. \\ \left. -\frac{1}{R_0^2}\left(\frac{1}{q}\right)'\frac{r^2 \xi \xi^*}{q} \left(\frac{1}{r^2 F} \pdd{(r^2 F g^{rr})}{r} + 2\pdd{g^{r\theta}}{\theta}\right) \right\}
\label{eq:dWbar_3}
\end{multline}
with prime indicating derivatives against $r$. Please note that the integration is now over $r$ and $\theta$ only.

\subsection{Magnetic field compressibility}
\label{sec:magn-field-compr}

The term of lowest order in equation (\ref{eq:dWbar_3}) is the term corresponding to the compressibility of magnetic field lines, it is noted $\overline{\delta W}_0$.
\begin{equation}
  \overline{\delta W}_0 = \pi \frac{F_0^2}{\mu_0 R_0} \int_P r \dd{r}\dd{\theta} \frac{1}{r^2}\left|\pdd{(r\xi)}{r} + \pdd{\chi}{\theta}\right|^2
\end{equation}
In comparison, other terms in $\overline{\delta W}$ are $\Oe{2}$ and are noted $\overline{\delta W}_2$. 

The quantity $\Xi$ is defined as 
\begin{equation}
  \Xi = \frac{1}{r}\left(\pdd{(r\xi)}{r} + \pdd{\chi}{\theta}\right).
\end{equation}
If one solves the Euler equations corresponding to the minimization of $\overline{\delta W}$, with the condition that $\xi$ and $\chi$ are vanishing at the plasma surface, one then finds that the quantity $\Xi$ is in fact only of order $\Oe{2}$ over the whole plasma. It then follows that the displacement $\vb{\xi}_p$ can be decomposed as $\vb{\xi}_p^{(0)} + \vb{\xi}_p^{(2)}$ where $\vb{\xi}_p^{(2)}$ is of order $\Oe{2}$ in comparison to $\vb{\xi}_p^{(0)}$ and $\vb{\xi}_p^{(0)}$ verifies
\begin{equation}
  \frac{1}{r}\left(\pdd{(r\xi^{(0)})}{r} + \pdd{\chi^{(0)}}{\theta}\right) = 0.
  \label{eq:xi_Bincomp}
\end{equation}
The decomposition is chosen such that the average over $\theta$ of $\chi^{(2)}$ vanishes, this leads to 
\begin{equation*}
  \frac{1}{r}\pdd{(r\xi^{(2)})}{r} = \langle\Xi\rangle, \qquad
  \frac{1}{r}\pdd{\chi^{(2)}}{\theta} = \Xi - \langle\Xi\rangle,
\end{equation*}
where the notation $\langle \mathcal{A} \rangle = \int \mathcal{A} \dd{\theta}/2\pi$ have been introduced.

As a result of this, it can be shown that the minimized expression for $\overline{\delta W}$ can be written
\begin{equation}
  \overline{\delta W}(\vb{\xi}_p,\vb{\xi}_p^{*}) = \overline{\delta W}_2(\vb{\xi}_p^{(0)},\vb{\xi}_p^{(0)*}) - \pi \frac{F_0^2}{\mu_0 R_0} \int r \dd{r}\dd{\theta} \left|\Xi\right|^2 + \Oe{6}.
\end{equation}

\subsection{Decomposition in poloidal Fourier harmonics}
\label{sec:decomp-polo-four}

Since the $\mm \neq 0$ components of the equilibrium are at least one order smaller in $\varepsilon$ than the $m=0$ component, an ordering of the poloidal harmonics of the MHD displacement is available. The notation $\langle \ldots \rangle = \int \ldots \dd{\theta}/2\pi$ is used. Writing $\xi^{(0)}$ and $\chi^{(0)}$ as
\begin{equation*}
  \xi^{(0)}(r,\theta) = \sum_\mm \xi_\mm e^{i \mm \theta},\ \chi^{(0)}(r,\theta) = \sum_\mm \chi_\mm e^{i \mm \theta}
\end{equation*}
equation (\ref{eq:xi_Bincomp}) is still verified such that for all $\mm$,
\begin{equation}
  \pdd{}{r}(r\xi_{\mm}) + i \mm \chi_\mm = 0.
  \label{eq:xi_Bincomp_mm}
\end{equation}

The expression for $\delta\tilde{W}$ defined as $\overline{\delta W} = 2\pi^2 F_0^2 \delta\tilde{W}/\mu_0 R_0 $ as a function of the $(\xi_\mm,\chi_\mm)$ is obtained.
\begin{multline}
  \delta\tilde{W}(\vb{\xi}_p,\vb{\xi}_p^{*}) = \sum_{k,l,\mm} \delta_{k-l,\mm} \int r \dd{r} \left\{\frac{1}{R_0^2}\left(\frac{k}{q} - \nn\right)\left(\frac{l}{q} - \nn\right) \left(\vphantom{\frac{l}{q}} \xi_k\xi_l^* r^2 \langle g^{\theta\theta}e^{i\mm\theta}\rangle + \ldots \right. \right. \\ 
\left. \left. \vphantom{\frac{l}{q}} + \chi_k\chi_l^* \langle g^{rr}e^{i\mm\theta}\rangle\right)  - \frac{r}{R_0^2}\left[\left(\frac{l}{q} - \nn\right)^2\xi_k \chi_l^* + \left(\frac{k}{q} - \nn\right)^2 \chi_k \xi_l^*\right]\langle g^{r\theta} e^{i\mm\theta}\rangle + \ldots \right. \\ 
\left. -\frac{1}{q R_0^2}\left(\frac{1}{r^2 F} \pdd{}{r}\left(\langle r^2 F g^{rr} e^{i \mm \theta}\rangle\right) - 2 i \mm \langle g^{r\theta} e^{i \mm \theta}\rangle\right)\left(\left(\frac{1}{q}\right)' r^2\xi_k \xi_l^* - i r \left(\frac{k}{q} - \nn\right) \chi_k \xi_l^*  + \ldots \right. \right. \\ 
\left. \left. + i r \left(\frac{l}{q} - \nn\right) \xi_k \chi_l^*\right) + \frac{\mu_0 p'}{F} \pdd{}{r}\left(\bigg\langle \frac{R^2}{F} e^{i \mm \theta} \bigg\rangle \right) \xi_k\xi_l^* \right\} \\ 
- \delta_{\mm,0} \int r \dd{r} \Xi_k \Xi_l^* + \Oe{6}
\label{eq:deltaWhat_Fourier}
\end{multline}

Equation (\ref{eq:deltaWhat_Fourier}) shows that the term in $\delta\tilde{W}$ coming from the coupling between the $k$ and $l$ components of the plasma displacement is proportional to the $\mm = k-l$ component of the metric tensor. The metric tensor has a dominant $\mm=0$ component and its $\mm = \pm 1$ components are $\Oe{}$ in comparison (if $|\mm| > 2$ they are even smaller). Therefore terms coming from the coupling of a given harmonic to its sidebands will be one order smaller in $\varepsilon$ than the term coming from the coupling to itself.

Terms with lowest order in the above expression are $\Oe{2}$ and one wishes to obtain an expression correct to order $\Oe{6}$. Therefore for the $\mm = 0$ component of the metric tensor, only terms up to $\Oe{2}$ are needed (there is no $\Oe{3}$ terms for the $\mm=0$ component). For $\mm = \pm 1$, $\Oe{}$ terms of the metric tensor are needed as well as $\Oe{3}$ terms if there are two adjacent harmonics of order $O(1)$.

\subsection{The case with a dominant poloidal harmonic}

If one makes the additional assumption that the MHD displacement is dominated by a single poloidal mode number $\mm$, a further ordering of $\delta\tilde{W}$ is possible. Because of the coupling terms due to toroidal geometry present in (\ref{eq:deltaWhat_Fourier}), other harmonics are also present.

In particular harmonics with poloidal mode number $(\mm \pm 1)$ are coupled to the main harmonic through a $\Oe{3}$ term while they are coupled to themselves through a term which is $\Oe{2}$, such that it is possible to find a solution minimizing $\delta\tilde{W}$ with the sidebands $(\mm \pm 1)$ being $\Oe{}$ in comparison to the main harmonic. This way the two terms governing the Euler equation for the sidebands are of the same order $\Oe{4}$. For harmonics $(\mm \pm l)$ with $l > 2$, the same argument could be made and this would result to terms only $\Oe{6}$ or higher for $\delta\tilde{W}$, which is beyond the desired accuracy so that one can neglect their influence on $\delta\tilde{W}$.

In summary, the potential energy $\delta\tilde{W}$ can be decomposed in the following sum
\begin{multline}
  \delta\tilde{W}(\vb{\xi}_p,\vb{\xi}_p^{*}) = \delta\tilde{W}^{(2)}(\vb{\xi}_{\mm},\vb{\xi}_{\mm}^{*}) + \delta\tilde{W}^{(4)}(\vb{\xi}_{\mm},\vb{\xi}_{\mm}^{*}) + \ldots \\ \delta\tilde{W}^{(2)}(\vb{\xi}_{\mm+1},\vb{\xi}_{\mm+1}^{*}) + \left(\delta\tilde{W}^{(3)}(\vb{\xi}_{\mm},\vb{\xi}_{\mm+1}^{*}) + c.c.\right) + \ldots \\ \delta\tilde{W}^{(2)}(\vb{\xi}_{\mm-1},\vb{\xi}_{\mm-1}^{*}) + \left(\delta\tilde{W}^{(3)}(\vb{\xi}_{\mm},\vb{\xi}_{\mm-1}^{*}) + c.c.\right) + \Oe{6}
\end{multline}
where $\delta\tilde{W}^{(i)}$ groups all $\Oe{i}$ terms of equation (\ref{eq:deltaWhat_Fourier}) (taking into consideration the different contributions of the metric tensor).

The lowest order term in this sum is $\delta\tilde{W}^{(2)}(\vb{\xi}_{\mm},\vb{\xi}_{\mm}^{*})$ which is an $\Oe{2}$ term. The following expression for $\delta\tilde{W}^{(2)}(\vb{\xi}_{\mm},\vb{\xi}_{\mm}^{*})$ is obtained,
\begin{equation}
  \delta\tilde{W}^{(2)}(\vb{\xi}_{\mm},\vb{\xi}_{\mm}^{*}) = \int r \dd{r} \left\{\frac{1}{R_0^2}\left(\frac{1}{q} - \frac{\nn}{\mm}\right)^2 \bigg((\mm^2 - 1)\xi_\mm\xi_\mm^* +r^2\xi_\mm'\xi_\mm'^* \bigg)\right\}
  \label{eq:dW2_m}
\end{equation}
Note that this expression is not valid if $\mm = 0$, since in this case equation (\ref{eq:xi_Bincomp_mm}) gives $\xi_0 = 0$ everywhere. Instead one has
\begin{equation}
\delta\tilde{W}^{(2)}(\vb{\xi}_{0},\vb{\xi}_{0}^{*}) = \int r \dd{r} \left\{\frac{\nn^2}{R_0^2} \chi_0\chi_0^*\right\}
\end{equation}

\subsection{The case of the \texorpdfstring{$\mm = 1$}{m=1} mode}

For $\mm = 1$, the main harmonic is $\vb{\xi}_1$ with sidebands $\vb{\xi}_0$ and $\vb{\xi}_2$ of order $\Oe{1}$ with respect to $\vb{\xi}_1$. Using equation (\ref{eq:xi_Bincomp_mm}) the $\vb{\xi}_1$ and $\vb{\xi}_2$ harmonics are described using $\xi_1$ and $\xi_2$ only, while $\vb{\xi}_0$ is described by $\chi_0$ only since $\xi_0 = 0$.

\subsubsection*{The \texorpdfstring{$\mm = 1$}{m=1} component}

The dominant term for $\delta\tilde{W}$ is then 
\begin{equation}
\delta\tilde{W}^{(2)}(\vb{\xi}_{1},\vb{\xi}_{1}^{*}) = \int \dd{r} \left\{\frac{r^3}{R_0^2}\left(\frac{1}{q} - \nn\right)^2 |\xi_1'|^2 \bigg)\right\},
\end{equation}
the equation for the minimization of $\delta\tilde{W}$ is then 
\begin{equation}
  \ddr{}{r} \left[\frac{r^3}{R_0^2}\left(\frac{1}{q} - \nn\right)^2 \xi_1'\right] = \Oe{2}.
  \label{eq:xi_FLBending}
\end{equation}
This equation implies that outside of regions where $(\nn q - 1) = \Oe{}$, the radial derivative of $\xi_1$ is a quantity of order $\Oe{2}$ such that $\xi_1$ is constant up to a quantity of order $\Oe{2}$. If $(\nn q - 1) = \Oe{}$, then other effects such as inertial effects can modify the structure of the mode.

\subsubsection*{The \texorpdfstring{$\mm = 0$}{m=0} component}

The structure of the $\mm=0$ component can be derived from expression (\ref{eq:deltaWhat_Fourier}). The Euler equation for $\chi_0$ is, to lowest order,
\begin{multline}
 \frac{\nn^2}{R_0^2} \chi_0 -\frac{1}{R_0^2}\left(\frac{1}{q} - \nn\right)i\nn\langle g^{rr}e^{i\theta}\rangle (r\xi_1)' + \frac{1}{q R_0^2} \frac{1}{rF} \langle r^2 F g^{rr} e^{i\theta}\rangle'(i\nn) \xi_1 + \ldots \\ 
\frac{r}{R_0^2} \langle g^{r\theta} e^{i \theta}\rangle \left(\frac{2}{q} - \nn\right) \nn\xi_1 = 0.
\label{eq:chi_0}
\end{multline}

If $\chi_0$ verifies equation (\ref{eq:chi_0}), then $\delta\tilde{W}^{(3)}(\xi_{1},\chi_{0}^{*}) = \delta\tilde{W}^{(3)}(\chi_{0},\xi_{1}^*) = - \delta\tilde{W}^{(2)}(\chi_{0},\chi_{0}^{*})$ such that the total of the $3$ terms where $\chi_0$ appears is 
\begin{equation}
\delta\tilde{W}^{(2)}(\chi_{0},\chi_{0}^{*}) + \delta\tilde{W}^{(3)}(\xi_{1},\chi_{0}^{*}) + \delta\tilde{W}^{(3)}(\chi_{0},\xi_{1}^*)= -\int r \dd{r} \left\{\frac{\nn^2}{R_0^2} \chi_0\chi_0^*\right\} + \Oe{6}.
\end{equation}

\subsection{Minimization against the parallel flow}
\label{sec:minim-against-parall}

This section will deal with the minimization of $K + \overline{\overline{\delta W}}$ to an expression of order $\Oe{4}$. The results of the previous section will be used , in particular the structure of the perpendicular flow which is characterized by a single toroidal mode number $\nn$, a dominant $\mm = 1$ harmonic with sidebands $\mm = 0,2$ of order $\Oe{1}$ with respect to the main harmonic.  equation \ref{eq:xi_Bincomp}, which expresses that to lowest order the perturbation does not compress magnetic field lines is recalled
\begin{equation*}
  \frac{1}{r}\left(\pdd{(r\xi)}{r} + \pdd{\chi}{\theta}\right) = R^2 \nabla \cdot \left(\frac{F}{R^2} \vb{\xi}_p\right) = \Oe{2},
\end{equation*}
as well as the expression for $K + \overline{\overline{\delta W}}$, which can be written
\begin{equation*}
  K + \overline{\overline{\delta W}} =  \frac{1}{2} \int_P \dd{\vc{x}} \left\{\gamma p \left|\nabla \cdot \vb{\xi}\right|^2 - \omega^2 \rho \left|\vb{\xi}\right|^2 \right\}
\end{equation*}
Introducing $\Gamma = -i \omega/\omega_A = -i \omega R_0^2\sqrt{\mu_0 \rho_0}/F_0$ ($\Gamma$ real and positive corresponds to a growing mode) and $\beta_c = \gamma \mu_0 p \rho_0 R_0^2/\rho F_0^2$, and considering a single toroidal mode number $\nn$, the Euler equation for $\alpha$ is then obtained
\begin{equation}
  \beta_c \frac{F^2}{R^4}\left(\frac{1}{q}\pdd{}{\theta} - i \nn \right)^2 \alpha  - \frac{\Gamma^2}{R_0^2}B^2 \alpha  = - \beta_c \frac{F}{R^2}\left(\frac{1}{q}\pdd{}{\theta} - i \nn \right) \left(\nabla \cdot \vb{\xi}_p\right) +\frac{\Gamma^2}{R_0^2} \left(\vc{B} \cdot \vb{\xi}_p\right)
  \label{eq:Euler_alpha}
\end{equation}

A Fourier analysis of the right-hand side of this equation is then performed to derive an ordering for $\alpha$. It turns out that the right hand side is dominated by the $\mm=0,2$ harmonics and the terms of next order have no $\mm=0,2$ harmonics. $\alpha$ is then written as
\begin{equation}
  \alpha = \alpha_0(r)  + \alpha_2(r) e^{2i\theta} + \Oe{}
\end{equation}
Equation (\ref{eq:Euler_alpha}) is then solved separately for the two harmonics.
\begin{align*}
  \left[\frac{\Gamma^2}{R_0^2} \frac{F_0^2}{R_0^2} + \beta_c \frac{F_0^2}{R_0^4} \nn^2\right] \alpha_0 &= \beta_c \frac{F_0}{R_0^2} (-i\nn) \frac{1}{R_0}\left(-r \xi_1' - 2 \xi_1\right) \\
  \left[\frac{\Gamma^2}{R_0^2} \frac{F_0^2}{R_0^2} + \beta_c \frac{F_0^2}{R_0^4} \left(\frac{2}{q} - \nn\right)^2\right] \alpha_2 &= \beta_c \frac{F_0}{R_0^2} i\left(\frac{2}{q}-\nn\right) \frac{1}{R_0}\left(r \xi_1 '\right)
\end{align*}
such that the solutions are
\begin{equation}
  \alpha_0 = \frac{i\nn}{\displaystyle\frac{\Gamma^2}{\beta_c} + \nn^2} \frac{R_0}{F_0} (r \xi_1' + 2 \xi_1), \qquad
  \alpha_2 = i\left(\frac{2}{q}-\nn\right)\frac{1}{\displaystyle\frac{\Gamma^2}{\beta_c} + \left(\frac{2}{q} - \nn\right)^2} \frac{R_0}{F_0} (r \xi_1').
\end{equation}

The corresponding expression for the sum $K + \overline{\overline{\delta W}}$ can be obtained after some algebra,
\begin{multline*}
  K + \overline{\overline{\delta W}} =  \frac{2\pi^2 F_0^2}{\mu_0 R_0} \int_P r \dd{r}\left\{\frac{\Gamma^2}{R_0^2} r^2 |\xi_1'|^2 \frac{\rho}{\rho_0} \left(1 + \frac{1}{\displaystyle\frac{\Gamma^2}{\beta_c} + \nn^2} + \frac{1}{\displaystyle\frac{\Gamma^2}{\beta_c} + \left(\frac{2}{q} - \nn\right)^2}\right) + \ldots \right. \\
\left.  - |\xi_1|^2  \ddr{}{r}\left(\frac{\Gamma^2}{R_0^2}\frac{\rho}{\rho_0} \left( 1 + \frac{2}{\displaystyle\frac{\Gamma^2}{\beta_c} + \nn^2}\right)\right)\right\} + \Oe{6}.
\end{multline*}
The second term is proportional to the gradient of $\rho$ and $\beta_c$ and can be neglected if these functions are supposed to be slowly varying functions of $r$. 

The quantity $M$, called the \emph{inertial enhancement factor}, is defined as
\begin{equation}
  \label{eq:defM}
 M =   \frac{\rho}{\rho_0} \left(1 + \frac{1}{\displaystyle\frac{\Gamma^2}{\beta_c} + \nn^2} + \frac{1}{\displaystyle\frac{\Gamma^2}{\beta_c} + \left(\frac{2}{q} - \nn\right)^2}\right),
\end{equation}
The quantity $\beta_c$ is of order $\Oe{2}$ and is proportional to the compressional energy. The assumption that $\Gamma^2$ is also a quantity of order $\Oe{2}$ has been made. Two limit cases are considered, if $\Gamma^2 \ll \beta_c$ the solution to equation \eqref{eq:Euler_alpha} verifies $\nabla \cdot \vb{\xi} = 0$ meaning that the plasma is incompressible, $M$ can then be approximated to
\begin{equation}
  M(r) = 1 + \frac{1}{\nn^2} + \frac{1}{\left(\frac{2}{q} - \nn\right)^2} \simeq 1+2q^2
\end{equation}
in the region where $q \simeq 1/\nn$, which gives $M \simeq 3$ for $\nn = 1$. Now if  $\beta_c \ll \Gamma^2$, the contribution of $\alpha$ to $K + \overline{\overline{\delta W}}$ is negligible and $M = 1$. In this case, the perturbed parallel flow vanishes and the predictions of the growth rate from the ideal MHD energy principle and the collisionless MHD energy principle match. Compared to the incompressible case, the growth rate will be larger by a factor $\sqrt{3}$. 

But these two models fail to reproduce correctly the ion inertia \cite{graPhD}. A drift-kinetic treatment of the thermal ions including kinetic contribution to inertia produces larger values of the inertial enhancement than the ideal MHD treatment \cite{graPhD,gra00}.

\section{The (\texorpdfstring{$\mm = 1, \nn = 1$}{m = 1, n=1}) internal kink mode}
\label{sec:m1_inter_kink_mode}

\subsubsection*{Total Energy}

It can be shown that in the case of $\nn = 1$ the total energy can be put in the form \cite{deb91}
\begin{multline}
  \tilde{E} = \int \frac{\dd{r}}{R_0^2} \left\{r^3\tilde{\Gamma}^2|\xi_1'|^2 + \frac{1}{2}\left(\frac{1}{q^2} - 1 \right)\frac{r^3}{R_0^2} |\xi_1|^2 - 3\left(\frac{1}{q} - \frac{1}{2}\right)^2 (r^2 |\xi_2|^2)' + \ldots \right. \\
  \left. r\left(\frac{1}{q} - \frac{1}{2}\right)^2\left|r \xi_2' + 3 \xi_2 + D \xi_1 - \Delta' r \left(\frac{1}{q^2} - 1 \right) \xi_1'\right|^2 + \ldots \right. \\
  \left. -\frac{3}{q}\left(\frac{1}{q} - 1 \right)\left(\xi_1^* \left[\left(\Delta' - \frac{r}{2 R_0}\right)r^2 \xi_2\right]' + c.c. + |\xi_1|^2 \left[r^2 \Delta'^2 - \frac{r^3}{R_0}\Delta'\right]'\right)\right\}
  \label{eq:Ehat_n1}
\end{multline}
where $D$ is defined by 
\begin{equation}
  \tilde{\Gamma}^2 = \left(\frac{1}{q} - 1\right)^2 + M \Gamma^2
\end{equation}
 $D$ is
\begin{equation}
  D = \left(r\Delta'' + 3 \Delta' - \frac{r}{R_0}\right)
\end{equation}
and $M$ is defined in the previous sections. The lowest order term is the first term and is $\Oe{2}$, all other terms are $\Oe{4}$.

The Euler equation for $\xi_1$ is then written
\begin{equation}
  \ddr{}{r} \left(\tilde{\Gamma}^2\xi_1' \right) = \Oe{2}
  \label{eq:Euler_FLB}
\end{equation}
such that $\tilde{\Gamma}^2\xi_1' = \Oe{2}$ over the whole minor radius. This equation implies equation \eqref{eq:xi_FLBending}.

\subsection{Structure of the mode}

\begin{figure}[!ht]
  \centering
  \includegraphics*[height = \figheight]{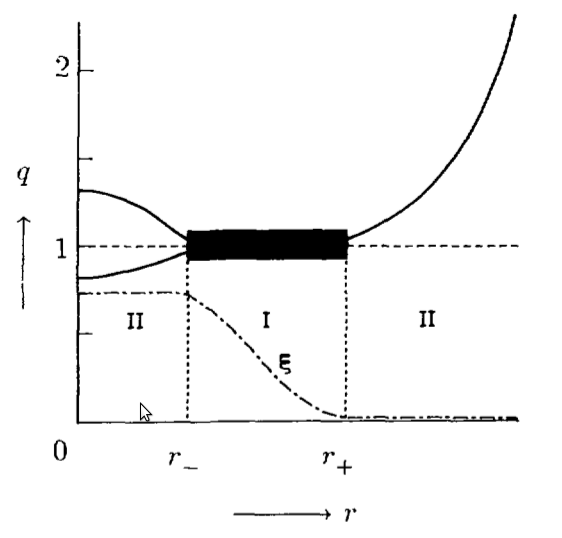}
  \caption[Sketches of the considered $q$ profiles]{Sketches of the considered $q$ profiles and of the radial displacement \cite{deb91}.}
  \label{fig:DeBlank_q_prof}
\end{figure}

Following \cite{deb91} the safety factor profile is assumed to be such that for $r \in [0,r_{-}]$ $q$ is not close to $1$ (but can be greater or lower than $1$), for $r \in [r_{-},r_+]$ $q-1 = \Oe{}$, and for $r \in [r_+,a]$ $q$ is greater than $1$ (see figure \ref{fig:DeBlank_q_prof} for an example).

\vspace{2mm}

The minimization of the total energy is then conducted separately in the three different regions of the plasmas.

\subsubsection*{Regions of constant \texorpdfstring{$\xi_1$}{xi}}

In the regions where $q-1 = O(1)$ (labeled $\mathrm{II}$ in figure \ref{fig:DeBlank_q_prof}), equation (\ref{eq:Euler_FLB}) implies $\xi_1' = \Oe{2}$ such that $\xi_1$ is almost constant on this region.

If one looks at the structure of the $\mm=2$ component of the mode then its structure is ruled by the following Euler equation 
\begin{equation}
  \ddr{}{r} \left(r^3 \left(\frac{1}{q} - \frac{1}{2}\right)^2\zeta' \right)= 3 r \left(\frac{1}{q} - \frac{1}{2}\right)^2 \zeta,
  \label{eq:Euler_zeta}
\end{equation}
with $\zeta$ defined as $\zeta = \xi_2 + (\Delta' + r/2R_0)\xi_c$  and $\xi_c$ is the approximate constant value of $\xi_1$ in this region. This equation is similar to the one obtained for the minimization of the MHD potential energy for a mode with a dominant $\mm=2$ component, see expression \eqref{eq:dW2_m}.

\subsubsection*{Regions where \texorpdfstring{$q\sim 1$}{q \~ 1}}

In the region $r \in [r_-,r_+]$ $q-1 = \Oe{}$ (labeled $\mathrm{I}$ in figure \ref{fig:DeBlank_q_prof}), and equation (\ref{eq:Euler_FLB}) does not impose the behavior of $\xi_1$ and contributions of inertia, field-line bending and magnetic field compressibility are all of the same order. The notation $q-1 = \varepsilon_q$ is adopted with $\varepsilon_q$ comparable to $\varepsilon$.

The equation for the evolution of $\xi_1$ can be obtained and integrated. It yields 
\begin{equation}
  \xi_1(r) = \xi_1(r_-) + c_1 \overline{S}_0(r)  + c_2 \overline{S}_1(r).
  \label{eq:xi_1_inert}
\end{equation}
with the integrals $\overline{S}_i$ defined as
\begin{equation}
\overline{S}_i(r) = \int_{r_{-}}^r \frac{\dd{x}}{x^3 \tilde{\Gamma}^2} \left[\frac{x^4}{R_0} \left(\beta_p + \hat{s} \right)\right]^i
\end{equation}
and $\overline{S}_i$ by $\overline{S}_i = \overline{S}_i(r_+)$. $c_1$ and $c_2$ are constants which characterize the solution in this region. The equation for $\xi_2$ is
\begin{equation}
  \left[r^3\xi_2 + \int_0^r x^2 D \dd{x} \xi_1 - c_2 r^4 - c_2 \overline{S}_2(r) - c_1 \overline{S}_1(r)\right]_{r_-}^r = 0.
  \label{eq:xi_2_inert}
\end{equation}

\subsubsection*{Intermediate layers}

These  intermediate layers are located in between the two previously cited regions, $q-1$ is neither $O(1)$ nor $\Oe{}$. In these layers $\xi_1'$ can be of order $O(1)$. Their width is typically of order $O(\varepsilon_q)$ such that the total contribution to $\tilde{E}$ is of order $O(\varepsilon^4 \varepsilon_q)$

For a layer located at $r_b$, the correction to $\xi_1$ is 
\begin{equation}
  \delta \xi_1 = c\int^{r_b}\frac{\dd{r}}{r^3 \tilde{\Gamma}^2} + O(\varepsilon\varepsilon_q)
  \label{eq:dxi_int_layer}
\end{equation}with $c$ being a constant.

\section{The dispersion relation}
\label{sec:dispersion-relation}

\subsection{Derivation}

In addition to the description of the $q$-profile of the previous section, it is assumed that a position $r_2$ where $q=2$ exists. Then from the previous section, the perturbation takes the following form \cite{deb91}
\begin{enumerate}
  \renewcommand{\theenumi}{\alph{enumi}}
  \itemsep0pt \parskip0pt \parsep0pt
\item $[0,r_-]$: $\xi_1(r) = \xi_c$ and $\xi_2(r) = \zeta_-(r) - (\Delta' + r/2R_0)\xi_c$ where $\zeta_-$ is the solution of (\ref{eq:Euler_zeta}) which is regular in $r=0$. This implies $\zeta_-(0) = 0$.
\item $[\ ,r_-]$: Correction $\delta \xi_1(r_-)$ to $\xi_1$ characterized by the constant $c_-$.
\item $[r_-,r_+]$: $\xi_1$ and $\xi_2$ are given by equations (\ref{eq:xi_1_inert}) and (\ref{eq:xi_2_inert}). They depend on $\xi_1(r_\pm)$ and $\xi_2(r_\pm)$, the constant $c_2$ is also expressed as a function of these four variables.
\item $[r_+,\ ]$: Correction $\delta \xi_1(r_+)$ to $\xi_1$ characterized by the constant $c_+$.
\item $[r_+,r_2]$: $\xi_1(r) = 0$ and $\xi_2(r) = \zeta_+(r)$ where $\zeta_+$ is the solution of (\ref{eq:Euler_zeta}) which is regular in $r=r_2$. This solution is characterized by $\zeta_+'(r_2) = 0$ but $\zeta_+(r_2)$ is not necessarily $0$. $\xi_2(r_+)$ determines the whole solution. 
\item $[r_2,b]$: $\xi_1(r) = 0$ and $\xi_2(r) = 0$. This solution is possible since equation (\ref{eq:Euler_zeta}) has a singular point in $r = r_2$. The value $\xi_2 = 0$ is chosen to minimize the total energy (this does not come from the Euler equation).
\end{enumerate}

The form of $\vb{\xi}$ is determined by the choice of 9 constants $\xi_c, \zeta_-(r_-)$, $c_-$, $\xi_1(r_-),\xi_2(r_-)$, $\xi_1(r_+),\xi_2(r_+)$, $c_+$ and $\zeta_+(r_+)$ (the constants $c_1$ and $c_2$ mentioned above are functions of $\xi_1(r_-),\xi_2(r_-)$, $\xi_1(r_+),\xi_2(r_+)$). These nine constants are bound by the continuity relations for $\xi_1$ and $\xi_2$ at $r_-$ and $r_+$, such that the number of independent variables is reduced to 5.

The final step to obtain the dispersion relation for the ($\mm=1, \nn=1$) internal kink mode is the minimization of the total energy against the remaining constants (4 of them at least, the last one which is $\xi_c$ in general is kept as a global scaling factor for the pertubation). This process is straightforward but elaborate and will not be shown here.

The dispersion relation is then obtained by setting $\tilde{E} R_0^2/|\xi_c|^2 = 0$, 
\begin{equation}
  \left(S_2 - \frac{S_1^2}{S_0} + A_- - A_+\right)^{-1}\left(\frac{S_1}{S_0} - \frac{3}{4} A_-\right)^2 + \frac{1}{2}\frac{r_-^4}{R_0^2} \hat{s}(r_-) - \frac{9}{16}\frac{A_-}{R_0^2} + \frac{1}{S_0} = 0,
\label{eq:KinkDispRel}
\end{equation}
the integrals $S_i$ have been defined as 
\begin{equation}
S_i = \int \frac{\dd{x}}{x^3 \tilde{\Gamma}^2} \left[\frac{x^4}{R_0} \left(\beta_p + \hat{s} \right)\right]^i
\end{equation}
where the integral covers both the singular layers and the $q \sim 1$ region. The quantities $A_\pm$ are linked to the solutions of equation \eqref{eq:Euler_zeta} by $A_\pm = r_\pm^4 (1 - a_\pm)/(3+a_\pm)$ and
\begin{equation}
  a_\pm = \left.4 \left(\frac{1}{q} - \frac{1}{2}\right)^2r\frac{\zeta'(r)}{\zeta(r)}\right|_{r = r_\pm}.
\end{equation}

\subsection{The thin singular layer case}

Let $r_{s} = |r_+ + r_-|/2$ and $w = |r_+ - r_-|/r_{s}$. If $w \ll 1$, the integrals $S_i$ can be expanded in $w$ and a simplified expression for the dispersion relation is found.
\begin{align*}
  S_2 - \frac{S_1^2}{S_0} &= O(w^3) \\
\frac{S_1}{S_0} &= \frac{r_{s}^4}{R_0} \left(\beta_p(r_{s}) + \hat{s}(r_{s}) \right) \left( 1 + O(w^2) \right)
\end{align*}

This gives the following expression for the dispersion relation  
\begin{equation}
  \left(A_- - A_+\right)^{-1}\left(\frac{r_{s}^4}{R_0} \left(\beta_p(r_{s}) + \hat{s}(r_{s}) \right) - \frac{3}{4} A_-\right)^2 + \frac{1}{2}\frac{r_{s}^4}{R_0^2} \hat{s}(r_{s}) - \frac{9}{16}\frac{A_-}{R_0^2} + \frac{1}{S_0} = 0,
\label{eq:KinkDispRel_thin}
\end{equation}
in the case of a thin singular layer ($w \ll 1$).

The major contribution to the integral $S_0$ will come from the $q\sim 1$ layer such that it can be evaluated in the following way
\begin{equation}
  S_0 = \frac{1}{r_{s}^2}\int_{-\infty}^{+\infty} \frac{\dd{y}}{\left(q^{-1}-1\right)^2 + M \Gamma^2}
\end{equation}
where $y = (r-r_{s})/r_{s}$. If $q(r_{s}) = 1$ and $q'(r_{s}) \neq 0$ then with $s = rq'/q$ the magnetic shear at $r_{s}$ one has 
\begin{equation}
  S_0 = \frac{\pi}{|s| \Gamma \sqrt{M} r_{s}^2}
\end{equation}
such that the dispersion relation can be written, remembering that $\Gamma = -i\omega/\omega_A$,
\begin{equation}
  \frac{\pi \,\delta \tilde{W} R_0^2}{r_{s}^2 |\xi_c|^2} - |s| \frac{i\omega}{\omega_A}\sqrt{M} = 0
  \label{eq:Kink_disprel_mon}
\end{equation}
where $\delta \tilde{W}$ has been defined by 
\begin{equation}
  \frac{\,\delta \tilde{W} R_0^2}{|\xi_c|^2} = \left(A_- - A_+\right)^{-1}\left(\frac{r_{s}^4}{R_0} \left(\beta_p(r_{s}) + \hat{s}(r_{s}) \right) - \frac{3}{4} A_-\right)^2 + \frac{1}{2}\frac{r_{s}^4}{R_0^2} \hat{s}(r_{s}) - \frac{9}{16}\frac{A_-}{R_0^2},
  \label{eq:deltaW_hastie}
\end{equation}
and corresponds to the minimized value of the MHD potential energy. This expression is equivalent to the one found in Bussac et al \cite{bus75} or Hastie et al \cite{has87}. If now $q(r_{s}) = 1 + \delta q$ with $|\delta q| \ll 1$ and $q'(r_{s}) = 0$ but $S^2 = r_{s}^2 q''(r_{s})/q(r_{s})^2$ is non-vanishing the following dispersion relation is obtained \cite{has87}
\begin{equation}
  \frac{\pi \,\delta \tilde{W} R_0^2}{r_{s}^2 |\xi_c|^2} + S \sqrt{\delta q^2 - M \frac{\omega^2}{\omega_A^2}} \sqrt{ \delta q^2 + \sqrt{\delta q^2 - M \frac{\omega^2}{\omega_A^2}}} = 0
  \label{eq:Kink_disprel_rev}
\end{equation}
The normalized potential energy $\dW$ is then defined by
\begin{equation}
  \label{eq:deltaWhat}
  \dW = \frac{\pi \,\delta \tilde{W} R_0^2}{r_{s}^2 |\xi_c|^2} = \frac{\pi \mu_0 \,\delta W }{2 \pi^2 B_0^2 R_0 r_{s}^2 |\xi_c|^2}.
\end{equation}

\vspace{2mm}

The value of $M$ depends on the considered model, for the ideal MHD model in the incompressible limit $M=3$, for the collisionless model $M=1$ and if one includes the kinetic effects of thermal ions one then has $M = 1 + (1.6/\sqrt{r_{s}/R_0} + 0.5)q^2$ \cite{graPhD,gra00}.

\section{Modification by resistivity}
\label{sec:res_disp_rel}

In the case of a thin singular layer, the dispersion relation \eqref{eq:KinkDispRel_thin} can also be obtained by separating the plasma in two regions, an MHD region where $q-1 = O(1)$ and an inertial region where $q-1$ is small. In the MHD region, the structure of the MHD displacement is obtained by minimizing only $\delta W$ (inertial effects are neglected). In the inertial region the structure of the mode obeys an Euler equation which expresses the competition between field-line bending and inertia. Both solutions are then asymptotically matched to obtain the dispersion relation. See for example Rosenbluth et al \cite{ros73} in the case of cylindrical geometry (note that in this case $M$ was set to 1 because parallel inertia was neglected).

The same approach can be used to study the influence of other effects. In ideal MHD, Ohm's Law is simply written $\vc{E} + \vc{v} \times \vc{B} = 0$, but according to the previous section, the $\vc{v} \times \vc{B}$ term vanishes at the inertial layer such that other terms that were neglected before can play an important role. This section deals with the addition of finite resistivity such that now $\vc{E} + \vc{v} \times \vc{B} = \eta \vc{J}$. The results presented here, although the formulas are derived in cylindrical geometry only, persist in toroidal geometry \cite{cop76,ara78}.

\subsection{Resistive equations}

It is assumed that the equilibrium perturbation has a single toroidal mode number $\nn$ and a single poloidal mode number $\mm$. This means that the perturbed part of any scalar quantity $\mathcal{A}$ can be written $\mathcal{A}_1(r,\theta,\varphi) = \mathcal{A}_1(r) \exp (i\mm\theta - i \nn \varphi)$.

\vspace{2mm}

Following Ara et al. \cite{ara78}, $2 \pi \Psi$ is the magnetic flux through the helical ribbon defined by the axis and the helix intersecting the point $(r,\theta,\varphi)$. This flux can be linked to the component of the vector potential in the direction of the helical perturbation by $\Psi = - \vc{A} \cdot \vc{e}_h$ with $\vc{e}_h = \vc{e}_\varphi + (\mm/\nn) \vc{e}_\theta$ and $\vc{e}_\varphi,\,\vc{e}_\theta$ are the vectors of the covariant basis.

The equation for the evolution of $\Psi$ is obtained by taking the dot product of Ohm's law and $\vc{e_h}$, namely
\begin{equation}
  \label{eq:OhmL_res}
  \vc{e}_h \cdot \left(\vc{E} + \vc{v} \times \vc{B} - \eta \vc{J}\right) = 0
\end{equation}
In cylindrical geometry (or in the high aspect ratio approximation), this yields the following equation for the evolution of $\Psi_h$
\begin{equation}
  \ddr{\Psi}{t} = \frac{\eta}{\mu_0} \left(2 \frac{\nn }{\mm} B_0 + \nabla_\perp^2 \Psi\right).
  \label{eq:Psi_res}
\end{equation}

\vspace{2mm}

If one assumes that the flow is incompressible and that the aspect ratio is high, then the velocity can be written $\vc{v} = \nabla U \times (R_0 \nabla \varphi)$ such that $U$ is the stream function of the flow $\vc{v}$. Taking the curl of the momentum conservation equation and projecting it on the toroidal direction, one obtains the equation for the evolution of $U$.
\begin{equation}
  \mu_0 \ddr{}{t}(\rho \, \nabla_\perp^2 U) = - \frac{\nabla \varphi}{R_0} \cdot (\nabla \Psi \times \nabla (\nabla_\perp^2 \Psi))
  \label{eq:U_res}
\end{equation}

Equations (\ref{eq:Psi_res},\ref{eq:U_res}) form a system of coupled differential equations ruling the evolution of the two variables $U$ and $\Psi$. 

\subsection{Equilibrium}

The linearization of equations \eqref{eq:Psi_res} and \eqref{eq:U_res} with the assumption that $\Psi$ has an equilibrium part $\Psi_{0}$ as well as a perturbed part $\Psi_{1}$, and that the equilibrium part of $U$ is null (since the calculation is valid only locally it can be done in the plasma rest frame where the equilibrium radial electric field is null). Then the equilibrium equations are 
\begin{align}
  &\nabla_\perp^2 \Psi_0 + 2\frac{\nn}{\mm} B_0 = 0, \\
  &\nabla \varphi \cdot (\nabla \Psi_0 \times \nabla (\nabla_\perp^2 \Psi_0)) = 0.
\end{align}
Because $\Psi_0$ depends only on $r$, the second equation is trivially verified. The definition of $\Psi$ allows us to obtain the expression for the derivative of the equilibrium flux $\Psi_0 = -\psi_p + (\nn/\mm) \psi_t$ giving with $\psi_{t,0} = B_0 r^2 / 2 $,
\begin{equation}
\Psi_0' = B_0 r\left(\frac{\nn}{\mm} - \frac{1}{q}\right).
\end{equation}

\subsection{Linearization}

The linearized equations for the perturbations are then obtained
\begin{align*}
  &\pdd{\Psi_1}{t} + (\nabla U_1 \times R_0 \nabla \varphi)\cdot \nabla \Psi_0 = \frac{\eta}{\mu_0} \nabla_\perp^2 \Psi_1, \\
  &\mu_0 \pdd{}{t}(\rho \, \nabla_\perp^2 U_1) = - \frac{\nabla \varphi}{R_0} \cdot \left( \nabla \Psi_0 \times \nabla (\nabla_\perp^2 \Psi_1) + \nabla \Psi_1 \times \nabla (\nabla_\perp^2 \Psi_0) \right).
\end{align*}
The perturbation is supposed to have the form $\Psi_1 = \Psi_1(r) \exp(\gamma t + i (\mm \theta - \nn \varphi))$ and  $U_1(r) = r \gamma \xi/i \mm$ (which gives $\vc{v}_1 = \gamma \xi \vc{e}_r - \gamma/ i\mm \cdot \partial/\partial r (r \xi) \vc{e}_\theta$ and this is consistent with the usual definition of the plasma displacement $\vc{v} = \partial \vb{\xi}/ \partial t$ with $\vc{v}\cdot \vc{e_z} = 0$ and $\nabla \cdot \vc{v} = 0$), one obtains
\begin{align}
  \label{eq:Psi_lin}
  &\Psi_1(r) +  \xi(r) \Psi_0' = \frac{\eta}{\gamma \mu_0} \nabla_\perp^2 \Psi_1,  \\
  \label{eq:U_lin}
  &\gamma^2 \rho \,\mu_0 \, \nabla_\perp^2 (r \xi) = \frac{\mm^2 \Psi_0'}{r R_0^2}\left(\nabla_\perp^2 \Psi_1 -\Psi_1 \nabla (\nabla_\perp^2 \Psi_0) \right),
\end{align}
where prime denotes derivation against $r$.

Outside the resistive layer ($q = \mm/\nn$ or $\Psi_0' = 0$), the effects of resistivity and inertia are negligible (these conditions correspond to $\gamma \, \tau_R \gg 1$ and $\gamma \, \tau_A \ll 1$ where $\tau_R,\tau_A$ are the resistive and Alfvén time), this yields the following solutions
\begin{align}
  \Psi_1 = -\xi\, \Psi_0', \\
  \nabla_\perp^2 \Psi_1 = \nabla (\nabla_\perp^2 \Psi_0) \Psi_1.
\end{align}

Inside the layer, the flux $\Psi_1$ is continuous but the derivative $\Psi_1'$ is allowed to change very rapidly $\nabla_\perp^2 \Psi_1 \simeq \Psi_1''$ so that the right-hand side of equation \eqref{eq:U_lin} is dominated by the $\nabla_\perp^2 \Psi_1$ term. Writing $r_{s}$ the location of the resonant layer, $s = r q'/q$ the magnetic shear at $r_{s}$ and $x = (r - r_{s})/r_{s}$ the derivative of $\Psi_0$ can be approximated by 
\begin{equation*}
\Psi_0' = B_0 r\left(\frac{\nn}{\mm} - \frac{1}{q}\right) \simeq B_0 r_{s} \frac{\nn}{\mm} s x,
\end{equation*}
one obtains the following equations inside the resonant layer (where now prime denotes derivative against $x$)
\begin{align}
  &\Psi_1 + \xi B_0 r_{s} \frac{\nn}{\mm} s x  = \frac{\eta}{\gamma \mu_0 r_{s}^2} {\Psi_1}'',  \label{eq:Psi_lin_layer}\\
  &\gamma^2 \rho \, \mu_0 \, r_{s} \xi'' = \mm \frac{B_0}{R_0^2} \nn  s x {\Psi_1}''.\label{eq:U_lin_layer}
\end{align}
Writing 
\[
\Psi_1 = B_0 r_{s}\frac{\nn}{\mm} s \, \psi, \quad 
\lambda = \gamma \frac{q R_0\sqrt{\rho \, \mu_0}}{B_0 \mm s} = \gamma \frac{\tau_A}{\mm s} = \gamma \tau_H, \quad
\tau_R = \frac{\mu_0\,r_{s}^2}{\eta}\ \mbox{and}\ 
\epsilon = \frac{\tau_H}{\tau_R},
\]
equations \eqref{eq:Psi_lin_layer} and \eqref{eq:U_lin_layer} become 
\begin{equation}
  \left\{
    \begin{array}{l}
      \displaystyle \psi = - x \xi + \frac{\epsilon}{\lambda}\mdd{\psi}{x}{2} \\
      \displaystyle \lambda^2 \mdd{\xi}{x}{2} = x \mdd{\psi}{x}{2}
    \end{array}
  \right.
  \label{eq:reduced_system}
\end{equation}

\subsection{Asymptotic matching}

The final step is to match asymptotically the solutions in the two regions. The complete steps of this matching are technical and are not presented here. In order to match the outer solution for the $(\mm=1,\, \nn = 1)$ mode, the inner solution must verify
\[
\left\{
  \begin{array}{lcl}
    \displaystyle \xi \sim \xi_c &\mbox{when} &x \rightarrow - \infty \\
    \displaystyle \xi \sim 0 &\mbox{when} &x \rightarrow + \infty \\
    \displaystyle \xi' \sim \frac{\xi_c}{\pi s^2 x^2} \dW &\mbox{when} &|x| \rightarrow + \infty
  \end{array}
\right.
\]
with $\dW$ the normalized MHD potential energy defined in \eqref{eq:deltaWhat}. 

The details of the asymptotic matching can be found in \cite{ara78} and are reproduced in appendix \ref{cha:asympt-match-resist}. It gives the following expression (with $\hat{\lambda} = \lambda /\epsilon^{1/3}$):
\begin{equation}
  -\frac{ \pi s^2 }{\dW} = \frac{\pi}{8 \, \epsilon^{1/3}} \hat{\lambda}^{5/4}  \frac{\Gamma((\hat{\lambda}^{3/2} - 1)/4)}{\Gamma((\hat{\lambda}^{3/2} + 5)/4)}
\end{equation}
This result is often expressed as
\begin{equation}
  \hat{\lambda} = \hat{\lambda}_H \frac{\hat{\lambda}^{9/4}\Gamma((\hat{\lambda}^{3/2} - 1)/4)}{8 \, \Gamma((\hat{\lambda}^{3/2} + 5)/4)},
  \label{eq:res_kink_disprel}
\end{equation}
with $\hat{\lambda}_H = \lambda_H/\epsilon^{1/3}$ and $\lambda_H = -\dW/s^2 )$. 

\subsection{Consequences}

In the ideal limit ($\tau_R \rightarrow 0$, $\epsilon \rightarrow 0$, $\hat{\lambda} \rightarrow + \infty$), using the Stirling formula for the Gamma functions,the ideal result is recovered
\begin{equation}
  \lambda = \lambda_H,
\end{equation}
which can be written as
\begin{equation}
  \gamma_I \tau_A s = - \dW,
  \label{eq:Id_kinkdisprel_Zonca}
\end{equation}
and which is identical to equation \eqref{eq:Kink_disprel_mon} in the incompressible case where $M \sim 1$.

\begin{figure}[!ht]
  \centering
  \includegraphics*[height = \figheight]{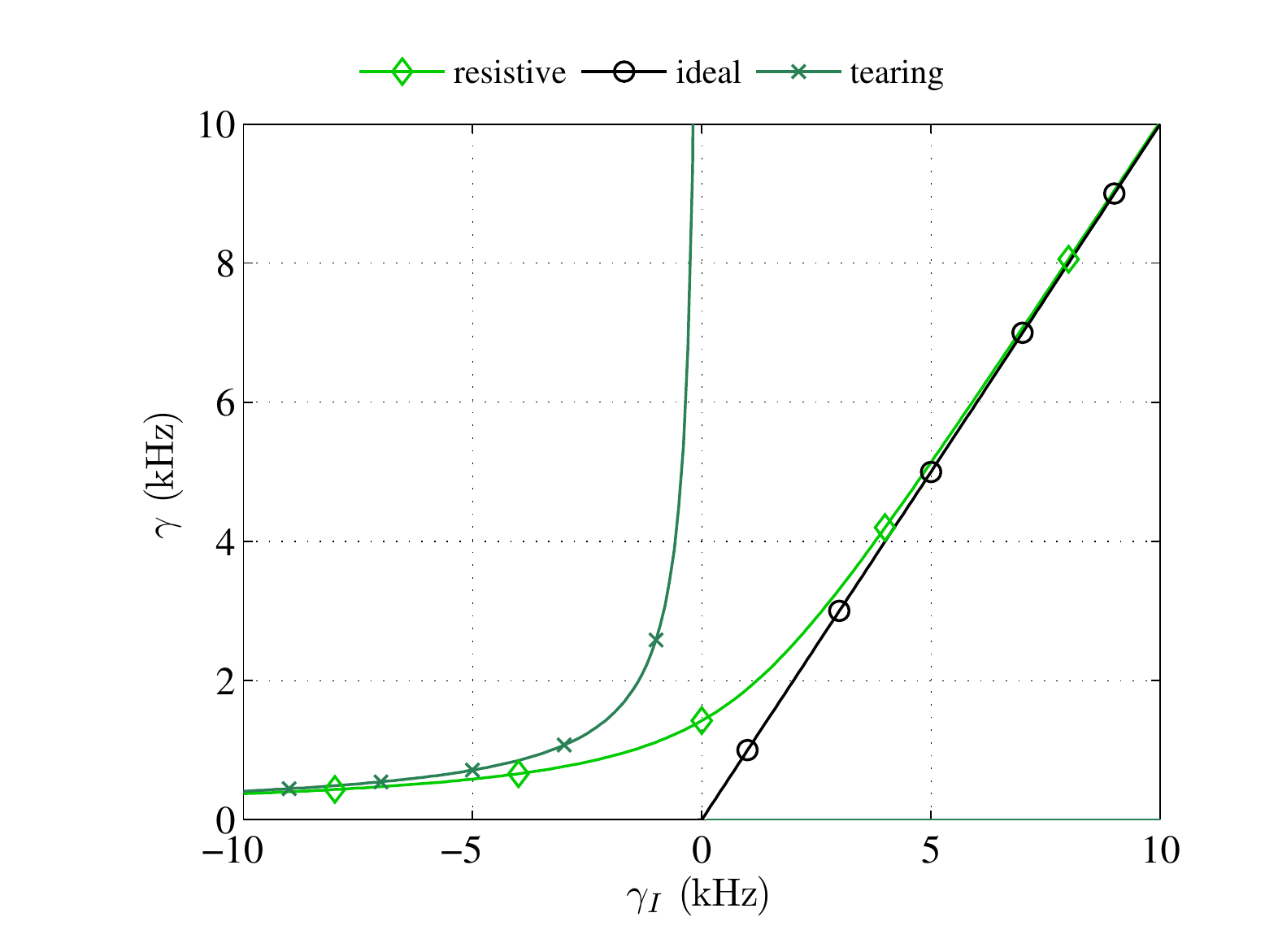}
  \caption[Internal kink growth rate with resistive effects]{Solution of the dispersion relation \eqref{eq:res_kink_disprel} in light green. The solution in the ideal limit ($\hat{\lambda}_H,\,\gamma_I \rightarrow + \infty$) is represented in black. The solution in the tearing limit ($\hat{\lambda}_H,\,\gamma_I \rightarrow - \infty$) is represented in dark green.}
  \label{fig:gamma_vs_gammaI_res}
\end{figure}

The main effect of resistivity is that the internal kink mode is always unstable. As can be seen on figure \ref{fig:gamma_vs_gammaI_res}, for all values of the ideal growth rate $\gamma_I$ (negative values of $\gamma_I$ correspond to positive values of $\delta W$) the growth rate is positive. At marginal stability in the ideal case $\delta W = 0$ ($\hat{\lambda}_H = 0$), the solution is $\hat{\lambda} = 1$. The resistive growth rate $\gamma_R$ is defined by $\hat{\lambda} = \gamma/\gamma_R$,
\begin{equation}
  \label{eq:res_gamma}
  \gamma_R = \left( \frac{s^2}{\tau_A^{2} \tau_R}\right)^{1/3} = \frac{1}{\tau_A }\left(\frac{s^2}{S}\right)^{1/3} 
\end{equation}
where the Lundquist number is defined as $S = \tau_R/\tau_A$.

In the regime where the ideal solution is stable ($\hat{\lambda}_H < 0$), the solution verifies $\hat{\lambda} < 1$ such that one can study the limit $\hat{\lambda} \ll 1$ of equation \eqref{eq:res_kink_disprel}. This gives
\begin{equation}
  \label{eq:res_tear_disprel}
  \gamma \frac{\tau_A}{s} \simeq \left(\frac{\sqrt{2}}{\pi} \Gamma\left(\frac{3}{4}\right)\right)^{-4/5} s^{-3/5} S^{-3/5}  \lambda_H^{-4/5},
\end{equation}
this solution is represented in dark green in figure \ref{fig:gamma_vs_gammaI_res}. This regime is often called the tearing regime, since in this case the shape of the radial displacement in the inertial layer matches the one from usual $|\mm| > 1$ tearing modes. Moreover the scaling $\gamma \propto S^{-3/5}$ is also recovered.

\section{Bi-fluid effects}
\label{sec:bifluid_disp_rel}

Furthermore bi-fluid effects can be considered and the following form of Ohm's Law used
\begin{equation}
  \label{eq:OhmsLaw_Ara_1}
  \vc{E} + \vc{v} \times \vc{B} = \frac{1}{e n} \left(\vc{J} \times \vc{B} - \nabla p_e + \vc{R}_e \right),
\end{equation}
which corresponds to equation \eqref{eq:MHD_OhmsLaw} where the contributions from electron inertia and from the electron viscous tensor have been neglected (this is valid if times much longer than the electron-ion collision time are considered). $\vc{R}_e$ can be approximated by $e n (\eta \vc{J}) - 0.71 n \nabla_\Vert T_e$ (see Braginskii \cite{bra58}). Equation \eqref{eq:OhmsLaw_Ara} then becomes 
\begin{equation}
  \label{eq:OhmsLaw_Ara}
  \vc{E} + \vc{v} \times \vc{B} = \eta \vc{J} + \frac{1}{e n} \left(\vc{J} \times \vc{B} - \nabla p_e  - 0.71 n \nabla_\Vert T_e \right).
\end{equation}
Another consequence of the inclusion of bi-fluid effects is the fact that the equilibrium ion and electron flows are dominated by the diamagnetic velocities $\vc{v}_{*s} = - \nabla p_s \times \vc{B}_0 / e_s n B_0^2$ (the $\vc{E} \times \vc{B}$ velocity is absent since in the plasma rest frame, the radial electric field is null). This will result in the introduction of the diamagnetic frequencies $\omega_{*s} = \vc{k} \cdot \vc{v}_{*s} \simeq (\mm/n e_s r B_0) \dd{p_s}/\dd{r}$.

\subsection{Bi-fluid layer equations}

The derivation of the linearized equations in this case is very similar to the purely resistive case (see \cite{ara78}). The ion momentum conservation equation is used to obtain an equation for the perturbed ion radial velocity (which is related to $\xi$ by $\vc{v}_{i,1}\cdot \vc{e}_r = -i(\omega - \wi)\xi$) and the generalised Ohm's Law to obtain an equation for the perturbed radial magnetic field (related to $\Psi_1$ by $i \mm \Psi_1 = r B_{1,r}$). The form of the equations obtained is the same as the one from section \ref{sec:res_disp_rel}.

\subsection{Bi-fluid dispersion relation}

The new dispersion relation is
\begin{equation}
  \left(\hat{\lambda}(\hat{\lambda} - i \hat{\lambda}_i)\right)^{1/2} = \hat{\lambda}_H \frac{\Lambda^{9/4}\Gamma((\Lambda^{3/2} - 1)/4)}{8 \, \Gamma((\Lambda^{3/2} + 5)/4)},
  \label{eq:BF_kinkdisprel}
\end{equation}
with $\Lambda = \left(\hat{\lambda}(\hat{\lambda} - i \hat{\lambda}_i)(\hat{\lambda} - i \hat{\lambda}_e)\right)^{1/3}$, $\hat{\lambda}_s = \lambda_s/\epsilon^{1/3}$, $\lambda = - i \omega' \tau_H$, $\lambda_i = - \omega_{*i} \tau_H$, $\lambda_e = - \hat{\omega}_{*e} \tau_H$, $\hat{\omega}_{*e} = \omega_{*e} + (0.71/e B_0 r) \, \dd{T_{e}}/\dd{r}$. If $\hat{\lambda}_i = \hat{\lambda}_e = 0$, equation \eqref{eq:BF_kinkdisprel} is equivalent to equation \eqref{eq:res_kink_disprel}.

\subsection{Solution properties}

In the ideal limit, letting $\epsilon \rightarrow 0, \, \hat{\lambda} \rightarrow + \infty$ one obtains
\begin{equation}
  \left(\lambda(\lambda - i \lambda_i)\right)^{1/2} = \lambda_H,
\end{equation}
which can be written as
\begin{equation}
  i \left(\omega (\omega-\omega_{*i})\right)^{1/2} = -\gamma_I,
  \label{eq:BF_kinkdisprel_id}
\end{equation}
the solutions to these equations is
\begin{equation}
  \omega = \frac{\wi}{2} \pm \sqrt{\frac{\wi^2}{4} - \gamma_I^2},
  \label{eq:BF_kinksol_id}
\end{equation}
 If $\gamma_I > \wi / 2$, the mode is unstable and the frequency is $\wi/2$, the mode growth rate is smaller than in the ideal case. If $- \wi/2 < \gamma_I < \wi/2$ there is two solutions both of them marginally stable, the first solution has a frequency between $0$ and $\wi/2$ depending on the value of $\gamma_I$, the second one between $\wi/2$ and $\wi$. In the ideal case, the bi-fluid effects are globally stabilizing and the mode rotates in the ion diamagnetic direction.

\begin{figure}[!ht]
  \centering
  \includegraphics*[height = \figheight]{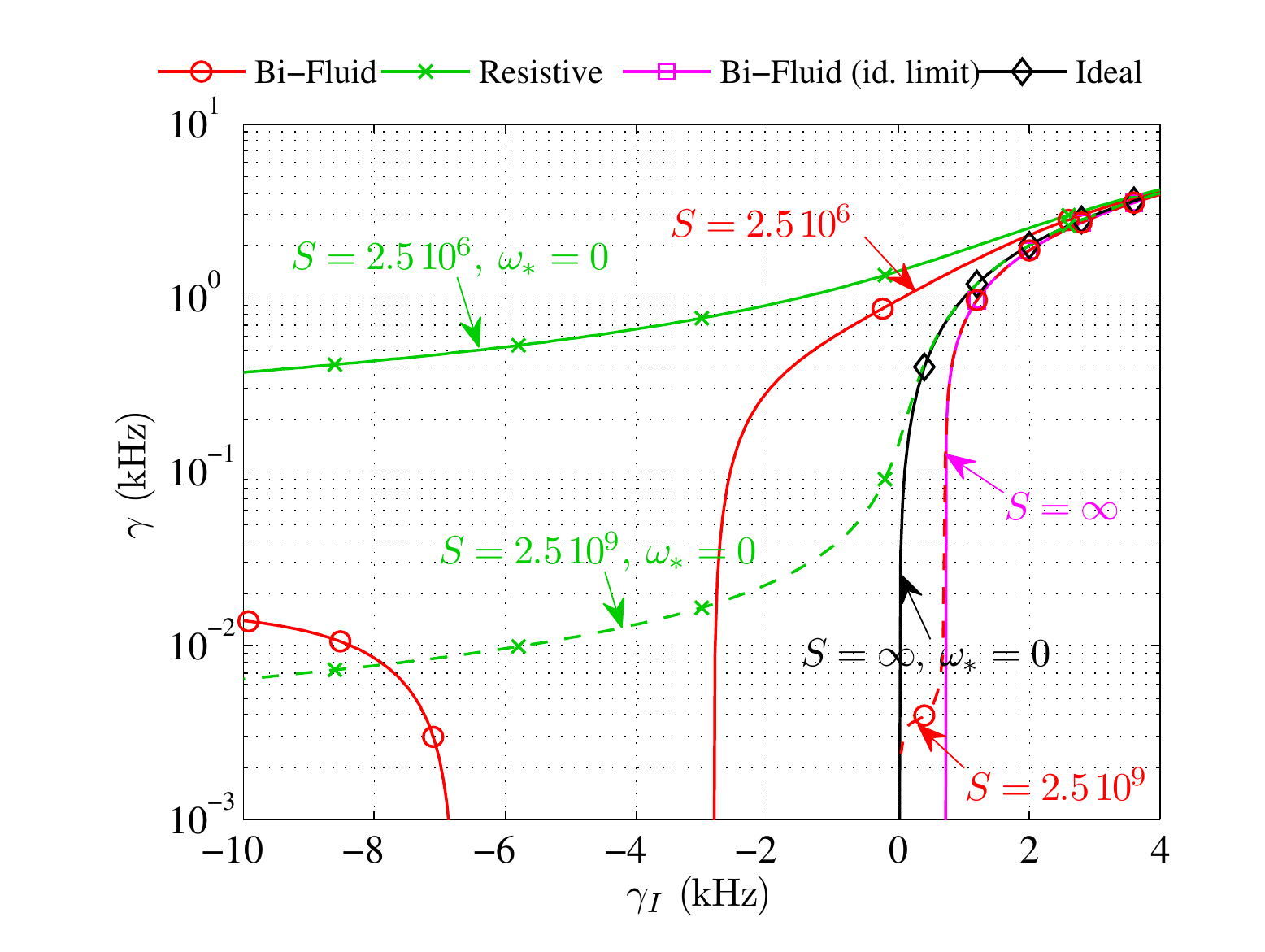}
  \caption[Internal kink growth rate with bi-fluid effects]{Growth rate of the solution of equation  \eqref{eq:BF_kinkdisprel} as a function of the ideal growth rate. The parameters are $\tau_A^{-1} = 0.56 \ \mathrm{kHz}$, $s = 0.2$, $\wi = - \hat{\omega}_{*e} = 1.42 \ \mathrm{kHz}$.}
  \label{fig:gamma_vs_gammaI_log_lab}
\end{figure}

\begin{figure}[!ht]
  \centering
  \includegraphics*[height = \figheight]{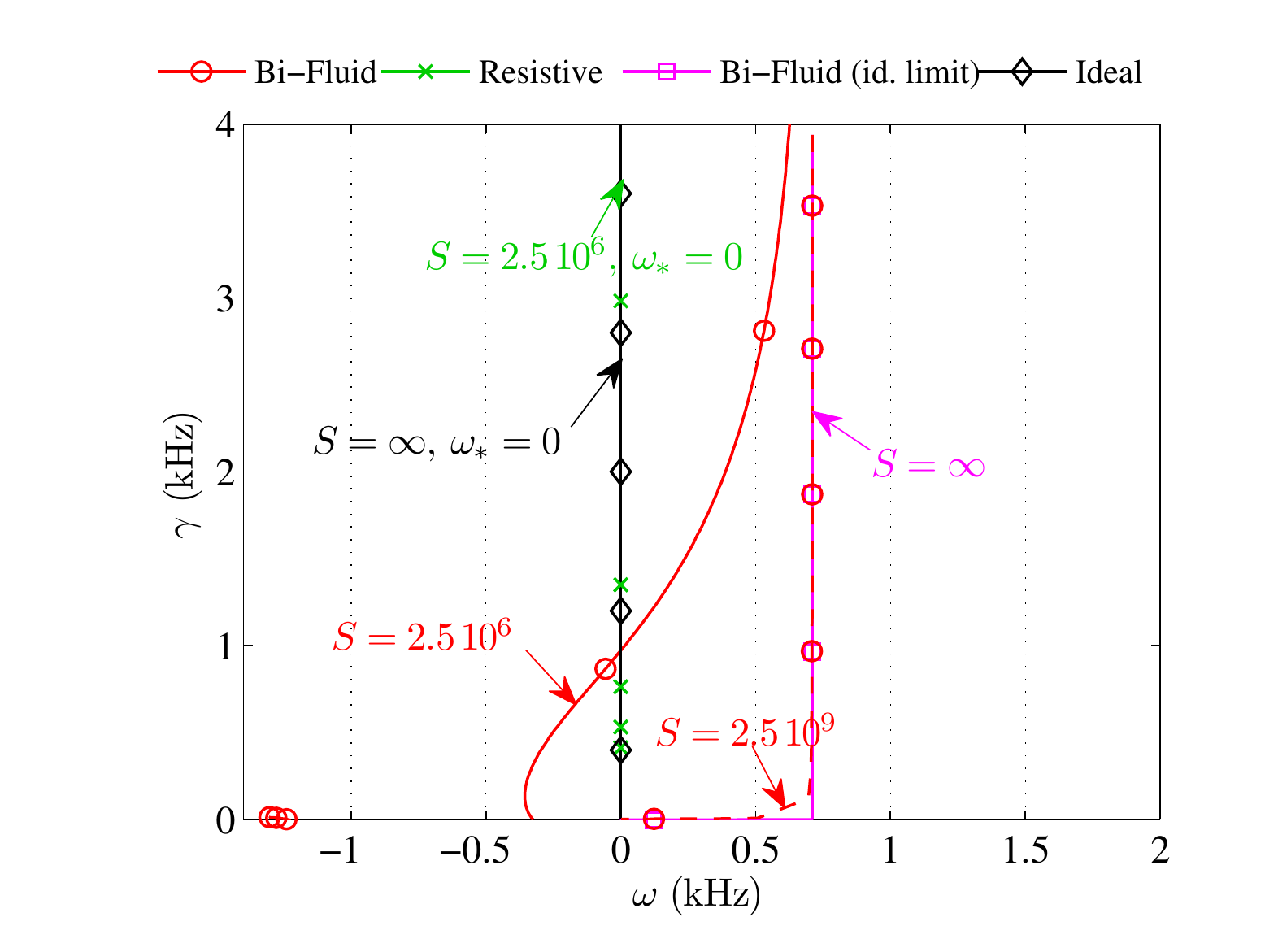}
  \caption[Internal kink growth rate and frequency with bi-fluid effects]{Growth rate and frequency of the solution of equation \eqref{eq:BF_kinkdisprel}. The parameters are the same as in figure \ref{fig:gamma_vs_gammaI_log_lab}.}
  \label{fig:gamma_vs_omega_S_lab}
\end{figure}

If one considers the effect of finite resistivity, the solutions of equation \eqref{eq:BF_kinkdisprel} can be studied numerically. In figure \ref{fig:gamma_vs_gammaI_log_lab} the growth rate of the solution in log scale is plotted versus the ideal growth rate for two different values of the resistivity. The solutions in the cases where diamagnetic effects or resistive effects are neglected have been added. In figure \ref{fig:gamma_vs_omega_S_lab} the growth rate is plotted versus the frequency of the same solutions.

The growth rate of the mode with resistivity and diamagnetic effects is higher than in the ideal case (with $\wi$) but stays lower than in the purely resistive case. In the high resistivity case ($S = 2.5 \,10^6$) $\gamma_R = \wi = - \hat{\omega}_{*e} = 1.42 \ \mathrm{kHz}$ and there is two unstable branches, one at positive or slightly negative $\gamma_I$ where the frequency of the mode is close to $\wi/2$ except close to the marginal stability where the frequency can be negative, the other one at negative $\gamma_I$ with a frequency comparable to $\hat{\omega}_{*e}$. The latter branch is reminiscent from the tearing branch in the purely resistive case, it is sometimes called the \emph{electron} branch since it rotates in the electron diamagnetic direction (see White et al. \cite{whi90}). In the low resistivity case ($S = 2.5 \,10^9$) $\gamma_R = 0.14 \ \mathrm{kHz}$ while $\wi = - \hat{\omega}_{*e} = 1.42 \ \mathrm{kHz}$, the electron branch is now stable and the solution is very close to the ideal case. 

\section{Summary}

 The method described by De Blank et al. \cite{deb91} to derive the dispersion relation for the $\mm=1,\nn=1$ internal kink mode in the case of high-aspect ratio equilibria with circular cross-sections is reproduced. In section \ref{sec:high-aspect-ratio} the coordinate system as well as an approximate solution to the dispersion relation are described. In section \ref{sec:preliminary-steps} the first steps of the potential energy minimization are carried out, in particular the perturbation minimizes the compressibility of the magnetic field, and the mode is dominated by its $\mm=1$ harmonic. The characteristic structure of the internal kink mode follows from the final steps of the minimization presented in section \ref{sec:m1_inter_kink_mode}: the radial MHD displacement is constant in regions where $q-1$ is not small. The dispersion relation is derived in section \ref{sec:dispersion-relation} and is applied to the case of a single $q=1$ layer recovering the results from Bussac et al. \cite{bus75} in the case of monotonic $q$-profiles and those of Hastie et al. \cite{has87} in the case of reversed $q$-profiles. If one adds the effects of finite resistivity the internal kink mode is always unstable (see section \ref{sec:res_disp_rel}). In section \ref{sec:bifluid_disp_rel} it is showed that diamagnetic effects have a double influence: the mode growth rate is reduced such that the mode is stabilized in some cases and the mode frequency tends to favor the ion diamagnetic frequency especially at low resistivity.

The stability of the internal kink mode for non-circular flux-surfaces has been investigated by Edery et al. \cite{ede76}, Bondesson et al. \cite{bon92} or L\"utjens et al. \cite{lut92}. The effects of ion-ion collisions are discussed in a paper by Ara et al. \cite{ara79}, the ones of ion viscosity and ion finite Larmor radius in a paper by Porcelli et al. \cite{por86}. Finally it is worth mentioning that the dispersion relation derived here in section \ref{sec:dispersion-relation} can be used to study profiles with very low shear and a wide region where $q$ is close to $1$. In this case the results are similar to the ones obtained by Wesson on the ``quasi-interchange mode'' \cite{wes86} or the ones of Hastie et al. \cite{has88}.


\chapter{Derivation of the Fishbone Dispersion Relation}
\label{cha:FDR_derivation}

The variational formalism used here for the derivation of the dispersion relation of \(\nn=1\) internal kink modes is the one introduced by Edery et al. \cite{ede92} and used later for the study of BAEs and GAMs by Nguyen et al. \cite{ngu08}. The dispersion relation provides explicit expression for the contributions of both the fluid (thermal) part and the kinetic (fast) part of the particle population. In particular, the kinetic term is identical to the one obtained by Chen et al. \cite{che84} in the case of ions or Zonca et al. \cite{zon07} in the case of electrons.

\section{The electromagnetic lagrangian}

We start from the electromagnetic lagrangian which is the variational form of the Maxwell equations in their linear version. We express it in function of the scalar potential $\Phi$ and the vector potential $\vc{A}$ for a perturbation of a single fourier time-harmonic $\omega$ (all quantities with an $\omega$ subscript will denote a single Fourier mode at frequency $\omega$, $M(\sim,t) = M_\omega(\sim) \exp (-i\omega t)$) :
\begin{multline}
  \mathcal{L}_\omega(\Phi_\omega,\vc{A}_\omega,\Phi_\omega^*,\vc{A}_\omega^*) = \varepsilon_0 \int \mathrm{d}^3\vc{x}\,\left(i\omega \vc{A}_\omega - \nabla \Phi_\omega\right)\cdot\left(-i\omega {\vc{A}_\omega}^* - \nabla {\Phi_\omega}^*\right) \ldots\\ 
  - \frac{1}{\mu_0}\int \mathrm{d}^3\vc{x} \,\left(\nabla \times \vc{A}_\omega\right)\cdot\left(\nabla \times {\vc{A}_\omega}^*\right) + \sum_s \int \mathrm{d}^3\vc{x} \left(\vc{J}_{s,\omega} \cdot \vc{A}_\omega^* - \rho_{s,\omega} \Phi_\omega^*\right)
\end{multline}
where $\rho_{s,\omega} = \int \mathrm{d}^3\vc{p}\ e_s f_{s,\omega}$ is the perturbed charge density associated with $f_{s,\omega}$ the perturbed distribution function for particles of type $s$ with charge $e_s$ and mass $m_s$; $\vc{J}_s = \int \mathrm{d}^3\vc{p}\ e_s (\vc{v}(F_s+f_s))_\omega$ is the perturbed current. The expression $(\vc{v}(F_s+f_s))_\omega$ denotes the harmonic of frequency $\omega$ of the product of the particle velocity and total particle distribution function,
\begin{equation}
  \vc{j}_{s,\omega} = \int \mathrm{d}^3\vc{p}\ e_s (\vc{v}f_{s,\omega} - \frac{1}{m_s} \vc{A}_\omega F_s).
\end{equation}
The extremalization of $\mathcal{L}_\omega$ with respect to the virtual fields ${\Phi_\omega}^*$ and ${\vc{A}_\omega}^*$ yields Maxwell equations.

Writing $\mathcal{L}_\omega = \mathcal{L}_{\omega,\varnothing} + \sum_s\mathcal{L}_s$, where $ \mathcal{L}_{\omega,\varnothing}$ is the lagrangian for the vacuum fields and $\mathcal{L}_s$ contains the particle-field interactions, one has 
\begin{equation}
  \mathcal{L}_s = - \int \mathrm{d}^3\vc{x} \mathrm{d}^3\vc{p} F_s \frac{{e_s}^2}{m_s} \vc{A}_\omega \cdot {\vc{A}_\omega}^* + e_s \int \mathrm{d}^3\vc{x} \mathrm{d}^3\vc{p} f_{s,\omega} \left(\vc{v}\cdot\vc{A}_\omega^* - \Phi_\omega^*\right)
  \label{eq:L_s}
\end{equation}
The last integral is linked to the perturbed hamiltonian for canonical coordinates
\begin{equation}
  h_{s,\omega} = e_s(\Phi_\omega - \vc{v}\cdot\vc{A}_\omega).
\end{equation}
which appears in the linear Vlasov equation which we need to solve in order to get the perturbed distribution function $f_{s,\omega}$ and which writes ($H_{s,0}$ being the unperturbed hamiltonian) :
\begin{equation}
  \pdd{f_s}{t} - [H_{s,0},f_s] = [h_s,F_s].
\end{equation}

\section{Action-angle variables}
\label{sec:AAvar}

Action-angle variables which can be derived from the unperturbed hamiltonian, are noted \((\vb{\alpha},\vc{J})\) where \(\vb{\alpha}\) are the angles ($\alpha_1$ is linked to the gyromotion, $\alpha_2$ is linked to the poloidal motion, and $\alpha_3$ to the toroidal motion) and \(\vc{J}\) the corresponding actions (in particular, $J_1$ is linked to the magnetic momentum $\mu$ and $J_3$ to the toroidal angular momentum $P_\varphi$).The geometrical angles $\theta$ and $\varphi$ are expressed as functions of $\alpha_2$ and $\alpha_3$ \cite{gar09a}
\begin{align}
  \theta &= \hat{\theta}(\alpha_2,\vc{J}) + \delta_P \, \alpha_2, \label{eq:theta_AA}\\
  \varphi &= \alpha_3 + q \,\hat{\theta}(\alpha_2,\vc{J}) + \hat{\varphi}(\alpha_2,\vc{J}), \label{eq:phi_AA}
\end{align}
where $\hat{\theta}$ and $\hat{\varphi}$ are periodic functions of $\alpha_2$ and have vanishing mean values,  and $\delta_ {P} = 1$ for passing particles only. 
The unperturbed motion has frequencies denoted by \(\vb{\Omega}\). $\Omega_1$ is the gyrofrequency, $\Omega_2$ is the poloidal transit frequency $\omega_b$ and $\Omega_3$ is the toroidal transit frequency which can be written
\begin{equation}
  \Omega_3 = \delta_P \, q \, \omega_b + \omega_d,
  \label{eq:tor_transit_freq}
\end{equation}
where $\delta_P$ is equal to $1$ for passing particles and $0$ for trapped particles, $\omega_d$ is the toroidal drift frequency; the ratio \(\omega_d/\omega_b\) is usually of the order of $\rho^* \ll 1$.

\section{Solving the linear Vlasov equation}

This particular set of variables allows us to solve the linear Vlasov equation in a very simple and elegant way. Performing a fourier transform in time and all 3 angles for any dynamical variable $V$,
\[V(\vc{x},\vc{p}) = \sum_{\vc{n} = (n_1,n_2,n_3)} V_\vc{n}(\vc{J})\exp i(\vc{n} \cdot \vb{\alpha}),\]
the following holds,
\begin{align}
  \int V(\vc{x},\vc{p}) W^*(\vc{x},\vc{p}) \mathrm{d}^3\vc{x} \mathrm{d}^3\vc{p} &= \sum_{\vc{n} = (n_1,n_2,n_3)}  \int \mathrm{d}^3\vb{\alpha} \mathrm{d}^3\vc{J} \, V_\vc{n}(\vc{J}) W_\vc{n}(\vc{J})^* \label{eq:Fourier_int}\\
  \{V(\vc{x},\vc{p}),K(\vc{J})\}_\vc{n} &= i \, \vc{n} \cdot \pdd{K}{\vc{J}} \\
  \{V(\vc{x},\vc{p}),H(\vc{J})\}_\vc{n} &= i \, \vc{n} \cdot \vb{\Omega}.
\end{align}
One then has
\begin{equation}
  f_{s,\vc{n},\omega} = - \frac{\vc{n} \cdot \partial {F_s}/ \partial \vc{J}}{\omega - \vc{n}\cdot\vc{\Omega}} h_{s,\vc{n},\omega}.
  \label{eq:Sol_Vlasov_linear}
\end{equation}
Note that by using the Vlasov equation, the effect of collisions on the perturbed distribution function is neglected. If these collisions were modeled using a simple Krook operator with an effective collision frequency $\nu_{eff}$, the denominator in equation (\ref{eq:Sol_Vlasov_linear}) would be replaced by $\omega - \vc{n}\cdot\vc{\Omega} + i\,\nu_{eff}$ such that if $\nu_{eff}$ is small compared to the frequency $\omega$ or compared to the frequencies of motion the effect of collisions can be neglected. A discussion for the case of electron-driven fishbones can be found in chapter \ref{sec:EFB_stability_circ}.

Instead of using the 3 actions as variables, we will use an equilibrium distribution function which will depend on $(J_1,E,J_3)$ where $E$ is the energy. The numerator of equation \eqref{eq:Sol_Vlasov_linear} now becomes
\begin{equation}
  \vc{n} \cdot \pdd{F_s}{\vc{J}} = \vc{n}\cdot\vb{\Omega}\left.\pdd{F_s}{E}\right|_{J_1,J_3} + n_1 \left.\pdd{F_s}{J_1}\right|_{E,J_3} + n_3 \left.\pdd{F_s}{J_3}\right|_{E,J_1},
  \label{eq:J_to_E}
\end{equation}
and the subscripts for the derivatives will be dropped from now on.

\section{The resonant lagrangian}

\subsection{Resonances at the cyclotron frequency}

Substituting expressions \eqref{eq:Sol_Vlasov_linear} and \eqref{eq:J_to_E} in \eqref{eq:L_s} one obtains 
\begin{align}
  \mathcal{L}_s &= - \int \mathrm{d}^3\vc{x} \mathrm{d}^3\vc{p} F_s \frac{{e_s}^2}{m_s} \vc{A}_\omega \cdot {\vc{A}_\omega}^* \ldots \nonumber\\
  & \hspace{2.5cm} + \sum_{\vc{n} = (n_1,n_2,n_3)} e_s \int \mathrm{d}^3\vb{\alpha} \mathrm{d}^3\vc{J} \frac{\vc{n}\cdot\vb{\Omega}\pdd{F_s}{E} + n_1 \pdd{F_s}{J_1} + n_3 \pdd{F_s}{J_3}}{\omega - \vc{n}\cdot\vc{\Omega}} h_{s,\vc{n},\omega} {h_{s,\vc{n},\omega}}^* \nonumber\\
  \mathcal{L}_s &= \int \mathrm{d}^3\vc{p}\mathrm{d}^3\vc{x} \left( - F_s \frac{{e_s}^2}{m_s} \left|\vc{A}_\omega\right|^2 - \pdd{F_s}{E}{e_s}^2 \left|\Phi_\omega - \vc{v} \cdot\vc{A}_\omega\right|^2 \right) \ldots \label{eq:L_adiabatic}\\
  & \hspace{2.5cm} + \sum_{\vc{n} = (n_1 \neq 0,n_2,n_3)} \int \mathrm{d}^3\vb{\alpha} \mathrm{d}^3\vc{J}  \frac{\omega \pdd{F_s}{E} + n_1 \pdd{F_s}{J_1} + n_3 \pdd{F_s}{J_3}}{\omega - \vc{n}\cdot\vc{\Omega}} h_{s,\vc{n},\omega} {h_{s,\vc{n},\omega}}^* \ldots \label{eq:L_mu}\\
  & \hspace{2.5cm} + \sum_{\vc{n} = (n_1 = 0,n_2,n_3)} \int \mathrm{d}^3\vb{\alpha} \mathrm{d}^3\vc{J}  \frac{\omega \pdd{F_s}{E} + n_3 \pdd{F_s}{J_3}}{\omega - \vc{n}\cdot\vc{\Omega}} h_{s,\vc{n},\omega} {h_{s,\vc{n},\omega}}^* \label{eq:L_diamagnetic}
\end{align}
where we obtained the second equality by making use of expression \eqref{eq:Fourier_int}. One can rewrite the previous equation in the following form,
\begin{equation}
  \mathcal{L}_s = \mathcal{L}_{adiabatic} + \mathcal{L}_\mu + \mathcal{L}_{diamagnetic}
\end{equation}
Because we will consider only perturbations with time-scales much slower than the cyclotron frequency $\Omega_1$, the resonances will all be included in the last term $\mathcal{L}_{diamagnetic}$. Indeed, the denominator of $\mathcal{L}_\mu$ will be dominated by $n_1 \Omega_1$.

\subsection{MHD modes}

In the low-beta limit, MHD modes can be described by a perturbation with a perturbed electrostatic potential $\Phi_\omega$ defined by 
\begin{equation}
  \vb{\xi}_\perp = \frac{1}{-i\omega} \frac{\nabla \Phi_\omega \times \vc{B}_0}{B_0^2}
\label{eq:Phi_pert_MHD}
\end{equation}
and a perturbed vector potential with a vanishing perpendicular component and the parallel component is such that the perturbed parallel electric field vanishes
\begin{equation}
  A_{\Vert,\omega} =\frac{1}{i\omega} \nabla_\Vert \Phi_\omega
\label{eq:Apar_pert_MHD}
\end{equation}
This description is consistent with the structure of the internal kink mode as described in chapter \ref{cha:internal-kink-mode} since it corresponds to a perturbation with a vanishing parallel magnetic field.

\subsection{The resonant lagrangian}

Noting $J_0$ the gyro-average operator (averaging over the cyclotronic period) and $\varphi$ the gyrophase, one has:
\begin{align}
  J_0 V = \frac{1}{2\pi} \int_0^{2 \pi} h(\vc{x},\vc{p})\mathrm{d} \varphi, \\
  \left\{ \begin{array}{lc} (J_0 V)_\vc{n} = 0 &\quad \mbox{if $n_1 \neq 0$} \\ (J_0 V)_\vc{n} = V_\vc{n} &\quad \mbox{if $n_1 = 0$} \end{array} \right.
\end{align}
The consequence for $\mathcal{L}_{diamagnetic}$ is that one can replace the reduced hamiltonian $h_{s,\vc{n},\omega}$ by its gyro-averaged value $(J_0 h_{s,\omega})_\vc{n}$. One can further modify expression \eqref{eq:L_diamagnetic} making it more relevant in the case of large-scale perturbations (like MHD modes).
\begin{equation}
  (J_0\,h_{s,\omega})_\vc{n} = e_s J_0\,\left(\frac{i\omega\Phi_\omega - \vc{v}\cdot\nabla \Phi_\omega}{i\omega} + \frac{\vc{v}_\perp \cdot \nabla \Phi_\omega}{i \omega} - v_\Vert \left(A_{\Vert,\omega} - \frac{\nabla_\Vert \Phi_\omega}{i \omega}\right)\right)
\end{equation}
The last parenthesis is proportional to the parallel electric field $E_{\Vert,\omega} = i \omega A_{\Vert,\omega} - \nabla_\Vert \Phi_\omega$. The $\vc{v}\cdot \nabla$ operator corresponds to the full time-derivative operator $\vc{v}\cdot \nabla V = \mathrm{d}V/\mathrm{d}t$ (for time-independent variables) and so $(\vc{v}\cdot \nabla \Phi_\omega)_\vc{n} = i \vc{n} \cdot \vb{\Omega} \, \Phi_{\vc{n},\omega}$.
\begin{equation}
  (J_0\,h_{s,\omega})_\vc{n} = e_s \left(\frac{\omega - \vc{n} \cdot \vb{\Omega}}{\omega} (J_0 \Phi_\omega)_\vc{n}\right) + e_s \left(J_0\left( \frac{\vc{v}_\perp \cdot \nabla \Phi_\omega}{i \omega} - v_\Vert E_{\Vert,\omega}\right) \right)_\vc{n}
\end{equation}
When combined with expression \eqref{eq:L_diamagnetic}, it is clear that the only resonant part of $\mathcal{L}_{diamagnetic}$ comes from the last term of the previous equation and one can write $\mathcal{L}_{diamagnetic} = \mathcal{L}' + \sum_s \mathcal{L}_{res}^{s}$ with 
\begin{equation}
  \mathcal{L}_{res}^{s} = \sum_{\vc{n} = (n_1 = 0,n_2,n_3)} \int \mathrm{d}^3\vb{\alpha} \mathrm{d}^3\vc{J}  \frac{\omega \pdd{F_s}{E} + n_3 \pdd{F_s}{J_3}}{\omega - \vc{n}\cdot\vc{\Omega}} ({h'}_{s,\omega})_\vc{n}  {({h'}_{s,\omega})_\vc{n}}^*
  \label{eq:L_res}
\end{equation}
\begin{equation}
  {h'}_{s,\omega} = e_s J_0\left( \frac{\vc{v}_\perp \cdot \nabla \Phi_\omega}{i \omega} - v_\Vert E_{\Vert,\omega}\right)  = e_s J_0\left( \frac{\vc{v}_{g,\perp} \cdot \nabla \Phi_\omega}{i \omega} - v_{g,\Vert} E_{\Vert,\omega} - \frac{\mu}{e} \left(\nabla \times \frac{\nabla \Phi_\omega}{i \omega}\right)_\Vert\right)
\label{eq:h_prime}
\end{equation}
A demonstration of equation \eqref{eq:h_prime} is given in appendix \ref{sec:h_prime_demo}. The last term of (\ref{eq:h_prime}) has been left purposely even though it is vanishing in this case. Indeed if one would want to keep the perpendicular part of the perturbed vector potential, then it can be shown that the only modification to $\mathcal{L}_{res}$ is through the addition of $\vc{A}_{\perp,\omega}$ to $\nabla \Phi_\omega/i \omega$ in this last term.

\vspace{2mm}

In the case of MHD perturbations, one obtains simply
\begin{equation}
  {h'}_{s,\omega} =  e_s \frac{\vc{v}_{g,\perp} \cdot \nabla \Phi_\omega}{i \omega} 
\label{eq:h_prime_MHD}
\end{equation}

\section{The extended energy principle}

\subsection{Other terms in the Lagrangian}

In our derivation of equation \eqref{eq:L_res} for the resonant part of the lagrangian, we have left out a few terms which can then be recombined to obtain the following form for the total electromagnetic lagrangian \cite{ede92},
\begin{equation}
  \label{eq:EMLag_decomp}
  \mathcal{L} = \mathcal{L}_{mag} + \mathcal{L}_{inertia} + \mathcal{L}_{tear} + \mathcal{L}_{int} + \sum_s \mathcal{L}^{s}_{res}.
\end{equation}
In this expression we have neglected terms present in reference \cite{ede92} corresponding to non-MHD perturbations, this means terms proportional to the perturbed parallel electric field or to the parallel magnetic field. The expressions of the different terms are
\begin{align}
  \mathcal{L}_{mag} &= -\frac{1}{\mu_0} \int \dd^3{\vc{x}}\,\left(\left|\nabla \times A_{\Vert,\omega} \vc{b}\right|^2 + \sum_s \mu_0 \frac{p_{\perp,s} - p_{\Vert,s}}{B^2} |\nabla_\perp A_{\Vert,\omega}|^2\right) \\
  \mathcal{L}_{inertia} &= \sum_s \Re \int \dd^3{\vc{x}}\,\frac{n_s m_s}{B^2} \left(|\nabla_\perp \Phi_\omega|^2 - \frac{\nabla p_{\perp,s} \times \vc{B}}{e_s n_s B^2}\cdot \frac{\nabla_\perp\Phi_\omega}{i\omega} \Delta \Phi_\omega^{*}\right) \\
  \mathcal{L}_{tear} &= \sum_s \Re \int \dd^3{\vc{x}}\, \left(\nabla \frac{J_{\Vert,s}}{B}\times \vc{b} \right)\cdot \frac{\nabla_\perp \Phi_\omega}{i\omega} \left(\frac{\nabla_\Vert \Phi_\omega}{i\omega}\right)^* \\
  \mathcal{L}_{int} &= -\sum_s \Re \int \dd^3{\vc{x}}\dd^3{\vc{p}}\, e_s \pdd{F_s}{J_3} \left(\nabla \psi_p \times \frac{\vc{B}}{B^2}\right) \cdot \frac{\nabla_\perp \Phi_\omega}{i\omega}\, {h'_{s,\omega}}^* \\
\end{align}
where we defined the quantities ($n_s,p_{\perp,s},p_{\Vert,s},J_{\Vert,s}$) as
\begin{align}
 \{n_s,p_{\perp,s},p_{\Vert,s},J_{\Vert,s}\} &= \int \dd^3{\vc{p}}\, \{1, m_s v_\perp^2/2, m_s v_\Vert^2, e_s v_\Vert\} F_s.
\end{align}

\subsection{Link with the MHD energy principle}

We suppose that the plasma is constituted of electrons and ions and that only one of these has a distribution that differs significantly from a Maxwellian distribution. We denote by $e$ and $i$ the thermal populations of electrons and electrons and by $h$ the population of fast particles.

Edery et al. \cite{ede92} and Nguyen et al. \cite{ngu08} have shown that the combination $\mathcal{L}_{mag} + \mathcal{L}_{tear} + \mathcal{L}_{int}$ for thermal species (electrons and ions) is compatible with the potential energy $\delta W$ found in the MHD energy principle (\ref{eq:dWp}) or (\ref{eq:dWp-int}) with the relation $\delta W = -2 \mathcal{L}$. $\mathcal{L}_{inertia}$ corresponds with the inertial term of the energy principle $(-\omega^2 K)$, and resonant effects with the thermal population are negligible.

Thus, assuming that the kinetic effects of the fast component are negligible as a first approximation, the potentials which minimize the action $\int \mathcal{L} \dd{t}$ are similar to those obtained using the MHD energy principle.

\subsection{From the Lagrangian to \texorpdfstring{$\dWk$}{dWk}}

Starting from the resonant part of the linear electro-magnetic lagrangian for a particle population of type $s$,
\begin{equation}
  \mathcal{L}_{res}^{s} = \sum_{\vc{n} = (n_1 = 0,n_2,n_3)} \int \mathrm{d}^3\vb{\alpha} \mathrm{d}^3\vc{J}  \frac{\omega \pdd{F_s}{E} + n_3 \pdd{F_s}{J_3}}{\omega - \vc{n}\cdot\vc{\Omega}} ({h'}_{s,\omega})_\vc{n}  {({h'}_{s,\omega})_\vc{n}}^*.
  \label{eq:L_res2}
\end{equation}

Even in the case of a single species, one still has to compute the integral in (\ref{eq:L_res2}) for every combination of mode numbers $n_2$ and $n_3$. But if the perturbation is assumed to have a single toroidal mode number and a single poloidal mode number, then only a few of those terms contribute significantly. The perturbed electrostatic potential is assumed to have the following form:
\begin{equation}
  \Phi_\omega(r,\theta,\varphi) = \Phi_\omega(r) e^{-i \mm \theta + i \nn \varphi}
\end{equation}
Then $(h'_{s\omega})_{\vc{n}}$ corresponds to the following integral
\begin{equation}
  h'_{s\vc{n}\omega} = \frac{1}{(2 \pi)^3} \int \dd^3{\vb{\alpha}} \frac{\left(\vc{v}_{g} \cdot i\vc{k}_\perp\right) \Phi_\omega}{i \omega} \,  e^{i (-\mm \theta +\nn \varphi - \vc{n} \cdot \vb{\alpha})}.
\end{equation}

The integral over $\alpha_1$ corresponds to a gyroaverage operator because $n_1 = 0$. One can approximate this integral by simply replacing the particle position by the gyrocenter position.

The integral over $\alpha_3$ corresponds to an integral around the torus axis of symmetry. Since only $\varphi$ is changing, the integral vanishes unless $n_3 = \nn$.

The expression for $h'_{s\vc{n}\omega}$ then reduces to:
\begin{equation}
  h'_{s\vc{n}\omega} = \frac{1}{2 \pi} \int \dd{\alpha_2} \frac{\left(\vc{v}_{g} \cdot i\vc{k}_\perp\right) \Phi_\omega}{i \omega} \, e^{i (-\mm \theta + \nn \varphi - n \alpha_3 - n_2 \alpha_2)}.
  \label{eq:hn}
\end{equation}
and, using equations (\ref{eq:theta_AA}) and (\ref{eq:phi_AA}), the term in the exponential can be written (up to a factor $i$) as
\begin{equation}
  \mm \left(\hat{\theta}(\alpha_2) + \delta_{P} \, \alpha_2 \right) + \nn \left(q \,\hat{\theta}(\alpha_2) + \hat{\varphi}(\alpha_2)\right) - n_2 \alpha_2.
\end{equation}

\subsubsection*{Trapped particles}

For deeply trapped particles, both functions $\hat{\theta}$ and $\hat{\varphi}$ are negligible compared to $\alpha_{2}$; furthermore the other term in the integral is almost independent of $\alpha_{2}$. Thus, the only significant contribution to $\mathcal{L}_{res}$ comes from $\vc{n}=(0,0,\nn)$. For weekly trapped particles, the choice of the mode number will be dictated by the resonance condition $\omega = n_{2}\omega_{b} + \nn\omega_{d}$. Indeed, in the present work, we consider only modes with frequencies much lower than the typical thermal poloidal orbit frequency, so that resonance for $n_{2} \neq 0$ is negligible.

\subsubsection*{Passing particles}

For deeply passing particles, the same argument as the one used for deeply trapped particles would lead us to choose $n_{2}=\mm$. Furthermore, from the resonance condition $\omega=(n_{2}+q\,\nn)\omega_{b}+\nn\omega_{d}$ we must choose $n_{2}$ such that the factor in front of $\omega_{b}$ is as low as possible; which yields $n_{2}=-\mm$. Since $\nn q - \mm \simeq k_{\Vert} R_{0} q$, mode resonance with circulating particles is restricted to the region where $k_{\Vert}$ is small (around $q = \mm/\nn$).

\subsection{The dispersion relation}

\subsubsection*{Derivation}

\label{sec:DR_derivation}

As described in chapter \ref{cha:internal-kink-mode}, for the ($\nn=1$, $\mm = 1$) internal kink mode the minimization is a two-scale problem. If the safety factor profile is such that $q=1$ at a radius $r_{s}$, a thin layer exists around $r_{s}$ where the gradients of the potentials are very strong and inertia plays a significant role.

Outside this layer, the ideal MHD solution is recovered, and the potentials can be derived from the radial MHD displacement : $\xi_{r}(r)=\xi_{c}$ inside the $q=1$ surface and $\xi_{r}(r)=0$ outside. If one chooses the gauge such that the electric field is purely non-inductive, then the electric potential is linked to the MHD displacement by $\partial_t \vb{\xi} = - i \vb{k}_\perp \Phi_\omega \times \vc{B}/B^2$ such that, to first order in $\varepsilon$, $\Phi_\omega = -\omega B_0 r \xi_c$. Note that this choice is consistent with the previous derivation since it implies that the perturbed vector potential is parallel to the equilibrium magnetic field.

For the inertial layer, the form of the solution depends on the different effects that one wishes to take into account. In any case, the asymptotic matching between the inertial layer solution and the MHD solution yields a dispersion relation in the following form 
\begin{equation}
  \dWf+\dWh=\delta I,
  \label{eq:FDR_app}
\end{equation}
where $\dWf$ and $\dWh$ represent the respective contributions of the plasma thermal bulk and of the plasma hot component ($\mathcal{L}=-2\delta W$ and $\delta\hat{W}=\hat{C}\delta W$ with $\hat{C}=\mu_{0}R_{0}/B_0^{2}2\pi r_{s}^{2}{\xi_c}^{2}$ consistently with \cite{zon07}).

\subsubsection*{Contribution of fast particles}

For the fast particle component (represented by the subscript ${}_h$), the contribution to $\mathcal{L}_{inertia}$ is usually negligible due to the very low density of fast particles compared to the thermal species. The contribution to $\mathcal{L}_{magn}$ is limited to the second term and is proportional to $|A_{\Vert,\omega}|^2$ and therefore to $k_\Vert^2$. The $\mathcal{L}_{tear}$ term is proportional to $\nabla_\Vert \Phi_\omega$ and therefore to $k_\Vert$ while $\mathcal{L}_{int}$ does not depend on $k_\Vert$. Here we will only consider $q$-profiles where $|q-1|$ stays usually of the order of $10^{-1}$ or smaller inside the $q=1$ surface such that one can only consider the contributions coming from $\mathcal{L}_{int}$ and $\mathcal{L}_{res}$ for the fast particle population.

In the following sections we will denote by $\dWk$ the contribution of fast particles due to $\mathcal{L}_{res}$ and by $\dW_{f,h}$ the one due to $\mathcal{L}_{int}$ such that we can write 
\begin{equation}
  \label{eq:dWh}
  \dWh = \dWk + \dW_{f,h}.
\end{equation}

\subsubsection*{Expression in guiding-center coordinates in the limit of zero-orbit width}

Since the equilibrium distribution function depends only on the invariants of motion we will use the following set of guiding-center coordinates: $(\bar{\psi}_p, p , \xi_0)$, where $\bar{\psi}_p$ is the orbit-averaged radial position of the particle, $p$ is the particle's momentum and $\xi_0$ is a pitch angle coordinate defined by
\begin{equation}
  \mu = \frac{E(1-\xi_0^2)}{B_m(\bar{\psi}_p)},
\end{equation}
with $B_{m}(\psi_p)$ the minimum amplitude of the magnetic field on the flux-surface indexed by $\psi_p$. It is equivalent to say that $\xi_0$ is the ratio of the parallel velocity to the total velocity at the point of minimum magnetic field amplitude ($\theta = 0$ for circular plasmas).

\comment{
Since the equilibrium distribution function depends only on the invariants of motion we will use the following set of guiding-center coordinates: $(\bar{\psi}_p, p , \hat{\lambda}, \sigma)$, where $\bar{\psi}_p$ is the orbit-averaged radial position of the particle, $p$ is the particle's momentum, $\hat{\lambda}$ is a pitch angle coordinate defined by
\begin{equation}
  \hat{\lambda} = \frac{\mu B_0}{E}
\end{equation}
and $\sigma$ is the sign of $v_\Vert$ at the point of minimum magnetic field amplitude along its orbit ($\theta = 0$ for circular plasmas).
}

\vspace{2mm}

In the case of a general axisymmetric magnetic equilibrium, in the limit of zero orbit width and to first order in $\rho_s L_B$ (where $L_B = \nabla B/B$ is the gradient length of the magnetic field amplitude), the jacobian of the transformation from action-angle coordinates to guiding-center coordinates is \cite{bri09}:
\begin{equation}
  \mathcal{J}(\psi_p, p , \xi_0) = (2\pi)^3 p^2 \frac{\tilde{q}(\psi_p) R_0 }{B_m(\psi_p)} |\xi_0| \bar{\tau}_b(\psi_p,\xi_0)
\end{equation}
where the identity $\bar{\psi}_p \simeq \psi_p$ has been assumed since the difference between those two quantities is typically of the order of the orbit width\footnote{In the remaining of this thesis, no distinction between $\bar{\psi}_p$ and $\psi_p$ will be made.}. $\bar{\tau}_b(\psi_p,\xi_0)$ is defined by 
\begin{equation}
  \bar{\tau}_b(\psi_p,\xi_0) = \frac{1}{\tilde{q}(\psi_p)} \int_{\theta_{min}}^{\theta_{max}} \frac{\dd{\theta}}{2 \pi} \frac{B}{R_0 B^\theta}\frac{1}{|\xi|},
\end{equation}
where $\xi = v_\Vert/v$ is the cosine of the particle's pitch-angle at the poloidal angle $\theta$; $\theta_{min}$ and $\theta_{max}$ are the poloidal angles corresponding to the maximum excursion of the particles of a given type. For passing particles $\theta_{min} = -\pi$ and $\theta_{max} = \pi$ while for trapped particles these quantities depend both on $\psi_p$ and $\xi_0$. $\tilde{q}$ is defined by equation (\ref{eq:qhat_def}). $\bar{\tau}_b$ corresponds to the normalized bounce time since
\begin{equation}
  \tau_b(\psi_p,\xi_0) = \frac{2 \pi m_h \tilde{q}(\psi_p) R_0}{p} \, \bar{\tau}_b(\psi_p,\xi_0).
\end{equation}
This expression of the bounce time corresponds to a complete orbit for passing particles and to a single leg of the banana orbit for trapped particles.

\vspace{2mm}

Then $\dWk$ is derived from equation (\ref{eq:L_res}), noting that in the limit of zero-orbit width the toroidal angular momentum $J_3 \simeq -e_h \bar{\psi}_p$,
\begin{equation}
  \dWk = - 4\pi^3 \hat{C} \int \dd{\psi_p} \dd{\xi_0} \dd{p} \, \frac{\tilde{q} R_0}{B_m} |\xi_0| \bar{\tau}_b \, p^2 \frac{\omega \, \partial_E F_h - e_h^{-1} \, \partial_{\psi_p} F_h}{\omega - \delta_P (q-1) \omega_b - \omega_d}\left|h'_{h\vc{n}\omega}\right|^2.
  \label{eq:dWk_GC_Gen}
\end{equation}
The reader will note that in deriving this expression we have used the fact that all quantities present in the integral are functions of the particle invariants only and are therefore independent of the poloidal angle $\theta$, in particular the equilibrium distribution function $F_h$, the motion frequencies $\omega_b$ and $\omega_d$ as well as $h'_{h\vc{n}\omega}$.

\vspace{2mm}

For $\dW_{f,h}$, the situation is different due to the fact that its expression contains $h_{h,\omega}$ which depends on $\theta$ through $\vc{v}_{g}$. The expression then becomes
\begin{equation}
  \dW_{f,h} = 4\pi^3 \hat{C} \int \dd{\psi_p} \dd{\xi_0} \dd{p} \, \frac{\tilde{q} R_0}{B_m} |\xi_0| \bar{\tau}_b \, p^2 \, \pdd{F_h}{\psi_p} \left(\vb{\xi}\cdot\nabla \psi_p\right) \left(\Re {\overline{h}'_{h,\omega}}^*\right)
  \label{eq:dWint_GC_Gen}
\end{equation}
where $\overline{h'}_{h,\omega}$ can be computed as 
\begin{equation}
  \label{eq:h_sw_orbit_av}
  \overline{h}'_{h,\omega} = \frac{1}{\bar{\tau}_b\tilde{q}} \int_{\theta_{min}}^{\theta_{max}} \frac{\dd{\theta}}{2\pi} \frac{B}{ R_0 B^\theta} \frac{1}{|\xi|} h'_{h,\omega}
\end{equation}
and corresponds to the orbit-averaged value of $h'_{h,\omega}$. The link between $\vb{\xi}$ and $\Phi_\omega$ yields the following relation
\begin{equation*}
  \nabla \psi_p \times \frac{\vc{B}}{B^2} \cdot \frac{i\vc{k}_\perp \Phi_\omega}{i\omega} = - \vb{\xi} \cdot \nabla\psi_p
\end{equation*}
which was used to derive equation (\ref{eq:dWint_GC_Gen}).

\paragraph*{Further approximations}

In the remaining of this thesis, the radial component of the MHD displacement for the $\mm=1$ internal kink is assumed to be a top-hat function such that $\xi_r = \xi_c$ inside the $q=1$ surface and $\xi_r = 0$ outside, as was pointed out by the analysis of chapter \ref{cha:internal-kink-mode}. As a consequence one obtains the following approximation
\begin{equation}
  - \vb{\xi} \cdot \nabla\psi_p = \xi_c \frac{r B_0}{q},
  \label{eq:top_hat_app}
\end{equation}
with $r$ being defined in equation (\ref{eq:def_r_DeB}).

\vspace{2mm}

For simplicity, $\overline{h}'_{h,\omega}$ and $\left|h'_{h\vc{n}\omega}\right|$ can be replaced by the expressions obtained by integration using the ballooning transform in the case of large aspect ratio (see \cite{che84} or \cite{zon07}) and which is exact for deeply trapped particles or in the case of vanishing magnetic shear
\begin{equation}
  \overline{h}'_{h,\omega} \simeq h'_{h\vc{n}\omega} \simeq \frac{e_h B_0 r}{q} \, \omega_d \xi_c = \frac{E}{R_0} \, \bar{\Omega}_d \xi_c.
  \label{eq:hsnw_app}
\end{equation}

Using these two approximations, the following expressions are obtained for $\dWk$ and $\dW_{f,h}$
\begin{align}
  \dWk &= - 4\pi^3 \hat{C} \int \dd{\psi_p} \dd{\xi_0} \dd{p} \, \frac{\tilde{q} R_0}{B_m} |\xi_0| \bar{\tau}_b \, p^2 \, \dd{p} \frac{\omega \, \partial_E F_h - e_h^{-1} \, \partial_{\psi_p} F_h}{\omega - \delta_P (q-1) \omega_b - \omega_d}\left(\frac{E}{R_0} \bar{\Omega}_d \xi_c\right)^2,
  \label{eq:dWk_gen_app}
\\
  \dW_{f,h} &= 4\pi^3 \hat{C} \int \dd{\psi_p} \dd{\xi_0} \dd{p} \, \frac{\tilde{q} R_0}{B_m} |\xi_0| \bar{\tau}_b \, p^2 \, \pdd{F_h}{\psi_p} \left(\frac{r B_0}{q}\xi_c\right) \left(\frac{E}{R_0} \bar{\Omega}_d \xi_c\right).
  \label{eq:dWint_gen_app}
\end{align}

\comment{
\vspace{2mm}

We introduce the notation \(\langle \mathcal{A} \rangle_{\vc{x},\vc{p}} = V^{-1}\int \dd^3 \vc{x} \dd^3 \vc{p} \mathcal{A} F_h \), $\vc{x}$ is the position in real space and $\vc{x}$ in momentum space, the integral is limited to the space inside the $q=1$ surface of total volume $V = 2 \pi^2 r_s^2 R_0$ and $F_h$ is the distribution function of fast particles. Then the expression of $\dWk$ becomes
\begin{equation}
  \dWk = - \frac{\pi}{2} \frac{\mu_0}{{B_0}^2} \bigg{\langle} E^2 \bar{\Omega}_d^2 \frac{\omega \, \partial_E\ln F_h  - e_h^{-1} \, \partial_{\psi_p} \ln F_h }{\omega - (q-1)\omega_b\delta_P - \omega_d}\bigg{\rangle}_{\vc{x},\vc{p}}.
  \label{eq:dWk_gen2}
\end{equation}
The same operation is done for $\dW_{f,h}$, this gives an expression compatible with \cite{whi85},
\begin{equation}
  \delta\hat{W}_{int,h} = \frac{\pi}{2} \frac{\mu_0}{{B_0}^2} \bigg{\langle} R_0 \bar{\Omega}_d E \, \partial_{r} \ln F_h \bigg{\rangle}_{\vc{x},\vc{p}}.
  \label{eq:dWint_gen2}
\end{equation}
}

\section{Summary}

A lagrangian formalism is used to derive the contribution of a population of fast particles to the dispersion relation of the internal kink mode. In the ideal MHD limit the energy principle derived in chapter \ref{cha:MHD-Energy-Principle} is recovered from the kinetic contributions of the thermal component of the plasma. The contribution of the plasma hot component is assumed to be small compared to the one of the plasma bulk such that the structure of the ideal MHD internal kink mode is also recovered with the radial MHD-displacement being approximated by a ``top-hat'' function. The modified dispersion relation obtained is compatible with the one found in White et al. \cite{whi85} or Zonca et al. \cite{zon07}, but the resonance condition for energetic passing particles has been modified to account for the term due to the parallel motion of passing particles. This term can be significant for energetic electrons and the effects of this modification on the stability of electron-driven fishbones will be discussed in chapter \ref{sec:EFB_stability_circ}.


\chapter{MIKE : solving the fishbone dispersion relation}
\label{cha:MIKE_solver}

The MIKE code has been designed to compute all the terms of the linear dispersion relation of the fishbone mode and solve it.

It can be used with analytical distributions, which provide means to verify that the code is in agreement with the linear theory developed in the previous chapter but also to study the general influence of any parameter, like the equilibrium shape for example. In chapter \ref{sec:EFB_stability_circ}, the code will be used with different types of analytical distributions to study the effect of finite $k_\Vert$ on the solutions of the dispersion relation.

It can also be used with distribution functions and equilibrium profiles reconstructed from actual experiments on tokamaks and provide a tool to compare the theory with the experiment. To this end, the MIKE code has been coupled to the Fokker-Planck code C3PO/LUKE \cite{dec08} to study electron-driven fishbone experiments in the Tore-Supra tokamak.

In this chapter, the physical content of the code is described in section \ref{sec:structure-mike-code}, then the normalization of the distribution functions and other parameters in the code is explained in section \ref{sec:normalization-mike}. The next two sections deal with some major features of the code, namely the computation of the resonant integral (section \ref{sec:ResIntComp}) and a method to find the zeros of a complex function (section \ref{sec:solv-disp-relat}). Finally the verification of the MIKE code is tackled in section \ref{sec:verif-mike-code}.

\section{Structure of the MIKE code}
\label{sec:structure-mike-code}

\subsection{Model and approximations}

\subsubsection*{Dispersion relation}

The dispersion relation is written
\begin{equation}
  \dWf + \delta \hat{W}_{int,h} + \dWk(\omega) = \delta I(\omega).
  \label{eq:FDR_MIKE}
\end{equation}

\subsubsection*{Fast particle contributions}

The contribution of fast particles to the dispersion relation comes from the $\dWk$ term $\dWk$ which includes all resonant effects between the particles and the mode and from the $\delta \hat{W}_{int,h}$ term which includes the modification of the fluid contribution to first order in $k_\Vert = (1-q)/qR_0$. Their expressions in the limit of zero-orbit width and using approximations (\ref{eq:top_hat_app}) and (\ref{eq:hsnw_app}) are
\begin{align}
  \dWk &= - 4\pi^3 \hat{C} \int \dd{\psi_p} \dd{\xi_0} \dd{p} \, \frac{\tilde{q} R_0}{B_m} |\xi_0| \bar{\tau}_b \, p^2 \, \dd{p} \frac{\omega \, \partial_E F_h - e_h^{-1} \, \partial_{\psi_p} F_h}{\omega - \delta_P (q-1) \omega_b - \omega_d}\left(\frac{E}{R_0} \bar{\Omega}_d \xi_c\right)^2,
  \label{eq:dWk_gen_MIKE}
\\
  \dW_{f,h} &= 4\pi^3 \hat{C} \int \dd{\psi_p} \dd{\xi_0} \dd{p} \, \frac{\tilde{q} R_0}{B_m} |\xi_0| \bar{\tau}_b \, p^2 \, \pdd{F_h}{\psi_p} \left(\frac{r B_0}{q}\xi_c\right) \left(\frac{E}{R_0} \bar{\Omega}_d \xi_c\right).
  \label{eq:dWint_gen_MIKE}
\end{align}

\subsubsection*{Fluid contribution}

$\dWf$ is the fluid term, it can be calculated from the expression derived by Bussac et al. \cite{bus75} for the $\nn = 1$ kink mode in toroidal geometry for the case of large aspect ratio and monotonic $q$-profiles, 
\begin{equation}
  \dWf = 3 \pi (1 - q_{min}) \frac{r_{s}^2}{R_0^2}\left(\frac{13}{144} - {\beta_{ps}}^2 \right).
\end{equation}

\subsubsection*{Inertia term}

$\delta I$ is called the inertial term and its form depends on the relevant physics inside the inertial ($q=1$) layer. The details for the different forms of $\delta I$ can be found in appendix \ref{cha:inert-term-fishb}.

\subsubsection*{Orbit characteristics}

The characteristics of the unperturbed particle orbits, $\omega_b$, $\omega_d$, $\overline{h}'_{s,\omega}$ and $h'_{s\vc{n}\omega}$, are calculated in the limit of zero-orbit width giving simple dependence over the particle momentum:
\begin{align*}
  \omega_b (\psi_p,\xi_0,p) &= p \, \tilde{\omega}_b (\psi_p,\xi_0), \\
  \omega_d (\psi_p,\xi_0,p) &= p^2 \, \tilde{\omega}_d (\psi_p,\xi_0), \\
  h'_{s\vc{n}\omega} (\psi_p,\xi_0,p) &\simeq p^2 \, \tilde{h}'_{s\vc{n}\omega} (\psi_p,\xi_0).
\end{align*}
This approximation, which is appropriate for electrons, results in significant computation time reduction.

\section{Normalization in MIKE}
\label{sec:normalization-mike}

\subsection{Link with density}

The MIKE code uses the same normalization for the distribution function as the Fokker-Planck code LUKE. The distribution functions calculated by the code LUKE have the following normalization \cite{pey08a}. The total number of electrons $N_e$ in the closed field-line region of the plasma (for $\psi_p \in [0,\psi_a]$) is 
\begin{equation}
  N_e = (2\pi)^3 \int_0^{\psi_a} \dd{\psi_p}  \int_{-1}^{1} \dd{\xi_0}  \int_0^{\infty} \dd{p} \, \frac{\tilde{q} R_0}{B_m} |\xi_0| \bar{\tau}_b \, p^{2} \, F_e(\psi_p,\xi_0,p)
  \label{eq:Ntot_Fe}
\end{equation}
where $F_e$ is the electron distribution function, $p$ is the momentum, and $\xi_{0}$ is the cosine of the pitch-angle taken the position $\theta_{0}$ of minimum $B$ field on the flux-surface, $B(\psi_p,\theta_0) = B_m(\psi_p)$. 

If one defines $n_e$ as the flux-surface averaged electronic density, such that $n_e$ does not depend either on $\theta$ or $\varphi$, then $N_e$ can also be computed as
\begin{equation}
  N_e = \int_0^{\psi_a} \dd{\psi_p} \int_0^{2\pi} \dd{\theta} \int_0^{2\pi} \dd{\varphi} \,  \frac{n_e(\psi_p)}{B^\theta(\psi_p,\theta)}
\end{equation}
which can also be written as 
\begin{equation}
  N_e = (2\pi)^2 \int_0^{\psi_a} \dd{\psi_p}\, \frac{n_e(\psi_p) R_0}{B_m(\psi_p)}\, \hat{q}(\psi_p)
  \label{eq:Ntot_ne}
\end{equation}
where $\hat{q}(\psi_p)$ is defined by equation (\ref{eq:qtilde_def}). Combining equations (\ref{eq:Ntot_Fe}) and (\ref{eq:Ntot_ne}), one obtains the following expression 
\begin{equation}
  n_e(\psi_p) = 2\pi \frac{\tilde{q}}{\hat{q}} \int_0^{\psi_a} \dd{\psi_p}  \int_{-1}^{1} \dd{\xi_0}  \int_0^{\infty} \dd{p} \, \frac{\tilde{q} R_0}{B_m} |\xi_0| \bar{\tau}_b \, p^{2} \, F_e(\psi_p,\xi_0,p)
  \label{eq:ne_Fe}
\end{equation}

\subsection{Non-dimensional variables}

\subsubsection*{Momentum space}

Numerically, the momentum is normalized to $\bar{p}=p/p_{\textrm{ref}}$ where $p_{\textrm{ref}}$ is some reference momentum value related to some reference energy $E_{\textrm{ref}}$ by $p_{\textrm{ref}}=\sqrt{m_h E_{\textrm{ref}}}$ such that the energy can always be written $E/E_{ref} = \bar{p}^2/2$.

\subsubsection*{Radial variable}

The radial variable $\rho$ used in MIKE (but also present in LUKE) is the distance of the flux-surface to the magnetic axis on the outboard midplane normalized to its value at the plasma edge located at $\psi_p = \psi_a$,
\begin{equation}
  \label{eq:def_rho_MIKE}
  \rho(\psi_p) = \frac{R(\psi_p,0) - R(0,0)}{R(\psi_a,0) - R(0,0)}.
\end{equation}
The transformation from $\psi_p$ to $\rho$ can be computed from the equilibrium geometry. In the case of a circular concentric equilibrium, one has simply $\rho = r/a$ where $r$ is the minor radius of the flux-surface.

\subsubsection*{Pitch-angle variable}

We have introduced $\xi_0$ the pitch-angle variable used in LUKE. The value representing the boundary between trapped particles and passing particles for $\xi_0$ is noted $\xi_{0T}$. This quantity tends to $0$ as $\rho$ (or $\psi_p$) approaches $0$. And so for a fixed grid, the number of points in the trapped region decreases strongly around the plasma center. This would not be a problem if the mode-particle interaction and in particular the imaginary value of $\dWk$ did not exhibit a strong dependence over $\xi_0$ when one approaches the trapped-passing boundary for electron-driven fishbones. This strong dependence is due to the presence of the region where the precession drift frequency $\omega_d$ reverses. This led to the introduction of a new variable $\hat{\iota}$ which is defined as
\begin{equation}
  \label{eq:iota_def_MIKE}
  \hat{\iota}^2 = \frac{1-\xi_{0T}^{-2}}{1-\xi_{0}^{-2}}.
\end{equation}
In the trapped region $\hat{\iota}$ ranges from $0$ for deeply trapped articles to $1$ at the trapped-passing boundary, and in the passing domain $\hat{\iota}$ ranges from $1$ to $+\infty$ for well passing particles.

\subsubsection*{Expressions for $\bar{n}_e$}

 We define a non-dimensional distribution function $\bar{F}_e = F_e\,p_{\textrm{ref}}^{3}/n_{\textrm{ref}}$ such that 
\begin{equation}
  \bar{n}_e(\psi_p) = 2\pi \frac{\tilde{q}(\psi_p)}{\hat{q}(\psi_p)} \int_{-1}^{1} \dd{\xi_0}  \int_0^{\infty} \dd{\bar{p}} \, \frac{\tilde{q} R_0}{B_m} |\xi_0| \bar{\tau}_b \, \bar{p}^2 \, \bar{F}_e(\psi_p,\xi_0,p)
  \label{eq:ne_Fe_ND}
\end{equation}
where $\bar{n}_e(\psi_p)=n_e(\psi_p)/n_{\textrm{ref}}$ is the normalized electron density.

\vspace{2mm}

Since $\hat{\iota}$ does not discriminate particles with $\xi_0 < 0$ from particles with $\xi_0 > 0$, we represent particles with $\xi_0 < 0$ by a triplet $(\rho,\hat{\iota},\bar{p})$ with $\bar{p} < 0$.

\vspace{2mm}

The normalized density is now computed by
\begin{equation}
  \bar{n}_e(\rho) = 2\pi \frac{\tilde{q}(\rho)}{\hat{q}(\rho)} \int_{0}^{\infty} \dd{\hat{\iota}} \int_{-\infty}^{\infty}\dd{\bar{p}}\left.\pdd{\xi_0}{\hat{\iota}}\right|_{\rho} \, |\xi_0| \bar{\tau}_b \, \bar{p}^{2} \, \bar{F}_e(\rho,\hat{\iota},\bar{p})
  \label{eq:ne_Fe_MIKE}
\end{equation}
Expressions for $\dWk$ and $\dW_{int,h}$ can easily be derived (see appendix \ref{sec:app_MIKE_Solver}).

\section{Resonant Integral Computation}
\label{sec:ResIntComp}

The first step in the computation of $\dWk$ is the integral over $\bar{p}$, which is the most challenging. Because the denominator of the integrand vanishes when particles do resonate with the wave, the calculation of this integral with a classic trapezoidal approximation can lead to dramatic errors.

In MIKE, particles going in both directions (co- and counter-current) are treated simultaneously. The dependence of $\omega_{b}$ and $\omega_{d}$ over $\bar{p}$ implies that the denominator is a second degree polynomial, and the integral over $\bar{p}$ can always be written in the following form
\begin{equation}
  J(g,b_{ref},c_{ref}) = \frac{1}{\sqrt{2}} \int_{-\infty}^{+\infty} \frac{\bar{p}^4 g(\bar{p})}{\bar{p}^2 + \sqrt{2} b_{ref} \bar{p} - 2 c_{ref}} \dd{\bar{p}}.
  \label{eq:Eint_def_MIKE}
\end{equation}

\subsection{Integration contour}

Let $\alpha_+$ and $\alpha_-$ be the roots of the denominator,
\begin{equation}
  \alpha_{\pm} = - \frac{b_{ref}}{\sqrt{2}} \pm \sqrt{\frac{b_{ref}^2}{2} + 2 c_{ref}},
  \label{eq:alpha_pm}
\end{equation}
because $b_{ref}$ is real in all cases, the sum of $\alpha_+$ and $\alpha_{-}$ is also real and their imaginary parts have opposite signs. $\alpha_+$ is chosen to be the root with the positive imaginary part when $\Im \omega > 0$, and when $\Im \omega \leq 0$, $\alpha_+$ is chosen such that its dependence over $\omega$ is analytic. The integration contour is defined to be the real axis $[-\infty,+\infty]$ when $\Im \omega > 0$, and the integral is analytically continued for $\Im \omega \leq 0$. This is equivalent to keeping the integration contour going below $\alpha_+$ and above $\alpha_-$.

\subsection{Case of near-Maxwellian distributions}

When the distribution is close to a Maxwellian distribution, the function $g$ decreases exponentially fast with $\bar{E} = \bar{p}^2/2$,  and the plasma dispersion function $Z$ is used to solve the singular integral. It is defined as
\begin{equation}
  Z\left(u\right) = \frac{1}{\sqrt{\pi}} \int_{-\infty}^{+\infty} \frac{e^{-u^2}\dd{u}}{u - z},
\end{equation}
where the integration contour goes below the pole located at $u = z$.

Defining $G(\bar{p}) = \bar{p}^4 e^{\bar{p}^2/2} g(\bar{p})$, the integral is expanded by decomposing the fraction in simple elements,
\begin{multline}
  J = \frac{1}{\sqrt{2}}\frac{1}{\alpha_+ - \alpha_-} \left[ \int_{-\infty}^{+\infty} \frac{G(\bar{p}) - G(\alpha_+)}{\bar{p} - \alpha_+} e^{-\bar{p}^2/2} \dd{\bar{p}} - \int_{-\infty}^{+\infty} \frac{G(\bar{p}) - G(\alpha_-)}{\bar{p} - \alpha_-} e^{-\bar{p}^2/2} \dd{\bar{p}} + \ldots \right. \\ \left. G(\alpha_+) \int_{-\infty}^{+\infty} \frac{e^{-\bar{p}^2/2}\dd{\bar{p}}}{\bar{p} -\alpha_+} - G(\alpha_-) \int_{-\infty}^{+\infty} \frac{e^{-\bar{p}^2/2} \dd{\bar{p}}}{\bar{p} -\alpha_-} \right]
\end{multline}
The two last integrals contain all the singularities, and $J$ can be expressed as,
\begin{multline}
  J = \frac{1}{\sqrt{2}}\frac{1}{\alpha_+ - \alpha_-} \left[ \int_{-\infty}^{+\infty} \frac{G(\bar{p}) - G(\alpha_+)}{\bar{p} - \alpha_+} e^{-\bar{p}^2/2} \dd{\bar{p}} - \int_{-\infty}^{+\infty} \frac{G(\bar{p}) - G(\alpha_-)}{\bar{p} - \alpha_-} e^{-\bar{p}^2/2} \dd{\bar{p}} + \ldots \right. \\ \left. \sqrt{\pi} G(\alpha_+) Z\left(\frac{\alpha_+}{\sqrt{2}}\right) - \sqrt{\pi} G(\alpha_-) \left(-Z\left(-\frac{\alpha_-}{\sqrt{2}}\right)\right) \right].
  \label{eq:ZEnInt}
\end{multline}
The first two integrals are regular and can be dealt with by using a trapezoidal approximation.

This method is most effective when using near-Maxwellian distributions. Otherwise, the same method is used without the exponential factor in the definition of $g$ such that the plasma dispersion function is replaced by a logarithm.

\subsection{General case}

If $g$ does not decrease exponentially fast with $p^2/2$, then the method of decomposition in simple elements can still be used but the singularity in the integral will be handled using the complex logarithm. Supposing that the function $g$ is non-zero on a finite interval $[0,p_{max}]$, let the function $G$ be defined as $G(p) = p^4 g(p)$. We then expand the integral by decomposing the fraction in simple elements,
\begin{multline}
  J = \frac{1}{\sqrt{2}}\frac{1}{\alpha_+ - \alpha_-} \left[ \int_{-p_{max}}^{+p_{max}} \frac{G(p) - G(\alpha_+)}{p - \alpha_+} \dd{p} - \int_{-p_{max}}^{+p_{max}} \frac{G(p) - G(\alpha_-)}{p - \alpha_-} \dd{p} + \ldots \right. \\ \left. G(\alpha_+) \int_{-p_{max}}^{+p_{max}} \frac{\dd{p}}{p -\alpha_+} - G(\alpha_-) \int_{-p_{max}}^{+p_{max}} \frac{\dd{p}}{p -\alpha_-} \right]
\end{multline}
The two last integrals contain all the singularities, and $J$ can be expressed as,
\begin{multline}
  J = \frac{1}{\sqrt{2}}\frac{1}{\alpha_+ - \alpha_-} \left[ \int_{-p_{max}}^{+p_{max}} \frac{G(p) - G(\alpha_+)}{p - \alpha_+} \dd{p} - \int_{-p_{max}}^{+p_{max}} \frac{G(p) - G(\alpha_-)}{p - \alpha_-} \dd{p} + \ldots \right. \\ \left. G(\alpha_+) \ln\left(\frac{\alpha_+ - p_{max}}{\alpha_+ + p_{max}}\right) - G(\alpha_-) \ln\left(\frac{\alpha_- - p_{max}}{\alpha_- + p_{max}}\right) \right].
  \label{eq:LogEnInt}
\end{multline}

\subsection{Application to real distributions}\label{sec:EnReal}

If the distribution function of the particles is obtained from a Fokker-Planck code, then we only know the function $g$ for a finite number of values of $\bar{p}$ along the real axis and the evaluation of $G(\alpha_\pm)$ requires the interpolation of the distribution function to the whole complex plane. This is a very complex problem and the obtained accuracy for the interpolation is often poor.

In order to avoid this, we use the following method:
\begin{itemize}
\item We define the function $\tilde{G}$ as $\tilde{G}(\bar{p}) = G(\Re\, \bar{p})$, $\tilde{G}$ can then be computed numerically in the complex plane by simple interpolation to the real axis with acceptable accuracy. 
\item $J$ is then computed according to equation (\ref{eq:ZEnInt}) by replacing $G(\alpha_{\pm})$ by $\tilde{G}(\alpha_{\pm})$.
\item If $\Im \omega > 0$ then the result of the calculation of $J$ by equation (\ref{eq:ZEnInt}) is correct with a simple real integration contour.
\item But if $\Im \omega \leq 0$, then the residue of the first integral at $\alpha_{\pm}$ is not zero (due to the function $\Re$ which is not analytical) and it must be added to the result in order to get the correct expression for $J$. Equation (\ref{eq:ZEnInt}) then becomes
  \begin{multline}
  J = \frac{1}{\sqrt{2}}\frac{1}{\alpha_+ - \alpha_-} \left[ \int_{-\infty}^{+\infty} \frac{G(\bar{p}) - G(\Re \alpha_+)}{\bar{p} - \alpha_+} e^{-\bar{p}^2/2} \dd{\bar{p}} - \ldots \right. \\ \left. \int_{-\infty}^{+\infty} \frac{G(\bar{p}) - G(\Re \alpha_-)}{\bar{p} - \alpha_-} e^{-\bar{p}^2/2} \dd{\bar{p}} + \ldots \right. \\ \left. \sqrt{\pi} G(\alpha_+) Z\left(\frac{\alpha_+}{\sqrt{2}}\right) - \sqrt{\pi} G(\alpha_-) \left(-Z\left(-\frac{\alpha_-}{\sqrt{2}}\right)\right) + \ldots \right. \\ \left. + \vphantom{\int_{-\infty}^{+\infty}} 2 i \pi \left(G(\alpha_+) - G(\Re \alpha_+)\right)e^{-\alpha_+^2/2} - 2 i \pi \left(G(\alpha_-) - G(\Re \alpha_-)\right)e^{-\alpha_-^2/2} \right]
    \label{eq:fullZEnInt}
  \end{multline}
  where $G(\alpha_{\pm})$ is calculated by performing a polynomial interpolation of $G$. A similar expression can be derived to replace equation (\ref{eq:LogEnInt}).
\end{itemize}

Thus the loss of accuracy is limited to the lower mid-plane, which is of lesser interest if we are looking only for unstable modes.

\subsection{Additional Remarks}

The definition of the function $G$ (both for the near-maxwellian case and the general case) is not unique. Since the weakest link in the computation of the integral is the interpolation of the function $G$ to $\alpha_\pm$, it can be advantageous, if one has a good idea of the dependence of $g$ over $\bar{p}$, to choose the function $G$ such that its interpolation is the simplest possible. However this choice should always allow one to compute the singular part of the integral analytically.

For example if $g(p) = \exp(-\bar{p}^2/2)$, then the best choice for the function $G$ is $G(\bar{p}) = \exp(\bar{p}^2/2) g(\bar{p}) = 1$ which can be trivially interpolated. In this case one obtains
\begin{equation}
  J = \sqrt{\frac{\pi}{2}}\frac{1}{\alpha_+ - \alpha_-}\left[\left(2\alpha_+ + 2 \alpha_+^3 + \alpha_+^4 Z\left(\frac{\alpha_+}{\sqrt{2}}\right)\right) - \left(2\alpha_- + 2 \alpha_-^3 - \alpha_-^4 Z\left(-\frac{\alpha_-}{\sqrt{2}}\right)\right)\right].
  \label{eq:ZEnInt_exact}
\end{equation}

\vspace{2mm}

The method described in the case of near-maxwellian distribution can easily be adapted to distributions whose energy dependence are close to $\exp(-\bar{p}^2/2 T)$ with any value for $T$. If one then defines $G(\bar{p}) = \exp(\bar{p}^2/2 T) g(\bar{p})$, one can obtain an expression similar to equation (\ref{eq:ZEnInt}). In the numerical implementation of this method one should be careful that the choice of $T$ is compatible with the grid in $\bar{p}$ since the exponential factor can lead to numerical errors due to infinite values.

\comment{
\vspace{2mm}

The definition of $\dWk$ requires that the function $\omega \rightarrow \dWk$ is analytical in the upper ($\Im \omega > 0$) midplane. $\omega$ enters the definition of $J$ through $c_{ref}$ which is proportional to $\omega$. Then because of the presence of the radical in equation (\ref{eq:alpha_pm}), we know that $\alpha_\pm$ are not analytical functions of $\omega$ in the whole complex plane. Still it is possible to obtain an expression analytical in the upper midplane by a careful choice of the branch-cut in the definition of $\alpha_\pm$ (for example.
}

\subsection{Accuracy test}

The accuracy of both methods is now tested and compared to the accuracy obtained by a simple trapezoidal approximation of the integral. Focus will be made on the potential accuracy enhancement near the real axis.

\begin{figure}[!ht]
  \centering
  \begin{subfigure}[b]{0.49\textwidth}
    \centering
    \includegraphics*[width=\figwidth]{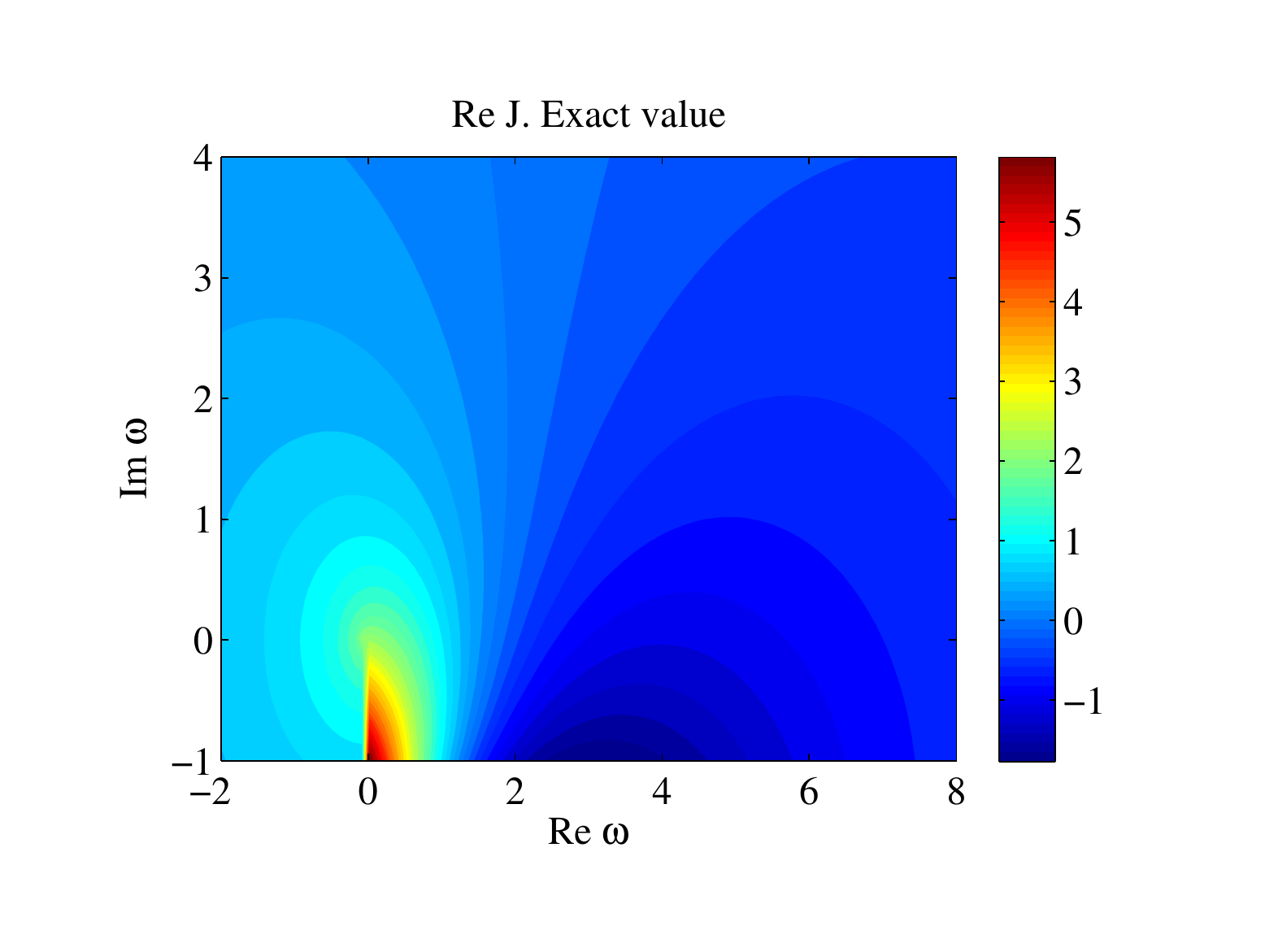}
    \caption{Real part of $J$}
    \label{fig:Eint_test_exp_Re}
  \end{subfigure}
  \begin{subfigure}[b]{0.49\textwidth}
    \centering
    \includegraphics*[width=\figwidth]{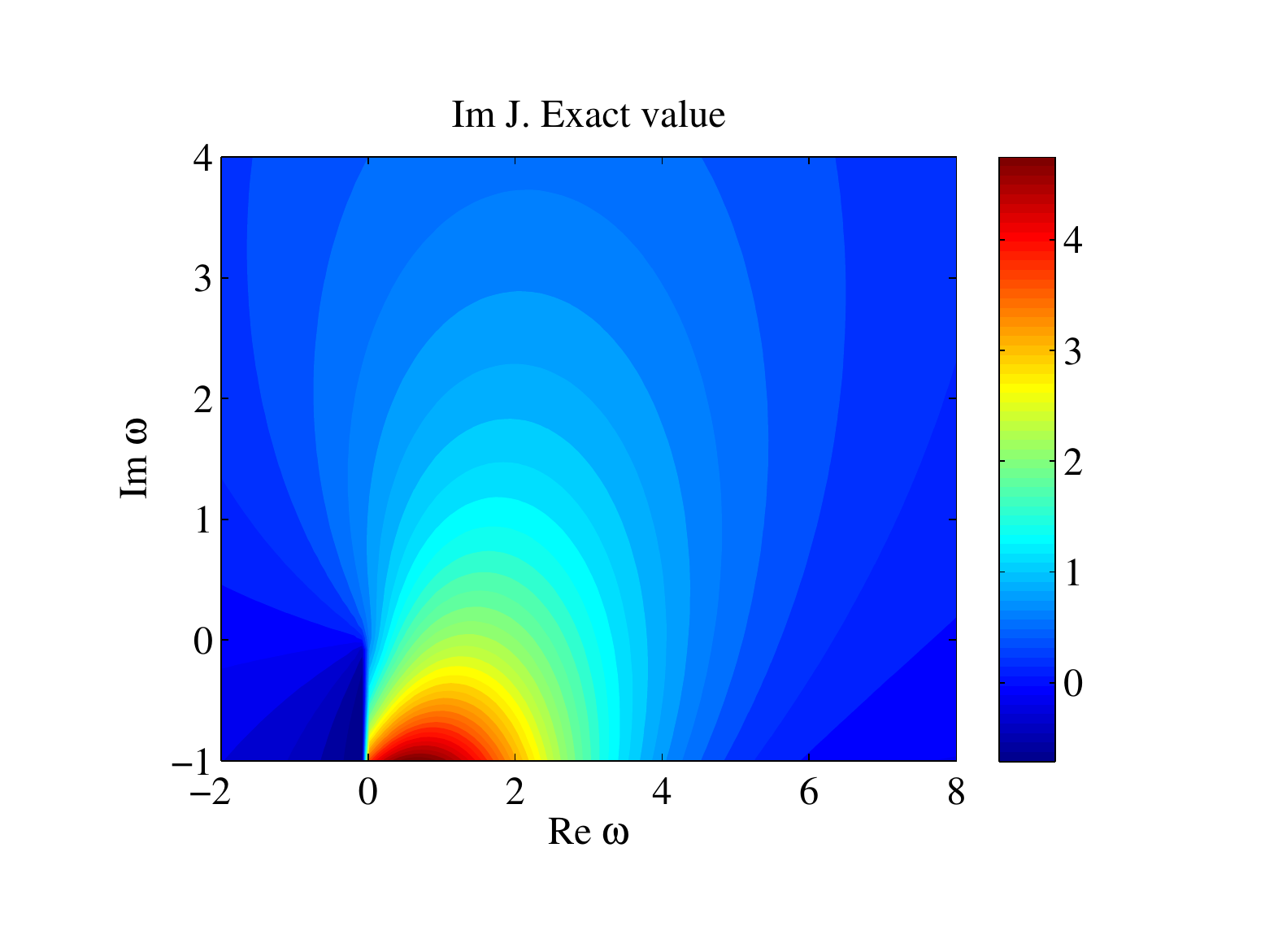}
    \caption{Imaginary part of $J$}
    \label{fig:Eint_test_exp_Im}
  \end{subfigure}
  \caption[Complex value of the resonant integral $J$]{Real part (a) and imaginary part (b) of the value of the resonant integral $J$ for $g(\bar{p}) = \exp(-\bar{p}^2/2)$, $b_{ref} = -0.5$, $c_{ref} = \omega$, $\Re \omega \in [-2,8]$ and $\Im \omega \in [-1, 4]$.}
  \label{fig:Eint_test_exp}
\end{figure}

The following parameters are set $g(\bar{p}) = \exp(-\bar{p}^2/2)$, $b_{ref} = -0.5$, $c_{ref} = \omega$. The integral $J$ defined in (\ref{eq:Eint_def_MIKE}) is then computed analytically (see equation (\ref{eq:ZEnInt_exact})) for a wide range of values of $\omega$ namely $\Re \omega \in [-2,8]$ and $\Im \omega \in [-1, 4]$. The results are represented on figure \ref{fig:Eint_test_exp}.

\begin{figure}[!ht]
  \centering
  \includegraphics*[width=\figwidth]{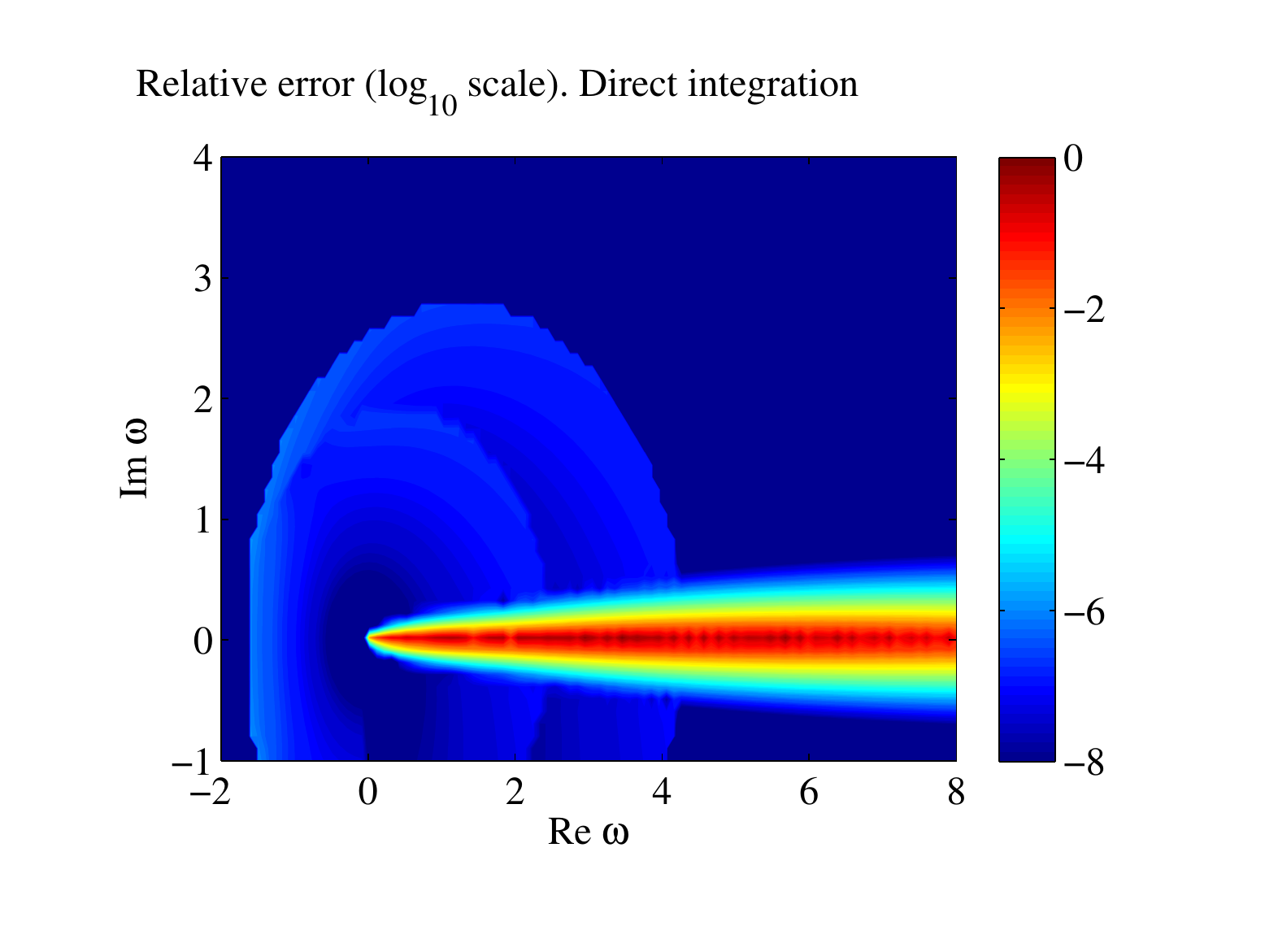}
  \caption[Relative error of the numerical computation of $J$ (trapezoidal approximation)]{Relative error of the value of $J$ for a trapezoidal approximation on a 300 points grid. The parameters are the same as for figure \ref{fig:Eint_test_exp}.}
  \label{fig:Eint_test_exp_Dir}
\end{figure}

A numerical approximation of this integral has then been computed using the trapezoidal approximation on a grid of 300 points located between $\bar{p} = -30$ and $\bar{p} = 30$. The relative error of the value of the integral is shown on figure \ref{fig:Eint_test_exp_Dir} in log scale.
As was expected, the accuracy drops substantially near the real axis.

\vspace{2mm}

\begin{figure}[!ht]
  \centering
  \includegraphics*[width=\figwidth]{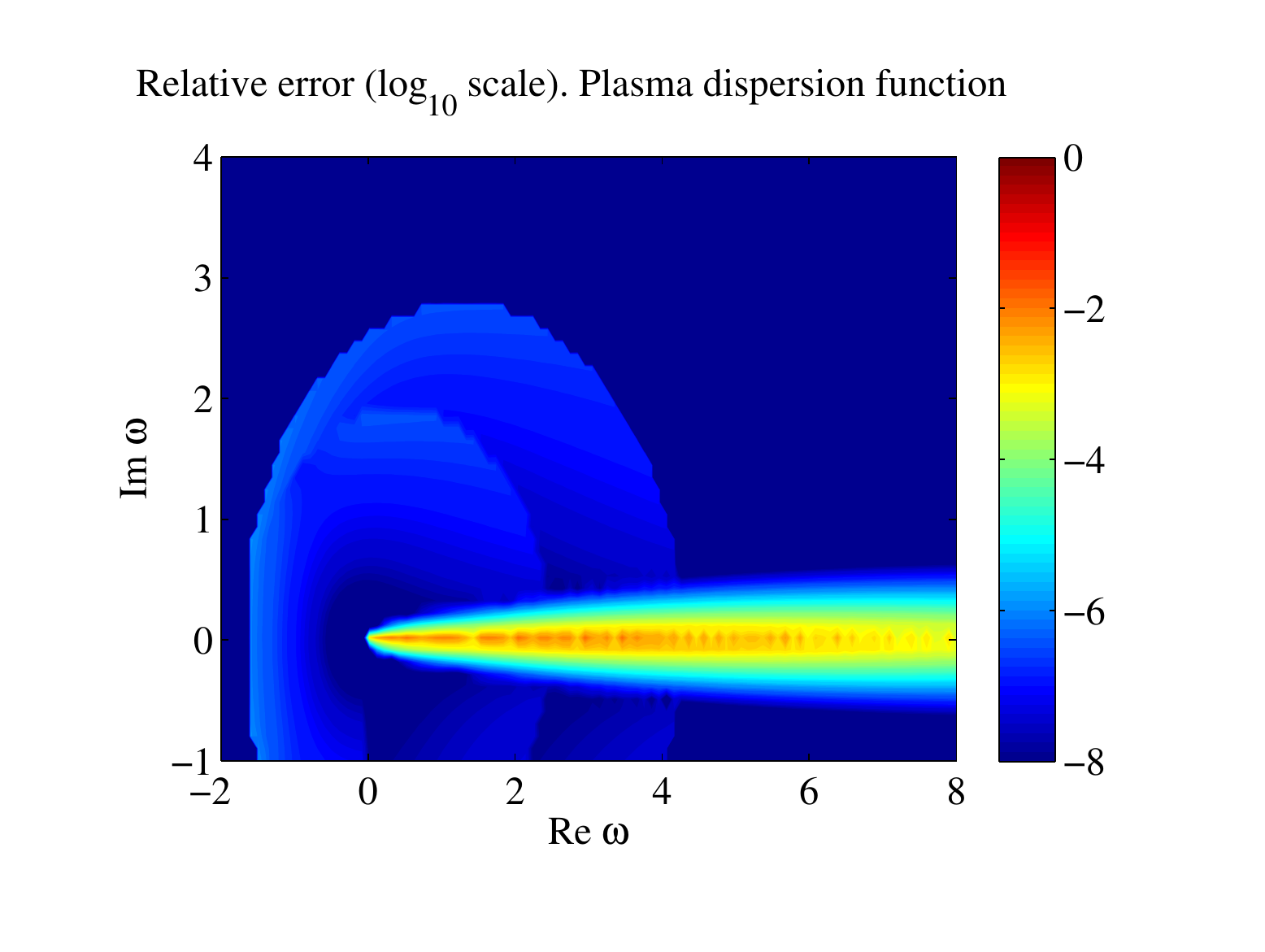}
  \caption[Relative error of the numerical computation of $J$ (plasma dispersion function)]{Relative error of the value of $J$ computed using the plasma dispersion function. The parameters are the same as for figure \ref{fig:Eint_test_exp}.}
  \label{fig:Eint_test_exp_Z}
\end{figure}

A second approximation of the value of $J$ is then computed using the plasma dispersion function through equation (\ref{eq:ZEnInt}) where the two non-singular integrals are approximated using the trapezoidal approximation with the same grid as mentioned above. The relative error is again represented in log scale in figure \ref{fig:Eint_test_exp_Z}.

In this computation and in the case where $\Im \omega$ was negative, the interpolation of the function $G$ (or $g$) to the complex plane at a position $z$ with $\Im z \neq 0$ was replaced by a simple evaluation of the function at the considered point. The negative values of $\Im \omega$ were only included to give a better perspective at the variations of the relative error near the real axis. This explains why the error is approximately symmetric with respect to the real axis.

If one compares figures \ref{fig:Eint_test_exp_Dir} and \ref{fig:Eint_test_exp_Z}, it is clear that the accuracy near the real axis has been enhanced by treating analytically the singular integral. In this region the error which could reach values of the order of $1$ is now below $10^{-2}$.

\vspace{2mm}

\begin{figure}[!ht]
  \centering
  \includegraphics*[width=\figwidth]{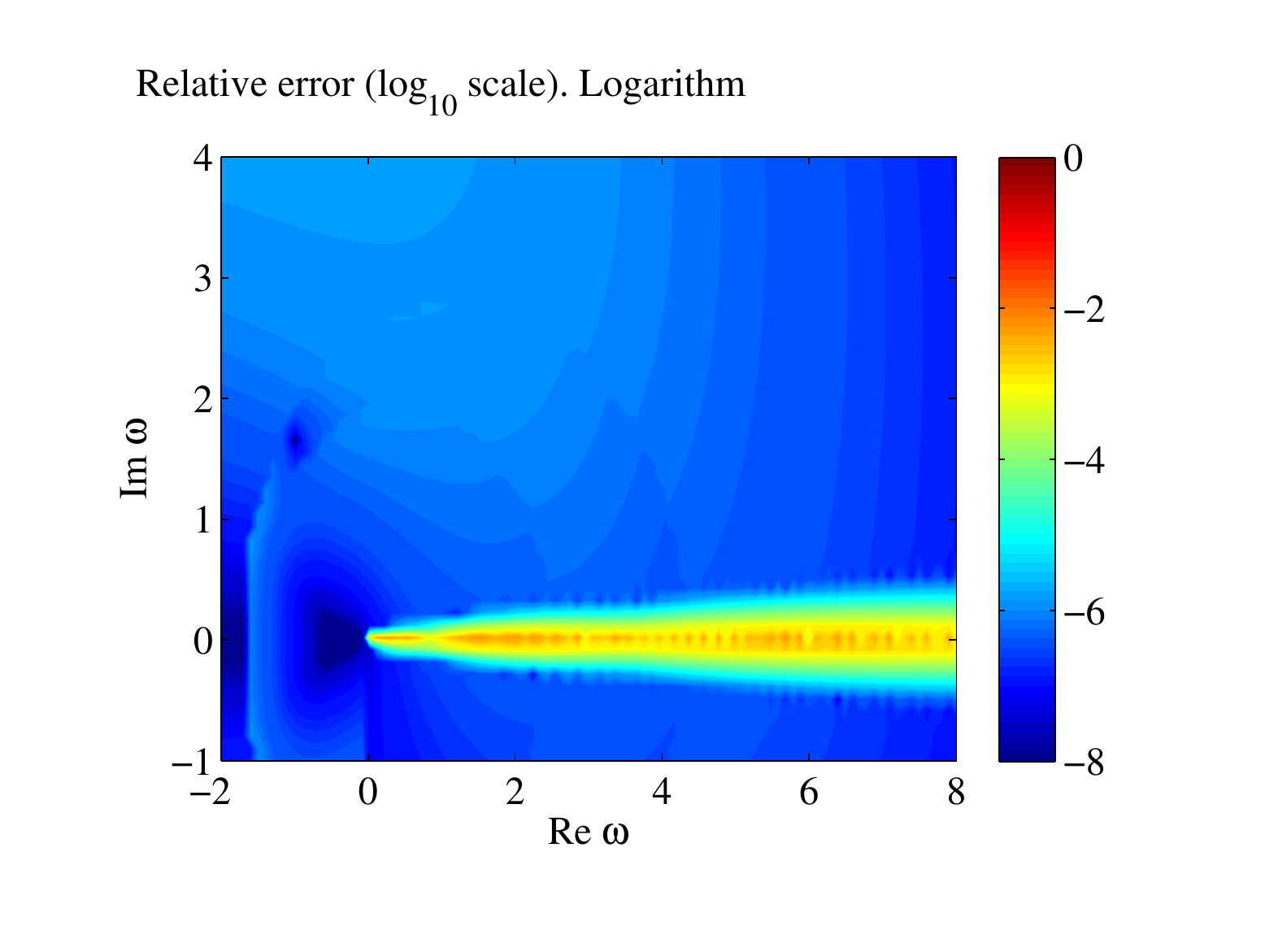}
  \caption[Relative error of the numerical computation of $J$ (complex logarithm)]{Relative error of the value of $J$ computed using the complex logarithm. The parameters are the same as for figure \ref{fig:Eint_test_exp}.}
  \label{fig:Eint_test_exp_Log}
\end{figure}

Finally the integral $J$ is computed using the complex logarithm through equation  (\ref{eq:LogEnInt}). The same remarks on the computation of the integral for values of $\omega$ with $\Im \omega < 0$ mentioned above apply here. The relative error is represented in log scale in figure \ref{fig:Eint_test_exp_Log}.

Once again comparing figures \ref{fig:Eint_test_exp_Dir} and \ref{fig:Eint_test_exp_Log}, an enhancement of the accuracy of the integration method is observed near the real axis using the logarithm method. This is not true further from the axis but the accuracy stays in the acceptable range. The accuracy of the logarithm method and of the plasma dispersion function method are comparable near the real axis.

\subsection{Conclusion}

The methods developed to obtain a better numerical approximation of the integral $J$ defined in equation (\ref{eq:Eint_def_MIKE}) for the MIKE code were able to reduce the error down to acceptable values. In particular, the method using the plasma dispersion function (\ref{eq:ZEnInt}) is very effective for distributions with a maxwellian energy dependence. For more general distributions, the method using the complex logarithm (\ref{eq:LogEnInt}) is effective and very reliable.

\section{Solving the dispersion relation}
\label{sec:solv-disp-relat}

The code MIKE solves the linear dispersion relation for fishbone-like modes using a method first described by Davies \cite{dav86}.

\subsection{Overview}

This method is based on the residue theorem to compute the zeros of an analytic function. The details of the method can be found in the original article by Davies \cite{dav86}. The method is described for the case of searching the zeros inside the unit circle centered at $z=0$ but it can be trivially extended to any circle with any radius and any center.

The problem initially formulated by Davies is to find the zeros of a given complex function $z \rightarrow h(z)$ inside the unit circle $C$ ($|z| < 1$). The integrals $S_k$ are defined as 
\begin{equation}
  \label{eq:Dav_Sn_def}
  S_k = \frac{1}{2i\pi} \int_C z^k \frac{h'(z)}{h(z)} \dd{z}.
\end{equation}
If $h$ has $N_r$ zeros $(z_i)_{i=1 \ldots N_r}$ inside the circle $C$ then these integrals can be computed using the residue theorem which gives
\begin{equation}
  \label{eq:Dav_Sn_val}
  S_n = \sum_{i=1}^{N_r} (z_i)^k.
\end{equation}
It appears that $S_k$ are symmetric functions of the roots $z_i$ such that if the values of $S_k$ for $k \in [0,N_r]$ are known a polynomial function whose roots are $(z_i)_{i=1 \ldots N_r}$ can be easily constructed. And the search for the zeros of any complex function $h$ has been transformed into the search for the roots of a polynomial function, for which efficient algorithms already exist.

\subsection{Implementation}

The first step is to compute the number of zeros inside the unit circle. For this $h$ is evaluated on $n$ equally spaced points along this circle, and $n$ is increased until the change in modulus between two consecutive values of $h$ lies between $1/r_{max}$ and $r_{max}$ and the change in argument is bounded by $\Phi_{max}$ ($\Phi_{max}$ and $r_{max}$ are predefined values which ensure a good accuracy of the method). The overall change of argument of $h$ along the circle is equal to the number of zeros times $2\pi$.

The next step is the computation of an approximate value $\tilde{S}_k$ of the integrals $S_k$ (using those $n$ points), an efficient method is described in \cite{dav86}. One then constructs the polynomial function $P$ associated to the $\tilde{S}_k$, which then gives approximate values $\tilde{z}_i$ for the zeros of the function $h$.

The accuracy of the approximate values of the $z_i$ can be evaluated by computing the values $h(\tilde{z}_i)$. If the accuracy level is not satisfactory, one can either increase the number $n$ of evaluations of $h$ or iterate the process by reducing the radius of the circle and translating its center at the estimated values $\tilde{z}_i$.

\subsection{Testing}

In this section, the numerical implementation of the algorithm described above to find in the complex plane the zeros of an analytic function is tested. The following function is used
\begin{equation}
  \label{eq:Davies_test_f}
  f:z \longrightarrow \sin \left( 5 \pi z - i \frac{\pi}{2}\right) \cos\left( 5 \pi z + i \frac{\pi}{2}\right),
\end{equation}
it possesses many zeros located at $z = (2k + i)/10$ or $z = (2k+1-i)/10$ with $k \in \mathbf{Z}$. The ability of the Davies algorithm to find a large number of zeros at once is described in the original article of Davies \cite{dav86}. It is of little interest here since the solutions of the fishbone dispersion relation are usually well separated.

\vspace{2mm}

The numerical evaluation of all terms of the fishbone dispersion relation on a large number of frequencies can be time-consuming. It is then of interest to look for the best strategy to obtain a solution with the maximum accuracy with a given number of evaluations of the dispersion relation.

The evolution of the accuracy of the algorithm when the number of function evaluations is increased is tested. In the first test only the number $n_{eval}$ of points where the function $f$ is evaluated is modified. In the second test, a first evaluation $\tilde{z}_{1}$ of the position of the zero is obtained using $n_{eval}/2$ points. The solution is then refined by using $n_{eval}/2$ points along the circle of center $\tilde{z}_1$ and radius $r/2$.
\begin{figure}[!ht]
  \centering
  \includegraphics*[width=\figwidth]{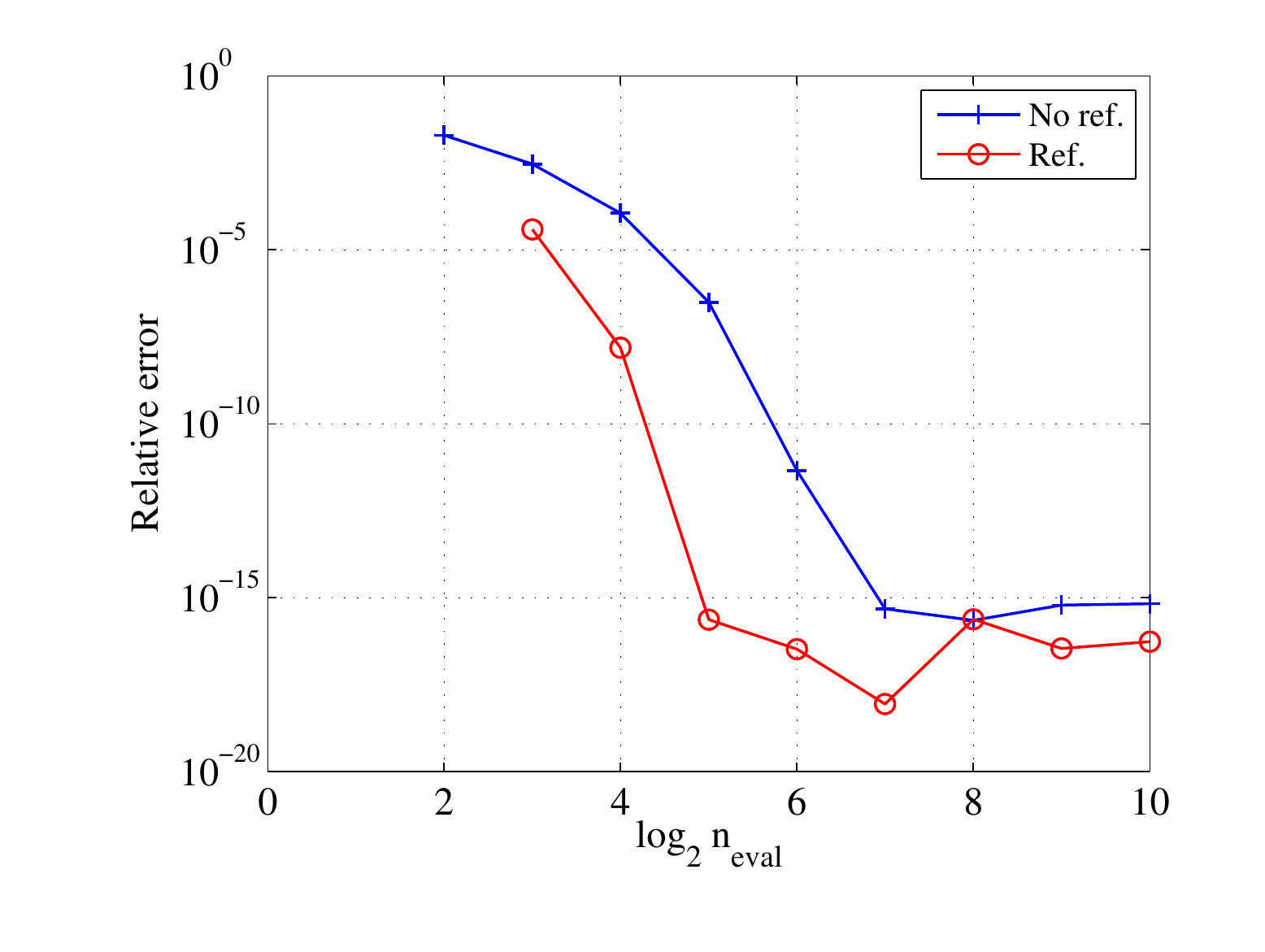}
  \caption[Relative error for the numerical solution of $f(z)=0$]{$\log-\log$ plot of the relative error on the value of the solution of $f(z) = 0$ versus the total number of function evaluations. The following parameters are set, $r_{max} = 6.1$, $\Phi_{max} = 3\pi/4$. Solutions are looked inside the circle of center $z_g = -0.03 + i\,0.12$ and radius $r = 0.05$. See text for an explanation of the difference between the two curves.}
  \label{fig:Davies_test_neval}
\end{figure}
The results of the two tests are compared in figure \ref{fig:Davies_test_neval} with the blue curve with the crosses corresponding to the first test while the red curve with the circles corresponds to the second test. The accuracy of the numerical solution is indeed enhanced when the number of function evaluations is increased. It appears also that at a given number of function evaluations, the strategy which consists in refining the solution by reducing step by step the radius of the circle has a better accuracy than the one which consists in maximizing the number of function evaluations on a given circle.

\comment{

\begin{figure}[ht!]
  \centering
  \includegraphics*[width=\figwidth]{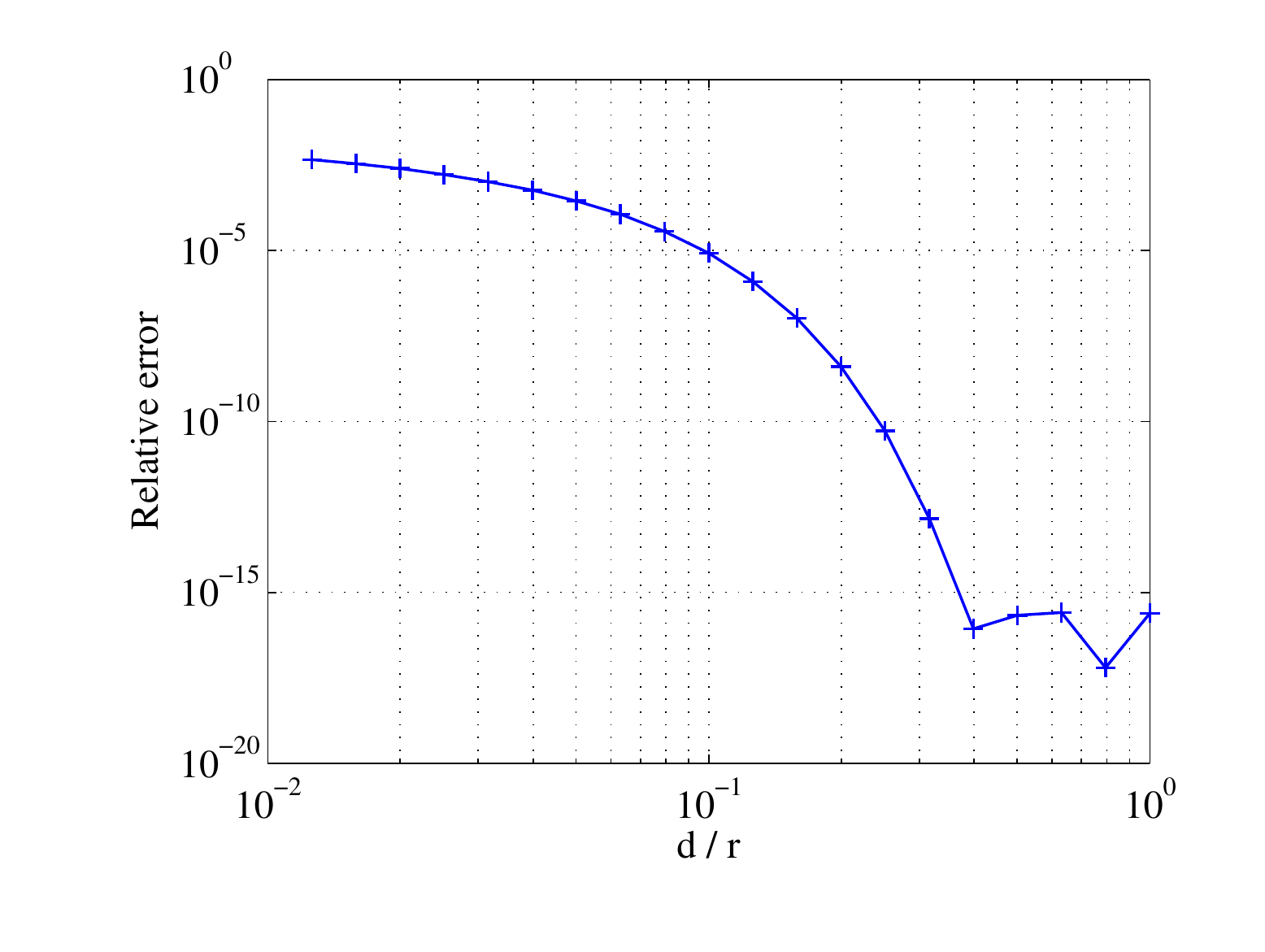}
  \caption{}
  \label{fig:Davies_test_dist}
\end{figure}

\subsection{Remarks}

It is important to point out that
\begin{itemize}
\item the number $n$ influences the precision of the computation of the integrals and thus the precision in the computation of the zeros, so that choosing more restrictive conditions for the change in modulus and argument will improve the accuracy of the method;
\item because the method is applicable to any analytical function, $h$ can be composed with any conformal map of the complex plane, effectively changing the shape of the circle to any regular contour. This can be used to look for zeros inside circles with radii bigger, or smaller, than $1$ and centered at any point in the complex plane, but also to use shapes that are more complex, like an ellipse.
\end{itemize}
}

\section{Verification of the MIKE code}
\label{sec:verif-mike-code}

This section deals with the series of tests that were performed on the $\dWk$ module to ensure the verification of the code. The distribution functions that are used are the ones that are described in a series of article by White (see for example reference \cite{whi89}). In these articles the analytical expressions of the kinetic term are also found. To get those expressions, some approximations were made and were incorporated in the MIKE code as options for the sake of benchmarking.

\subsection{Analytical expressions}

\subsubsection{Test distributions}

We consider an electronic distribution with an energy dependence corresponding to a slowing down distribution  such that $F_e$ is proportional to $p^{-3}$ for energies in the range $[E_{0}, E_{m}]$. The distribution is strongly anisotropic such that there is a single value for the variable $\hat{\lambda} = \mu B_0/E$, this value $\hat{\lambda}_0$ is chosen such that electrons near the $q=1$ surface are in the barely trapped region and have a reversed drift frequency. Finally the distribution function is linearly increasing with radius to provide a positive radial gradient necessary for electron-driven fishbone destabilization.
\begin{equation}
  \label{eq:SD_dist_test}
  \bar{F}_e(\rho,\hat{\iota},\bar{p}) = \alpha_n \rho \, \delta(\hat{\lambda} - \hat{\lambda}_0) \, \bar{p}^{-3} \left(\log \frac{E_m}{E_0}\right)^{-1}, \quad \mbox{for $E_0 < E < E_m$},
\end{equation}
where $\alpha_n$ is a parameter to control the fast electron density. The normalized electronic density is then
\begin{equation}
  \label{eq:SD_nebar_test}
  \bar{n}_e(\rho) = 2 \pi \left(\frac{\tilde{q}(\rho) B_m(\rho)}{\hat{q}(\rho) B_0}\right) \alpha_n \rho \: \bar{\tau}_b(\rho,\hat{\lambda}_0).
\end{equation}

\subsubsection{Equilibrium}

We consider a low-beta high aspect ratio equilibrium such that the flux surfaces are circular and concentric. In this configuration, the coordinate $\rho$ in the MIKE code corresponds to the minor radius of the flux surfaces normalized to its value on the $q=1$ surface. Only lowest order terms in $\varepsilon_s = r_s/R_0$ are retained in the expressions for $\dWk$, $\dW_{int,h}$, $\omega_b$ and $\omega_d$. Therefore the energy derivative term in $\dWk$ is neglected since it contributes to the sum $\dWk + \dW_{int,h}$ only to order $\varepsilon_s$ compared to the radial derivative term. The approximate expression for the normalized density is
\begin{equation*}
  \bar{n}_e(\rho) = 2 \pi \alpha_n \rho \: \bar{\tau}_b(\rho,\hat{\lambda}_0).
\end{equation*}

The safety factor profile is assumed to be parabolic and monotonically increasing with a $q=1$ surface located at $r=r_s$. Therefore equation (\ref{eq:inertia_finite_s}) is used to compute the inertia term $\delta I$. The pressure profile is monotonically decreasing, has a parabolic dependence over $r$ and the central pressure is chosen such that the ion-diamagnetic frequency at the $q=1$ surface verifies
\begin{equation*}
  \wi = \frac{1}{10}\, \omega_d(r_s,\hat{\lambda}_0,E_m)
\end{equation*}

\subsubsection{Additional approximation}

The radial dependence of both $\omega_d$ and $\bar{\Omega}_d$ are neglected in the expressions of $\dWk$ and $\dW_{int,h}$. This strong approximation (in particular for $\omega_d$ which is proportional to $1/r$) is present in reference \cite{whi89} and can be understood as the fact that the mode interacts only with particles located in a region where the radial gradient is strongly enhanced. In this study, the linear radial dependence for $\bar{F}_e$ is retained since it is less numerically demanding.

\subsubsection{Expressions for \texorpdfstring{$\dWh$}{dWh}}

One then obtains the following expression for $\dWh = \dWk + \dW_{int,h}$,

\begin{equation}
\label{eq:SD_dWh_test}
  \dWh = \sqrt{2} \pi^2 \alpha_n \frac{2\mu_0 n_{ref} E_{ref}}{B_0^2} C \left(-\frac{\bar{\Omega}_{d,s}}{\varepsilon_s}\right) \frac{\omega}{\omega_{dref}} \log\left(\frac{1-\omega_{dm}/\omega}{1-\omega_{d0}/\omega}\right),
\end{equation}
where we have defined $\bar{\Omega}_{d,s} = \bar{\Omega}_{d}(r_s,\hat{\lambda}_0)$, $\omega_{dm} = \omega_{d}(r_s,\hat{\lambda}_0,E_m)$, $\omega_{dref} = \omega_{d}(r_s,\hat{\lambda}_0,E_{ref})$, $\omega_{dm} = \omega_{dref} (E_{m}/E_{ref})$, $\omega_{d0} = \omega_{dref} (E_{0}/E_{ref})$ and the constant $C$ corresponds to the following integral
\begin{equation*}
  C = \int_0^1 \rho \, \bar{\tau}_b(\rho,\hat{\lambda}_0) \dd{\rho}.
\end{equation*}

\subsection{Comparison with numerical results}

To compare the results of the computation of $\dWh$ with MIKE and the analytical expression \eqref{eq:SD_dWh_test}, the integral $C$ was first numerically computed in double precision.

\subsubsection{Computation of \texorpdfstring{$\dWh$}{dWh}}

 The three integrals ($\rho$, $\iota$ and $E$) were tested separately by incorporating the analytical expressions of the integrals in the code. The code was successively tested with $1$, $2$ or $3$ integrals computed numerically at the same time. All tests showed a good convergence for the computation of the integral, the approximation error decreasing as the total number of grid points was increased.
\begin{figure}[ht!]
  \centering
  \begin{subfigure}{0.495\textwidth}
    \includegraphics*[width=\figwidth]{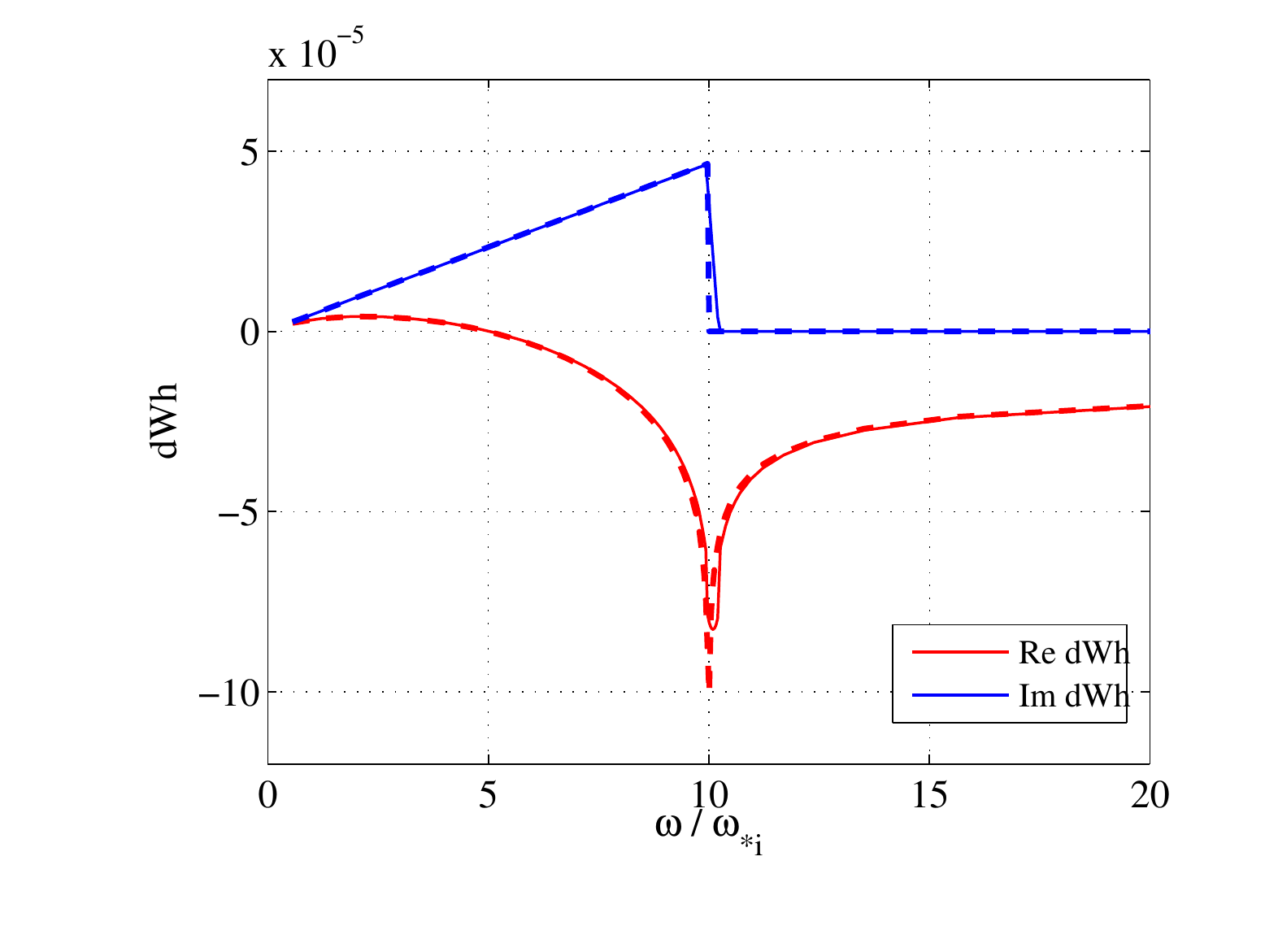}
    \caption{Real and imaginary part of $\dWh$}
    \label{Bench_SD_l0_riE_omg}
  \end{subfigure}
  \begin{subfigure}{0.495\textwidth}
    \includegraphics*[width=\figwidth]{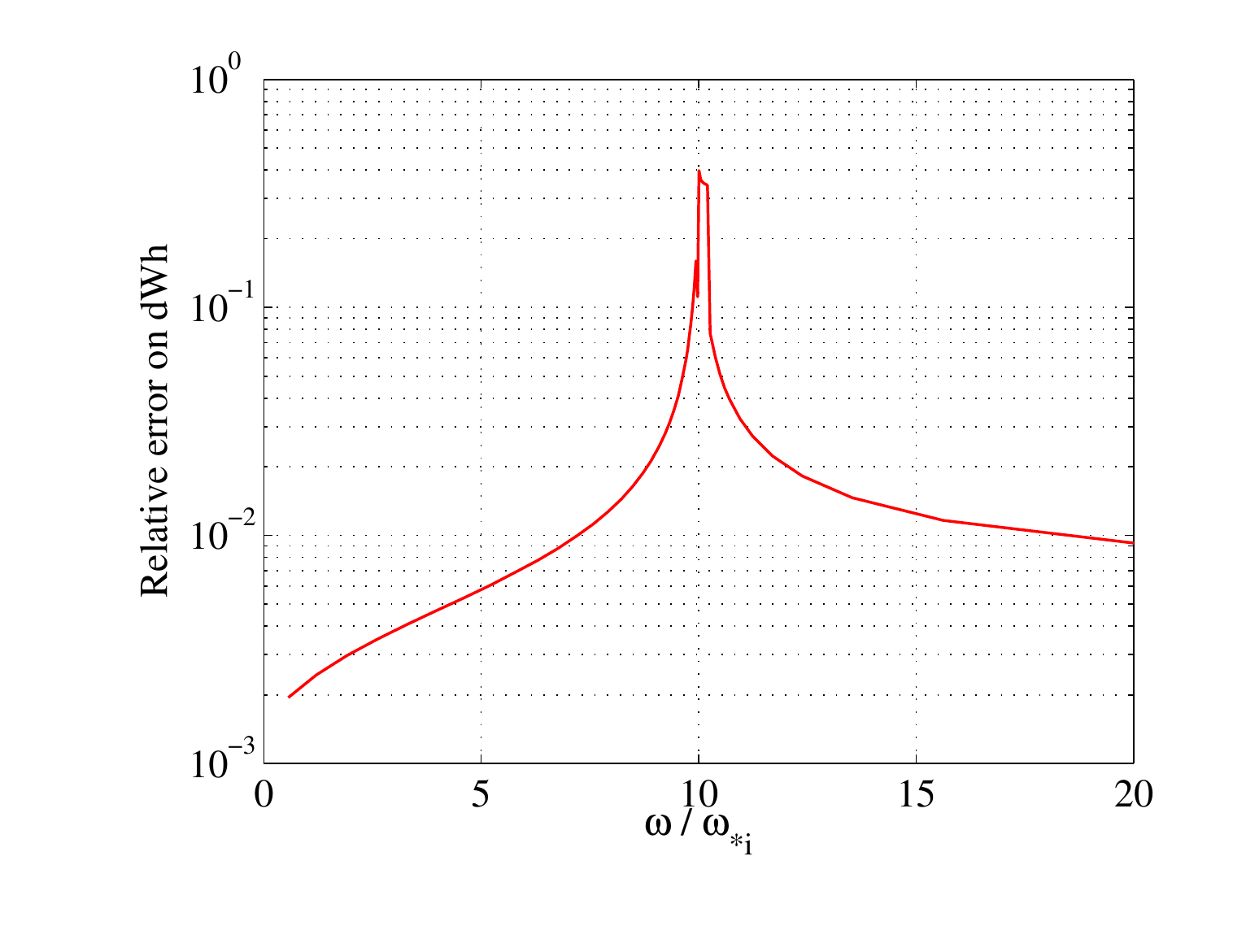}
    \caption{Relative error on the value of $\dWh$}
    \label{Bench_SD_l0_riE_err}
  \end{subfigure}
  \caption[Comparison of the numerical value of $\dWh$ with its analytical expression]{Comparison between the results from the computation of $\dWh$ with MIKE and its analytical expression. The computation is done for a range of real frequencies $\omega$ between $0$ and $20\, \omega_{*i}$. All integrals are computed numerically using a total number of $2.0 \,10^6$ grid points. The parameter $\alpha_n$ was set to $1$.}
  \label{Bench_SD_l0_riE}
\end{figure}
On figure \ref{Bench_SD_l0_riE} is shown on the left hand side the results of the fully numerical computation of $\dWh$ for different frequencies compared to the analytical expression and on the right hand side the relative error between the two values. The agreement is very good except in the region where the expression for $\dWh$ is discontinuous. This discontinuity comes in fact from the discontinuity in the distribution function at $E = E_m$, since distributions reconstructed from experimental conditions generally do not exhibit such singularities this region is of no particular interest. 

\subsubsection{Solution of the fishbone dispersion relation}

The whole MIKE code was then tested by studying the evolution of the solution of the fishbone dispersion relation when the parameter $\alpha_n$ was increased. The ``analytical'' solution was computed numerically up to double precision by using a standard solver. The MIKE solution was obtained using the solver based on the Davies method.
\begin{figure}[ht!]
  \centering
  \begin{subfigure}{0.495\textwidth}
    \includegraphics*[width=\figwidth]{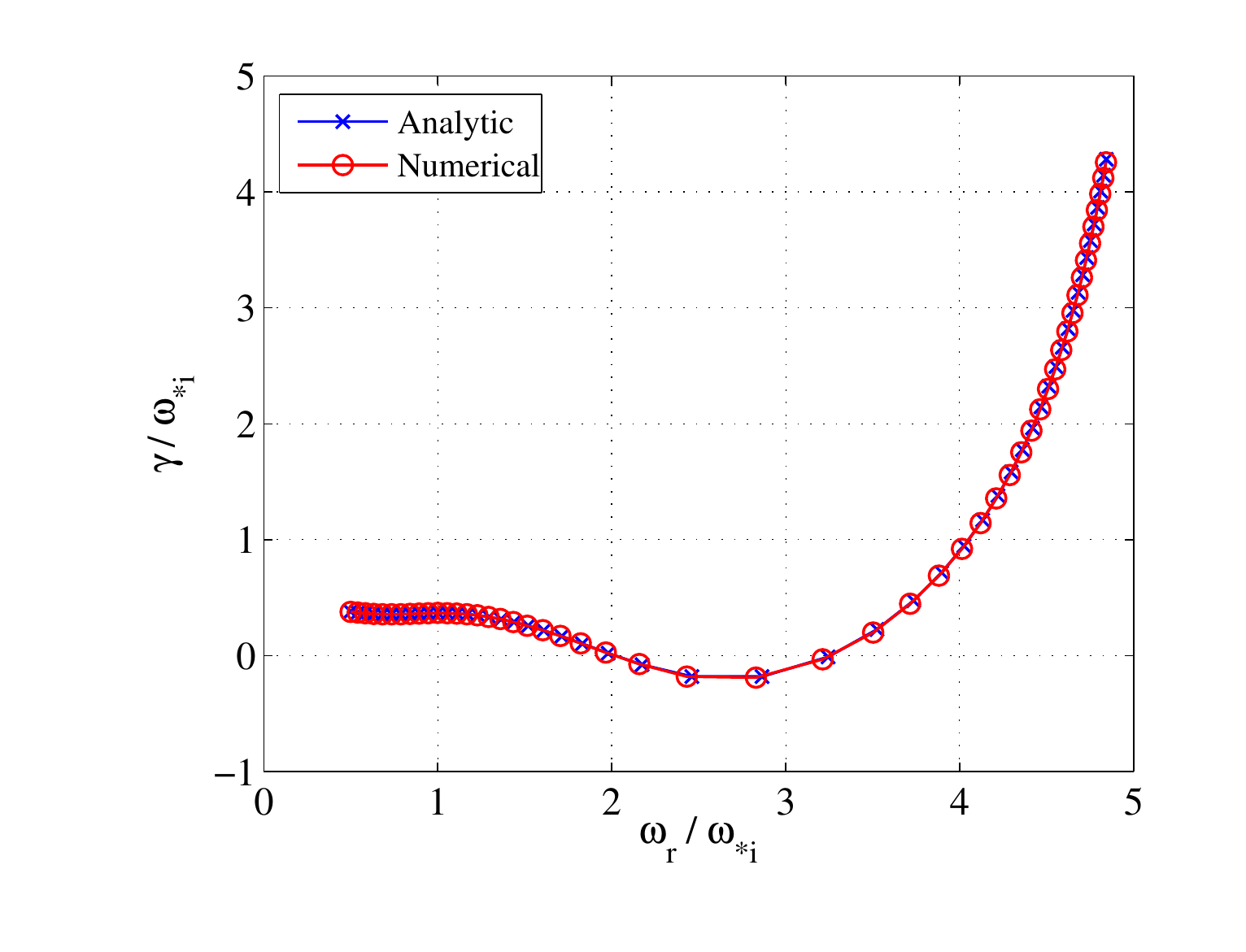}
    \caption{Evolution of $\omega_r/\omega_{dm}$ and $\gamma/\omega_{dm}$.}
    \label{Bench_SD_l0_nh_scan_omg}
  \end{subfigure}
  \begin{subfigure}{0.495\textwidth}
    \includegraphics*[width=\figwidth]{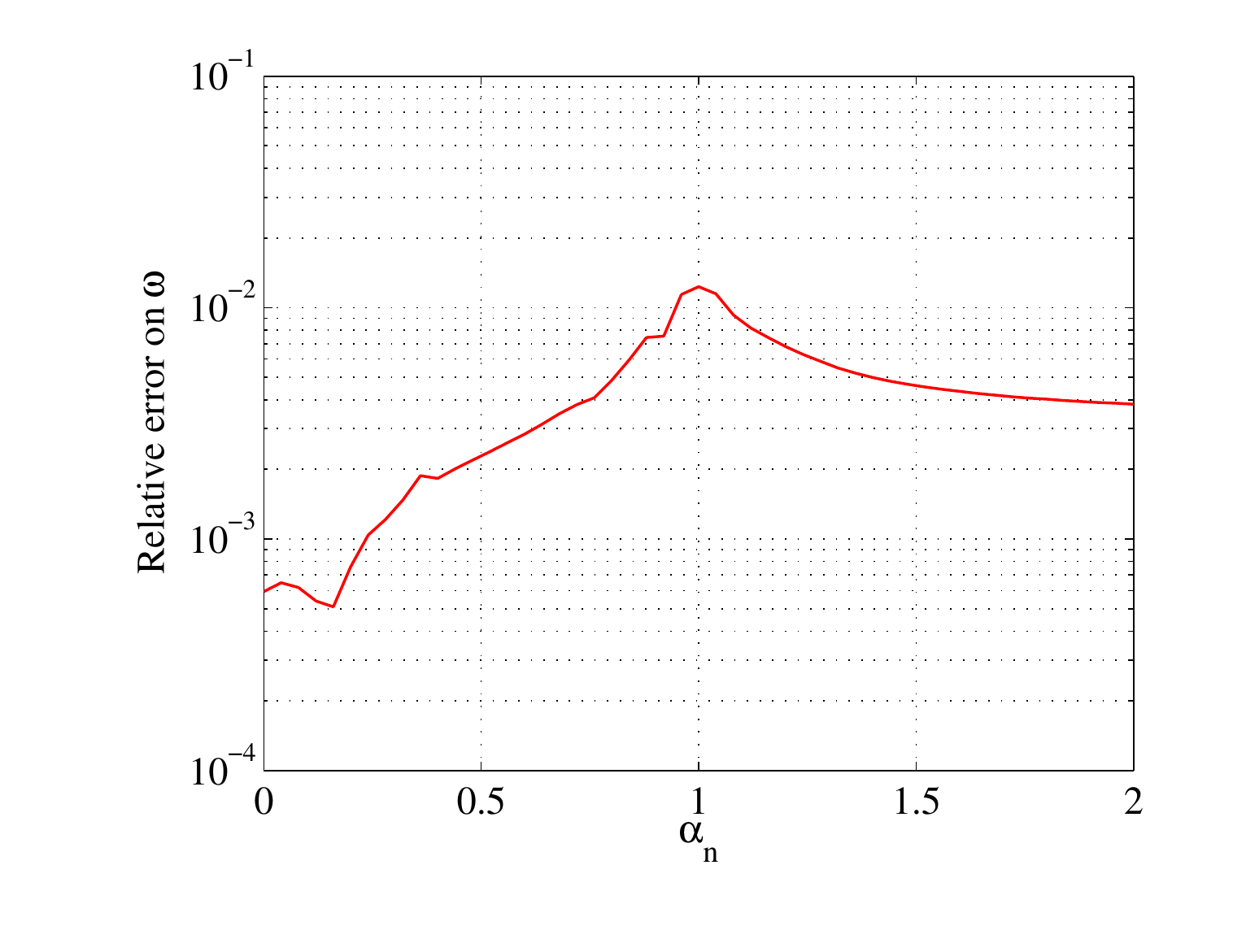}
    \caption{Relative error}
    \label{Bench_SD_l0_nh_scan_err}
  \end{subfigure}
  \caption[Comparison of the numerical solution with its analytical expression]{Comparison of the solution given by the MIKE code (red circles) and the analytical solution (blue crosses), the relative error between the 2 solutions is shown on the right panel.}
  \label{Bench_SD_l0_nh_scan}
\end{figure}
The results are shown on figure \ref{Bench_SD_l0_nh_scan}. The relative difference between the two solutions is less than $1\%$ for the whole scan. In this simulation, the integration in pitch-angle was computed analytically in order to save computing time. 

\subsection{Conclusion}

The MIKE code was successfully verified using analytical distributions. These distributions were designed to obtain simple analytical expressions but they were also very demanding for the code because of the presence of many singularities. Using the code with more standard distributions (for which simple analytical expressions are not available) showed that the convergence of the integral computation is much faster and does not need as many grid points.

\section{Summary}

The MIKE code which has been developed to study the stability of electron-driven fishbone modes is described. The implementation of the model recalled in section \ref{sec:structure-mike-code} is described in section \ref{sec:normalization-mike}. The different techniques used to compute the resonant integrals are showed in section \ref{sec:ResIntComp}. The accuracy of the different techniques is also tested. In section \ref{sec:solv-disp-relat} we describe an implementation of the method to find the zeros of the dispersion relation which was originally developed by Davies \cite{dav86}. Finally we show in section \ref{sec:verif-mike-code} how the MIKE code has been successfully benchmarked against simplistic analytical distributions.


\chapter{Finite \texorpdfstring{$k_\Vert$}{k\_par} effects on the stability of electron-driven fishbones}
\label{sec:EFB_stability_circ}

In this chapter, we show how the modification of the fishbone dispersion relation to take into account the parallel motion in the resonance with passing particles derived in chapter \ref{cha:FDR_derivation} affects the stability of electron-driven fishbones.

The resonance condition of trapped particles $\omega = \nn \omega_d$ becomes $\omega = \nn(q - 1)\omega_b + \nn \omega_d$ for passing particles ($\omega_d$ is the toroidal precession frequency and $\omega_b$ is the bounce frequency of particles). In previous works this additional term was usually neglected by assuming $q \sim 1$ \cite{zon07,wan07}. For energetic electrons $\omega_b$ is much larger than $\omega_d$ such that if $q$ gets close to $1$ then all terms of the resonance condition can be of similar weight $\omega \sim \nn\omega_d \sim \nn(q-1)\omega_b$. It is somewhat different from the work of Fredrickson et al. \cite{fre03} for ion fishbones where $\omega - \omega_d \sim \omega_b \ll \omega$.

We first compare the influence of trapped and passing particles on the linear stability of electron-driven fishbones using analytical distributions found in previous works by White et al.\cite{whi89}, Sun et al. \cite{sun05} or Wang et al. \cite{wan07}. The analysis was performed using the code MIKE which implements this model. It shows that energetic barely circulating electrons can resonantly interact with the internal kink even at low frequency ($\omega < \nn\omega_d$). This seems in agreement with a recent analysis of observations on the Tore Supra tokamak where electron-driven fishbones with a low $\omega/\nn$ ratio were measured \cite{gui12,gui11}.

The same analysis was then performed using a family of more realistic analytical distributions which were chosen to model those obtained in discharges heated with ECRH using a minimum number of parameters. The choice of ECRH over LHCD is justified by the fact that in the case of ECRH the energetic electrons do not generate any toroidal current and therefore the stability of the internal kink is only modified by the addition of the energetic particle term $\dWh$ to the dispersion relation and not by a modification of the $q$-profile. The influence of the safety factor profile is investigated separately.

\section{Linear theory of electron-driven fishbones}

\subsection{The dispersion relation}

The dispersion relation of the internal kink mode in the presence of energetic particles has been derived in chapter \ref{cha:FDR_derivation}. It can be written in the following form
\begin{equation}
  \label{eq:FDR}
  \delta I = \dWf + \dWh(\omega),
\end{equation}
where $\dWf$ and $\dWh$ are the respective contribution of the thermal bulk and of the energetic component of the plasma. $\delta I$ is called the inertia term which accounts mainly for the contribution of the so-called inertial layer and can take several forms (see appendix \ref{cha:inert-term-fishb}). In this chapter we consider expressions for $\delta I$ including bi-fluid effects in the limit of vanishing resistivity \cite{cop66,ara78} and of kinetic effects of thermal ions \cite{gra00} in a single inertial layer for low-frequency modes \cite{zon06a}. The $q$-profile is either monotonic with the existence of a $q=1$ surface at $r=r_s$ or inversed in the central region with a minimum value $q_{min} \sim 1$ located at $r=r_s$.

\subsection{Fast particle contribution}

The fast particle contribution to the fishbone dispersion relation can be written
\[ \dWh(\omega) = \dWk(\omega) + \delta\hat{W}_{f,h}\]
where $\dWk$ accounts for all resonant effects between the fast particles and the mode and $\delta\hat{W}_{f,h}$ is the contribution of fast particles to the $(\vb{\xi}\cdot \nabla p)$ term of the usual MHD energy. Their expressions in the limit of zero-orbit width and for a low-beta circular equilibrium are equations \eqref{eq:dWk_CCFS_app_l2} and \eqref{eq:dWint_CCFS_app_l2} which can be written as
\begin{equation}
  \dWk = - \frac{\pi}{2} \frac{\mu_0}{{B_0}^2} \bigg{\langle} E^2 \bar{\Omega}_d^2 \frac{\omega \, \partial_E \ln F_h - \omega_* \, \partial_{r} \ln F_h }{\omega - (q-1)\omega_b\delta_P - \omega_d}\bigg{\rangle}_{\vc{x},\vc{p}}
  \label{eq:dWk_gen}
\end{equation}
and 
\begin{equation}
  \delta\hat{W}_{f,h} = \frac{\pi}{2} \frac{\mu_0}{{B_0}^2} \bigg{\langle} R_0 \bar{\Omega}_d E \, \partial_{r} \ln F_h \bigg{\rangle}_{\vc{x},\vc{p}}.
  \label{eq:dWint_gen}
\end{equation}
with \(\langle \mathcal{A} \rangle_{\vc{x},\vc{p}} = V^{-1}\int \dd^3 \vc{x} \dd^3 \vc{p} \mathcal{A} F_h \), $\vc{x}$ is the position in real space and $\vc{p}$ in momentum space, the integral is limited to the space inside the $q=1$ surface of total volume $V = 2 \pi^2 r_s^2 R_0$, $F_h$ is the distribution function of fast particles of mass $m_h$ and charge $e_h$, $\omega_b$ and $\omega_d$ are the bounce-frequency and toroidal drift frequency of fast particles (see appendix \ref{sec:part_dynamics} for expressions of $\omega_b$ and $\omega_d$ in circular concentric geometry and zero orbit width limit), $\bar{\Omega}_d$ is defined by $\omega_d = (q E \bar{\Omega}_d)/(e_h B_0 R_0 r)$ where $E$ is the energy of the particle, $\omega_* = q/(e_h B_0 r)$. Finally $\delta_P$ is equal to $1$ for passing particles and $0$ for trapped particles. Expression (\ref{eq:dWk_gen}) was obtained by neglecting the effect of collisions on the perturbed electronic distribution. This is valid if $\nu_{\rm{dt}} \ll \omega,\omega_d$, where $\nu_{\rm{dt}}$ is the de-trapping frequency of energetic particles and is given by \cite{hin76}
\begin{equation}
  \nu_{\rm{dt}} = \frac{ e^4 Z_{eff} n_e \ln \Lambda}{4 \pi \varepsilon_0^2 m_e^{1/2} E^{3/2}}\frac{R_0}{r},
\end{equation}
For typical parameters used in the simulations presented in the section \ref{sec:ECRH_dis}, with particles of energy about $100 \ \mathrm{keV}$, one has $\nu_{\rm{dt}} \sim 0.5 \mathrm{kHz}$ which is much smaller than the typical drift frequency at the same energy $\omega_d \sim 10 \mathrm{kHz}$.

Expression (\ref{eq:dWk_gen}) implies that the resonance happens when the frequency of the mode is close to the toroidal drift frequency of the particles and that the source of the instability lies in the radial gradient of the distribution function. Unlike the ion case, electron-driven fishbones need an inversed radial profile of the electronic distribution function $\partial_r F_s > 0$. Resonant electrons must have a reversed toroidal drift $\bar{\Omega}_d < 0$. Hence only barely trapped or passing electrons can resonate \cite{zon07}.

\begin{figure}[!ht]
  \centering
  \includegraphics*[width=\figwidth]{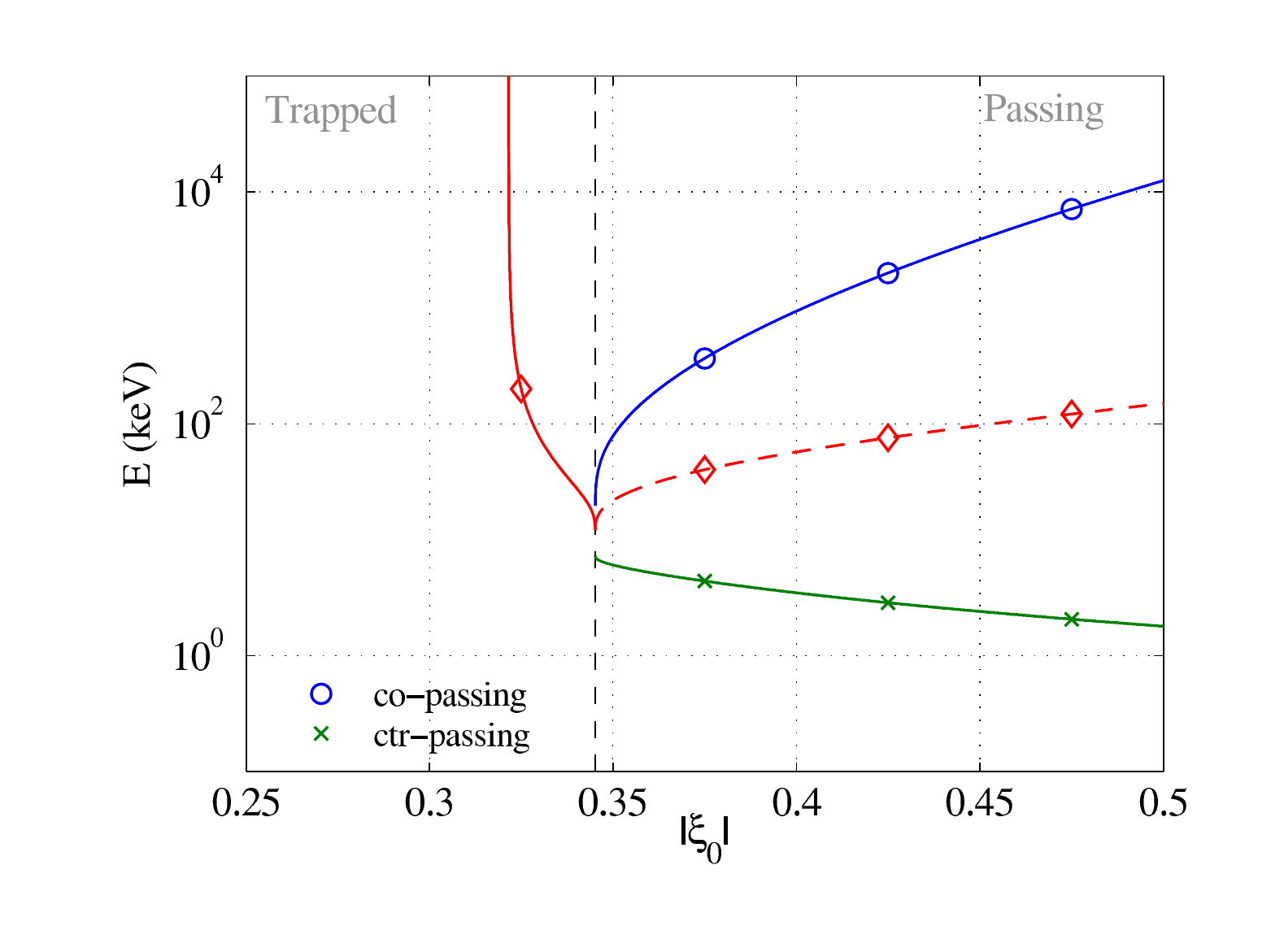}
  \caption[Energy of resonant particles versus pitch-angle]{Energy of resonant particles versus $|\xi_0| = \left|v_\Vert/v\right|_{\theta = 0}$. Circles correspond to co-passing particles while crosses correspond to counter-passing particles. The dotted line corresponds to the resonant energy of passing particles using $\omega = \omega_d$ as the resonance condition. Parameters are $B_0 \simeq 3.1\ \mathrm{T}$, $R_0 \simeq 2.46\ \mathrm{m}$, $r \simeq 0.16\ \mathrm{m}$, $1-q = 2.\ 10^{-3}$, $s = 4.\ 10^{-2} $, $\omega = 2.0\, \mathrm{kHz}$. If $1-q$ changes sign, then curves for co-passing  and counter-passing particles are switched.}
  \label{fig:Resonant_Energies_art}
\end{figure}

The term associated to $\delta_P$ comes from the parallel part of  the usual resonance condition of the Landau effect $\omega = \langle \vc{k}\cdot \vc{v} \rangle$ where the brackets stand for orbit-averaging. For trapped particles, the parallel velocity averages to $0$ over one poloidal orbit, while for passing particles one has $\langle v_\Vert \rangle = q R_0 \omega_b$ and $k_\Vert = (q - 1)/(q R_0)$. This term was usually neglected in previous studies using the argument that $q$ is close to $1$ ($k_\Vert$ small) or restricting the study to barely passing particles ($\omega_b$ small). Since $\omega_d/\omega_b \sim \rho_L/r$, the argument could stand for passing fast ions which have a ratio $\rho_L/r$ of the order of unity, but for passing fast electrons $\rho_L/r \ll 1$ and $(q - 1) \omega_b$ can be of the order of $\omega_d$ or much greater than $\omega_d$ for well-passing electrons. As we will see later on, it turns out that this term does have a significant influence on the linear stability of the fast electron driven fishbone mode. It breaks the symmetry of the resonance condition between co-passing and counter-passing particles, producing a branch at low energies and a branch at high energies. 
This can be seen on figure \ref{fig:Resonant_Energies_art} where the energy of resonant particles has been plotted versus pitch-angle for a standard case. The dependence of the energy of resonant particles over frequency is also weakened. 

At $\omega = 0$, the total contribution of fast particles is 
\[
  \dWh(0) = \frac{\pi}{2} \frac{\mu_0}{{B_0}^2} \bigg{\langle} \frac{(q-1)\omega_b\delta_P}{(q-1)\omega_b\delta_P + \omega_d} R_0 \bar{\Omega}_d E \, \partial_{r} \ln F_h \bigg{\rangle}_{\vc{x},\vc{p}}
\]
such that the contribution of trapped particles vanishes at low frequency.

\subsection{Solving the dispersion relation}

In the absence of fast particles ($\dWh = 0$) and neglecting $\wi$, according to equations (\ref{eq:FDR}) and (\ref{eq:inertia_omi}), the internal kink is unstable for $\dWf < 0$ and stable for $\dWf > 0$. The effect of $\wi$ on the growth rate is stabilizing, and creates a window around $\dWf = 0$ where the mode is marginally stable. For the frequency, there is a global shift toward frequencies in the ion diamagnetic direction ($\omega$ has the same sign as $\wi$).

If fast particles are present, global trends can still be identified. The real part of $\dWh$ will mainly influence the stability and growth rate of the mode, in a similar way to $\dWf$, a negative value being destabilizing. The imaginary part of $\dWh$ will mainly influence the frequency of the mode. It can be linked to the power exchanged between the particles and the mode and is balanced by the imaginary part of $\delta_I$ which is linked to the damping of the mode by coupling to the Alfv\'en continuum. A bigger value for $\Im\, \dWh$ corresponds to a higher frequency. Due to the form of (\ref{eq:inertia_omi}), the ion diamagnetic direction is the direction favored by the mode as it experiences less damping.

\section{Unidirectional distributions}
\label{sec:Drift_Bounce_Res}

Intrinsic properties of the electron-driven fishbone mode are studied using analytic unidirectional distributions. These distributions were used to verify the code MIKE \cite{mer10} against analytical results \cite{whi88}. Although they do not reflect realistic distributions, they are of interest to determine the specific contributions of various classes of electrons.

Let us consider a model maxwellian distribution
\[F_h(p,\hat{\lambda},r) = n_h H(r-r_h) \delta_{\hat{\lambda}_h}(\hat{\lambda}) \frac{\exp \left(-p^2/2m_hk_BT\right)}{{m_h}^{3/2} (k_BT)^{3/2}}\]
with $\hat{\lambda} = \mu B_0/E$ ($\mu$ is the magnetic moment of the particle), $\delta_{z}(x) = \delta(x-z)$, $\delta$ being the Dirac distribution and $H(r-r_h) = 1$ if $r>r_h$, $0$ otherwise. The total contribution coming from the energy derivative of the distribution function to $\dWh$ is of the order of the contribution from its radial derivative multiplied by $\varepsilon = r/R_0$, so we will neglect this term in the computation of $\dWh$. The radial derivative of $F_h$ (with $E$ and $\mu$ kept constant) takes the form of a double Dirac distribution both in radial position and pitch-angle.
\[\partial_r F_h(p,\hat{\lambda},r) = n_h \delta_{r_h}(r) \delta_{\hat{\lambda}_h}(\hat{\lambda}) \frac{\exp \left(-p^2/2m_hk_BT\right)}{{m_h}^{3/2} (k_BT)^{3/2}}.\]
We now define $\omega_{dT}$ and $\omega_{bT}$ the drift and bounce frequency of the particles located at $r=r_h$ with energy $k_B T$ and $\hat{\lambda} = \hat{\lambda}_h$. Introducing $\beta_h = \langle E \rangle_{\vc{x},\vc{p}}$, one has 
\begin{equation}
  \dWh = \hat{C}_0 \beta_h \frac{G(\bar{p}_+) + G(-\bar{p}_-)}{\bar{p}_+ - \bar{p}_-}
\end{equation}
with  $\hat{C}_0$ a normalization constant, $G(x) = x^3 \left(1/2 + x^2 + x^{3} Z(x)\right)$ and $\bar{p}_{\pm}$ are the two roots of the second degree polynomial $\omega - \delta_P(q(r_h)-1)\omega_{bT} \, \bar{p} - \omega_{dT} \, \bar{p}^2$ with $\bar{p} = p/\sqrt{2 m_h k_B T}$.

The expression for trapped particles found in \cite{whi88} is recovered since in this case $\bar{p}_+ = - \bar{p}_- = \sqrt{\omega/\omega_{dT}}$. As a result of the symmetry breaking between co- and counter-passing particles, their contribution to $\dWh$ is shifted toward lower frequencies. The difference in energy between the two branches and the strength of the frequency dependence are related to the ratio $(q-1)\omega_{bT}/\omega_{dT}$. If it is much lower than $1$, then the same behavior as for trapped particles is recovered. If it is comparable to $1$, then the same behavior is expected at zero frequency than for trapped particles with a frequency approaching $\omega_{dT}$. Finally if the ratio is much larger than $1$ ($> 2$ is enough), then for frequencies in the range of $\omega_{dT}$, the high energy branch is much larger than $k_BT$ and the low energy branch is much lower than $k_B T$ so that $\dWh$ will be almost real.

\begin{figure}[!ht]
  \centering
  \includegraphics*[width=\figwidth]{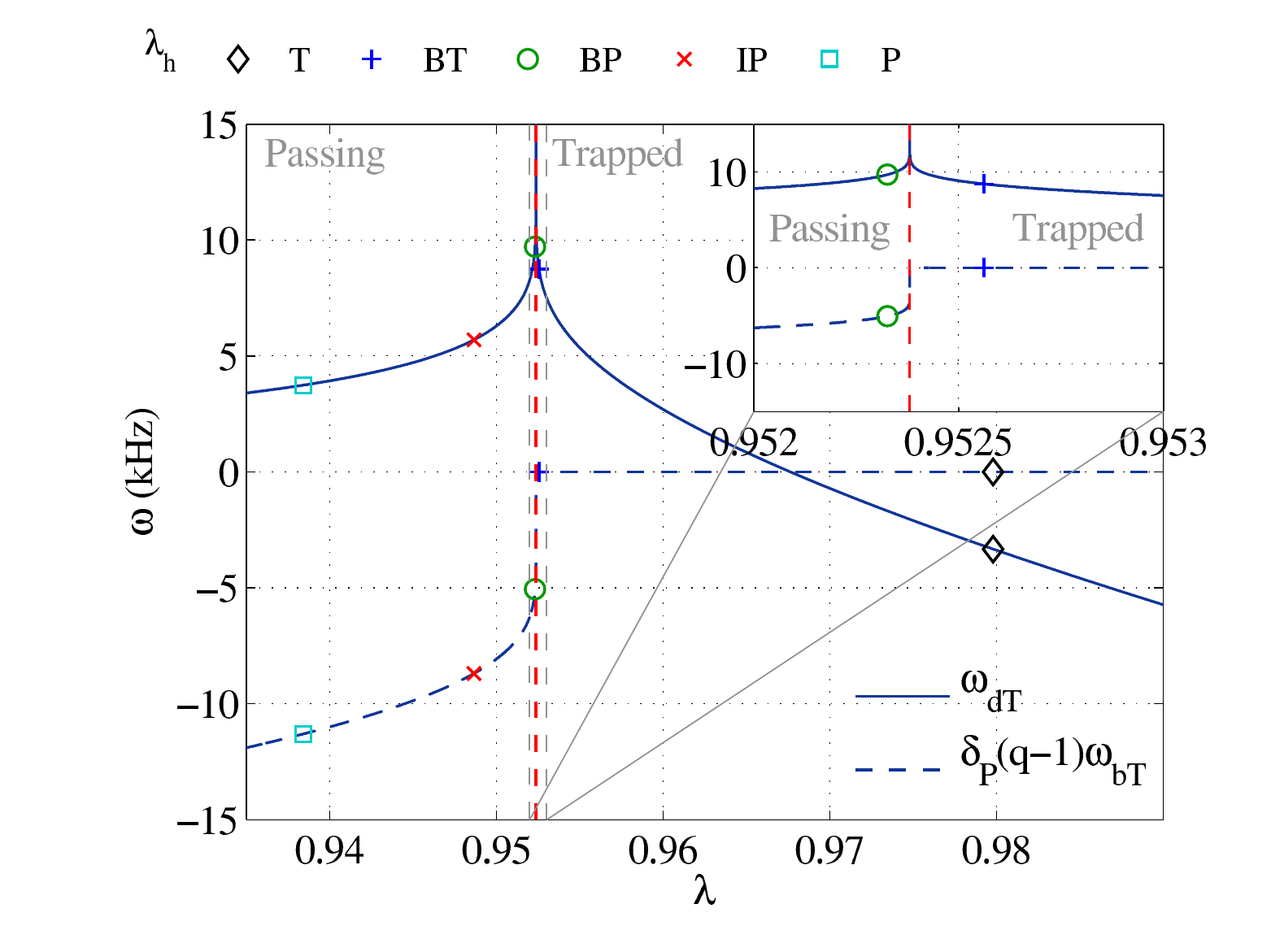}
  \caption[Dependence of $\omega_{dT}$ and $\delta_P(q-1)\omega_{bT}$ and chosen values of $\hat{\lambda}$]{Dependence of $\omega_{dT}$ and $\delta_P(q-1)\omega_{bT}$ over $\hat{\lambda}$. The $5$ values of $\hat{\lambda}$ retained for this study are marked. The vertical dotted line indicates the position of the trapped-passing boundary. Parameters are $B_0 \simeq 3.1\ \mathrm{T}$, $R_0 \simeq 2.46 \ \mathrm{m}$, $k_B T = 100\ \mathrm{keV}$, $r_h \simeq 0.12\ \mathrm{m}$, $1-q(r_h) \sim 4.\ 10^{-3}$, $s(r_h) \sim 1.\ 10^{-2}$.}
  \label{fig:White_Max_pitch}
\end{figure}

\begin{figure}[!ht]
  \centering
  \includegraphics*[width=\figwidth]{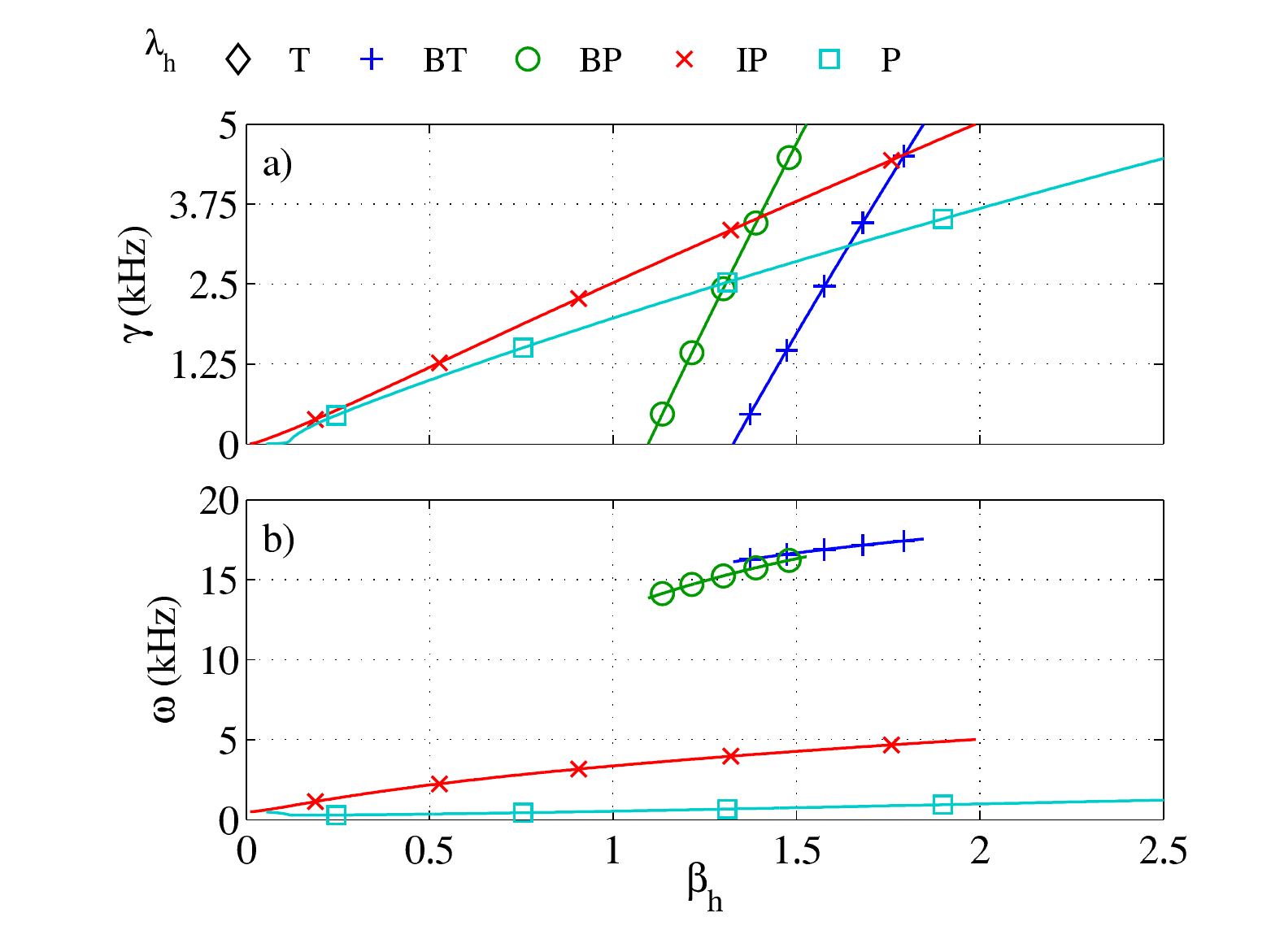}
  \caption[Frequency and growth rate versus $\beta_h$]{Evolution of $\omega$ (a) and $\gamma$ (b) against $\beta_h$. Each color and symbol corresponds to a value of $\hat{\lambda}_h$ according to figure \ref{fig:White_Max_pitch}. The diamond curve(s) lies entirely in the stable domain ($\gamma < 0$) and is not displayed here.}
  \label{fig:White_Max_vsbetah}
\end{figure}

\begin{figure}[!ht]
  \centering
  \includegraphics*[width=\figwidth]{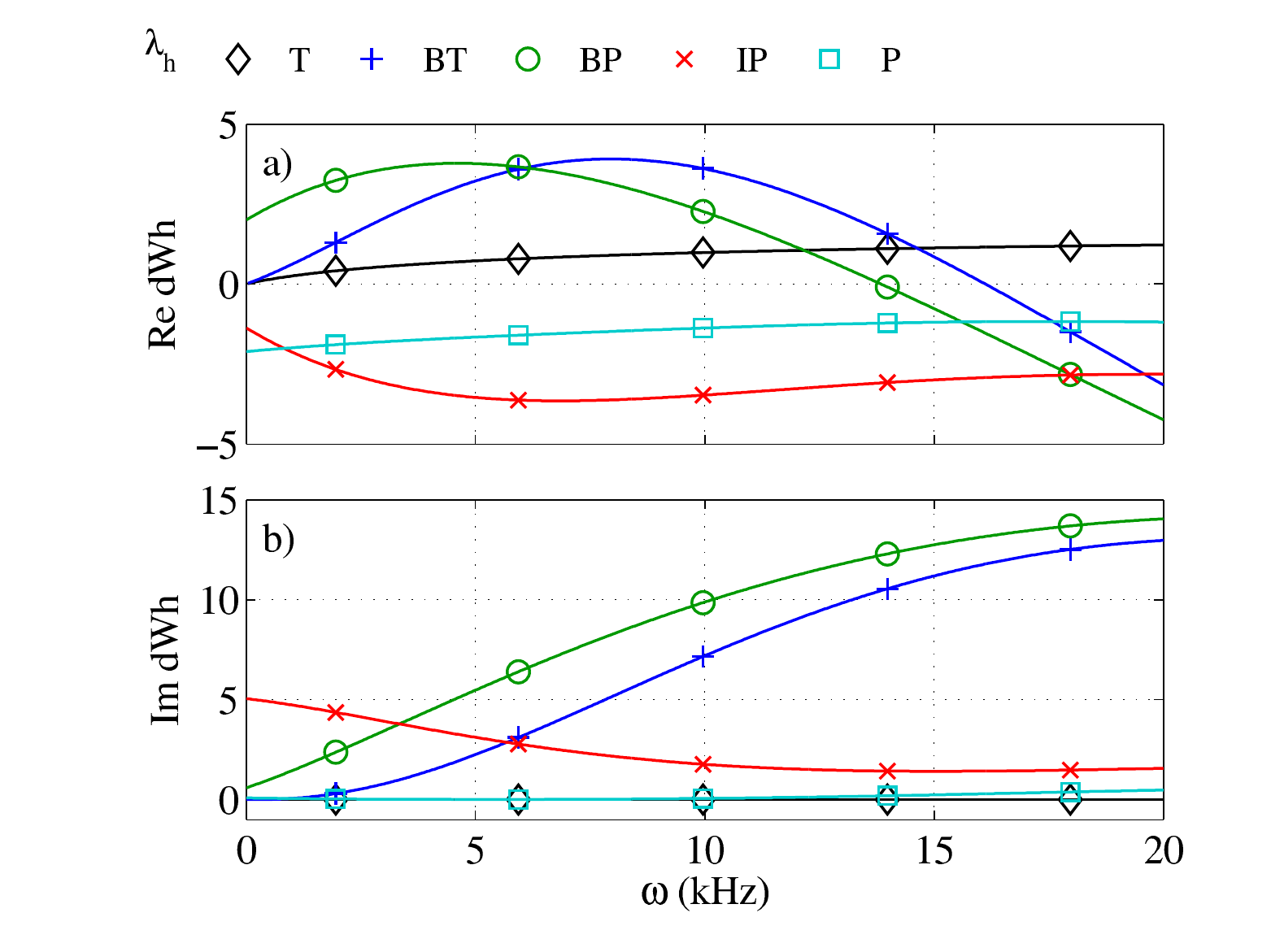}
  \caption[Evolution of $\dWh$ for real frequencies]{Evolution of the real part (a) and the imaginary part (b) of $\dWh$ for real frequencies. Each color and symbol corresponds to a value of $\hat{\lambda}_h$ according to figure \ref{fig:White_Max_pitch}.}
  \label{fig:White_Max_dWk}
\end{figure}

We study the solution of the fishbone dispersion relation with this model distribution function using electrons as fast particles and standard plasma parameters, taken from the Tore Supra discharge number 40816 where modes identified as electron-driven fishbones were observed, ($B_0 \simeq 3.1\ \mathrm{T}$, $R_0 \simeq 2.46 \ \mathrm{m}$, $k_B T = 100\ \mathrm{keV}$, $r_h \simeq 0.12\ \mathrm{m}$, $1-q(r_h) \sim 4.\ 10^{-3}$, $s(r_h) \sim 1.\ 10^{-2} $). We are interested in the behavior of the solution at $\dWf = 0$ when the fast particle beta $\beta_h$ is increased. We study distributions with different values of $\hat{\lambda}_h$ around the trapped-passing boundary but keeping $T$ constant, the values of $\hat{\lambda}_h$ are chosen to be representative of their class of particles. The first one, $\hat{\lambda}_h = 0.9798$, corresponds to the trapped case and is noted ``T''; $\hat{\lambda}_h = 0.9526$ corresponds to barely trapped particles and is noted ``BT'', $\hat{\lambda}_h = 0.9523$ corresponds to barely passing particles and is noted ``BP'', $\hat{\lambda}_h = 0.9384$ corresponds to well passing particles and is noted ``P'', finally in between those two values $\hat{\lambda}_h = 0.9487$ is noted ``IP''. 
They are shown on figure \ref{fig:White_Max_pitch} where we have also plotted $\omega_{dT}$ and $(q-1)\omega_{bT} \delta_P$ as a function of $\hat{\lambda}_h$. Results are shown on figure \ref{fig:White_Max_vsbetah} where we have plotted $\omega$ and $\gamma$ versus $\beta_h$ when $\gamma$ is positive, 
and on figure \ref{fig:White_Max_dWk} where the dependence of $\dWh$ over $\omega$ at $\gamma = 0$ is displayed.

For the ``T'' case, electrons at $r_h$ are trapped and precess in the electron diamagnetic direction $\bar{\Omega}_d > 0$. Their contribution to the fishbone dispersion relation is purely non-resonant since $\Im \dWh = 0$ and their influence on the mode is stabilizing because $\Re \dWh > 0$, especially at higher frequencies since $\dWh = 0$ at $\omega = 0$. This is consistent with the fact that $\gamma$ stays below $0$ when $\beta_h$ is increased, see figure \ref{fig:White_Max_vsbetah}.
The ``BT'' case corresponds to distributions where resonant particles are barely trapped electrons with $\bar{\Omega}_d < 0$. As $\beta_h$ increases, the energy transfer from the particles to the mode due to the resonance ($\Im \dWh$) increases and this is compensated by an increase in the real frequency which increases the continuum damping ($\Im \delta I$). However the influence of the fast particles is stabilizing at low frequency since $\Re \dWh > 0$ as can be seen on figure \ref{fig:White_Max_dWk}. The mode is then driven unstable when the total potential energy enters the ideally unstable region $\dWf + \Re \dWh < 0$, this requires a region where $\Re \dWh$ is decreasing with real frequency. According to other simulations, the threshold frequency varies almost proportionally to $\omega_{dT}$ such that the energy of resonant electrons at the excitation threshold is about $1.8\ k_B T$. Since $\Im \dWh$ is proportional to $\bar{\Omega}_d(r_h,\hat{\lambda}_h)$, the energy transfer is more effective and the threshold value for $\beta_h$ is lower with particles of higher $\bar{\Omega}_d$.

Let us now consider the ``BP'' case, corresponding to passing particles very close to the trapped region. $\omega_{dT}$ is comparable to the previous case ($9.70\ \mathrm{kHz}$ versus $8.74\ \mathrm{kHz}$) but $(1-q)\omega_{bT}/\omega_{dT} \sim 0.5$ so that even at frequencies close to $1\ \mathrm{kHz}$ the energy of resonant electrons are close to $2 k_B T$ and the conditions for the mode to be destabilized ($\dWf + \Re \dWh < 0$, with $\Im \dWh > 0$) are met. This frequency is also much closer to the frequency of the mode at low $\beta_h$ (which is close to $\wi$), allowing for a much lower $\beta_h$ value.
When the particles are further away from the trapped-passing boundary, the parameter $(1-q)\omega_{bT}/\omega_{dT}$ increases, the last 2 curves correspond to this parameter equal to $1.5$ for the ``IP'' case and $3.0$ for the ``P'' case. At this level, the energy of resonant electrons in the considered range is greater than $5 k_B T$ for the high energy branch and lower than $0.05 k_B T$ for the low energy branch. Thus the imaginary part of $\dWh$ is very small and the real part is almost constant and negative. Increasing the density of fast particles acts almost exactly like making the plasma more and more ideally unstable, therefore the mode growth rate will increase while the frequency will not change much. This is the opposite case from deeply trapped electrons which provide a stabilizing influence ($\Re \dWh > 0$ and $\Im \dWh = 0$). As we get further away from the trapped-passing boundary, the destabilizing effect gets weaker.

In summary, deeply trapped electrons are stabilizing. Barely trapped electrons are able to destabilize a mode at frequencies close to $\omega_{dT}$. The effect of barely passing electrons is similar but the frequency of the mode at the excitation threshold is lower than for barely trapped electrons. Well-passing electrons have a global destabilizing influence, this influence decreases as they are further from the trapped-passing boundary.

\section{ECRH-like distributions}
\label{sec:ECRH_dis}

In this section, the MIKE code is used to study the stability of the $\nn = 1$ internal kink mode in the presence of fast electrons using analytical distribution functions of fast electrons that are characteristic of those obtained in ECRH-experiments.

\subsection{Parameters}

\subsubsection{Distribution function}

To model ECRH-heated plasmas, the fast electron distribution function is chosen to have a Maxwellian momentum dependence with an anisotropic temperature,
\begin{equation}
  f(p,\xi_{0},r)=\tilde{f}(r)\exp\left(-\frac{p^{2}}{2m_{e}k_{B}T(\xi_{0})}\right)
\end{equation}
with $\xi_0 = {v_\Vert}_{\theta=0}/v$ such that $\hat{\lambda} = (1-\xi_0^2)/(1-\varepsilon)$. For the function $T(\xi_{0})$, the 2-temperature model ($1/T(\xi_{0})={\xi_{0}}^{2}/T_{\Vert}+(1-{\xi_{0}}^{2})/T_{\perp})$ is modified to include a third temperature $T_{t}=T(\xi_{0,T})$ where $\xi_{0,T}$ is the position of the trapped-passing boundary at the $q=1$ surface. $T$ has a power-law dependence on $\xi_{0}$ for trapped and passing domains, the exponent $\alpha_T$ is used to control the width of the peak in temperature, the dependence over $\xi_0$ becoming more peaked as $\alpha_T$ is increased.
\begin{equation}
  T(\xi_{0}) = \left\{ 
    \begin{array}{cl} 
      \displaystyle T_\perp + (T_t-T_\perp)\left|\frac{\xi_0}{\xi_{0,T}}\right|^{\alpha_T} & \mbox{ if } \xi_0 < \xi_{0,T}, \\
      \displaystyle T_\Vert + (T_t-T_\Vert)\left|\frac{\xi_0-1}{\xi_{0,T}-1}\right|^{\alpha_T} & \mbox{ if } \xi_0 > \xi_{0,T}.
      \end{array}
    \right.
  \label{eq:T_xi0}
\end{equation}
 To further reduce the number of parameters, the 3 temperatures are linked, by the relation 
\begin{equation}
  T_{\perp}=2\xi_{0,T}^2 T_{\Vert} + (1-2\xi_{0,T}^{2})^{\left(T_t/T_\Vert\right)^{1/3}}T_t.
  \label{eq:T_perp}
\end{equation}
 In this way, when $\xi_{0,T}=0$, $T_{\perp}=T_{t}$ and when $\xi_{0,T}^{2}=1/2$, $T_{\perp}=T_{\Vert}$. Also $T_\Vert/T_t \rightarrow 0$ implies $T_\perp/T_t \rightarrow 0$; and $T_\Vert = T_t$ implies $T_\perp = T_\Vert = T_t$. The $1/3$ exponent has been chosen as a best fit to experimental conditions on various machines using ECRH.
\begin{figure}[!ht]
  \centering
  \includegraphics*[width=\figwidth]{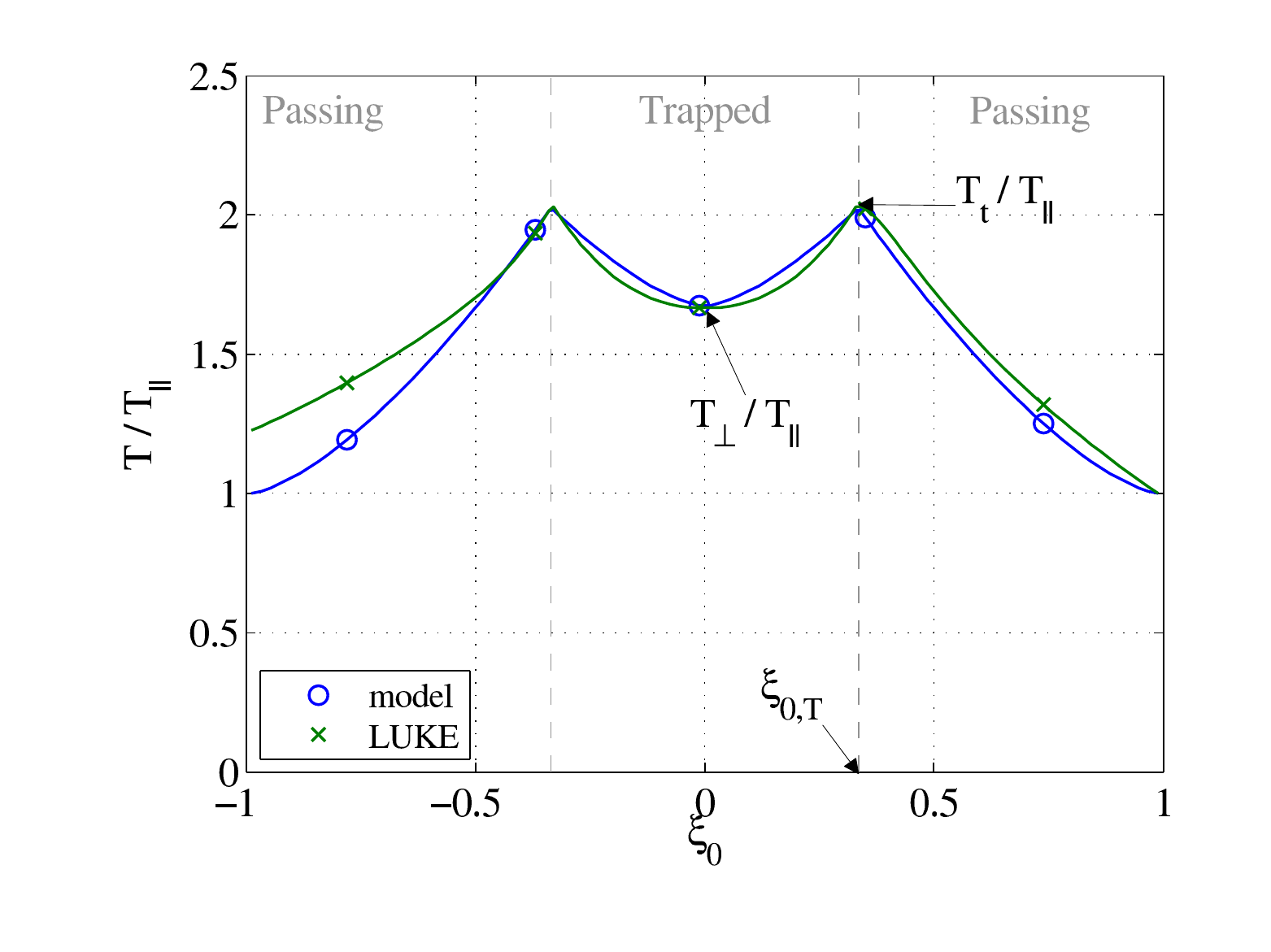}
  \caption[Anisotropic temperature model and comparison to a TCV experiment]{Temperature (normalized to $T_{\Vert}$) dependence over $\xi_0$ for the TCV discharge number 31737 with an additional power of 1MW of ECRH. Crosses correspond to results of the simulation using C3P0/LUKE, circles to the best fit using the model given by (\ref{eq:T_xi0}) and (\ref{eq:T_perp}). The value of $\alpha_T$ used was $1.5$.}
  \label{fig:T_xi_0}
\end{figure}
On figure \ref{fig:T_xi_0} is given an example taken from the TCV discharge number 31737. The Fokker-Planck code LUKE/C3P0 has been used to compute the electronic distribution function created by $1\ \mathrm{MW}$ of ECRH. The temperature of the fast-particle component is plotted along with the best fit using the model described by equations (\ref{eq:T_xi0}) and (\ref{eq:T_perp}).

A reasonable first approximation for the flux-surface averaged fast electron density, consistent with off-axis ECRH, is to choose a linear function between $0$ and $r_{s}$, $n(r)=n_{h}r/r_{s}$ where $n_h$ is the density of fast electrons at the $q=1$ surface. This implies $\tilde{f}(r) = n_h \tilde{I}(r) r/r_{s}$, where
\begin{equation}
  \tilde{I}(r) = \frac{1}{2} \frac{1}{1-\varepsilon} \int_{-1}^{1} \left|\xi_0\right|\bar{\tau}_b(r,\xi_0) \, \mathrm{d}\xi_{0} \left(2\pi m_e k_B T(\xi_0)\right)^{3/2}.
\end{equation}
and $\bar{\tau}_b = \omega_b^{-1} v/q R_0$ is the normalized bounce time.

The fast particle density $n_h$ and the height of the temperature peak $T_t$ are related to RF power density, while the ratio $T_t/T_\Vert$ or $\alpha_T$, which will influence the width of the peak of temperature, can be linked to $Z_{eff}$, the effective ion charge of the plasma.

\vspace{2mm}

\begin{figure}[!ht]
  \centering
  \includegraphics*[width=\figwidth]{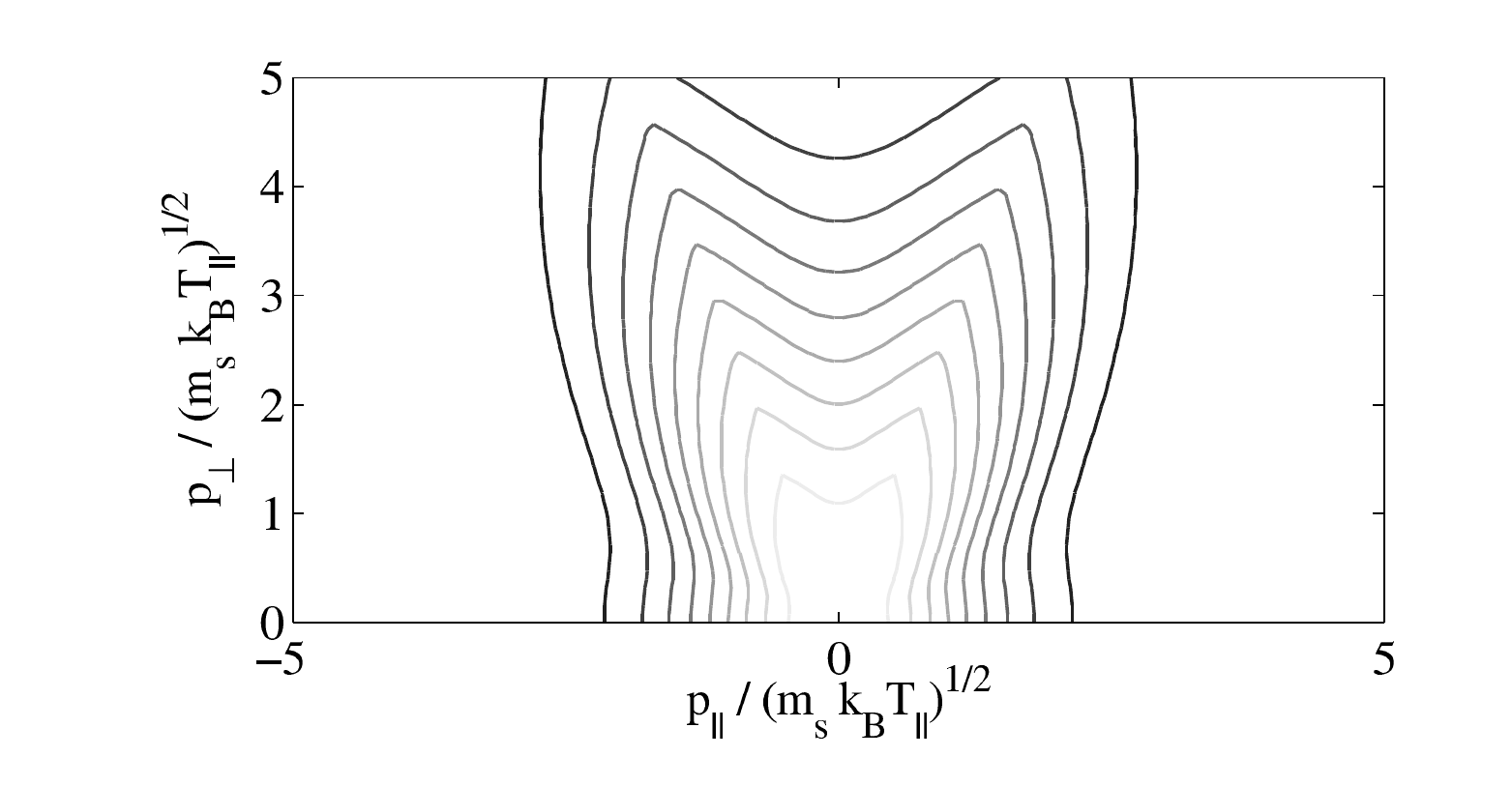}
  \caption[Contours of the distribution function in momentum space]{Contours of the distribution function in $(p_{\Vert},p_\perp)$ space for $T_t/T_\Vert = 10$, $\alpha_T = 2$ (increasing levels of gray indicate decreasing levels of $F_h$.}
  \label{fig:f_contours}
\end{figure}

The distribution function is described by only three parameters, the density at $q=1$, noted $n_h$ and the temperatures $T_t$ and $T_\Vert$. An example of the distribution function is displayed in figure \ref{fig:f_contours}.

\subsubsection{Equilibrium}

For the safety factor profile, we chose 2 types of profiles which are generally associated with the internal kink mode.

\begin{figure}[!ht]
  \centering
  \includegraphics*[width=\figwidth]{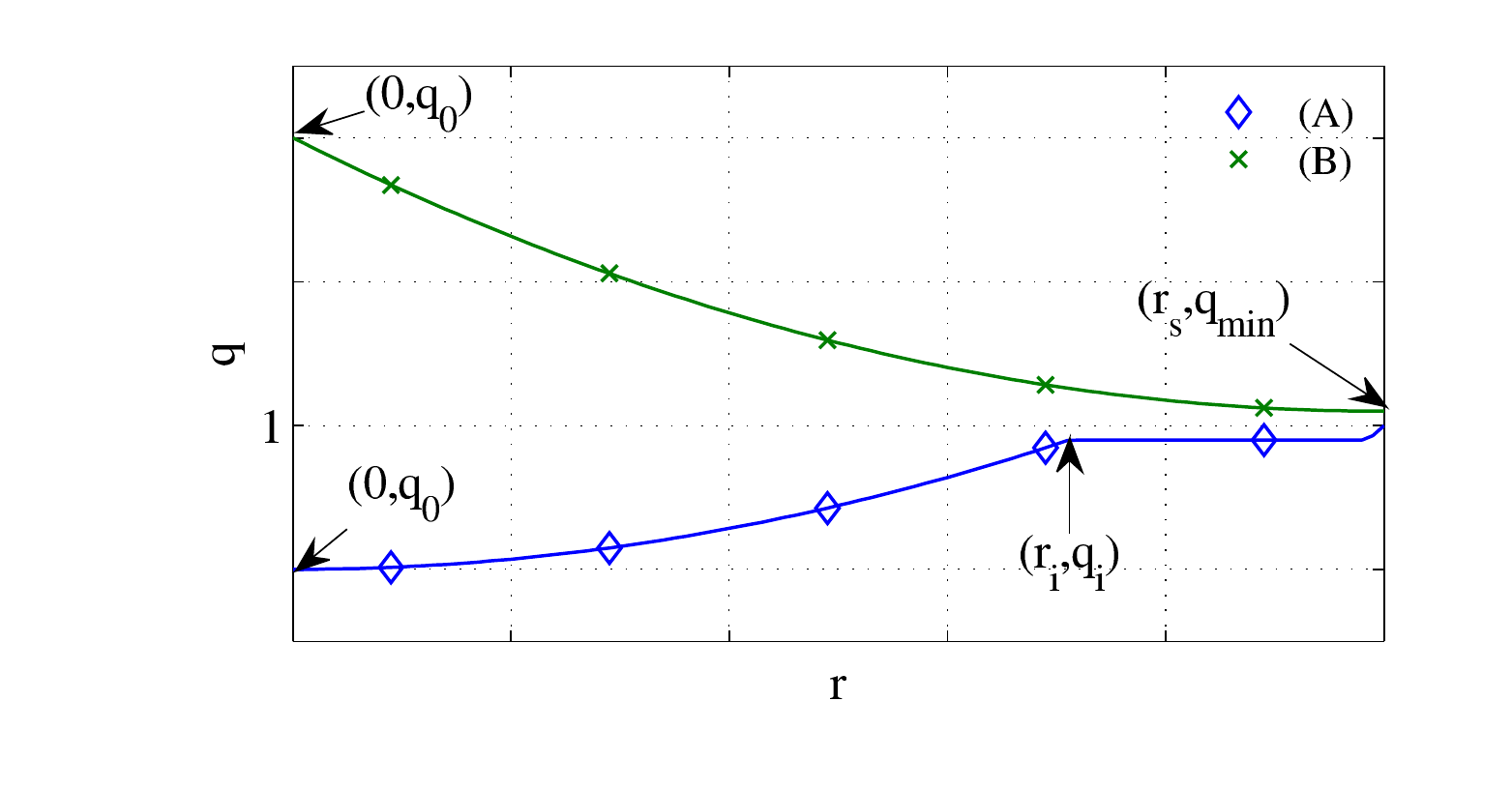}
  \caption[Safety factor profiles used for the parametric study]{The two types of safety factor profiles chosen for this study. Profiles corresponding to type (A) (see text) are represented with diamonds, profiles of type (B) with crosses.}
  \label{fig:qprof_TS}
\end{figure}

  (A) The first one has been proposed for sawtoothing plasmas where partial reconnection can occur and a plateau in $q$ appear near $q=1$. It starts with a parabolic profile between $r=0$ and $r = r_i$, followed by a plateau between $r=r_i$ and $r=r_s$. The value at the center is noted $q_0$, the value of the plateau $q_i$ and is close to $1$. For $r > r_s$, $q$ rises up to the edge, the magnetic shear at $r={r_s}^+$ is noted $s$ (see the curve with diamonds on figure \ref{fig:qprof_TS} for an example). It is of interest in this study since electron-fishbones have been obsereved in-between sawteeth on various tokamaks such as HL-2A\cite{che09}. In this case, inertial effects should be important in the whole region where $q \simeq 1$, but De Blank showed \cite{deb91} that if the width $r_s - r_i$ of the plateau is small compared to $r_s$ then the structure of the dispersion relation given by equations (\ref{eq:FDR}) and (\ref{eq:inertia_finite_s}) are correct up to order $(1-r_i/r_s)^2$.

  (B) For the second type, the q-profiles are reversed in the center and the point of minimum $q$ is located at $r=r_s$ with $q_{min} \simeq 1$. The value at the center is noted $q_0$. This corresponds to the case of DIII-D \cite{won00} or FTU \cite{zon07} where electron fishbones have been observed in discharges where the q-profile was reversed in the center. With these profiles, the dispersion relation is given by equations (\ref{eq:FDR}) and (\ref{eq:inertia_zero_s}).

These two types of profiles are represented in figure \ref{fig:qprof_TS}.

\subsection{Results}

We now study the evolution of the frequency and growth rate of the mode when the fraction of fast particles $n_h/n_e$ is increased from $0$.

\subsubsection{Influence of the shape of the distribution function}

In this part, we choose an equilibrium corresponding to sawtoothing plasmas with the following parameters $q_0 = 0.9$, $1-q_i = 5.\,10^{-3}$, $r_s = R_0/15$, $r_i = 0.75 r_s$, the magnetic shear at $r=r_s$ is $s = 0.1$, other parameters of the equilibrium are the same as the one chosen in section \ref{sec:Drift_Bounce_Res}.

If the equilibrium is kept fixed, then the characteristic frequencies of the particle orbits are also fixed and so are the characteristics of the particle-mode interaction. It all comes down to knowing the respective population of each category described in section \ref{sec:Drift_Bounce_Res}.

\begin{figure}[!ht]
  \centering
  \includegraphics*[width=\figwidth]{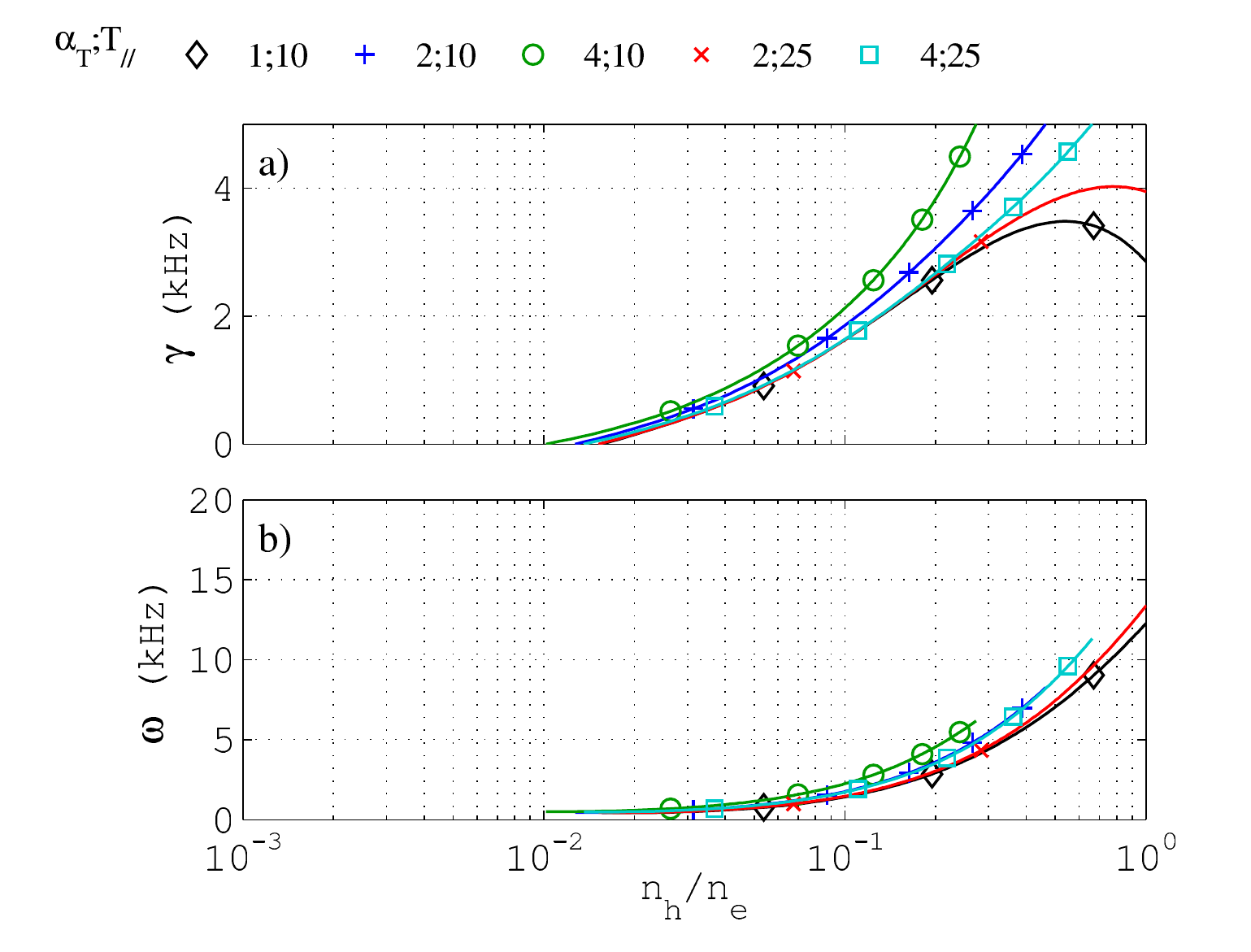}
  \caption[Solutions for different values of $\alpha_T$ and $T_{\Vert}$]{Evolution of $\omega$ (a) and $\gamma$ (b) against $n_h/n_e$. The parameters are $q_0 = 0.9$, $1-q_i = 5.\,10^{-3}$, $r_s = R_0/15$, $r_i = 0.75 r_s$, $k_B T_{t} = 100\, \mathrm{keV}$ and $\dWf = 10^{-4}$.}
  \label{fig:RE2_art_alpha_Eref_scan_vsnh}
\end{figure}

We performed several simulations with different values for $\alpha_T$ and $T_{\Vert}$, keeping $k_B T_{t} = 100\, \mathrm{keV}$. The dependence of $\gamma$ and $\omega$ over $n_h/n_e$ at $\dWf = 10^{-4}$ are presented in figure \ref{fig:RE2_art_alpha_Eref_scan_vsnh}.
The first 3 cases correspond to a fixed value of $k_B T_\Vert = 10\,\mathrm{keV}$ with $\alpha_T = 1,\,2,\,4$. The curve corresponding to $\alpha_T = 1$ exhibits the competition between resonant barely passing particles and non-resonant trapped particles. The barely passing particles provide the drive for the destabilization of the mode at low-frequency (between $1$ and $2\,\mathrm{kHz}$). At higher frequency (higher $n_h/n_e$), the drive by barely trapped particles takes over the decreasing drive by barely passing particles, but the damping by non-resonant trapped particles becomes also more effective, such that the mode is re-stabilized.

When $\alpha_T$ increases, the temperature dependence over $\xi_0$ gets more peaked and the amount of trapped particles and passing particles decreases especially in the region around $\xi_{0,T}$. As trapped particles (except for barely trapped particles) have a stabilizing influence, the growth rate of the mode is stronger and the threshold value of $n_h$ is lower for higher $\alpha_T$. Moreover, figure \ref{fig:RE2_art_alpha_Eref_scan_vsnh} shows that for $k_B T_{\Vert} = 10\ \mathrm{keV}$ and $\alpha_T \geq 2$ the growth rate is a monotonically increasing function of $n_h/n_e$ even at high fast particle fraction ($n_h/n_e$ of the order of $1$). The lack of passing particles does not have such a strong effect, the frequency of the mode is slightly higher for higher $\alpha_T$.

The second set of simulations is performed at $k_B T_{\Vert} = 25\, \mathrm{keV}$, so that at $\alpha_T$ fixed the most affected regions are at $\xi_0$ close to $1$ (deeply passing) but also $0$ (deeply trapped) since $T_\perp$ is linked to $T_\Vert$. Those 2 regions are more populated at higher $T_\Vert$. Once again the stabilizing influence of deeply trapped particles is recovered by comparing the curves corresponding to $\alpha_T = 2$ ($+$ and $\times$) or $\alpha_T = 4$ ($\circ$ and $\square$). The fact that the mode has a higher frequency at lower $T_\Vert$ can again be explained by the higher population of passing particles.

\subsubsection{Influence of the equilibrium}

\begin{figure}[!ht]
  \centering
  \includegraphics*[width=\figwidth]{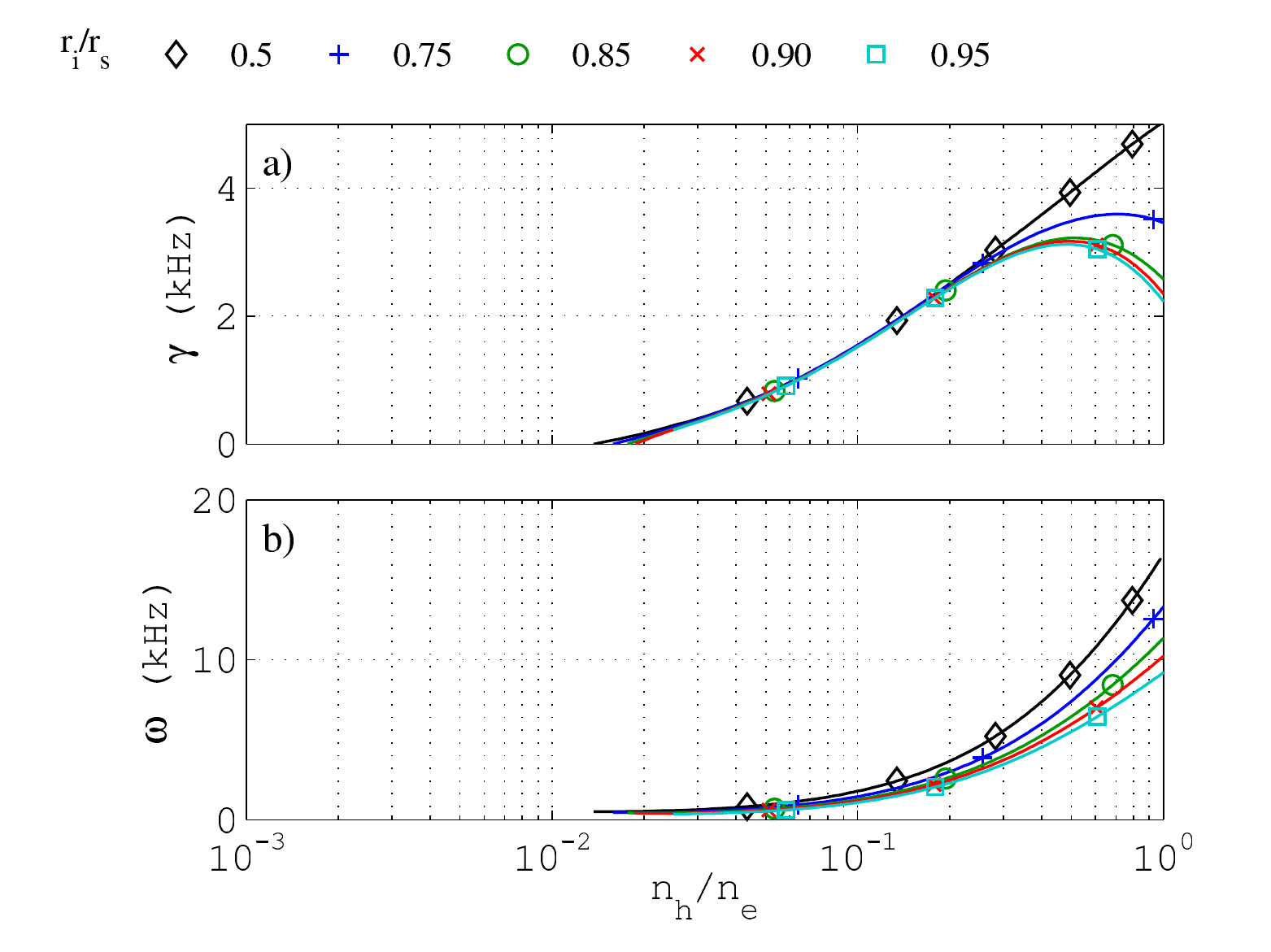}
  \caption[Solutions for different values of $r_i/r_s$]{Evolution of $\omega$ (a) and $\gamma$ (b) against $n_h/n_e$ for different values of $r_i/r_s$. The parameters are $q_0 = 0.9$, $1-q_i = 5.\,10^{-3}$, $r_s = R_0/15$, $\alpha_T = 2$, $k_B T_{\Vert} = 25\, \mathrm{keV}$, $k_B T_{t} = 100\, \mathrm{keV}$ and $\dWf = 10^{-4}$.}
  \label{fig:RE2_art_rhoi_scan_vsnh}
\end{figure}

We then perform a scan in the parameter $r_i/r_s$ to check that the results obtained in the previous section do not strongly depend on the size of the plateau in the q-profile. Parameters for the distribution function are $\alpha_T = 2$, $k_B T_\Vert = 25 \,\mathrm{keV}$, all other parameters are kept constant. Results are shown in figure \ref{fig:RE2_art_rhoi_scan_vsnh}.
It appears that the width of the plateau has a limited effect on the frequency and growth rate of the mode. Except for large plateaus ($r_i/r_s > 0.25$), the dependence of $(\omega,\gamma)$ over $n_h$ is globally conserved. When $r_i>r_s$ is increased, the main effect is an increase of the ratio $(1-q)\omega_{bT}/\omega_{dT}$ for the particles located where the q-profile is changed. Such that there is a depletion of particles with low values of this parameter, whereas the population of particles with intermediate values does not change much. Hence a drop in growth rate and frequency is observed.

\begin{figure}[!ht]
  \centering
  \includegraphics*[width=\figwidth]{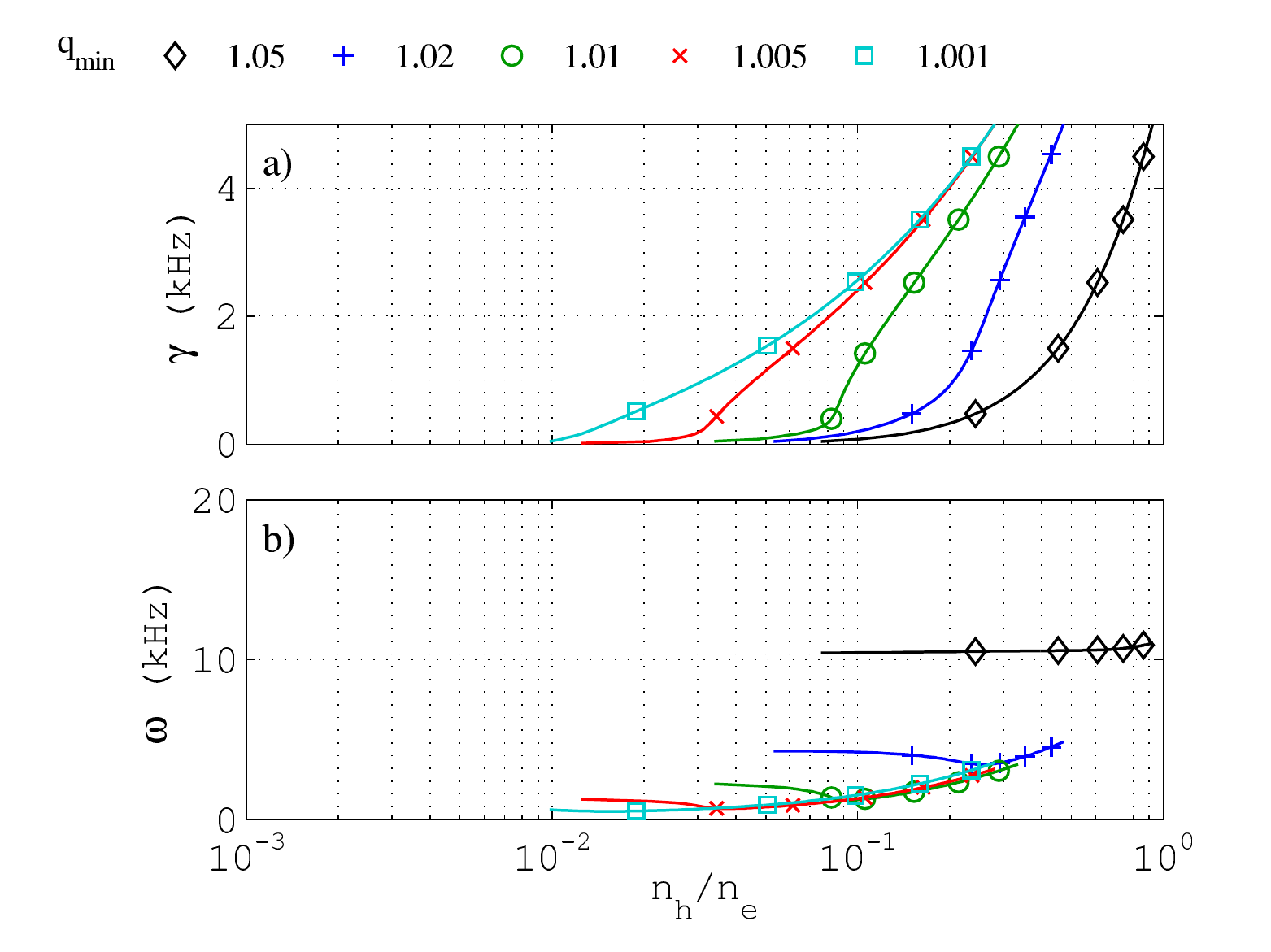}
  \caption[Solutions for different values of $q_{min}$]{Evolution of $\omega$ (a) and $\gamma$ (b) against $n_h/n_e$ for different values of $q_{min}$. The parameters are $q_0 = 1.2$, $r_s = R_0/15$, $\alpha_T = 2$, $k_B T_{\Vert} = 25\, \mathrm{keV}$, $k_B T_{t} = 100\, \mathrm{keV}$ and $\dWf = 10^{-4}$}
  \label{fig:RE2_art_qmin_scan_vsnh}
\end{figure}

We consider also the case of a reversed q-profile as described in the previous section (case (B)). At fixed parameters for the distribution function, we choose $r_s = R_0/15$ and $q_0 = 1.2$ and we vary $q_{min}$.  As $q_{min}$ drops toward unity, the continuum damping gets weaker as is implied by equation (\ref{eq:inertia_zero_s}) and the resonance with passing particles is more effective. This is confirmed by the results presented in figure \ref{fig:RE2_art_qmin_scan_vsnh}.
If $q_{min} = 1.05$, the continuum damping is too strong and the mode is driven unstable only at a very high fast particle fraction ($n_h/n_e \sim 0.1$) but if $q_{min}$ decreases the value of $n_h$ at threshold decreases and for $|1-q_{min}| < 10^{-2}$, $n_h/n_e$ at threshold is of the order of $10^{-2}$. It should be noted that the frequency of the mode decreases when $q_{min}$ decreases. Further study of this case shows that the dominant effect is the reduction of the continuum damping and not the increased resonance with passing particles.

\subsubsection{Influence of the resonance condition}

\begin{figure}[!ht]
  \centering
  \includegraphics*[width=\figwidth]{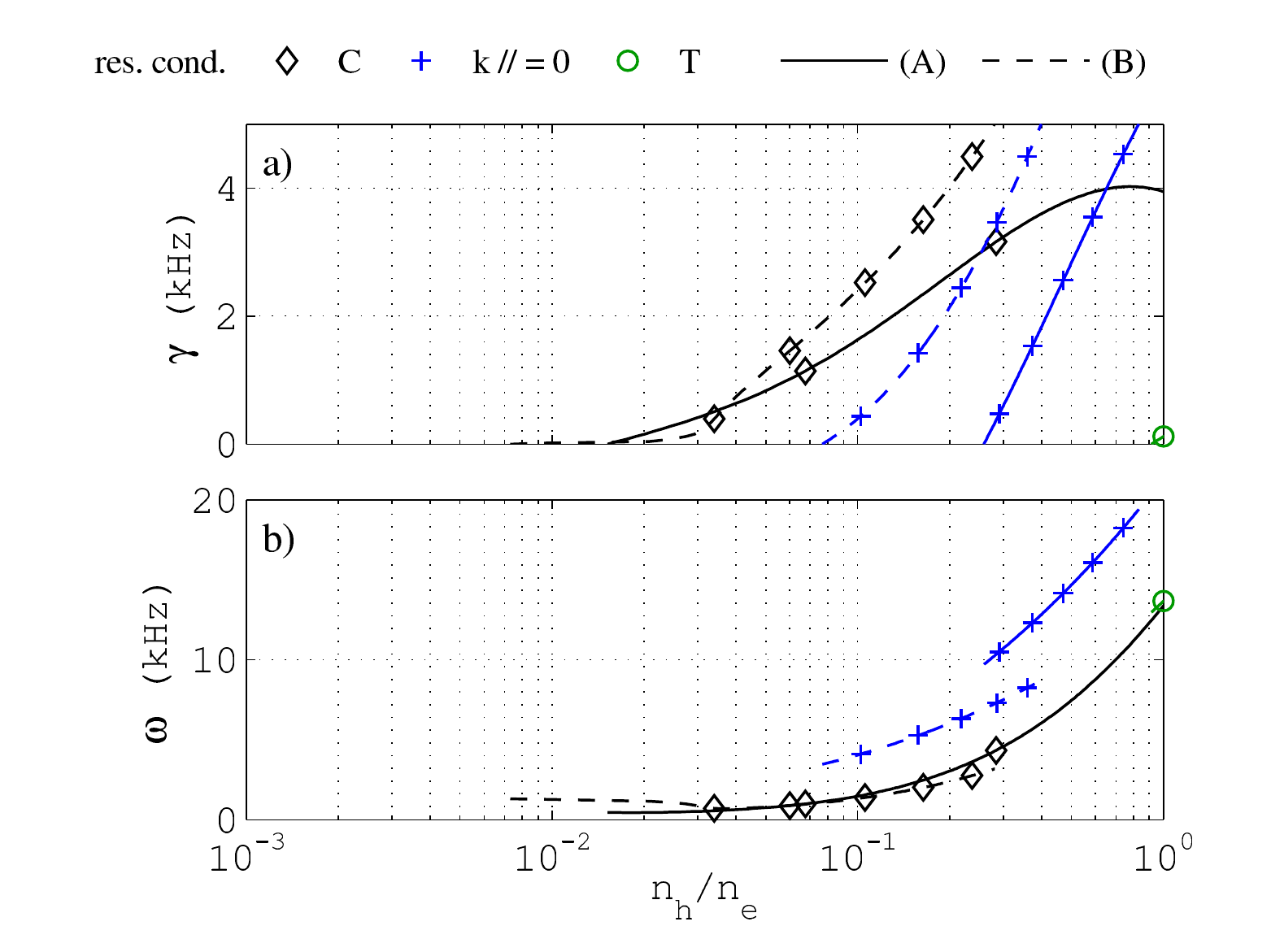}
  \caption[Solutions for different resonance conditions]{Evolution of $\omega$ (a) and $\gamma$ (b) against $n_h/n_e$ for different resonance conditions. Solid curves correspond to the case of figure \ref{fig:RE2_art_rhoi_scan_vsnh} with $r_i/r_s = 0.75$. Dashed curves correspond to the case of figure \ref{fig:RE2_art_qmin_scan_vsnh} with $q_{min} = 1.005$.}
  \label{fig:RE2_art_ResCond_comp_vsnh}
\end{figure}

To highlight the effect of the $\langle k_\Vert v_\Vert\rangle$ term in the resonance condition, we compare the results of the previous simulations with the ones obtained by setting $k_{\Vert} = 0$ or by including only trapped particles in the computation of $\dWh$. 2 reference simulations are chosen, one with each type of q-profile. Figure ~\ref{fig:RE2_art_ResCond_comp_vsnh}
 shows the results. As was expected, when $k_\Vert$ is set to $0$, the value of $n_h$ at threshold is strongly increased, as well as the frequency of the mode. If only the trapped particles contribution is retained in $\dWh$ then the mode is stable for $n_h/n_e < 1$.

\vspace{2mm}

This result, along with the previous discussion about the influence of the safety factor profile, points out that the critical contribution of passing particles is the one of particles where $(q-1)\omega_b$ is of the same order as $\omega_d$, labeled ''IP'' in section ~\ref{sec:Drift_Bounce_Res}.

\section{Summary}

In this chapter the original fishbone dispersion relation was extended to account for the transit frequency in the resonance with passing particles in the zero-orbit width limit. The inclusion of the term due to the parallel motion of particles breaks the symmetry of the resonance condition for passing particles. The resonance with energetic passing particles is limited to regions where $q$ is close to $1$. Using the MIKE code with analytical unidirectional distributions, we confirm that the internal kink mode can be driven unstable by barely trapped electrons resonating at $\omega = \omega_d$. More deeply trapped electrons have a stabilizing influence ($\dWh$ real and positive). We also show that it can be driven by barely passing electrons even if $\omega < \omega_d$, provided that $1-q$ is small enough. Passing electrons further from the trapped-passing boundary have a destabilizing influence ($\dWh$ mostly real and negative). This destabilizing effect quickly decreases away from the trapped-passing boundary. It is also shown that the linear stability of electron-driven fishbones exhibit different characteristics from the ion-driven fishbone \cite{whi89}, such as a lower frequency. Using more realistic distribution functions close to those created in ECRH-experiments, we find as expected that the destabilization of electron-driven fishbones is favored by a more densely populated region around the trapped-passing boundary which provides more resonant particles. The influence of the safety factor profile was also investigated and we show that, if the profile includes a plateau near $q=1$, then the frequency and the growth rate of the mode do not depend much on the width of this plateau . We also show that for reversed-shear profiles with $q_{min} > 1$, the dominant effect when $q_{min}$ decreases is the reduction of the continuum damping of the mode. The contribution of energetic passing electrons to the dispersion relation of the electron-driven fishbone allows both for a reduced mode frequency and a reduced threshold value for the density of energetic electrons. This effect could help understanding of the observations of low-frequency modes during lower hybrid current-drive in the Tore Supra tokamak \cite{gui12,gui11}.


\chapter{Conclusion}

The observations of electron-driven fishbones in Tore Supra were in apparent contradiction with the standard theory of the electron-driven fishbone stability since the observed frequency was much lower than the toroidal precession frequency of the energetic electrons created by the Lower Hybrid wave in those discharges \cite{gui12,gui11}. 

In this thesis we have generalized the original fishbone dispersion relation to account for the transit frequency in the resonance with passing particles (see chapter \ref{cha:FDR_derivation}). In particular, a term due to the parallel motion of passing particles has been added while it was neglected in previous studies \cite{sun05,zon07,wan07}. In the regions where the safety factor is close to $1$, the value of this term is of particular importance for passing electrons due to their large transit frequency. 

We developed the code MIKE to solve the generalized fishbone dispersion relation with arbitrary distribution functions and study the stability of electron-driven fishbones. In chapter 8 we have investigated the influence of the different classes of electrons using the code MIKE with simple analytical distribution functions. We have shown that, unlike barely trapped electrons which can drive the internal kink mode unstable at frequencies close to their precession frequency, barely passing electrons are destabilizing at a lower frequency. For such particles all three terms in the resonance condition have a similar weight. For passing electrons further from the trapped-passing boundary, the term due to the parallel motion of the particles dominates the other terms; , the contribution of energetic particles remains destabilizing but mainly as a non-resonant effect. The MIKE code was also used with realistic distribution functions based on the modeling of ECRH experiments using the code C3PO/LUKE \cite{dec04a}. Whether in experiments or modeling, using ECRH rather than LHCD simplifies the interpretation as the electron distribution can be significantly modified with minimal effect on the current profile. With this analysis we showed that the modification of the resonance condition for passing electrons reduces both the energetic electron density threshold for the mode stability and the frequency of the mode. By extension, the relatively low frequency of the electron-driven fishbones observed in the Tore Supra tokamak could be explained by this effect. 

\begin{figure}[!ht]
  \centering
  \begin{subfigure}[b]{0.495\textwidth}
    \includegraphics[width = \figwidth]{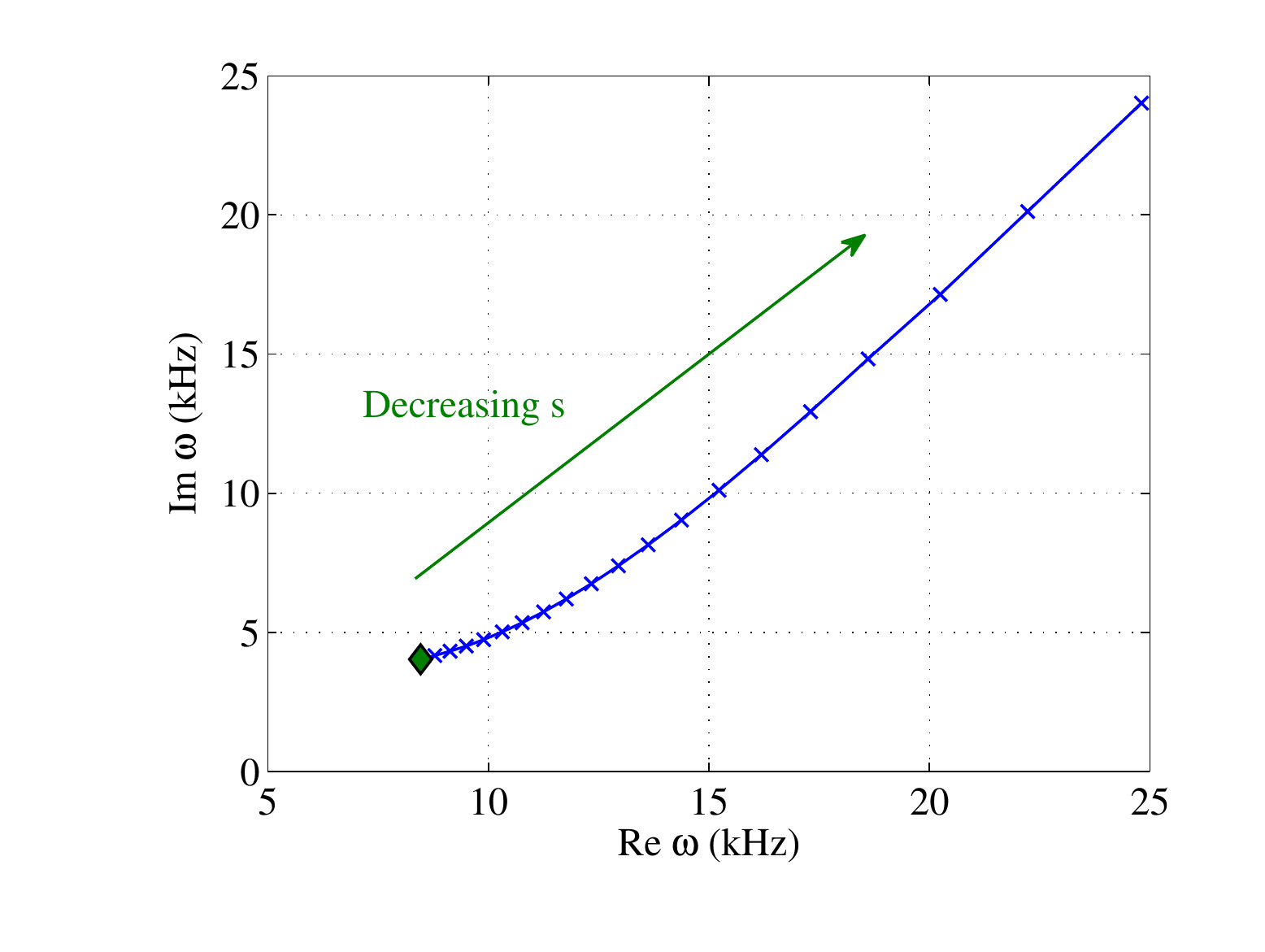}
    \caption{}
    \label{fig:40816_Dr0_0_s1_effect}
  \end{subfigure}
  \begin{subfigure}[b]{0.495\textwidth}
    \includegraphics[width = \figwidth]{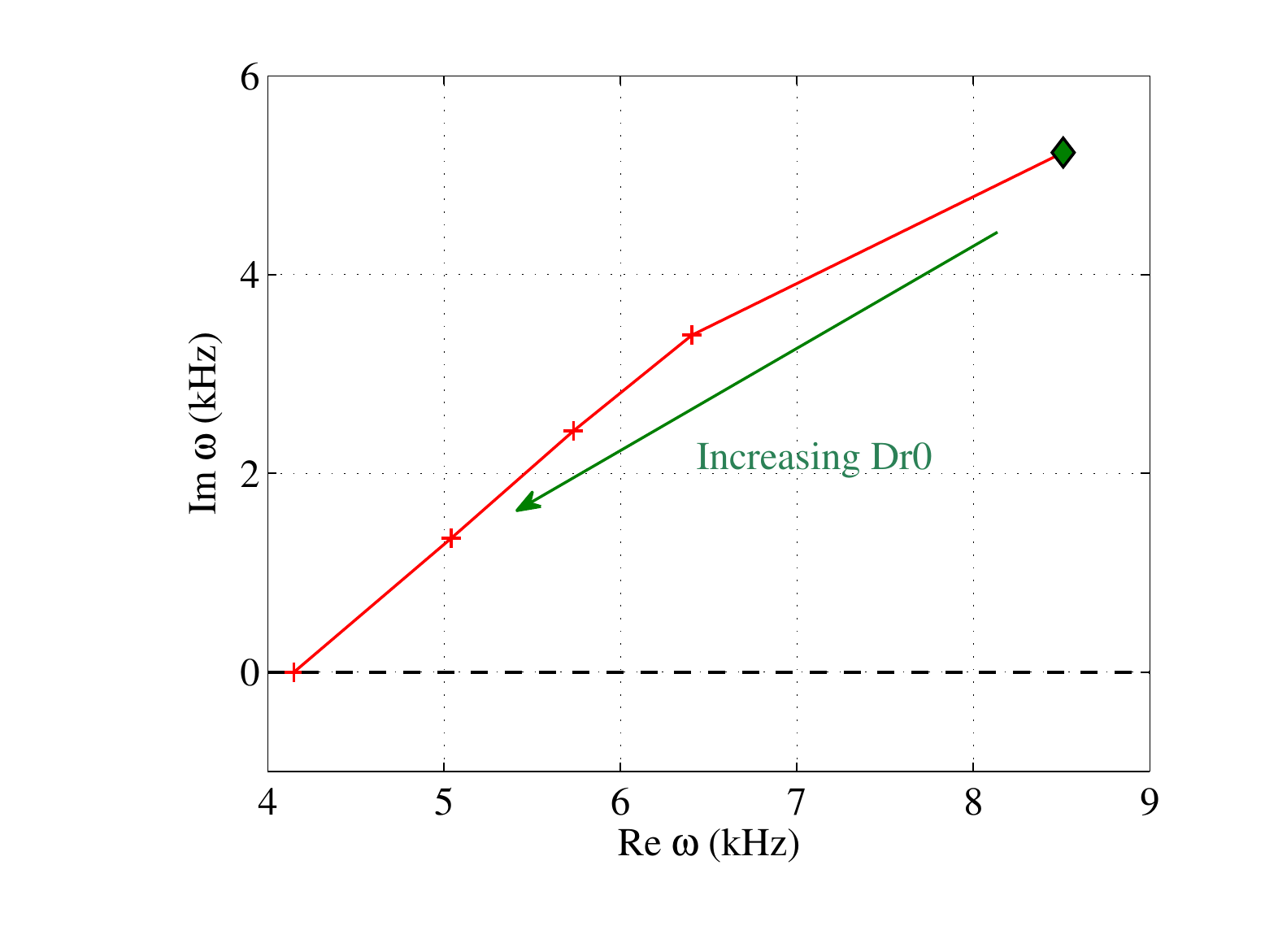}
    \caption{}
    \label{fig:40816_Dr0scan_fg}
  \end{subfigure}
  \caption[Sensitivity of the solution to the parameters of $q$ and $F_h$]{Evolution of the mode frequency and growth rate for the Tore Supra discharge number 40816 (\ref{fig:40816_Dr0_0_s1_effect}) when the magnetic shear $s$ at the $q=1$ surface is decreased from $0.1$ to $0.02$ and (\ref{fig:40816_Dr0scan_fg}) when the radial diffusion coefficient $D_{r0}$ due to the turbulent anomalous transport of electrons used to reconstruct the electronic distribution function is increased from $0.0 \ \mathrm{m^2 \, s^{-1}}$ to $0.05 \ \mathrm{m^2 \, s^{-1}}$. In these simulations, finite $k_\Vert$ effects were neglected.}
  \label{fig:40816_scans}
\end{figure}

The development of the MIKE code, which is described in chapter 7, has required the development of new techniques to overcome several difficulties, such as the computation of resonant integrals, the use of arbitrary distribution functions, or the search for the solutions of the dispersion relation in the complex plane. The development of the code MIKE, which is extensively benchmarked against analytical results obtained with simplistic distribution functions, served two purposes. The first objective, developed in this thesis, is to study the intrinsic properties of electron fishbone modes and determine the role of passing electrons. The second objective is to use the MIKE code for a comparison of theory and experiment, by coupling the code MIKE to the transport code CRONOS \cite{bas03}, which provides the equilibrium profiles, and to the relativistic Fokker-Planck code C3PO/LUKE \cite{dec04a}, which is able to reconstruct the electronic distribution function. MIKE could also be inserted as an element of integrated modeling platforms. So far, our attempts to compare the results of the MIKE code with the observations on Tore Supra have not been convincing due to the very high sensitivity of the solution to the details of both the safety factor profile and the distribution function. An example is provided in figure \ref{fig:40816_scans}, where the evolution of the mode frequency and growth rate is shown as a function of the magnetic shear at the $q=1$ surface, or the level of diffusion due to the anomalous transport of electrons by turbulence.
The sensitivity to the details of the safety factor profile is particularly problematic in Tore Supra where the reconstruction of the q-profile by transport codes cannot be compared to experimental measurements such as those that could be provided by the motional Stark effect diagnostic.

The latest version of the MIKE code accounts for the effect of arbitrary flux-surface geometry in the computation of the equilibrium frequencies of motion (see chapter \ref{sec:GC_motion}) and in the computation of the contribution of fast particles to the fishbone dispersion relation (see chapter \ref{cha:FDR_derivation}), as well as relativistic effects \cite{wan07}. We hope this will help obtain a qualitative agreement between the results of the MIKE computations and the Tore Supra observations. 

In addition, the most unstable modes in Tore Supra have usually poloidal and toroidal mode numbers above $1$ and the $q$-profiles in these discharges have a very low shear in the central region such that $q$ is close to $1$ over a wide region, which enhances the resonance with passing electrons. Additional work is needed so that MIKE can calculate the stability for higher poloidal mode numbers. Indeed, in deriving the expressions implemented in the MIKE code (equations \eqref{eq:dWk_gen_MIKE} and \eqref{eq:dWint_gen_MIKE}) we have assumed that the radial MHD- displacement is a simple top-hat function, while the analysis reproduced in chapter \ref{cha:internal-kink-mode} showed that for low-shear profiles the radial MHD-displacement can differ significantly from the top-hat function, and depends on the mode growth rate (and frequency).



\appendix


\chapter{The inertia term for the fishbone dispersion relation}
\label{cha:inert-term-fishb}

The form of the inertia term $\delta I$ in the $(\mm=1,\nn = 1)$ internal kink dispersion relation or to the fishbone dispersion relation depends on the relevant physics inside the inertial ($q=1$) layer. Several different forms are used in this thesis but all concern the case of a single singular layer, we recall them here. In any case, the dispersion relation is written
\begin{equation}
  \delta I = \dW
\end{equation}
with $\dW$ defined in equation \eqref{eq:deltaWhat}.

\section{Shape of the safety factor profile}

 If the magnetic shear $s = rq'/q$ at $r=r_{s}$  does not vanish then the expression for $\delta I$  is,
\begin{equation}
  \delta I=i|s|\Lambda_I.
  \label{eq:inertia_finite_s}
\end{equation}
If $s=0$ but $S=r^{2}q''/q$ does not vanish then one has \cite{has87}
\begin{equation}
  \delta I=-S\left(\Delta q_{s}^{2}-\Lambda_I^{2}\right)^{3/4}\left(1+\frac{\Delta q_{s}}{\sqrt{\Delta q_{s}^{2}-\Lambda_I^{2}}}\right)^{1/2},
  \label{eq:inertia_zero_s}
\end{equation}
with $\Delta q_{s}=1-q(r_{s})$. $\Lambda_I$ in equations \eqref{eq:inertia_finite_s} and \eqref{eq:inertia_zero_s} is the generalized inertia term, introduced in \cite{zon06a}. Its expressions are detailed in the next section.

\section{Physical model}

\subsection{Ideal MHD}

In the case of low-frequency modes with the inclusion of diamagnetic effects in the limit of vanishing resistivity \cite{cop66,ara78} a general expression for $\Lambda_I$ is 
\begin{equation}
  \Lambda_I=\frac{\sqrt{\omega(\omega-\wi)}}{\omega_{A}}\sqrt{M}
  \label{eq:inertia_omi}
\end{equation}
where $\wi = (\dd{p_i}/\dd{r})/(e_i n r B_0)$ is the ion diamagnetic frequency, $\omega_{A} = B_0/(qR_0\sqrt{\mu_0 m_i n})$ the Alfv\'en frequency, all those quantities being evaluated at the position $r_{s}$ of the inertial layer. $M$ is the inertial enhancement factor defined in equation \eqref{eq:defM} recalled here
\begin{equation}
  \label{eq:defM_}
 M =  \frac{\rho}{\rho_0} \left(1 + \frac{1}{\displaystyle\frac{\Gamma^2}{\beta_c} + \nn^2} + \frac{1}{\displaystyle\frac{\Gamma^2}{\beta_c} + \left(\frac{2}{q} - \nn\right)^2}\right),
\end{equation}
In the incompressible limit, $\Gamma^2 \ll \beta_c$ and $M = 1+2q^2$. The other limit case is $\Gamma^2 \ll \beta_c$ where the parallel inertia is negligible and $M = 1$.

\subsection{Collisionless MHD}

In the collisionless MHD model, $\Lambda_I$ is still given by equation (\ref{eq:inertia_omi}) but $M = 1$ due to the absence of the parallel inertia in the energy principle.

\subsection{Drift-kinetic thermal ions}

If one includes the kinetic effects of thermal ions $\Lambda_I$ is still given by equation (\ref{eq:inertia_omi}) but $M$ is given by \cite{graPhD,gra00}
\begin{equation}
M = 1 + (1.6/\sqrt{r_s/R_0} + 0.5)q^2.
\end{equation}

\subsection{Bi-fluid Resistive MHD}

The form of the inertia term in the bi-fluid resistive MHD model has been derived only for monotonic q-profiles with
\begin{equation}
  \Lambda_I=\frac{\sqrt{\omega(\omega-\wi)}}{\omega_{A}}\frac{8\Gamma\left(\frac{\Lambda^{3/2}+5}{4}\right)}{\Lambda^{9/4}\Gamma\left(\frac{\Lambda^{3/2}-1}{4}\right)}
  \label{eq:inertia_bif}
\end{equation}
with $\Lambda = (\omega(\omega-\wi)(\omega - \hat{\omega}_{*e})^{1/3} \, (\tau_R \tau_A/s)^{1/3}$, see section \ref{sec:bifluid_disp_rel} for the definitions of $\hat{\omega}_{*e},\tau_R$. 

In the limit of vanishing resistivity, one recovers equation (\ref{eq:inertia_omi}) with $M=1$ since the parallel inertia was neglected.

\chapter{Asymptotic matching in the resistive layer for the internal kink mode}
\label{cha:asympt-match-resist}

\section{Solving the layer equations}

Integrating the second equation of sytem \eqref{eq:reduced_system}, one has $\lambda^2 \xi' + \mathrm{const} = x \psi' - \psi$. Introducing 
\begin{equation}
  \chi(x) = \lambda^2 \xi' + \chi_\infty = x \psi' - \psi,
  \label{eq:chi_def}
\end{equation}
which gives (assuming $\psi,\xi \sim 0$ when $x \rightarrow +\infty$)
\begin{align}
  \xi(x) &= - \lambda^{-2}\int_x^{+\infty}\mathrm{d}y \, (\chi(y) - \chi_\infty), \label{eq:xi_chi}\\
  \psi(x) &= - \chi(x) - x \int_x^{+\infty}\frac{\mathrm{d}y}{y} \ddr{\chi}{y}, \label{eq:psi_chi}
\end{align}
Dividing the first equation of \eqref{eq:reduced_system} by $x$, one obtains,
\[
-\frac{\chi}{x} - \int_x^{+\infty}\frac{\mathrm{d}y}{y} \ddr{\chi}{y} - \lambda^{-2}\int_x^{+\infty}\mathrm{d}y \, (\chi(y) - \chi_\infty) - \frac{\epsilon}{\lambda \, x^2} \ddr{\chi}{x} = 0,
\]
then differentiating,
\begin{align*}
  \frac{\chi}{x^2} +\lambda^{-2}(\chi - \chi_\infty) + \frac{2\epsilon}{\lambda \, x^3} \ddr{\chi}{x} - \frac{\epsilon}{\lambda \, x^2} \mdd{\chi}{x}{2} = 0, \\
  \mdd{\chi}{x}{2} - \frac{2}{x} \ddr{\chi}{x} - \left(\frac{x^2}{\epsilon \lambda} + \frac{ \lambda}{\epsilon}\right) \chi = -\frac{x^2}{\epsilon \lambda} \chi_\infty,
\end{align*}
then introducing $x = \epsilon^{1/4} \lambda^{1/4} u$,
\begin{equation}
  \mdd{\chi}{u}{2} - \frac{2}{u} \ddr{\chi}{u} - \left(u^2 + \sqrt{\frac{\lambda^3}{\epsilon}}\right) \chi = - u^2 \chi_\infty.
  \label{eq:chi}
\end{equation}
The solution of this equation can be put in the form (for $\Re (\lambda^{3/2}\epsilon^{-1/2}) > 1$)
\begin{equation}
  \frac{\chi}{\chi_\infty} = 1 - \frac{1}{2} \sqrt{\frac{\lambda^3}{\epsilon}} \int_0^1 \mathrm{d}t (1-t)^{(\lambda^{3/2}\epsilon^{-1/2} - 5)/4} (1+t)^{-(\lambda^{3/2}\epsilon^{-1/2} + 5)/4} \exp\left(-\frac{1}{2}t \, u^2\right)
  \label{eq:chi_sol}
\end{equation}

\begin{proof}
The evaluation of the left hand side of \eqref{eq:chi}, gives (writing $f(t)\exp(-t \, u^2/2)$ the integrand in \eqref{eq:chi_sol})
\[
  -\chi_\infty \left(u^2 + \sqrt{\frac{\lambda^3}{\epsilon}}\right) - \frac{\chi_\infty}{2} \sqrt{\frac{\lambda^3}{\epsilon}} \int_0^1 \mathrm{d}t f(t) \exp\left(-\frac{1}{2}t \, u^2\right) \left(t - \sqrt{\frac{\lambda^3}{\epsilon}} + u^2(t^2-1)\right).
\]
Integrating by parts the $u^2 (t^2-1)$ term,
\begin{multline*}
   u^2 \int_0^1 \mathrm{d}t f_u(t) \,(t^2-1) = -u^2 \int_0^1 \mathrm{d}t f(t) (1-t)(1+t) \exp\left(-\frac{1}{2}t \, u^2\right) \\
  = -u^2\left(\left[-\frac{2}{u^2} f(t) (1-t)(1+t) \exp\left(-\frac{1}{2}t \, u^2\right)\right]_0^1 + \right. \ldots \\ \left. \frac{2}{u^2} \int_0^1 \mathrm{d}t \left(f(t)(1-t)(1+t)\right)' \exp\left(-\frac{1}{2}t \, u^2\right) \right).
\end{multline*}
The computation of the derivative gives,
\[
u^2 \int_0^1 \mathrm{d}t f_u(t) \,(t^2-1) = -u^2 \left( \frac{2}{u^2}  + \frac{2}{u^2}\int_0^1 f(t) \exp\left(-\frac{1}{2}t \, u^2\right) \frac{t-\lambda^{3/2}\epsilon^{1/2}}{2}\right),
\]
\[
- \frac{\chi_\infty}{2} \sqrt{\frac{\lambda^3}{\epsilon}} \int_0^1 \mathrm{d}t f_u(t) \left(t - \sqrt{\frac{\lambda^3}{\epsilon}} + u^2(t^2-1)\right) = \chi_\infty \sqrt{\frac{\lambda^3}{\epsilon}}.
\]
This proves that if $\chi$ is given by \eqref{eq:chi_sol}, then it verifies equation \eqref{eq:chi}.
\end{proof}

\subsubsection{Asymptotic matching}

Recalling from the ideal calculation that the solution in the resonant layer must verify 
\begin{equation*}
\left\{
  \begin{array}{lcl}
    \displaystyle \xi \sim \xi_\infty &\mbox{when} &x \rightarrow - \infty \\
    \displaystyle \xi \sim 0 &\mbox{when} &x \rightarrow + \infty \\
    \displaystyle \xi' \sim \frac{\xi_c\,\delta\hat{W}}{\pi s^2 x^2} &\mbox{when} &|x| \rightarrow + \infty
  \end{array}
\right.
\end{equation*}

Remembering that outside the resonant layer $\psi = - x \xi$ and thus $\psi' = - \xi - x \xi'$, we have the following conditions for $\psi$
\[
\left\{
  \begin{array}{lcl}
    \displaystyle \psi \sim - x \xi_\infty &\mbox{when} &x \rightarrow - \infty \\
    \displaystyle \psi' \sim - \xi_\infty -  \frac{\xi_c \, \delta\hat{W}}{\pi s^2} \frac{1}{x} &\mbox{when} &x \rightarrow - \infty \\
    \displaystyle \psi' \sim  - \frac{\xi_c\,\delta\hat{W}}{\pi s^2} \frac{1}{x} &\mbox{when} &x \rightarrow + \infty
  \end{array}
\right.
\]
From the definiton of $\chi$ \eqref{eq:chi_def} and the previous conditions on $\psi$ and $\xi$, we have
\[
\mbox{for $x \rightarrow +\infty$, } \chi \sim \chi_\infty \mbox{ and } \chi \sim -\frac{\xi_c\, \delta\hat{W}}{\pi s^2}
\]
where the first comes from the relation between $\chi$ and $\xi$, and the second one between $\chi$ and $\psi$, and the results is exactly the same for $x \rightarrow - \infty$ ($\chi$ is an even function). Therefore
\[
\chi_\infty = -\frac{\xi_c\, \delta\hat{W}}{\pi s^2}.
\]
From \eqref{eq:psi_chi} we also have 
\[
\psi \sim - x \int_{-\infty}^{+\infty} \frac{\mathrm{d}y}{y} \ddr{\chi}{y} \mbox{ for } x \rightarrow - \infty,
\]
giving
\[
\xi_\infty = \int_{-\infty}^{+\infty} \frac{\mathrm{d}y}{y} \ddr{\chi}{y}.
\]
Finally the condition for asymtotic matching reduces to 
\begin{equation}
  -\frac{\pi s^2}{\delta \hat{W}} = \int_{-\infty}^{+\infty} \frac{1}{\chi_\infty} \ddr{\chi}{y} \frac{\mathrm{d}y}{y}.
  \label{eq:asymptotic_matching}
\end{equation}
The computation of this integral is treated in the next sections. 

\subsection{Another expression for  \texorpdfstring{$\chi$}{chi}}

Starting from expression \eqref{eq:chi_sol}, we perform the change of variable $z = (1-t)/(1+t)$ (which gives also $t = (1-z)/(1+z)$:
\begin{align}
  \frac{\chi}{\chi_\infty} &= 1 - \frac{1}{2} \sqrt{\frac{\lambda^3}{\epsilon}} \int_0^1 \mathrm{d}t (1-t)^{(\lambda^{3/2}\epsilon^{-1/2} - 5)/4} (1+t)^{-(\lambda^{3/2}\epsilon^{-1/2} + 5)/4} \exp\left(-\frac{x^2}{2 \sqrt{\lambda\epsilon}}t \right) \nonumber \\
  &= 1 - \frac{1}{2} \sqrt{\frac{\lambda^3}{\epsilon}} \int_0^1 \frac{2\mathrm{d}z}{(1+z)^2} z^{(\lambda^{3/2}\epsilon^{-1/2} - 5)/4} \left(\frac{2}{1+z}\right)^{-5/2} \exp\left(-\frac{x^2}{2 \sqrt{\lambda\epsilon}}\frac{1-z}{1+z} \right) \nonumber \\
   \frac{\chi}{\chi_\infty} &=1 - 2^{-5/2} \sqrt{\frac{\lambda^3}{\epsilon}} \int_0^1 \mathrm{d}z \, (1+z)^{1/2} z^{(\lambda^{3/2}\epsilon^{-1/2} - 5)/4} \exp\left(-\frac{x^2}{2 \sqrt{\lambda\epsilon}}\frac{1-z}{1+z} \right) \label{eq:chi_sol_alt}
\end{align}

This gives the following expression for $\dd{\chi}/\dd{x}$,
\begin{equation}
  \frac{1}{x \, \chi_\infty} \ddr{\chi}{x} = 2^{-5/2} \frac{\lambda}{\epsilon} \int_0^1 \mathrm{d}z \, \frac{1-z}{\sqrt{1+z}} z^{(\lambda^{3/2}\epsilon^{-1/2} - 5)/4} \exp\left(-\frac{\, x^2}{2 \sqrt{\lambda\epsilon}}\frac{1-z}{1+z} \right)
  \label{eq:dchi_alt}
\end{equation}

\subsection{Computation of the integral}

Remembering that
\begin{equation*}
\int_{-\infty}^{+\infty} \exp(-y^2/2\sigma^2) \mathrm{d}y = \sqrt{2 \pi \sigma^2},
\end{equation*}
integration over $x$ is done easily by inverting the two integrals
\begin{align*}
  \int_{-\infty}^{+\infty} \frac{1}{\chi_\infty} \ddr{\chi}{y} \frac{\mathrm{d}y}{y} &= 2^{-5/2} \frac{\lambda}{\epsilon} \int_{-\infty}^{+\infty} \int_0^1 \frac{1-z}{\sqrt{1+z}} z^{(\lambda^{3/2}\epsilon^{-1/2} - 5)/4} \exp\left(-\frac{y^2}{2 \sqrt{\lambda\epsilon}}\frac{1-z}{1+z} \right)\mathrm{d}y \, \mathrm{d}z \\
  &= 2^{-5/2} \frac{\lambda}{\epsilon} \int_0^1 \frac{1-z}{\sqrt{1+z}} z^{(\lambda^{3/2}\epsilon^{-1/2} - 5)/4} \int_{-\infty}^{+\infty} \exp\left(-\frac{y^2}{2 \sqrt{\lambda\epsilon}}\frac{1-z}{1+z} \right)\mathrm{d}y \, \mathrm{d}z \\
  &= 2^{-5/2} \frac{\lambda}{\epsilon} \int_0^1 \frac{1-z}{\sqrt{1+z}} z^{(\lambda^{3/2}\epsilon^{-1/2} - 5)/4} \sqrt{\frac{2\pi(1+z)}{(1-z)\sqrt{\lambda\epsilon}}} \, \mathrm{d}z \\
  &= \frac{\sqrt{\pi \lambda^{5/2}\epsilon^{-3/2}}}{4} \int_0^1 \sqrt{1-z} \, z^{(\lambda^{3/2}\epsilon^{-1/2} - 5)/4} \, \mathrm{d}z \\
 &= \frac{\sqrt{\pi \lambda^{5/2}\epsilon^{-3/2}}}{4} \, \mathrm{B}\left(\frac{\lambda^{3/2}\epsilon^{-1/2} - 1}{4},\frac{3}{2}\right)
\end{align*}
where $\mathrm{B}$ is the Beta function and it is linked to the gamma function 
by \eqref{eq:beta_gamma}.
\begin{align}
  \int_{-\infty}^{+\infty} \frac{1}{\chi_\infty} \ddr{\chi}{y} \frac{\mathrm{d}y}{y} &= \frac{\sqrt{\pi \lambda^{5/2}\epsilon^{-3/2}}}{4}  \frac{\Gamma((\lambda^{3/2}\epsilon^{-1/2} - 1)/4)\Gamma(3/2)}{\Gamma((\lambda^{3/2}\epsilon^{-1/2} + 5)/4)} \nonumber \\
  &= \frac{\pi\sqrt{ \lambda^{5/2}\epsilon^{-3/2}}}{8}  \frac{\Gamma((\lambda^{3/2}\epsilon^{-1/2} - 1)/4)}{\Gamma((\lambda^{3/2}\epsilon^{-1/2} + 5)/4)} \nonumber \\
  &= \frac{\pi}{8 \, \epsilon^{1/3}} \hat{\lambda}^{5/4}  \frac{\Gamma((\hat{\lambda}^{3/2} - 1)/4)}{\Gamma((\hat{\lambda}^{3/2} + 5)/4)} \label{eq:int_dchi_x}
\end{align}
with $\hat{\lambda} = \lambda/\epsilon^{1/3}$.

Finally the asymptotic matching condition, equation \eqref{eq:asymptotic_matching}, reduces to (with $\hat{\lambda} = \lambda /\epsilon^{1/3}$):
\begin{equation}
  -\frac{\pi s^2}{\delta \hat{W}} = \frac{\pi}{8 \, \epsilon^{1/3}} \hat{\lambda}^{5/4}  \frac{\Gamma((\hat{\lambda}^{3/2} - 1)/4)}{\Gamma((\hat{\lambda}^{3/2} + 5)/4)}
\end{equation}

\section{Relationship between \texorpdfstring{$\Gamma$}{Gamma} and  \texorpdfstring{$\mathrm{B}$}{Beta} function}
To derive the integral representation of the $\mathrm{B}$ function, write the product of two $\Gamma$ functions as 
\[
 \Gamma(x)\Gamma(y) =
  \int_0^\infty\ e^{-u} u^{x-1}\,du \int_0^\infty\ e^{-v} v^{y-1}\,dv
=\int_0^\infty\int_0^\infty\ e^{-u-v} u^{x-1}v^{y-1}\,du  \,dv.
\]

Changing variables to $u=zt$, $v=z(1-t)$ shows that this is
\[
\int_{z=0}^\infty\int_{t=0}^1\ e^{-z} (zt)^{x-1}(z(1-t))^{y-1}z\,dt  \,dz
=\int_{z=0}^\infty \ e^{-z}z^{x+y-1} \,dz\int_{t=0}^1t^{x-1}(1-t)^{y-1}\,dt 
\]

Hence
\begin{equation}
  \Gamma(x) \, \Gamma(y)= \Gamma(x+y) \mathrm{B}(x,y).
  \label{eq:beta_gamma}
\end{equation}


\chapter{Appendices to the derivation of the fishbone dispersion relation}
\label{sec:app_FDR_derivation}

\section{A derivation of equation (\ref{eq:h_prime})}
\label{sec:h_prime_demo}

The goal here is to get an expression for ${h'}_{s,\vc{n},\omega}$ in terms of the guiding-center velocity (and not the particle velocity). The particle velocity $\vc{v}$ can be decomposed in the sum of a parallel velocity $v_\Vert = v_{g,\Vert}$ , a perpendicular guiding-center velocity $\vc{v}_{g,\perp}$ (which is the drift velocity coming from the curvature and grad-$B$ drifts) and the perpendicular velocity $\vc{\tilde{v}}$ associated to the gyration around the field-lines. The the particle's position $\vc{x}$ need to be expanded by writing $\vc{x} = \vc{X}_G + \vb{\rho}$ where $\vb{\rho}$ is the gyroradius and $\vc{X}_G$ is the position of the guiding-center. In the case of a vanishing first-order $\vc{E}\times\vc{B}$ drift (no equilibrium perpendicular electric field), the dependence over the gyrophase $\varphi$ remains only in $\vc{\tilde{v}}$ and $\vb{\rho}$. Moreover, these two quantities are (to first order) $2\pi$ periodic in $\varphi$ and are linked by 
\begin{align}
  \vb{\rho} &= \rho (\cos \varphi \: \vc{e}_1 + \sin \varphi \: \vc{e}_2) \\
  \vc{\tilde{v}} &= \omega_c \vc{b} \times \vb{\rho}
\end{align}
where $\vc{b}$ is the magnetic field unit vector and $\vc{e}_1$ and $\vc{e}_2$ are two unit vectors such that $(\vc{e}_1,\vc{e}_2,\vc{b})$ forms a right-handed basis of the euclidian space. With these expressions, one is then able to compute the gyroaverage of ${h'}_{s,\omega}$.
\begin{align}
  J_0 {h'}_{s,\omega} &= \frac{1}{2 \pi} \int_0^{2\pi} \frac{\vc{v}_\perp \cdot \nabla \Phi_\omega}{i \omega}  - v_\Vert E_{\Vert,\omega} \mathrm{d} \varphi \\
  &= \frac{1}{2 \pi} \int_0^{2\pi} \frac{(\vc{v}_{g,\perp} + \vc{\tilde{v}}) \cdot (\nabla \Phi_\omega)(\vc{X}_G + \vb{\rho})}{i \omega}  - v_{g,\Vert} E_{\Vert,\omega}(\vc{X}_G + \vb{\rho}) \mathrm{d} \varphi \\
  &\sim\frac{\vc{v}_{g,\perp} \cdot \nabla \Phi_\omega}{i \omega}  - v_{g,\Vert} E_{\Vert,\omega} + \frac{1}{2\pi}\int_0^{2 \pi} \frac{\vc{\tilde{v}} \cdot (\vb{\rho} \cdot \nabla) \nabla \Phi_\omega}{i \omega} \mathrm{d} \varphi.
\end{align}
The $\sim$ sign denotes the fact that we have neglected the terms with higher order in $\vb{\rho}$.

\begin{align}
  \frac{1}{2\pi}\int_0^{2 \pi} \vc{\tilde{v}} \cdot (\vb{\rho} \cdot \nabla) \vc{f} \: \mathrm{d} \varphi &= \frac{1}{2 \pi}\int_0^{2 \pi} \omega_c (\vc{b} \times \vb{\rho})\cdot (\vb{\rho} \cdot \nabla) \vc{f} \mathrm{d} \varphi \\
  &= \frac{1}{2 \pi}\int_0^{2 \pi} \omega_c \vc{b} \cdot (\vb{\rho} \times (\vb{\rho} \cdot \nabla) \vc{f} \mathrm{d} \varphi \\
  &= \frac{\omega_c \rho^2}{2}\, \vc{b} \cdot (\vc{e}_1 \times (\vc{e}_1 \cdot \nabla \vc{f}) + \vc{e}_2 \times (\vc{e}_2 \cdot \nabla \vc{f}) \\
  &= \frac{\mu}{e_s}\, \vc{b} \cdot (\nabla \times \vc{f} - \vc{b} \times (\vc{b} \cdot \nabla \vc{f})) \\
  \frac{1}{2\pi}\int_0^{2 \pi} \vc{\tilde{v}} \cdot (\vb{\rho} \cdot \nabla) \vc{f} \: \mathrm{d} \varphi &= \frac{\mu}{e_s} \vc{b} \cdot (\nabla \times \vc{f})
\end{align}
where we have used the fact that $\mu = m {v_\perp}^2/2 B = e_s \rho^2 \omega_c /2 $ and that for any right-handed basis $(e_i)_{i = 1,2,3}$, \[\nabla \times \vc{f} = \sum_{i=1,2,3} \vc{e}_i \times (\vc{e}_i \cdot \nabla \vc{f}).\]

Finally, one has,
\begin{equation}
  J_0 \, {h'}_{s,\omega} = \frac{\vc{v}_{g,\perp} \cdot \nabla \Phi_\omega}{i \omega}  - v_{g,\Vert} E_{\Vert,\omega} + \frac{\mu}{e_s} \vc{b} \cdot \left(\nabla \times \frac{\nabla\Phi_\omega}{i \omega}\right)
\end{equation}
which is exactly equation  \eqref{eq:h_prime}


\chapter{Contribution of energetic particles in different coordinate systems}
\label{sec:app_MIKE_Solver}

In this appendix, we will introduce different coordinate systems and exhibit the corresponding expressions for $\dWk$ and $\dW_{int,h}$.

The starting point will be the $(\psi_p,p,\xi_0)$ coordinate system introduced in chapter \ref{cha:FDR_derivation}. The expression for $\dWk$ is equation (\ref{eq:dWk_gen_app})
\begin{equation}
  \dWk = - 4\pi^3 \hat{C} \int \dd{\psi_p} \dd{\xi_0} \dd{p} \, \frac{\tilde{q} R_0}{B_m} |\xi_0| \bar{\tau}_b \, p^2 \, \dd{p} \frac{\omega \, \partial_E F_h - e_h^{-1} \, \partial_{\psi_p} F_h}{\omega - \delta_P (q-1) \omega_b - \omega_d}\left(\frac{E}{R_0} \bar{\Omega}_d \xi_c\right)^2,
\end{equation}
and the expression for $\dW_{f,h}$ is equation (\ref{eq:dWint_gen_app}) 
\begin{equation}
  \dW_{f,h} = 4\pi^3 \hat{C} \int \dd{\psi_p} \dd{\xi_0} \dd{p} \, \frac{\tilde{q} R_0}{B_m} |\xi_0| \bar{\tau}_b \, p^2 \, \pdd{F_h}{\psi_p} \left(\frac{r B_0}{q}\xi_c\right) \left(\frac{E}{R_0} \bar{\Omega}_d \xi_c\right).
\end{equation}

\section{Variables}

\subsection{Definition of \texorpdfstring{$\hat{\lambda}$}{lambda}}

The pitch angle variable $\hat{\lambda}$ is defined by 
\begin{equation}
  \label{eq:hat_lambda_def}
  \hat{\lambda}(\psi_p,\xi_0) = (1-\xi_0^2) \frac{B_0}{B_m(\psi_p)}.
\end{equation}
With this definition, the trapped domain corresponds to
\begin{equation*}
  \frac{B_0}{B_M(\psi_p)}  < \hat{\lambda} < \frac{B_0}{B_m(\psi_p)}
\end{equation*}
with $B_M$ and $B_m$ defined in chapter \ref{cha:Magn_config}, we have used the following identity $(1-\xi_{0T}^2) = B_m/B_M$. The circulating domain corresponds to 
\begin{equation*}
  0 < \hat{\lambda} < \frac{B_0}{B_M(\psi_p)} 
\end{equation*}

\subsection{Definition of \texorpdfstring{$\hat{\kappa}$}{kappa}}

The pitch angle variable $\hat{\kappa}$ is defined by 
\begin{equation}
  \label{eq:kappa_def_MIKE_bis}
  \hat{\kappa}^2(\psi_p,\xi_0) = \frac{1-\xi_{0}^{-2}}{1-\xi_{0T}^{-2}(\psi_p)}.
\end{equation}
the trapped domain corresponds to
\begin{equation*}
  1 < \hat{\kappa} < +\infty
\end{equation*}
The circulating domain corresponds to 
\begin{equation*}
  0  < \hat{\kappa} < 1.
\end{equation*}

\subsection{Definition of \texorpdfstring{$\hat{\iota}$}{iota}}

The pitch angle variable $\hat{\iota}$ is defined by 
\begin{equation}
  \label{eq:iota_def_MIKE_bis}
  \hat{\iota}^2(\psi_p,\xi_0) = \frac{1-\xi_{0T}^{-2}(\psi_p)}{1-\xi_{0}^{-2}} = \frac{1}{\hat{\kappa}^2}.
\end{equation}
the trapped domain corresponds to
\begin{equation*}
  0  < \hat{\iota} < 1.
\end{equation*}
The circulating domain corresponds to 
\begin{equation*}
  1 < \hat{\iota} < +\infty.
\end{equation*}

\subsection{Definition of \texorpdfstring{$\rho$}{rho}}

The radial variable $\rho$ is defined as
\begin{equation}
  \rho(\psi_p) = \frac{R(\psi_p,0) - R(0,0)}{R(\psi_a,0) - R(0,0)}.
\end{equation}

\section{Expressions for the fast particle contributions}

Since $\hat{\lambda},\hat{\kappa}$ or $\hat{\iota}$ do not discriminate particles with different signs for $\xi_0$, we introduce $\sigma$ the sign of $v_\Vert$ at the point of minimum magnetic field amplitude along its orbit ($\theta = 0$ for circular plasmas).

\subsection{Expressions with \texorpdfstring{$\hat{\lambda}$}{lambda}}

If one uses the set of coordinates $(\psi_p,\hat{\lambda},p,\sigma)$, the following expressions for $\dWk$ and $\dW_{f,h}$ can be used
\begin{align}
  \dWk &= - 4\pi^3 \hat{C} \sum_{\sigma = \pm 1} \int \frac{\dd{\psi_p}}{B_0} \frac{\dd{\hat{\lambda}}}{2} \dd{p} \, \tilde{q} R_0 \bar{\tau}_b \, p^2 \, \frac{\omega \, \partial_E F_h - e_h^{-1} \, \partial_{\psi_p} F_h}{\omega - \delta_P (q-1) \omega_b - \omega_d}\left(\frac{E}{R_0} \bar{\Omega}_d \xi_c\right)^2,
  \label{eq:dWk_gen_app_l}
\\
  \dW_{f,h} &= 4\pi^3 \hat{C} \sum_{\sigma = \pm 1} \int \frac{\dd{\psi_p}}{B_0} \frac{\dd{\hat{\lambda}}}{2} \dd{p} \, \tilde{q} R_0 \bar{\tau}_b \, p^2 \, \pdd{F_h}{\psi_p} \left(\frac{r B_0}{q}\xi_c\right) \left(\frac{E}{R_0} \bar{\Omega}_d \xi_c\right).
  \label{eq:dWint_gen_app_l}
\end{align}

\subsection{Expressions with \texorpdfstring{$\hat{\kappa}$}{kappa} or \texorpdfstring{$\hat{\iota}$}{iota}}

If one wants to use $\hat{\kappa}$ or $\hat{\iota}$ in place of $\hat{\lambda}$ then one just needs to add the $\partial \hat{\lambda}/\partial \hat{\kappa}$ or $\partial \hat{\lambda}/\partial \hat{\iota}$ factors in the integral, the derivatives being made at constant $\psi_p$. 

\subsection{Expressions in the MIKE code}

The MIKE code uses the $(\rho,\iota,\bar{p})$ coordinate system with particles corresponding to $\sigma = -1$ being represented by a negative value of $\bar{p}$. The quantity $\bar{p}$ is defined by $\bar{E} = E/E_{ref} = \bar{p}^2/2$ with $E_{ref}$ being a reference energy such that $p = \sqrt{m_h E_{ref}} \, \bar{p}$. The distribution function is then normalized by $F_h = n_{ref} \bar{F}_h /(m_e E_{ref})^{3/2} $ where $n_{ref}$ is a reference density. $\dWk$ can then be expressed as
\begin{multline}
  \dWk = -  \frac{\pi^2}{4} \beta_{ref} \left(\frac{m_h}{m_e}\right)^{3/2} \frac{1}{B_0 r_s^2} \int \ddr{\psi_p}{\rho} \, \tilde{q} \, \dd{\rho} \ldots \\ \int \frac{1}{2}\pdd{\hat{\lambda}}{\hat{\iota}} \bar{\tau}_b {\bar{\Omega}_d}^2 \, \dd{\hat{\iota}} \int \bar{p}^6 \, \frac{\omega \, \partial_{\bar{E}} \bar{F}_h - \hat{\omega}_{*h} \, \partial_{\rho} \bar{F}_h}{\omega - \delta_P (q-1) \omega_b - \omega_d} \dd{\bar{p}},
  \label{eq:dWk_gen_app_MIKE}
\end{multline}
where $\beta_{ref} = (2 \mu_0 n_{ref} E_{ref})/B_0^2$, $\hat{\omega}_{*h} = E_{ref}/(e_h \, \dd \psi_p / \dd \rho)$. The expression of $\dW_{f,h}$ can then be written as
\begin{equation}
  \dW_{f,h} = \frac{\pi^2}{4} \beta_{ref} \left(\frac{m_h}{m_e}\right)^{3/2} \frac{1}{B_0 r_s^2} \int \ddr{\psi_p}{\rho} \, \tilde{q} \, \dd{\rho} \int \frac{1}{2}\pdd{\hat{\lambda}}{\hat{\iota}} \bar{\tau}_b {\bar{\Omega}_d}^2 \, \dd{\hat{\iota}} \int \bar{p}^6 \frac{\hat{\omega}_{*h}}{\omega_d} \pdd{\bar{F}_h}{\rho}\dd{\bar{p}}.
  \label{eq:dWint_gen_app_MIKE}
\end{equation}

Note that the radial derivatives of $\bar{F}_h$ in the expressions of $\dWk$ and $\dW_{f,h}$ are done keeping $\mu$ and $E$ constant which is equivalent to keeping $\bar{p}$ and $\hat{\lambda}$ constant. This means that it is a combination of the $\rho$-derivative (at $\bar{p}$ and $\hat{\iota}$ constant) and of the $\hat{\iota}$-derivative (at $\rho$ and $\bar{p}$ constant).
\begin{equation}
  \left.\pdd{\bar{F}_h}{\rho}\right|_{\mu,E} = \left.\pdd{\bar{F}_h}{\rho}\right|_{\bar{p},\hat{\iota}} + \pdd{\hat{\iota}}{\rho} \left.\pdd{\bar{F}_h}{\hat{\iota}}\right|_{\bar{p},\rho}.
\end{equation}

\subsubsection*{Total contribution of energetic particles}

If one then neglects the finite $k_\Vert$ effects the expression for the sum of $\dWk$ and $\dW_{f,h}$ is
\begin{multline}
  \dWh = -  \frac{\pi^2}{4} \beta_{ref} \left(\frac{m_h}{m_e}\right)^{3/2} \frac{1}{B_0 r_s^2} \int \ddr{\psi_p}{\rho} \, \tilde{q} \, \dd{\rho} \ldots \\ \int \frac{1}{2}\pdd{\hat{\lambda}}{\hat{\iota}} \bar{\tau}_b {\bar{\Omega}_d}^2 \, \dd{\hat{\iota}} \int \bar{p}^6 \, \frac{\omega \left(\partial_{\bar{E}} \bar{F}_h - (\hat{\omega}_{*h}/\omega_d) \, \partial_{\rho} \bar{F}_h \right)}{\omega - \omega_d} \dd{\bar{p}},
  \label{eq:dWh_gen_app_MIKE}
\end{multline}

\subsubsection*{Energy integral}

In MIKE the integral over $\bar{p}$ is computed separately. We recall the expression for the integral $J$, equation \eqref{eq:Eint_def_MIKE} 
\begin{equation}
  J(g,b_{ref},c_{ref}) = \frac{1}{\sqrt{2}} \int_{-\infty}^{+\infty} \frac{\bar{p}^4 g(\bar{p})}{\bar{p}^2 + \sqrt{2} b_{ref} \bar{p} - 2 c_{ref}} \dd{\bar{p}}.
\end{equation}
In the case where $g$ is en even function of $\bar{p}$ and $b_{ref} = 0$ (which corresponds to trapped particles). One then has
\begin{equation}
  J(g,b_{ref},c_{ref}) = 2 \int_{-\infty}^{+\infty} \frac{\bar{E}^{3/2} g(\bar{p})}{\bar{E} - c_{ref}} \, \dd{\bar{E}}.
\end{equation}

In terms of $J$, $\dWk$ is written
\begin{multline}
  \dWk = \sqrt{2} \pi^2 \beta_{ref} \left(\frac{m_h}{m_e}\right)^{3/2} \frac{1}{B_0 r_s^2} \int \ddr{\psi_p}{\rho} \, \tilde{q} \, \dd{\rho} \ldots \\ \int \frac{1}{2}\pdd{\hat{\lambda}}{\hat{\iota}} \bar{\tau}_b {\bar{\Omega}_d}^2 \, \dd{\hat{\iota}} \, \frac{\omega}{\omega_{d,ref}} J\left(\bar{p}^2(\partial_{\bar{E}} \bar{F}_h - \hat{\omega}_{*h} \, \partial_{\rho} \bar{F}_h),\delta_P(q-1)\omega_{b,ref},\omega_{d,ref}\right),  
\end{multline}
where $\omega_{d,ref} = \omega_d E/E_{ref}$ and $\omega_{b,ref} = \omega_b \sqrt{E/E_{ref}}$.

\subsubsection*{Conventions in MIKE}

In MIKE we define $\Lambda(\rho)$ and $\lambda(\rho,\xi_0)$ as
\begin{align}
  \Lambda(\rho) &= \int_0^{2\pi} \frac{\dd{\theta}}{2 \pi} \frac{B}{B^\theta}, \\
  \lambda(\rho,\xi_0) &= \frac{1}{\Lambda} \int \frac{\dd{\theta}}{2 \pi} \frac{B}{B^\theta} \left| \frac{\xi_0}{\xi} \right|.
\end{align}
such that $\Lambda = R_0 \tilde{q}$ and $\lambda = |\xi_0| \bar{\tau}_b$.

In LUKE, all $q$-factors have an additional $R_0/a$ factor.


\chapter{The high aspect ratio low-beta equilibrium approximation}
\label{cha:app_circular}

The approximate expressions for a low-beta high aspect ratio equilibrium with circular concentric flux-surfaces are recalled. The flux-surface label used is $r$ the minor radius of a given flux-surface. Only first-order terms in $\varepsilon = r/R_0$ are retained.

\section{Equilibrium}

The magnitude of the magnetic field is equal to the one of the toroidal magnetic field $B_T$. Since $B_\varphi = R\,B_T$ is constant to order $\varepsilon^2$, one has
\begin{equation}
  B(r,\theta) = B_0(1-\varepsilon \cos \theta).
\end{equation}
One then has $\theta_m(r) = 0$, $\theta_M(r) = \pi$ which yields
\begin{align}
  B_m(r) &= B_0 (1 - \varepsilon), \\
  B_M(r) &= B_0 (1 + \varepsilon).
\end{align}

The toroidal flux $\psi_t$ is
\begin{equation}
  \psi_t(r) = B_0 \frac{r^2}{2}
\end{equation}
the poloidal flux verifies
\begin{equation}
  \ddr{\psi_p}{r} = B_0 \frac{r}{q(r)}.
\end{equation}

The approximate expressions for $\tilde{q}$ and $\hat{q}$ defined in section \ref{sec:eq_add_def} are
\begin{align}
  \frac{\tilde{q} (r)}{q(r)} &= 1, \\
  \frac{\hat{q} (r)}{q(r)} &= 1-\varepsilon.
\end{align}

\section{Particle Dynamics}
\label{sec:part_dynamics}

In this section we recall the expressions for the characteristic frequencies of the particle's gyrocenter motion. We consider a particle of mass $m_s$, charge $e_s$. Its trajectory is determined by its energy $E = p^2 /2m_s$, orbit-averaged radial position $r$, and its magnetic moment $\mu$.

\subsection{Pitch-angle variables}

$\xi_0$ the cosine of the pitch angle at $\theta = 0$ such that,
\begin{equation}
  \mu = \frac{E(1-\xi_0^2)}{B_0(1-\varepsilon)},
\end{equation}
the value of $\xi_0$ associated to the trapped-passing boundary is noted $\xi_{0T}$, its value is
\begin{equation}
  \xi_{0T}^2 = 1-\frac{B_m(r)}{B_M(r)} = \frac{2 \varepsilon}{1 + \varepsilon}.
\end{equation}
Then $\hat{\lambda}$ is defined as
 \begin{equation}
   \hat{\lambda} = \frac{\mu B_0}{E} = \frac{1-\xi_0^2}{1-\varepsilon}
 \end{equation}
and finally, $\hat{\kappa}$ is defined as 
\begin{equation}
  \hat{\kappa}^2 = \frac{\xi_0^{-2}-1}{\xi_{0T}^{-2}-1} = \frac{2 \varepsilon}{1-\varepsilon} \frac{1-\xi_0^2}{\xi_0^2}.
\end{equation}
The trapped-passing boundary in $\hat{\kappa}$ is independent of the position and is simply $\hat{\kappa} = 1$, with the passing particles corresponding to $\hat{\kappa} < 1$ and the trapped particles to $\hat{\kappa} > 1$. We define also $\hat{\iota}$ as the inverse of $\hat{\kappa}$
\begin{equation}
  \hat{\iota} = \hat{\kappa}^{-1}.
\end{equation}

\subsection{Frequencies of motion}

The gyro-frequency is 
\begin{equation}
  \omega_{c} = e_sB/m_s.
\end{equation}
The particle bounce-frequency $\omega_b$ (which for trapped particles corresponds to half an orbit only) can be expressed as 
\begin{equation}
  \omega_b = \frac{p}{m_s R_0 q} \sqrt{\frac{2 \varepsilon}{2\varepsilon + (1-\varepsilon)\hat{\kappa}^2}} \frac{\pi}{2} \left\{\begin{array}{cl}\displaystyle \frac{1}{\mathbb{K} \, (\hat{\kappa})} & \mbox{if } \hat{\kappa} < 1, \\ \\ \displaystyle \frac{\hat{\kappa}}{\mathbb{K} \, \left(\frac{1}{\hat{\kappa}}\right)}  & \mbox{if } \hat{\kappa} > 1. \end{array} \right.
\end{equation}
The expression for the toroidal drift-frequency $\omega_d$ is
\begin{equation}
  \omega_d = \frac{q E}{e_s B_0 R_0 r} \frac{\hat{\kappa}^2}{2\varepsilon + (1- \varepsilon) \hat{\kappa}^2} \left\{ 
    \begin{array}{cl} 
      \displaystyle \frac{4s}{\hat{\kappa}^2} \left(\frac{\mathbb{E} \, (\hat{\kappa})}{\mathbb{K} \, (\hat{\kappa})} - \left(\frac{\pi}{2 \mathbb{K}(\hat{\kappa})}\right)^2\right) + 1 - \frac{2}{\hat{\kappa}^2} + \frac{2}{\hat{\kappa}^2} \frac{\mathbb{E} \, (\hat{\kappa})}{\mathbb{K} \, (\hat{\kappa})} & \mbox{if } \hat{\kappa} < 1, \\
      \\ \displaystyle 4 s \left(\frac{\mathbb{E} \, (1/\hat{\kappa})}{\mathbb{K} \, (1/\hat{\kappa})} + \frac{1}{\hat{\kappa}^2} - 1 \right) + 2 \frac{\mathbb{E} \, (1/\hat{\kappa})}{\mathbb{K} \, (1/\hat{\kappa})} - 1 & \mbox{if } \hat{\kappa} > 1.
    \end{array}
  \right. 
  \label{eq:tor_drift_freq}
\end{equation}
In the definition of $\omega_d$, equation (\ref{eq:tor_transit_freq}), $q$ is evaluated at the position of the orbit-averaged poloidal flux, which yields a slightly different expression for (\ref{eq:tor_drift_freq}) than the one found in Ref. \cite{zon07}.

\section{The fast particle contribution to the fishbone dispersion relation}

If one uses the set of coordinates $(r,\hat{\lambda},p,\sigma)$, the following expressions for $\dWk$ and $\dW_{f,h}$ can be used
\begin{align}
  \dWk &= - 4\pi^3 \hat{C} \sum_{\sigma = \pm 1} \int r \dd{r} \frac{\dd{\hat{\lambda}}}{2} \dd{p} \, R_0 \bar{\tau}_b \, p^2 \, \frac{\omega \, \partial_E F_h - (q/e_h B_0 r) \, \partial_{\psi_p} F_h}{\omega - \delta_P (q-1) \omega_b - \omega_d}\left(\frac{E}{R_0} \bar{\Omega}_d \xi_c\right)^2,
  \label{eq:dWk_CCFS_app_l}
\\
  \dW_{f,h} &= 4\pi^3 \hat{C} \sum_{\sigma = \pm 1} \int r \dd{r} \frac{\dd{\hat{\lambda}}}{2} \dd{p} \, R_0 \bar{\tau}_b \, p^2 \, \left(\frac{q}{B_0 r}\right)\pdd{F_h}{r} \left(\frac{r B_0}{q}\xi_c\right) \left(\frac{E}{R_0} \bar{\Omega}_d \xi_c\right).
  \label{eq:dWint_CCFS_app_l}
\end{align}
which can be written,
\begin{align}
  \dWk &= - \pi^2 \frac{2 \mu_0}{B_0^2} \sum_{\sigma = \pm 1} \int \frac{r \dd{r}}{r_s^2} \frac{\dd{\hat{\lambda}}}{2} \dd{p} \, \bar{\tau}_b \, p^2 \, \frac{\omega \, \partial_E F_h - (q/e_h B_0 r) \, \partial_{r} F_h}{\omega - \delta_P (q-1) \omega_b - \omega_d}\, E^2 \bar{\Omega}_d^2,
  \label{eq:dWk_CCFS_app_l2}
\\
  \dW_{f,h} &= \pi^2 \frac{2 \mu_0}{B_0^2} \sum_{\sigma = \pm 1} \int \frac{r \dd{r}}{r_s}^2 \frac{\dd{\hat{\lambda}}}{2} \dd{p} \, \bar{\tau}_b \, p^2 \, \pdd{F_h}{r} \, E \bar{\Omega}_d R_0.
  \label{eq:dWint_CCFS_app_l2}
\end{align}


\renewcommand{\glossarypreamble}{\indent This is a non-exhaustive list of the variables used in this thesis. Bold variables, like $\vc{B}$, indicate vector or tensor quantities while plain variables, like $T$ indicate scalar quantities. Extensive use of the subscript $s$ will be made indicating that the quantity is specific to the species of type $s$, the subscript $h$ is dedicated to the energetic particles population.\\ \indent Some variable names have been defined twice, however the context should help identify which of the definitions is used, for example $p$ stands for both the plasma pressure and the particle's momentum.\\}
\printglossary[type=notation,style=index]

\cleardoublepage
\phantomsection
\addcontentsline{toc}{chapter}{Bibliography}

\bibliographystyle{ieeetr}
\bibliography{biblio}

\end{document}